\documentclass[12pt]{article}%
\usepackage[sort]{cite}
\usepackage{graphicx}
\usepackage{multicol}
\usepackage{amsfonts}
\usepackage{amssymb}
\usepackage{amsmath}
\usepackage{heck}
\usepackage{afterpage}
\usepackage{setspace}
\usepackage{verbatim}
\usepackage{color}
\usepackage{longtable}
\usepackage{standalone}
\usepackage{float}
\usepackage{subcaption}
\usepackage{epsfig}
\usepackage{epstopdf}
\usepackage{adjustbox}
\usepackage{tikz}
\usepackage[margin=1in]{geometry}
\usepackage{titletoc}
\usepackage{hyperref}%
\usepackage{braket}
 \usepackage{relsize}
 \usepackage{scalerel}
\usepackage{mathrsfs}
\usepackage{setspace}

\usepackage{enumitem}

\usepackage{caption}
 \providecommand{\U}[1]{\protect\rule{.1in}{.1in}}
\newsavebox{\mysavebox}

\hypersetup{colorlinks,citecolor=black,filecolor=black,linkcolor=black,urlcolor=black}

\numberwithin{equation}{section}

\usetikzlibrary{arrows,decorations.pathmorphing,backgrounds,positioning,fit,petri}
\usetikzlibrary{decorations.pathreplacing}
\usetikzlibrary{decorations.markings}
\usetikzlibrary{decorations.shapes}
\usetikzlibrary{arrows.meta}
\usetikzlibrary{quotes,angles}
\usetikzlibrary{positioning}
\usetikzlibrary{patterns}
\usetikzlibrary{chains}
\usetikzlibrary{hobby}

\usepackage{tikz}
\usetikzlibrary{arrows}
\usetikzlibrary{arrows.meta}
\usetikzlibrary{shapes.geometric,calc,arrows, positioning,shapes.misc,decorations.markings}
\tikzset{
  big arrow/.style={
    decoration={markings,mark=at position 1 with {\arrow[scale=2,#1]{>}}},
    postaction={decorate},
    shorten >=0.4pt},
  big arrow/.default=black}

\pgfdeclarelayer{edgelayer}
\pgfdeclarelayer{nodelayer}
\pgfsetlayers{edgelayer,nodelayer,main}
\tikzstyle{none}=[inner sep=0pt]

\tikzstyle{NodeCross}=[draw, shape=circle, cross out, inner sep=0pt, minimum size=6pt,line width=0.25mm]
\tikzstyle{Circle}=[draw, shape=circle, black, fill=black, inner sep=0pt, minimum size=6pt]
\tikzstyle{circle}=[draw, shape=circle, black, fill=black, inner sep=0pt, minimum size=10pt]
\tikzstyle{Star}=[draw, shape=star, fill=red, star points=10, inner sep=0pt, minimum size=8pt]
\tikzstyle{CircleRed}=[draw, shape=circle, black, fill=red, inner sep=0pt, minimum size=4pt]
\tikzstyle{MidCircleRed}=[draw, shape=circle, black, fill=red, inner sep=0pt, minimum size=8pt]
\tikzstyle{MidCircleBlue}=[draw, shape=circle, black, fill=blue, inner sep=0pt, minimum size=8pt]
\tikzstyle{StarP}=[draw={rgb,255: red,128; green,0; blue,128}, shape=star, fill={rgb,256: red,128; green,0; blue,128}, star points=8, inner sep=0pt, minimum size=12pt]
\tikzstyle{ShadedCircRed}=[draw=red, shape=circle, fill={rgb, 255: red,255; green,114; blue, 118}, inner sep=0pt, minimum size=80pt, line width=0.5mm, fill opacity=0.2]
\tikzstyle{ShadedCircRed2}=[draw=red, shape=circle, fill={rgb, 255: red,255; green,114; blue, 118}, inner sep=0pt, minimum size=10pt]
\tikzstyle{ShadedCircRed3}=[draw=black, shape=rectangle, fill={rgb, 255: red,255; green,114; blue, 118}, inner sep=0pt, minimum size=113pt, line width=0.25mm]
\tikzstyle{ShadedCirc}=[draw=red, shape=circle, fill=white, inner sep=0pt, minimum size=45pt,  fill opacity=1.0,  line width=0.5mm]
\tikzstyle{CircleBlue}=[draw, shape=circle, fill=blue, inner sep=0pt, minimum size=6pt]
\tikzstyle{BigCirclePurple}=[draw, shape=circle, fill={rgb,255: red,191; green,0; blue,191}, inner sep=0pt, minimum size=12pt]
\tikzstyle{CirclePurple}=[draw, shape=circle, fill={rgb,255: red,191; green,0; blue,191}, inner sep=0pt, minimum size=8pt]
\tikzstyle{EmptyCircle}=[draw, shape=circle, inner sep=0pt, minimum size=4pt]
\tikzstyle{GreenCircle}=[draw, shape=circle,  fill={rgb,255: red,80; green,200; blue,120}, inner sep=0pt, minimum size=8pt]
\tikzstyle{BrownCircle}=[draw, shape=circle,  fill={rgb,255: red,115; green,115; blue,115}, inner sep=0pt, minimum size=8pt]
\tikzstyle{CirclePurpleSmall}=[draw, shape=circle, fill={rgb,255: red,191; green,0; blue,191}, inner sep=0pt, minimum size=4pt]
\tikzstyle{BigCircleGreen}=[draw, shape=circle, fill={rgb,255: red,0; green,191; blue,0}, inner sep=0pt, minimum size=12pt]
\tikzstyle{BigCircleBlue}=[draw, shape=circle, fill={rgb,255: red,0; green,0; blue,191}, inner sep=0pt, minimum size=12pt]
\tikzstyle{BigCircleRed}=[draw, shape=circle, fill={rgb,255: red,191; green,0; blue,0}, inner sep=0pt, minimum size=12pt]
\tikzstyle{CircleBrown}=[draw, shape=circle, fill={rgb,255: red,210; green,105; blue,30}, inner sep=0pt, minimum size=8pt]

\tikzstyle{BigCircleGrey}=[shape=circle,  fill={rgb,255: red,120; green,120; blue,120}, inner sep=0pt, minimum size=10pt]
\tikzstyle{SmallCircleGrey}=[shape=circle,  fill={rgb,255: red,120; green,120; blue,120}, inner sep=0pt, minimum size=6pt]

\tikzstyle{DashedLine}=[-, densely dashed, line width=0.25mm]
\tikzstyle{DottedLine}=[-, dotted, line width=0.25mm]
\tikzstyle{ThickLine}=[-, line width=0.25mm]
\tikzstyle{ThickerLine}=[-, line width=0.4mm]
\tikzstyle{ArrowLineRight}=[-, -{Stealth[scale=1.25]}, line width=0.25mm, scale=5]
\tikzstyle{ArrowLineRed}=[-, draw={rgb,255: red,191; green,0; blue,0}, -{Stealth[scale=1.75]}, line width=0.1mm, scale=5]
\tikzstyle{RedLine}=[-, draw={rgb,255: red,191; green,0; blue,0}, fill=none, line width=0.5mm]
\tikzstyle{DashedLineThin}=[-, densely dashed, line width=0.125mm, fill=none, draw=black]
\tikzstyle{DottedRed}=[-, dotted, draw={rgb,255: red,191; green,0; blue,0}, fill=none, line width=0.25mm]
\tikzstyle{DashedRed}=[-, densely dashed, draw={rgb,255: red,191; green,0; blue,0}, fill=none, line width=0.25mm]
\tikzstyle{BlueLine}=[-, draw={rgb,255: red,0; green,0; blue,191}, fill=none, line width=0.5mm]
\tikzstyle{ArrowLineBlue}=[-, draw={rgb,255: red,0; green,0; blue,191}, -{Stealth[scale=1.75]}, line width=0.1mm, scale=5]
\tikzstyle{GreenDoubleArrow}=[<->, draw={rgb,155: red,0; green,255; blue,0},  line width= 0.5mm, scale=5]
\tikzstyle{RedDoubleArrow}=[<->, draw={rgb,255: red,255; green,0; blue,0},  line width= 0.5mm, scale=5]
\tikzstyle{BlueDottedLight}=[-, dotted, draw={rgb,255: red,0; green,0; blue,191}, fill=none, line width=0.3mm]
\tikzstyle{BrownLine}=[-, draw={rgb,255: red,210; green,105; blue,30}, fill=none, line width=0.5mm]
\tikzstyle{DottedBrownLine}=[-, dotted, draw={rgb,255: red,210; green,105; blue,30}, fill=none, line width=0.5mm]
\tikzstyle{DottedRed}=[-, dotted, draw={rgb,255: red,191; green,0; blue,0}, fill=none, dotted, line width=0.5mm]
\tikzstyle{DottedPurple}=[-, dotted, draw={rgb,255: red,191; green,0; blue,191}, fill=none, dotted, line width=0.5mm]
\tikzstyle{BlueDottedLight}=[-, dotted, draw={rgb,255: red,0; green,0; blue,191}, fill=none, line width=0.5mm]
\tikzstyle{ArrowLinePurple}=[-, draw={rgb,255: red,191; green,0; blue,191}, -{Stealth[scale=1.75]}, line width=0.5mm, scale=5]
\tikzstyle{DashedLineGreen}=[-, densely dashed, draw={rgb,255: red,74; green,103; blue,65}, line width=0.25mm]
\tikzstyle{LineGreen}=[-, draw={rgb,255: red, 74; green,200; blue,65}, line width=0.5mm]
\tikzstyle{ArrowLineGreen}=[-, draw={rgb,255: red,0; green,191; blue,0}, -{Stealth[scale=1.75]}, line width=0.5mm, scale=5]
\tikzstyle{GreenLine}=[-, draw={rgb,255: red,0; green,191; blue,0}, fill=none, line width=0.5mm]
\tikzstyle{PurpleLine}=[-, draw={rgb,255: red,191; green,0; blue,191}, fill=none, line width=0.5mm]
\tikzstyle{DPurpleLine}=[-, dotted, draw={rgb,255: red,191; green,0; blue,191}, fill=none, line width=0.5mm]
\tikzstyle{LightBlue}=[-, draw={rgb,255: red,0; green,150; blue,255}, fill=none, line width=0.5mm]
\tikzstyle{DLightBlue}=[-, dotted, draw={rgb,255: red,0; green,150; blue,255}, fill=none, line width=0.5mm]
\tikzstyle{ThickGreenLine}=[-, draw={rgb,255: red,0; green,255; blue,0}, fill=none, line width=0.75mm]
\tikzstyle{ThickDarkGreenLine}=[-, draw={rgb,255: red,80; green,200; blue,120}, fill=none, line width=0.75mm]

\tikzstyle{NodeCross}=[draw, shape=circle, cross out, inner sep=0pt, minimum size=10pt,line width=0.25mm]
\tikzstyle{SmallCircle}=[draw, shape=circle, black, fill=black, inner sep=0pt, minimum size=6pt]
\tikzstyle{BigCircle}=[draw, shape=circle, black, fill=black, inner sep=0pt, minimum size=10pt]
\tikzstyle{MidCircle}=[draw, shape=circle, black, fill=black, inner sep=0pt, minimum size=8pt]
\tikzstyle{SmallCircleRed}=[shape=circle, fill={rgb,255: red,191; green,0; blue,0}, inner sep=0pt, minimum size=6pt]
\tikzstyle{BigCircleRed}=[shape=circle, fill={rgb,255: red,191; green,0; blue,0}, inner sep=0pt, minimum size=10pt]
\tikzstyle{SmallCircleBlue}=[shape=circle, fill=blue, inner sep=0pt, minimum size=6pt]
\tikzstyle{BigCircleBlue}=[shape=circle, fill=blue, inner sep=0pt, minimum size=10pt]
\tikzstyle{SmallCirclePurple}=[shape=circle, fill={rgb,255: red,191; green,0; blue,191}, inner sep=0pt, minimum size=6pt]
\tikzstyle{BigCirclePurple}=[shape=circle, fill={rgb,255: red,191; green,0; blue,191}, inner sep=0pt, minimum size=10pt]
\tikzstyle{SmallCircleGreen}=[shape=circle, fill={rgb,255: red,80; green,200; blue,120}, inner sep=0pt, minimum size=6pt]
\tikzstyle{SmallCircleBrightGreen}=[shape=circle, fill={rgb,255: red,80; green,255; blue,120}, inner sep=0pt, minimum size=6pt]
\tikzstyle{BigCircleGreen}=[shape=circle,  fill={rgb,255: red,80; green,200; blue,120}, inner sep=0pt, minimum size=10pt]
\tikzstyle{SmallCircleBrown}=[shape=circle,  fill={rgb,255: red,210; green,105; blue,30}, inner sep=0pt, minimum size=6pt]
\tikzstyle{BigCircleBrown}=[shape=circle,  fill={rgb,255: red,210; green,105; blue,30}, inner sep=0pt, minimum size=10pt]
\tikzstyle{Star}=[draw, shape=star, fill=black, star points=8, inner sep=0pt, minimum size=10pt]
\tikzstyle{MidCircleGreen}=[shape=circle,  fill={rgb,255: red,80; green,200; blue,120}, inner sep=0pt, minimum size=8pt]
\tikzstyle{StarBlue}=[draw, shape=star, fill=blue, star points=8, inner sep=0pt, minimum size=10pt]

\tikzstyle{DashedLine}=[-, densely dashed, line width=0.25mm]
\tikzstyle{DottedLine}=[-, dotted, line width=0.25mm]
\tikzstyle{ThickLine}=[-, line width=0.25mm]
\tikzstyle{ArrowLineRight}=[-, -{Stealth[scale=1.75]}, line width=0.15mm, scale=5]
\tikzstyle{RedLine}=[-, draw={rgb,255: red,191; green,0; blue,0}, fill=none, line width=0.5mm]
\tikzstyle{DashedRedLine}=[-, densely dashed, draw={rgb,255: red,191; green,0; blue,0}, fill=none, line width=0.5mm]
\tikzstyle{DottedRed}=[-, dotted, draw={rgb,255: red,191; green,0; blue,0}, fill=none, dotted, line width=0.5mm]
\tikzstyle{BlueLine}=[-, draw=blue, fill=none, line width=0.5mm]
\tikzstyle{ThickBlueLine}=[-, draw=blue, fill=none, line width=0.75mm]
\tikzstyle{DashedBlueLine}=[-, densely dashed, draw=blue, fill=none, line width=0.5mm]
\tikzstyle{DottedBlue}=[-, dotted, draw=blue, fill=none, dotted, line width=0.5mm]
\tikzstyle{PurpleLine}=[-, draw={rgb,255: red,191; green,0; blue,191}, fill=none, line width=0.5mm]
\tikzstyle{DashedPurpleLine}=[-, densely dashed, draw={rgb,255: red,191; green,0; blue,191}, fill=none, line width=0.5mm]
\tikzstyle{DottedPurple}=[-, dotted, draw={rgb,255: red,191; green,0; blue,191}, fill=none, dotted, line width=0.5mm]\tikzstyle{GreenLine}=[-, draw={rgb,255: red,80; green,200; blue,120}, fill=none, line width=0.5mm]
\tikzstyle{DashedGreenLine}=[-, densely dashed, draw={rgb,255: red,80; green,200; blue,120}, fill=none, line width=0.5mm]
\tikzstyle{DottedGreen}=[-, dotted, draw={rgb,255: red,80; green,200; blue,120}, fill=none, dotted, line width=0.5mm]
\tikzstyle{BrownLine}=[-, draw={rgb,255: red,210; green,105; blue,30}, fill=none, line width=0.5mm]
\tikzstyle{DashedBrownLine}=[-, densely dashed, draw={rgb,255: red,210; green,105; blue,30}, fill=none, line width=0.5mm]
\tikzstyle{DottedBrown}=[-, dotted, draw={rgb,255: red,210; green,105; blue,30}, fill=none, dotted, line width=0.5mm]

\tikzset{snake it/.style={decorate, decoration=snake}}

\newcommand{\lb}{\left(}
\newcommand{\rb}{\right)}
\newcommand{\lbb}{\left[}
\newcommand{\rbb}{\right]}
\newcommand{\lbbb}{\left\{}
\newcommand{\rbbb}{\right\}}

\newcommand{\be}{\begin{equation}}
\newcommand{\ee}{\end{equation}}
\newcommand{\ba}{\begin{aligned}}
\newcommand{\ea}{\end{aligned}}
\newcommand{\Z}{{\mathbb Z}}
\newcommand{\R}{{\mathbb R}}
\newcommand{\C}{{\mathbb C}}
\newcommand{\Q}{{\mathbb Q}}
\renewcommand{\P}{{\mathbb P}}

\begin{document}

\date{August 2024}

\title{Cornering Relative Symmetry Theories}

\institution{PENN}{\centerline{$^{1}$Department of Physics and Astronomy, University of Pennsylvania, Philadelphia, PA 19104, USA}}
\institution{PENNmath}{\centerline{$^{2}$Department of Mathematics, University of Pennsylvania, Philadelphia, PA 19104, USA}}
\institution{Maribor}{\centerline{$^{3}$Center for Applied Mathematics and Theoretical Physics, University of Maribor, Maribor, Slovenia}}
\institution{CERN}{\centerline{${}^{4}$Theoretical Physics Department, CERN, 1211 Geneva 23, Switzerland}}

\authors{Mirjam Cveti\v{c}\worksat{\PENN,\PENNmath,\Maribor}\footnote{e-mail: \texttt{cvetic@physics.upenn.edu}}, Ron Donagi\worksat{\PENNmath, \PENN}\footnote{e-mail: \texttt{donagi@upenn.edu}}, Jonathan J. Heckman\worksat{\PENN,\PENNmath}\footnote{e-mail: \texttt{jheckman@sas.upenn.edu}},\\[3mm]
Max H\"ubner\worksat{\PENN}\footnote{e-mail: \texttt{hmax@sas.upenn.edu}}, and
Ethan Torres\worksat{\CERN}\footnote{e-mail: \texttt{ethan.martin.torres@cern.ch}}
}

\abstract{The symmetry data of a $d$-dimensional quantum field theory (QFT)
can often be captured in terms of a higher-dimensional symmetry topological field theory (SymTFT). In top down (i.e., stringy) realizations of this structure, the QFT in question is localized in a higher-dimensional bulk. In many cases of interest, however, the associated $(d+1)$-dimensional bulk is not fully gapped and one must instead consider a filtration of theories to reach a gapped bulk in $D = d+m$ dimensions. Overall, this leads us to a nested structure of relative symmetry theories which descend to coupled edge modes, with the original
QFT degrees of freedom localized at a corner of this $D$-dimensional bulk system. We present a bottom up characterization of this structure and also show how it naturally arises in a number of string-based constructions of QFTs with both finite and continuous symmetries.}

{\small \texttt{\hfill UPR-1331-TH}}

{\small \texttt{\hfill CERN-TH-2024-139}}

\maketitle

\enlargethispage{\baselineskip}

\setcounter{tocdepth}{2}

\tableofcontents

\newpage

\section{Introduction}

Symmetry principles impose important constraints on the dynamics of physical systems.
In the context of quantum field theory (QFT) symmetries specify selection rules
and constraints on renormalization group flows. Recently, the notion of symmetry
itself has undergone rapid developments, especially with regards to the interplay
between these physical principles and topological\,/\,categorical structures which are
just now being discovered and systematized.

When available, a particularly helpful tool in understanding the symmetries of a given $d$-dimensional
QFT involves the symmetry topological field theory (SymTFT) of the
QFT. A SymTFT$_{d+1}$ is a $(d+1)$-dimensional TFT which captures the global categorical symmetries of a $d$-dimensional  QFT$_d$.
In this framework, one places the SymTFT$_{d+1}$ on an interval which splits the choice of global structure of the QFT$_d$ into two boundary states: there is a relative
QFT $\vert \mathcal{T}_d \rangle$ at one end of the interval and a choice of topological boundary conditions $\langle \mathcal{B}_{d} \vert$ at the other end. Evaluating $\langle \mathcal{B}_{d} \vert \mathcal{T}_{d} \rangle$ specifies the partition function for the initial absolute QFT which is independent of the length of the interval as the SymTFT is topological.

This setup is best established for finite symmetries (see e.g., \cite{Reshetikhin:1991tc, Turaev:1992hq, Barrett:1993ab, Witten:1998wy, Fuchs:2002cm, Kirillov:2010nh, Kapustin:2010if, Kitaev:2011dxc, Fuchs:2012dt, Freed:2012bs, Freed:2018cec, Gaiotto:2020iye, Apruzzi:2021nmk, Freed:2022qnc, Kaidi:2022cpf}), and there have also been recent proposals on extending this framework to certain continuous symmetries (see e.g., \cite{Brennan:2024fgj, Heckman:2024oot, Antinucci:2024zjp, Bonetti:2024cjk, Apruzzi:2024htg}).

Indeed, this general picture resonates well with the extra-dimensional structures present in
string-based constructions of QFTs. To obtain a QFT decoupled from gravity, one considers a curvature singularity / stack of
probe branes localized at a small region in a non-compact extra-dimensional geometry $X$. Then, assuming that $X$ is topologically a cone, i.e., $X = \mathrm{Cone}(\partial X)$, there is a natural radial direction which begins at the singularity and extends out to the asymptotic boundary $\partial X$. Dimensional reduction of the stringy background on $\partial X$ and dropping dynamical modes which decouple in the infrared (IR)
results in a $(d+1)$-dimensional symmetry TFT (see \cite{Apruzzi:2021nmk} as well as \cite{Aharony:1998qu, Belov:2006jd, Heckman:2017uxe, Heckman:2022xgu,vanBeest:2022fss, Baume:2023kkf, Yu:2023nyn, Apruzzi:2023uma, Lawrie:2023tdz, DelZotto:2024tae}). This provides a beautiful match between top down and bottom up approaches to the construction of SymTFTs.

But string constructions suggest further generalizations of the SymTFT paradigm.\footnote{See e.g., \cite{Baume:2023kkf, Heckman:2024oot, Apruzzi:2024htg, Heckman:2024zdo} for some recent string-motivated generalizations of the SymTFT formalism.}
A common occurrence in many stringy realizations of QFTs
is the presence of singularities which are \textit{not} isolated at a single point of $X$. This provides a general way to
introduce various flavor symmetries in the $d$-dimensional QFT$_d$ via a higher-dimensional ``flavor brane,'' though more broadly this may simply be another higher-dimensional QFT$_D$ where $D>d$. These additional singularities specify relative QFTs
in their own right, and as such, tracking just the radial direction of $\mathrm{Cone}(\partial X)$ would naively result in a $(d+1)$-dimensional
system with gapless degrees of freedom.\footnote{Of course, in terms of the impact of this sector on the QFT$_d$, there is a
decoupling limit one can first take to remove the dynamics of this flavor brane in the QFT$_d$, similar to the discussion in \cite{Bonetti:2024cjk}. This becomes especially subtle when $d < 4$ since the dynamics of the flavor brane can a priori also be non-trivial in the IR. A careful treatment in this case then requires specifying
a suitable order of limits for decoupling all dynamics where the flavor brane is first decoupled. This can be achieved because the QFTs in question are still localized on subspaces in the ambient target space.} Such gapless degrees of freedom are an indication that one is not necessarily dealing with a bulk TFT, but a more general bulk\,/\,boundary system.

In a suitable scaling limit of a bulk theory $\mathcal{S}_{d+1}(g)$ in which various parameters / scales are tuned / decoupled, one can expect to either reach a gapped or free theory. We shall refer to the bulk system obtained after applying such a scaling limit as a symmetry theory (SymTh), for now remaining agnostic as to whether we have a fully gapped bulk.\footnote{A more restricted version of SymTh was introduced in \cite{Apruzzi:2024htg} which refers to the specific case of free fields in the bulk. This can often be traded for a characterization in terms of a formal topological field theory with non-compact gauge groups as in \cite{Brennan:2024fgj, Antinucci:2024zjp, Bonetti:2024cjk, Copetti:2024onh}. Here, we allow ourselves a more general perspective to cover the different limits which can in principle arise. An additional comment here is that this is also what one expects for a CFT$_d$ with a holographic dual, where the physical boundary condition of the symmetry theory is itself ``smeared out'' to a bulk AdS (see, e.g., \cite{Heckman:2024oot}).} Overall, one can summarize this structure as a $d$-dimensional relative theory $\mathcal{T}_{d}$ which sits at one end of an interval filled by the symmetry theory $\mathcal{S}_{d+1}$, with another boundary mode / boundary condition $\mathcal{B}_d$ at the other end (which may or may not be fully topological). This results in a decompression of the original QFT$_d$ in terms of the formal quiverlike structure:
\be \label{eq:TScombo}
\scalebox{0.9}{
\begin{tikzpicture}
	\begin{pgfonlayer}{nodelayer}
		\node [style=none] (1) at (-1, 0) {};
		\node [style=none] (2) at (0, 0) {};
		\node [style=Star] (3) at (-2.5, 0) {};
		\node [style=SmallCircle] (4) at (1.5, 0) {};
		\node [style=SmallCircleRed] (5) at (4.5, 0) {};
		\node [style=none] (6) at (-2.5, 0.5) {QFT$_d$};
		\node [style=none] (7) at (4.5, 0.5) {$\mathcal{T}_d$};
		\node [style=none] (8) at (1.5, 0.5) {$\mathcal{B}_d$};
		\node [style=none] (9) at (3, 0.5) {$\mathcal{S}_{d+1}$};
		\node [style=none] (10) at (0, 0.875) {};
	\end{pgfonlayer}
	\begin{pgfonlayer}{edgelayer}
		\draw [style=ArrowLineRight] (1.center) to (2.center);
		\draw [style=ThickLine] (4) to (5);
	\end{pgfonlayer}
\end{tikzpicture}}
\ee
In the special case where the symmetries are of finite type, $\mathcal{S}_{d+1}$ is a TFT, and $\mathcal{B}_d$ specifies gapped boundary conditions, but more generally, $\mathcal{B}_d$ might support gapless free fields (especially in the case of continuous symmetries).
Indeed, there is no guarantee that the $(d+1)$-dimensional bulk is fully gapped, as naturally arises in many string-based examples.
Another comment here is that even in these cases, the structure of this decompression is not unique; one can in principle make different choices for the bulk theory provided the resulting categorical structures reduced to $d$-dimensions (such as the Drinfeld center) all match. That being said, stringy constructions typically favor a particular canonical choice, and we leave these choices implicit in what follows.

In the context of top down motivated constructions, $\mathcal{S}_{d+1}$ is obtained from taking a limit in a family of QFTs which
we schematically write as $\mathcal{S}_{d+1}(g)$, i.e.:
\begin{equation}
\mathcal{S}_{d+1} = \underset{g \rightarrow 0}{\lim} \, \mathcal{S}_{d+1}(g).
\end{equation}
In particular, we can treat $\mathcal{S}_{d+1}(g)$ as a QFT in its own right. As such,
it is natural to ask whether it too has a non-trivial symmetry theory.
Put together, then, there is another symmetry theory governing the
combined system of line (\ref{eq:TScombo}).

Denoting by $\mathcal{T}_{d+1}$ the relative theory associated with the bulk theory $\mathcal{S}_{d+1}(g)$, we see that there can be a symmetry theory $\mathcal{S}_{d+2}$. Just as we decompressed the absolute QFT$_d$ in line \eqref{eq:TScombo}, we can now decompress $\mathcal{S}_{d+1}(g)$. In the absence of edge modes for $\mathcal{S}_{d+1}(g)$, i.e., considering this theory on a manifold without boundary, we have the decompression:
\be\label{eq:TScombo2}
\scalebox{0.95}{
\begin{tikzpicture}
	\begin{pgfonlayer}{nodelayer}
		\node [style=none] (0) at (0, 0) {};
		\node [style=none] (1) at (1, 0) {};
		\node [style=none] (5) at (-2, 0.5) {$\mathcal{S}_{d+1}(g)$};
		\node [style=none] (9) at (1, 1.25) {};
		\node [style=none] (10) at (-2.75, 0) {};
		\node [style=none] (11) at (-1.25, 0) {};
		\node [style=none] (12) at (-1, 0) {};
		\node [style=none] (13) at (-3, 0) {};
		\node [style=none] (14) at (2.25, 0.5) {};
		\node [style=none] (15) at (3.75, 0.5) {};
		\node [style=none] (16) at (4, 0.5) {};
		\node [style=none] (17) at (2, 0.5) {};
		\node [style=none] (18) at (2.25, -0.5) {};
		\node [style=none] (19) at (3.75, -0.5) {};
		\node [style=none] (20) at (4, -0.5) {};
		\node [style=none] (21) at (2, -0.5) {};
		\node [style=none] (22) at (3, -1) {$\mathcal{T}_{d+1}$};
		\node [style=none] (23) at (3, 1) {$\mathcal{B}_{d+1}$};
		\node [style=none] (24) at (3, 0) {$\mathcal{S}_{d+2}$};
	\end{pgfonlayer}
	\begin{pgfonlayer}{edgelayer}
		\filldraw[fill=gray!50, draw=gray!50]  (2.25, 0.5) -- (2.25, -0.5) -- (3.75, -0.5) -- (3.75, 0.5) -- cycle;
		\draw [style=ArrowLineRight] (0.center) to (1.center);
		\draw [style=ThickLine] (10.center) to (11.center);
		\draw [style=DottedLine] (11.center) to (12.center);
		\draw [style=DottedLine] (10.center) to (13.center);
		\draw [style=PurpleLine] (14.center) to (15.center);
		\draw [style=RedLine] (18.center) to (19.center);
		\draw [style=DottedRed] (20.center) to (19.center);
		\draw [style=DottedRed] (21.center) to (18.center);
		\draw [style=DottedPurple] (16.center) to (15.center);
		\draw [style=DottedPurple] (17.center) to (14.center);
	\end{pgfonlayer}
\end{tikzpicture}}
\ee\vspace{-30pt}

\noindent Continuing in this manner, we can continue to decompress the various $\mathcal{S}_{d+m}$'s for $m \geq 1$ until eventually we reach a fully gapped bulk. The existence of such a bulk theory
is in some sense guaranteed by the SymTFT formalism,
and in the context of string\,/\,M-theory backgrounds, $d+m$ is bounded above.

What happens when we combine the decompression of the QFT$_d$ and its symmetry theory $\mathcal{S}_{d+1}$? In this case, we can decompress the bulk theory $\mathcal{S}_{d+1}(g)$ at the expense of introducing an additional junction which connects to the original relative theories $\mathcal{B}_d$ and $\mathcal{T}_{d}$:
\be
\scalebox{0.95}{
\begin{tikzpicture}
	\begin{pgfonlayer}{nodelayer}
		\node [style=SmallCircle] (0) at (-5, 0) {};
		\node [style=SmallCircleRed] (1) at (-2, 0) {};
		\node [style=none] (2) at (-2, 0.5) {$\mathcal{T}_d$};
		\node [style=none] (3) at (-5, 0.5) {$\mathcal{B}_d$};
		\node [style=none] (4) at (3.25, 0.5) {$\mathcal{S}_{d+1}(g)$};
		\node [style=none] (5) at (-0.5, 0) {};
		\node [style=none] (6) at (0.5, 0) {};
		\node [style=SmallCircle] (7) at (2, 0) {};
		\node [style=SmallCircleRed] (8) at (9, 0) {};
		\node [style=none] (9) at (9, 0.5) {$\mathcal{T}_d$};
		\node [style=none] (10) at (2, 0.5) {$\mathcal{B}_d$};
		\node [style=none] (11) at (4.75, 0) {};
		\node [style=none] (12) at (6.25, 0) {};
		\node [style=none] (13) at (5.5, 0.75) {};
		\node [style=none] (14) at (5.5, -0.75) {};
		\node [style=none] (15) at (5.5, -1.25) {$\mathcal{T}_{d+1}$};
		\node [style=none] (16) at (7.75, 0.5) {$\mathcal{S}_{d+1}(g)$};
		\node [style=none] (17) at (5.5, 1.25) {$\mathcal{B}_{d+1}$};
		\node [style=none] (22) at (4.25, -0.45) {$\overline{\mathcal{J}}_d$};
		\node [style=none] (23) at (6.75, -0.5) {$\mathcal{J}_d$};
		\node [style=none] (24) at (-3.5, 0.5) {$\mathcal{S}_{d+1}(g)$};
		\node [style=none] (25) at (5.5, 0) {$\mathcal{S}_{d+2}$};
	\end{pgfonlayer}
	\begin{pgfonlayer}{edgelayer}
		\draw[fill=gray!50, draw=gray!50] (5.5,0) circle (4ex);
		\draw [style=ThickLine] (0) to (1);
		\draw [style=ArrowLineRight] (5.center) to (6.center);
		\draw [style=ThickLine] (7) to (11.center);
		\draw [style=ThickLine] (12.center) to (8);
		\draw [style=RedLine, bend right=45] (11.center) to (14.center);
		\draw [style=RedLine, bend right=45] (14.center) to (12.center);
		\draw [style=PurpleLine, bend left=45] (11.center) to (13.center);
		\draw [style=PurpleLine, bend left=45] (13.center) to (12.center);
	\end{pgfonlayer}
\end{tikzpicture}}
\ee
where here, we have the junction theories $\mathcal{J}_d$ and its orientation reversed counterpart $\overline{\mathcal{J}}_d$ which fuse the triple $\mathcal{B}_{d+1},\mathcal{T}_{d+1},\mathcal{S}_{d+1}$ with different orientations (see \cite{Baume:2023kkf} for a discussion of junctions). We comment that in the context of our string constructions, the junctions $\mathcal{J}_d,\overline{\mathcal{J}}_d$ turn out to be relatively innocuous. Now, we can contract the $(d+1)$-dimensional edges supporting $\mathcal{S}_{d+1}$ colliding the junctions into $\mathcal{B}_d,\mathcal{T}_d$ producing the corners $\mathcal{B}_d',\mathcal{T}_d'$. In the context of our constructions we will have $\mathcal{B}_d=\mathcal{B}_d'$ and $\mathcal{T}_d=\mathcal{T}_d'$, but in principle this operation can alter corners. We present the process as:
\be
\scalebox{0.9}{
\begin{tikzpicture}
	\begin{pgfonlayer}{nodelayer}
		\node [style=none] (0) at (-7.75, 0.5) {$\mathcal{S}_{d+1}(g)$};
		\node [style=SmallCircle] (1) at (-9, 0) {};
		\node [style=SmallCircleRed] (2) at (-2, 0) {};
		\node [style=none] (3) at (-2, 0.5) {$\mathcal{T}_d$};
		\node [style=none] (4) at (-9, 0.5) {$\mathcal{B}_d$};
		\node [style=none] (5) at (-6.25, 0) {};
		\node [style=none] (6) at (-4.75, 0) {};
		\node [style=none] (7) at (-5.5, 0.75) {};
		\node [style=none] (8) at (-5.5, -0.75) {};
		\node [style=none] (9) at (-5.5, -1.25) {$\mathcal{T}_{d+1}$};
		\node [style=none] (10) at (-3.25, 0.5) {$\mathcal{S}_{d+1}(g)$};
		\node [style=none] (11) at (-5.5, 1.25) {$\mathcal{B}_{d+1}$};
		\node [style=none] (12) at (-6.75, -0.45) {$\overline{\mathcal{J}}_d$};
		\node [style=none] (13) at (-4.25, -0.5) {$\mathcal{J}_d$};
		\node [style=none] (14) at (-5.5, 0) {$\mathcal{S}_{d+2}$};
		\node [style=none] (15) at (-0.5, 0) {};
		\node [style=none] (16) at (0.5, 0) {};
		\node [style=SmallCircle] (18) at (2.25, 0) {};
		\node [style=SmallCircleRed] (19) at (5.25, 0) {};
		\node [style=none] (22) at (3, 0.75) {};
		\node [style=none] (23) at (3, -0.75) {};
		\node [style=none] (24) at (4.5, -0.75) {};
		\node [style=none] (25) at (4.5, 0.75) {};
		\node [style=none] (28) at (3.75, 0) {$\mathcal{S}_{d+2}$};
		\node [style=none] (29) at (3.75, -1.5) {$\mathcal{T}_{d+1}$};
		\node [style=none] (30) at (5.625, 0.375) {$\mathcal{T}_d$};
		\node [style=none] (31) at (1.875, 0.375) {$\mathcal{B}_d$};
		\node [style=none] (32) at (3.75, 1.5) {$\mathcal{B}_{d+1}$};
	\end{pgfonlayer}
	\begin{pgfonlayer}{edgelayer}
		\draw[fill=gray!50, draw=gray!50] (-5.5,0) circle (4ex);
		\draw[fill=gray!50, draw=gray!50] (3.75,0) circle (6.04ex);
		\filldraw[fill=gray!50, draw=gray!50]  (2.25, 0) -- (3, 0.75) -- (3, -0.75)  -- cycle;
		\filldraw[fill=gray!50, draw=gray!50]  (5.25, 0) -- (4.5, 0.75) -- (4.5, -0.75) -- cycle;
		\draw [style=ThickLine] (1) to (5.center);
		\draw [style=ThickLine] (6.center) to (2);
		\draw [style=RedLine, bend right=45] (5.center) to (8.center);
		\draw [style=RedLine, bend right=45] (8.center) to (6.center);
		\draw [style=PurpleLine, bend left=45] (5.center) to (7.center);
		\draw [style=PurpleLine, bend left=45] (7.center) to (6.center);
		\draw [style=ArrowLineRight] (15.center) to (16.center);
		\draw [style=RedLine] (24.center) to (19);
		\draw [style=RedLine, bend right=45, looseness=1.25] (23.center) to (24.center);
		\draw [style=RedLine] (23.center) to (18);
		\draw [style=PurpleLine] (18) to (22.center);
		\draw [style=PurpleLine, bend left=45, looseness=1.25] (22.center) to (25.center);
		\draw [style=PurpleLine] (25.center) to (19);
	\end{pgfonlayer}
\end{tikzpicture}}
\ee
Observe, then, that the original $\mathcal{T}_{d}$ now specifies a corner mode of a higher-dimensional bulk. In this setting the relative theories $\mathcal{T}_{d+1}$ and $\mathcal{B}_{d+1}$ now serve as edge theories which merge at the $\mathcal{T}_{d}$ corner.\footnote{Of course corners are not topological. The topologically invariant feature here is that both $\mathcal{B}_d,\mathcal{T}_d$ are interfaces between $\mathcal{B}_{d+1}, \mathcal{T}_{d+1}$ which are edges for $\mathcal{S}_{d+2}$. Nonetheless we represent both $\mathcal{B}_d, \mathcal{T}_d$ as corners with an eye on their geometric origin in string constructions.} Finally, it will be useful to disperse  the original boundary condition $\mathcal{B}_d$ as a gapped / free system which spans the edges far away from the $\mathcal{T}_d$ corner. This will decompress various distinct boundary conditions which are all subsumed into $\mathcal{B}_d$. We decompress as:
\be\label{eq:cheesesteak}
\scalebox{0.8}{
\begin{tikzpicture}
	\begin{pgfonlayer}{nodelayer}
		\node [style=SmallCircle] (0) at (-5, 0) {};
		\node [style=SmallCircleRed] (1) at (-2, 0) {};
		\node [style=none] (2) at (-4.25, 0.75) {};
		\node [style=none] (3) at (-4.25, -0.75) {};
		\node [style=none] (4) at (-2.75, -0.75) {};
		\node [style=none] (5) at (-2.75, 0.75) {};
		\node [style=none] (6) at (-3.5, 0) {$\mathcal{S}_{d+2}$};
		\node [style=none] (7) at (-3.5, -1.5) {$\mathcal{T}_{d+1}$};
		\node [style=none] (8) at (-1.5, 0.5) {$\mathcal{T}_d$};
		\node [style=none] (9) at (-5.5, 0.5) {$\mathcal{B}_d$};
		\node [style=none] (10) at (-3.5, 1.5) {$\mathcal{B}_{d+1}$};
		\node [style=SmallCircleRed] (12) at (5, 0) {};
		\node [style=none] (15) at (2.5, -1) {};
		\node [style=none] (16) at (2.5, 1) {};
		\node [style=none] (17) at (3.375, 0) {$\mathcal{S}_{d+2}$};
		\node [style=none] (18) at (4, -1) {$\mathcal{T}_{d+1}$};
		\node [style=none] (19) at (5.75, 0) {$\mathcal{T}_d$};
		\node [style=none] (21) at (4, 1) {$\mathcal{B}_{d+1}$};
		\node [style=none] (22) at (-0.5, 0) {};
		\node [style=none] (23) at (0.5, 0) {};
		\node [style=none] (24) at (1.75, 0.035) {$\mathbb{B}$};
		\node [style=SmallCircleBrown] (25) at (2.5, -1) {};
		\node [style=SmallCircleGrey] (26) at (2.5, 1) {};
		\node [style=none] (27) at (1.8, 0.35) {};
		\node [style=none] (28) at (1.8, -0.35) {};
		\node [style=none] (29) at (2.25, -0.85) {};
		\node [style=none] (30) at (2.25, 0.85) {};
		\node [style=none] (142) at (2.25, 1) {};
		\node [style=none] (143) at (2, 0) {};
		\node [style=none] (144) at (2.25, -1) {};
	\end{pgfonlayer}
	\begin{pgfonlayer}{edgelayer}
		\draw[fill=gray!50, draw=gray!50] (-3.5,0) circle (6.04ex);
		\filldraw[fill=gray!50, draw=gray!50]  (-2, 0) -- (-2.75, 0.75) -- (-2.75, -0.75)  -- cycle;
		\filldraw[fill=gray!50, draw=gray!50]  (-5, 0) -- (-4.25, 0.75) -- (-4.25, -0.75) -- cycle;
		\filldraw[fill=gray!50, draw=gray!50]  (5, 0) -- (2.5, 1) -- (2.5, -1) -- cycle;
		\draw [style=RedLine] (4.center) to (1);
		\draw [style=RedLine, bend right=45, looseness=1.25] (3.center) to (4.center);
		\draw [style=RedLine] (3.center) to (0);
		\draw [style=PurpleLine] (0) to (2.center);
		\draw [style=PurpleLine, bend left=45, looseness=1.25] (2.center) to (5.center);
		\draw [style=PurpleLine] (5.center) to (1);
		\draw [style=RedLine] (15.center) to (12);
		\draw [style=PurpleLine] (16.center) to (12);
		\draw [style=ArrowLineRight] (22.center) to (23.center);
		\draw [style=GreenLine] (16.center) to (15.center);
		\draw [in=0, out=-180, looseness=0.75] (142.center) to (143.center);
		\draw [in=180, out=0, looseness=0.75] (143.center) to (144.center);
	\end{pgfonlayer}
\end{tikzpicture}}
\ee

Here, on the right, we have indicated the different relative theories appearing as edges and corners, and denoted the remaining boundary conditions collectively as $\mathbb{B}$ (a tuple of two corners and an edge) which can in principle either be gapped or free.
Compared with the original decompression in line \eqref{eq:TScombo} with $\mathcal{S}_{d+1}$ not gapped, we will have examples where now $\mathcal{S}_{d+2}$ is gapped.  As such, standard manipulations of defects and topological symmetry operators can now be carried out in this bigger system. We view this as a nested collection of relative symmetry theories with corners, or perhaps more colloquially as a ``cheesesteak'' construction.\footnote{See, e.g., \cite{Cheesesteak}.}

Now, once we have a fully gapped bulk symmetry theory, we can then proceed to track how defects and symmetry operators push down onto the different edges, as well as the corner theory $\mathcal{T}_d$. One consequence of this setup is that we can now explicitly track more subtle features such as higher-group structures\footnote{See also \cite{Baez:2005sn, Sati:2008eg, Sati:2009ic, Fiorenza:2010mh, Fiorenza:2012tb, Kapustin:2013uxa} and \cite{Pantev:2005zs, Pantev:2005rh,Pantev:2005wj, Sharpe:2015mja}.} involving entwinement between higher-form symmetries \cite{Benini:2018reh, Cordova:2018cvg, Cordova:2020tij}, even in situations where some of the constituent generalized symmetries have continuous factors.

Our discussion thus far has focused on the symmetry theory from the perspective of the relative theory $\mathcal{T}_d$. Alternatively, one can consider a bulk interacting theory such as $\mathcal{T}_{d+1}$ and then simply ask what happens if we consider a system with a defect inserted at the ``end of the world''. This defect will in general also support a non-trivial QFT, and so inevitably there will be a non-trivial interplay between the symmetries of the bulk and that of the defect. In this case, it will be useful to decompress line \eqref{eq:cheesesteak} further. Schematically, we have:
\be
\scalebox{0.9}{
\begin{tikzpicture}
	\begin{pgfonlayer}{nodelayer}
		\node [style=SmallCircleRed] (0) at (-2.5, 0) {};
		\node [style=none] (1) at (-5, -1) {};
		\node [style=none] (2) at (-5, 1) {};
		\node [style=none] (3) at (-4.25, 0) {$\mathcal{S}_{d+2}$};
		\node [style=none] (4) at (-3.5, -1) {$\mathcal{T}_{d+1}$};
		\node [style=none] (5) at (-1.75, 0) {$\mathcal{T}_d$};
		\node [style=none] (6) at (-3.5, 1) {$\mathcal{B}_{d+1}$};
		\node [style=none] (7) at (-0.5, 0) {};
		\node [style=none] (8) at (0.5, 0) {};
		\node [style=none] (9) at (-5.75, 0.0) {$\mathbb{B}$};
		\node [style=SmallCircleBrown] (10) at (-5, -1) {};
		\node [style=SmallCircleGrey] (11) at (-5, 1) {};
		\node [style=SmallCircleRed] (12) at (5.25, -1) {};
		\node [style=none] (13) at (2.75, -1) {};
		\node [style=none] (14) at (2.75, 1) {};
		\node [style=none] (15) at (4, 0) {$\mathcal{S}_{d+2}$};
		\node [style=none] (16) at (4, -1.5) {$\mathcal{T}_{d+1}$};
		\node [style=none] (17) at (5.75, -1.5) {$\mathcal{T}_d$};
		\node [style=none] (18) at (6, 0) {$\mathcal{B}_{d+1}$};
		\node [style=none] (19) at (1.75, 2.1) {$\widetilde{\mathbb{B}}$};
		\node [style=SmallCircleBrown] (20) at (2.75, -1) {};
		\node [style=none] (21) at (5.25, 1) {};
		\node [style=SmallCircleBlue] (22) at (5.25, 1) {};
		\node [style=SmallCircleGreen] (23) at (2.75, 1) {};
		\node [style=none] (24) at (2.25, 2) {};
		\node [style=none] (25) at (5, 1.25) {};
		\node [style=none] (26) at (1.75, 1.5) {};
		\node [style=none] (27) at (2.5, -0.75) {};
		\node [style=none] (28) at (-5.625, 0.35) {};
		\node [style=none] (29) at (-5.2, 0.875) {};
		\node [style=none] (30) at (-5.2, -0.875) {};
		\node [style=none] (31) at (-5.625, -0.35) {};
		\node [style=none] (142) at (-5.25, 1) {};
		\node [style=none] (143) at (-5.5, 0) {};
		\node [style=none] (144) at (-5.25, -1) {};
		\node [style=none] (148) at (5.375, 1.375) {};
		\node [style=none] (150) at (2.375, -1.125) {};
		\node [style=none] (151) at (2.25, -0.25) {};
		\node [style=none] (152) at (4.25, 1.5) {};
		\node [style=none] (153) at (2, 1.75) {};
	\end{pgfonlayer}
	\begin{pgfonlayer}{edgelayer}
		\filldraw[fill=gray!50, draw=gray!50]  (-2.5, 0) -- (-5, 1) -- (-5, -1) -- cycle;
		\filldraw[fill=gray!50, draw=gray!50]  (2.75, 1) -- (2.75, -1) -- (5.25, -1) -- (5.25, 1) -- cycle;
		\draw [style=RedLine] (1.center) to (0);
		\draw [style=PurpleLine] (2.center) to (0);
		\draw [style=ArrowLineRight] (7.center) to (8.center);
		\draw [style=GreenLine] (2.center) to (1.center);
		\draw [style=RedLine] (13.center) to (12);
		\draw [style=GreenLine] (14.center) to (13.center);
		\draw [style=BlueLine] (14.center) to (22);
		\draw [style=PurpleLine] (22) to (12);
		\draw [in=0, out=-180, looseness=0.75] (142.center) to (143.center);
		\draw [in=180, out=0, looseness=0.75] (143.center) to (144.center);
		\draw [in=90, out=0, looseness=0.75] (152.center) to (148.center);
		\draw [in=-180, out=-105, looseness=0.75] (151.center) to (150.center);
		\draw [in=-45, out=75, looseness=0.75] (151.center) to (153.center);
		\draw [in=-175, out=-45, looseness=0.50] (153.center) to (152.center);
	\end{pgfonlayer}
\end{tikzpicture}}
\ee

Here, in transitioning between the tuples ${\mathbb{B}}$ and $\widetilde{\mathbb{B}}$ we require that an interface between two edges can be equally presented as two $d$-dimensional interfaces connected by a new $(d+1)$-dimensional edge. The result of this final decompression are two symmetry sandwiches, an absolute sandwich in dimension $(d+1)$ and two relative sandwiches in dimension $d$ (reading structures vertically).

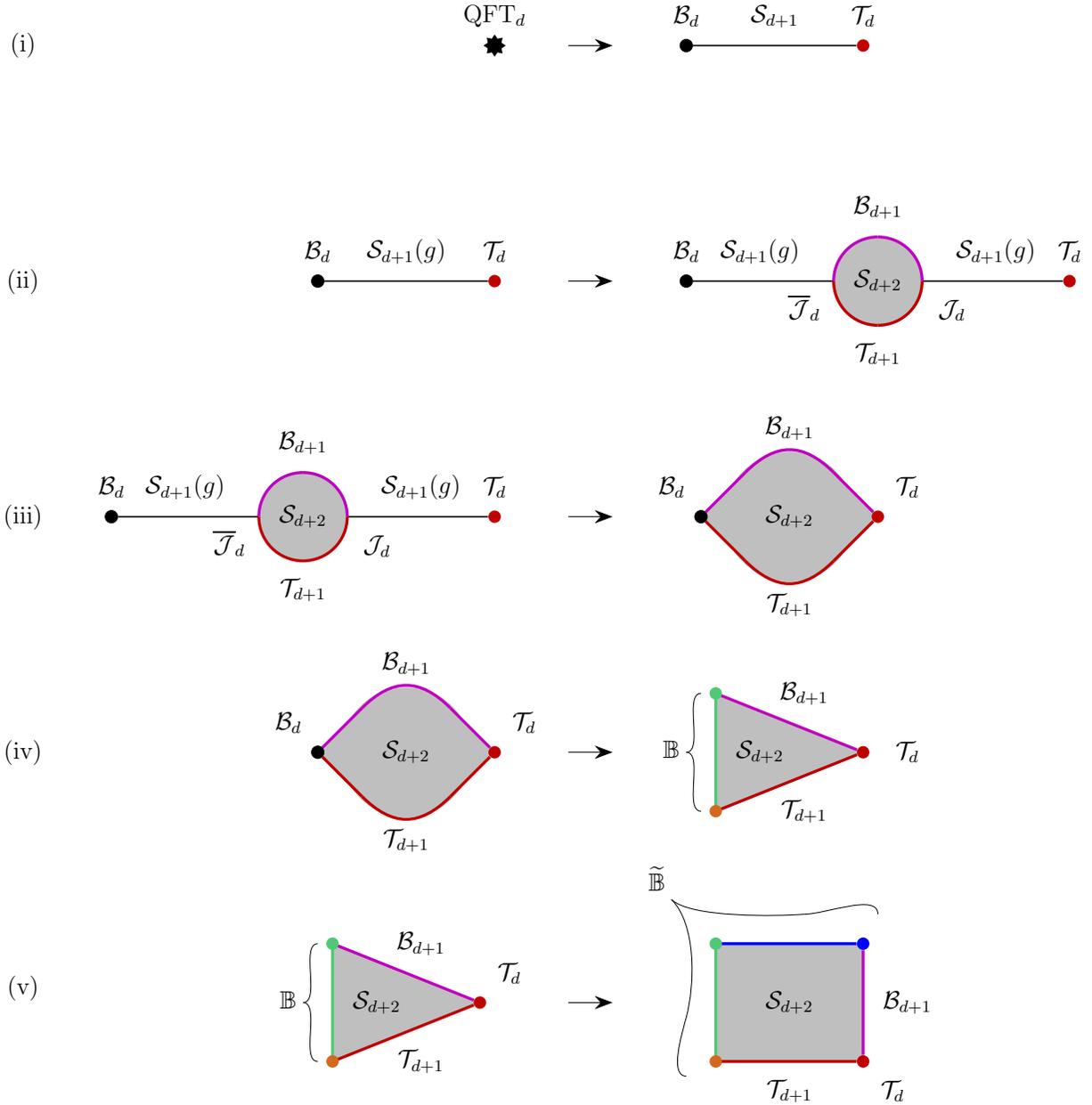
\begin{figure}
\centering
\scalebox{0.88}{
\begin{tikzpicture}
	\begin{pgfonlayer}{nodelayer}
		\node [style=SmallCircleRed] (0) at (-1.75, -6.75) {};
		\node [style=none] (1) at (-4.25, -7.75) {};
		\node [style=none] (2) at (-4.25, -5.75) {};
		\node [style=none] (3) at (-3.5, -6.75) {$\mathcal{S}_{d+2}$};
		\node [style=none] (4) at (-2.75, -7.75) {$\mathcal{T}_{d+1}$};
		\node [style=none] (5) at (-1.25, -6.25) {$\mathcal{T}_d$};
		\node [style=none] (6) at (-2.75, -5.75) {$\mathcal{B}_{d+1}$};
		\node [style=none] (7) at (-0.25, -6.75) {};
		\node [style=none] (8) at (0.5, -6.75) {};
		\node [style=none] (9) at (-5, -6.7) {$\mathbb{B}$};
		\node [style=SmallCircleBrown] (10) at (-4.25, -7.75) {};
		\node [style=SmallCircleGreen] (11) at (-4.25, -5.75) {};
		\node [style=SmallCircleRed] (12) at (4.75, -7.75) {};
		\node [style=none] (13) at (2.25, -7.75) {};
		\node [style=none] (14) at (2.25, -5.75) {};
		\node [style=none] (15) at (3.5, -6.75) {$\mathcal{S}_{d+2}$};
		\node [style=none] (16) at (3.5, -8.25) {$\mathcal{T}_{d+1}$};
		\node [style=none] (17) at (5.25, -8.25) {$\mathcal{T}_d$};
		\node [style=none] (18) at (5.5, -6.75) {$\mathcal{B}_{d+1}$};
		\node [style=none] (19) at (1.25, -4.65) {$\widetilde{\mathbb{B}}$};
		\node [style=SmallCircleBrown] (20) at (2.25, -7.75) {};
		\node [style=none] (21) at (4.75, -5.75) {};
		\node [style=SmallCircleBlue] (22) at (4.75, -5.75) {};
		\node [style=SmallCircleGreen] (23) at (2.25, -5.75) {};
		\node [style=none] (24) at (1.75, -4.75) {};
		\node [style=none] (25) at (4.5, -5.5) {};
		\node [style=none] (26) at (1.25, -5.25) {};
		\node [style=none] (27) at (2, -7.5) {};
		\node [style=SmallCircle] (28) at (-4.5, -2.5) {};
		\node [style=SmallCircleRed] (29) at (-1.5, -2.5) {};
		\node [style=none] (30) at (-3.75, -1.75) {};
		\node [style=none] (31) at (-3.75, -3.25) {};
		\node [style=none] (32) at (-2.25, -3.25) {};
		\node [style=none] (33) at (-2.25, -1.75) {};
		\node [style=none] (34) at (-3, -2.5) {$\mathcal{S}_{d+2}$};
		\node [style=none] (35) at (-3, -4) {$\mathcal{T}_{d+1}$};
		\node [style=none] (36) at (-1, -2) {$\mathcal{T}_d$};
		\node [style=none] (37) at (-5, -2) {$\mathcal{B}_d$};
		\node [style=none] (38) at (-3, -1) {$\mathcal{B}_{d+1}$};
		\node [style=SmallCircleRed] (39) at (4.75, -2.5) {};
		\node [style=none] (40) at (2.25, -3.5) {};
		\node [style=none] (41) at (2.25, -1.5) {};
		\node [style=none] (42) at (3, -2.5) {$\mathcal{S}_{d+2}$};
		\node [style=none] (43) at (3.75, -3.5) {$\mathcal{T}_{d+1}$};
		\node [style=none] (44) at (5.5, -2.5) {$\mathcal{T}_d$};
		\node [style=none] (45) at (3.75, -1.5) {$\mathcal{B}_{d+1}$};
		\node [style=none] (46) at (-0.25, -2.5) {};
		\node [style=none] (47) at (0.5, -2.5) {};
		\node [style=none] (48) at (1.5, -2.45) {$\mathbb{B}$};
		\node [style=SmallCircleBrown] (49) at (2.25, -3.5) {};
		\node [style=SmallCircleGreen] (50) at (2.25, -1.5) {};
		\node [style=none] (51) at (-6.75, 2) {$\mathcal{S}_{d+1}(g)$};
		\node [style=SmallCircle] (52) at (-8, 1.5) {};
		\node [style=SmallCircleRed] (53) at (-1.5, 1.5) {};
		\node [style=none] (54) at (-1.5, 2) {$\mathcal{T}_d$};
		\node [style=none] (55) at (-8, 2) {$\mathcal{B}_d$};
		\node [style=none] (56) at (-5.5, 1.5) {};
		\node [style=none] (57) at (-4, 1.5) {};
		\node [style=none] (58) at (-4.75, 2.25) {};
		\node [style=none] (59) at (-4.75, 0.75) {};
		\node [style=none] (60) at (-4.75, 0.25) {$\mathcal{T}_{d+1}$};
		\node [style=none] (61) at (-2.75, 2) {$\mathcal{S}_{d+1}(g)$};
		\node [style=none] (62) at (-4.75, 2.75) {$\mathcal{B}_{d+1}$};
		\node [style=none] (63) at (-6, 1.05) {$\overline{\mathcal{J}}_d$};
		\node [style=none] (64) at (-3.5, 1) {$\mathcal{J}_d$};
		\node [style=none] (65) at (-4.75, 1.5) {$\mathcal{S}_{d+2}$};
		\node [style=none] (66) at (-0.25, 1.5) {};
		\node [style=none] (67) at (0.5, 1.5) {};
		\node [style=SmallCircle] (68) at (2, 1.5) {};
		\node [style=SmallCircleRed] (69) at (5, 1.5) {};
		\node [style=none] (70) at (2.75, 2.25) {};
		\node [style=none] (71) at (2.75, 0.75) {};
		\node [style=none] (72) at (4.25, 0.75) {};
		\node [style=none] (73) at (4.25, 2.25) {};
		\node [style=none] (74) at (3.5, 1.5) {$\mathcal{S}_{d+2}$};
		\node [style=none] (75) at (3.5, 0) {$\mathcal{T}_{d+1}$};
		\node [style=none] (76) at (5.5, 2) {$\mathcal{T}_d$};
		\node [style=none] (77) at (1.5, 2) {$\mathcal{B}_d$};
		\node [style=none] (78) at (3.5, 3) {$\mathcal{B}_{d+1}$};
		\node [style=SmallCircle] (79) at (-4.5, 5.5) {};
		\node [style=SmallCircleRed] (80) at (-1.5, 5.5) {};
		\node [style=none] (81) at (-1.5, 6) {$\mathcal{T}_d$};
		\node [style=none] (82) at (-4.5, 6) {$\mathcal{B}_d$};
		\node [style=none] (84) at (-0.25, 5.5) {};
		\node [style=none] (85) at (0.5, 5.5) {};
		\node [style=none] (99) at (-3, 6) {$\mathcal{S}_{d+1}(g)$};
		\node [style=none] (101) at (3, 6) {$\mathcal{S}_{d+1}(g)$};
		\node [style=SmallCircle] (102) at (1.75, 5.5) {};
		\node [style=SmallCircleRed] (103) at (8.25, 5.5) {};
		\node [style=none] (104) at (8.25, 6) {$\mathcal{T}_d$};
		\node [style=none] (105) at (1.75, 6) {$\mathcal{B}_d$};
		\node [style=none] (106) at (4.25, 5.5) {};
		\node [style=none] (107) at (5.75, 5.5) {};
		\node [style=none] (108) at (5, 6.25) {};
		\node [style=none] (109) at (5, 4.75) {};
		\node [style=none] (110) at (5, 4.25) {$\mathcal{T}_{d+1}$};
		\node [style=none] (111) at (7, 6) {$\mathcal{S}_{d+1}(g)$};
		\node [style=none] (112) at (5, 6.75) {$\mathcal{B}_{d+1}$};
		\node [style=none] (113) at (3.75, 5.05) {$\overline{\mathcal{J}}_d$};
		\node [style=none] (114) at (6.25, 5) {$\mathcal{J}_d$};
		\node [style=none] (115) at (5, 5.5) {$\mathcal{S}_{d+2}$};
		\node [style=none] (116) at (-0.25, 9.5) {};
		\node [style=none] (117) at (0.5, 9.5) {};
		\node [style=Star] (118) at (-1.5, 9.5) {};
		\node [style=SmallCircle] (119) at (1.75, 9.5) {};
		\node [style=SmallCircleRed] (120) at (4.75, 9.5) {};
		\node [style=none] (121) at (-1.5, 10) {QFT$_d$};
		\node [style=none] (125) at (-9.5, 1.5) {(iii)};
		\node [style=none] (126) at (-9.5, 5.5) {(ii)};
		\node [style=none] (127) at (-9.5, 9.5) {(i)};
		\node [style=none] (128) at (-9.5, -2.5) {(iv)};
		\node [style=none] (129) at (-9.5, -6.5) {(v)};
		\node [style=none] (130) at (4.75, 10) {$\mathcal{T}_d$};
		\node [style=none] (131) at (1.75, 10) {$\mathcal{B}_d$};
		\node [style=none] (132) at (3.25, 10) {$\mathcal{S}_{d+1}$};
		\node [style=none] (133) at (0, -9) {};
		\node [style=none] (133) at (0, -9) {};
		
		\node [style=none] (134) at (-4.5, -5.875) {};
		\node [style=none] (135) at (-4.875, -6.4) {};
		\node [style=none] (136) at (-4.875, -7.1) {};
		\node [style=none] (137) at (-4.5, -7.625) {};
		
		\node [style=none] (138) at (2, -1.625) {};
		\node [style=none] (139) at (1.562, -2.15) {};
		\node [style=none] (140) at (1.625, -2.85) {};
		\node [style=none] (141) at (2, -3.375) {};
		\node [style=none] (142) at (2, -1.5) {};
		\node [style=none] (143) at (1.75, -2.5) {};
		\node [style=none] (144) at (2, -3.5) {};
		\node [style=none] (145) at (-4.5, -5.75) {};
		\node [style=none] (146) at (-4.75, -6.75) {};
		\node [style=none] (147) at (-4.5, -7.75) {};
		\node [style=none] (148) at (5, -5.25) {};
		\node [style=none] (150) at (1.75, -8) {};
		\node [style=none] (151) at (1.75, -7) {};
		\node [style=none] (152) at (3.75, -5.25) {};
		\node [style=none] (153) at (1.5, -5) {};
	\end{pgfonlayer}
	\begin{pgfonlayer}{edgelayer}
		\filldraw[fill=gray!50, draw=gray!50]  (-1.75, -6.75) -- (-4.25, -5.75) -- (-4.25, -7.75) -- cycle;
		\filldraw[fill=gray!50, draw=gray!50]  (4.75, -2.5) -- (2.25, -1.5) -- (2.25, -3.5) -- cycle;
		\filldraw[fill=gray!50, draw=gray!50]  (2.25, -7.75) -- (2.25, -5.75) -- (4.75, -5.75) -- (4.75, -7.75) -- cycle;
		\draw[fill=gray!50, draw=gray!50] (5,5.5) circle (4ex);
		\draw[fill=gray!50, draw=gray!50] (-4.75,1.5) circle (4ex);

		\draw[fill=gray!50, draw=gray!50] (3.5,1.5) circle (6.04ex);
		\filldraw[fill=gray!50, draw=gray!50]  (-1.5, -2.5) -- (-2.25, -1.75) -- (-2.25, -3.25)  -- cycle;
		\filldraw[fill=gray!50, draw=gray!50]  (-4.5, -2.5) -- (-3.75, -1.75) -- (-3.75, -3.25) -- cycle;
		
		\draw[fill=gray!50, draw=gray!50] (-3,-2.5) circle (6.04ex);
		\filldraw[fill=gray!50, draw=gray!50]  (5, 1.5) -- (4.25, 2.25) -- (4.25, 0.75)  -- cycle;
		\filldraw[fill=gray!50, draw=gray!50]  (2, 1.5) -- (2.75, 2.25) -- (2.75, 0.75) -- cycle;
		
		\draw [style=RedLine] (1.center) to (0);
		\draw [style=PurpleLine] (2.center) to (0);
		\draw [style=ArrowLineRight] (7.center) to (8.center);
		\draw [style=GreenLine] (2.center) to (1.center);
		\draw [style=RedLine] (13.center) to (12);
		\draw [style=GreenLine] (14.center) to (13.center);
		\draw [style=BlueLine] (14.center) to (22);
		\draw [style=PurpleLine] (22) to (12);
		\draw [style=RedLine] (32.center) to (29);
		\draw [style=RedLine, bend right=45, looseness=1.25] (31.center) to (32.center);
		\draw [style=RedLine] (31.center) to (28);
		\draw [style=PurpleLine] (28) to (30.center);
		\draw [style=PurpleLine, bend left=45, looseness=1.25] (30.center) to (33.center);
		\draw [style=PurpleLine] (33.center) to (29);
		\draw [style=RedLine] (40.center) to (39);
		\draw [style=PurpleLine] (41.center) to (39);
		\draw [style=ArrowLineRight] (46.center) to (47.center);
		\draw [style=GreenLine] (41.center) to (40.center);
		\draw [style=ThickLine] (52) to (56.center);
		\draw [style=ThickLine] (57.center) to (53);
		\draw [style=RedLine, bend right=45] (56.center) to (59.center);
		\draw [style=RedLine, bend right=45] (59.center) to (57.center);
		\draw [style=PurpleLine, bend left=45] (56.center) to (58.center);
		\draw [style=PurpleLine, bend left=45] (58.center) to (57.center);
		\draw [style=ArrowLineRight] (66.center) to (67.center);
		\draw [style=RedLine] (72.center) to (69);
		\draw [style=RedLine, bend right=45, looseness=1.25] (71.center) to (72.center);
		\draw [style=RedLine] (71.center) to (68);
		\draw [style=PurpleLine] (68) to (70.center);
		\draw [style=PurpleLine, bend left=45, looseness=1.25] (70.center) to (73.center);
		\draw [style=PurpleLine] (73.center) to (69);
		\draw [style=ThickLine] (79) to (80);
		\draw [style=ArrowLineRight] (84.center) to (85.center);
		\draw [style=ThickLine] (102) to (106.center);
		\draw [style=ThickLine] (107.center) to (103);
		\draw [style=RedLine, bend right=45] (106.center) to (109.center);
		\draw [style=RedLine, bend right=45] (109.center) to (107.center);
		\draw [style=PurpleLine, bend left=45] (106.center) to (108.center);
		\draw [style=PurpleLine, bend left=45] (108.center) to (107.center);
		\draw [style=ArrowLineRight] (116.center) to (117.center);
		\draw [style=ThickLine] (119) to (120);
		\draw [in=0, out=-180, looseness=0.75] (142.center) to (143.center);
		\draw [in=180, out=0, looseness=0.75] (143.center) to (144.center);
		\draw [in=0, out=-180, looseness=0.75] (145.center) to (146.center);
		\draw [in=180, out=0, looseness=0.75] (146.center) to (147.center);
		\draw [in=90, out=0, looseness=0.75] (152.center) to (148.center);
		\draw [in=-180, out=-105, looseness=0.75] (151.center) to (150.center);
		\draw [in=-45, out=75, looseness=0.75] (151.center) to (153.center);
		\draw [in=-175, out=-45, looseness=0.50] (153.center) to (152.center);
	\end{pgfonlayer}
\end{tikzpicture}
}
\caption{We summarize the five (de)compression steps we consider. In (iii) we have assumed simplifying assumptions for the junction theories.}
\label{fig:Summary}
\end{figure}

We summarize all (de)compression steps discussed in figure \ref{fig:Summary}.

This picture naturally arises in stringy constructions, and this was indeed the initial motivation for this work.
To further support the utility of this perspective, we turn to some explicit examples which illustrate these general features. A particularly prominent example is the case of 5D SCFTs realized by M-theory on the singular background $X = \mathbb{C}^{3} / \mathbb{Z}_{2n}$ with group action generated by $(z_1, z_2, z_3) \mapsto (\omega z_1, \omega z_2 , \omega^{-2} z_3)$ with $\omega = \exp(2 \pi i / 2n)$. This geometry has a 5D SCFT localized at the origin $z_1 = z_2 = z_3 = 0$, but also contains an $\mathfrak{su}(2)$ flavor symmetry factor along the locus $z_1 = z_2 = 0$. Observe that in this geometry, the radial direction of $X = \mathrm{Cone}(S^5 / \mathbb{Z}_{2n})$ contains a fixed locus, so a naive dimensional reduction along $\partial X$ would result in a gapless 6D theory in the candidate symmetry theory. Transverse to this flavor brane there is still a gapped system, but this instead would have been specified by reduction on a lower-dimensional space, i.e., $\partial^2 X^{\circ}$, the space obtained by first excising the flavor brane from $\partial X$ and considering the boundary of the resulting system. Performing the reductions on the appropriate boundaries and ``boundaries of boundaries'' all of the ingredients in line (\ref{eq:cheesesteak}) appear.

As an additional example, we also consider the relative symmetry theory of $N$ chiral multiplets as engineered from the collision of singularities in a local M-theory background \cite{Atiyah:2001qf, Acharya:2001gy, Witten:2001uq}. In this case, we view the 4D chiral multiplet as a defect of a bulk ``flavor brane'' system, which in principle has its own SymTFT. This case is especially subtle because of the large number of symmetries which can act on free fields. As such, it provides an interesting check on the formalism and the computation.

The geometric perspective also provides a systematic way to compute quantities of interest in the original QFT$_d$, including the structure
of various discrete and continuous anomalies. In the case of 5D SCFTs realized by singular M-theory backgrounds, we show how to account for excisions of singularities in calculating triple products of (co)homology classes. This in turn specifies more refined data on possible mixing structures present when dealing with 0-form and 1-form symmetries in these settings. We also illustrate these calculations in the case of our system of 4D chiral multiplets coupled to a bulk gauge theory.

The rest of this paper is organized as follows. Though our approach is motivated by string theory considerations,
we begin in section \ref{sec:cheesesteak} with a bottom up characterization of
relative symmetries theories, and in particular the nested structure which accommodates an eventual filtration. We follow this in section
\ref{sec:TopDown} with a top down construction of this structure. To illustrate these general considerations, we turn to some explicit examples in section \ref{sec:Illustrative}.
Section \ref{sec:CONC} contains our conclusions. We present some additional details of the geometric computations in appendix \ref{app:A}.

\section{Relative Symmetry Theories}\label{sec:cheesesteak}

In this section we introduce relative symmetry theories. The main idea is that the global symmetries of a $d$-dimensional QFT can
be captured in terms of a $(d+1)$-dimensional symmetry theory (SymTh). Treating the combined system as a $(d+1)$-dimensional bulk theory with
a $d$-dimensional edge mode, we can also analyze the symmetries of this combined system. The motivation for proceeding up in dimension in this way comes directly from string theory where one often encounters an intricate collection of intersecting branes and singularities which localize to produce a QFT$_d$ of interest. One can of course attempt to ``compress'' all of this data into a single extra dimension, but this can obscure various features of the bulk system, including global structures such as its spectrum of extended operators as well as higher-categorical structures. One could in principle anticipate further generalizations, but the construction we present matches well to expectations based on top down realizations of QFTs.

This section is organized as follows. We begin by briefly reviewing symmetry theories, and their use in specifying the absolute form of a QFT$_d$.
There is a natural uplift of this structure to stringy backgrounds on geometries of the form $X = \mathrm{Cone}(\partial X)$ where the
extra dimension of the symmetry theory is interpreted as the radial coordinate of the cone. Stringy considerations also include cases
where the radial direction is itself filled by an interacting QFT, which can have its own symmetry theory. Unpacking this, we show how to lift
the full system to a nested collection of relative symmetry theories. In particular, we explain how boundary conditions of the relative theories are correlated in this bigger nested structure.

\subsection{Symmetry Theories}

We now briefly review some aspects of symmetry theories. Our starting point will be to discuss the best established case with finite symmetries, in which case we have a symmetry topological field theory (SymTFT). We then turn to the case of continuous symmetries, as captured by a symmetry theory (SymTh). As mentioned our definition of a SymTh is simply a bulk system which captured the symmetries of the relative QFT localized on an edge. As such, we permit ourselves to consider both gapped and free field theories in the bulk, and we view both possibilities as obtained from a scaling limit of a possibly more complicated bulk QFT (as often happens in stringy constructions). The main aim of our approach is to filter this system further to produce a bulk theory which is fully gapped but which nonetheless encodes the structure of different symmetries, viewed as boundary modes in a possibly even bigger system.

To begin, we consider a QFT$_d$ with a collection of categorical symmetries.
In the case where these symmetries are finite there is a general construction available to capture the global form of the QFT$_d$ in terms
of an auxiliary $(d+1)$-dimensional symmetry topological field theory SymTFT$_{d+1}$ (see e.g., \cite{Reshetikhin:1991tc, Turaev:1992hq, Barrett:1993ab, Witten:1998wy, Fuchs:2002cm, Kirillov:2010nh, Kapustin:2010if, Kitaev:2011dxc, Fuchs:2012dt, Freed:2012bs, Freed:2018cec, Gaiotto:2020iye, Apruzzi:2021nmk, Freed:2022qnc, Kaidi:2022cpf}).
In this framework, the local\,/\,interacting degrees of freedom are separated from the global structure of the theory by introducing suitable boundary conditions on the SymTFT. One refers to the physical boundary conditions $\mathcal{T}_d$ as the relative QFT, and the gapped\,/\,topological boundary conditions as $\mathcal{B}_d$. The boundary conditions $\mathcal{B}_d$ dictate the global form of the theory, i.e.,
the spectrum of symmetry and defect operators of the absolute QFT$_d$. We can summarize this in terms of a decompression step:
\be \label{eq:decompress}
\scalebox{0.9}{
\begin{tikzpicture}
	\begin{pgfonlayer}{nodelayer}
		\node [style=none] (1) at (-1, 0) {};
		\node [style=none] (2) at (0, 0) {};
		\node [style=Star] (3) at (-2.5, 0) {};
		\node [style=SmallCircle] (4) at (1.5, 0) {};
		\node [style=SmallCircleRed] (5) at (4.5, 0) {};
		\node [style=none] (6) at (-2.5, 0.5) {QFT$_d$};
		\node [style=none] (7) at (4.5, 0.5) {$\mathcal{T}_d$};
		\node [style=none] (8) at (1.5, 0.5) {$\mathcal{B}_d$};
		\node [style=none] (9) at (3, 0.5) {$\mathcal{S}_{d+1}$};
		\node [style=none] (10) at (0, 0.875) {};
	\end{pgfonlayer}
	\begin{pgfonlayer}{edgelayer}
		\draw [style=ArrowLineRight] (1.center) to (2.center);
		\draw [style=ThickLine] (4) to (5);
	\end{pgfonlayer}
\end{tikzpicture}}
\ee
in the obvious notation.

In this setup, $q$-dimensional heavy defects of the QFT$_d$ lift to $(q+1)$-dimensional defects in the $(d+1)$-dimensional system which stretch between the two boundaries $\mathcal{B}_d$ and $\mathcal{T}_d$. Topological symmetry operators link\,/\,intersect with these defects and remain of the same dimension when pulled from the $d$-dimensional boundary out to the bulk. Observe that linking\,/\,intersection of defect operators with symmetry operators is consistent between the QFT$_d$ and the bulk symmetry theory because the heavy defects have support along the extra spatial direction.

One can in principle generalize this basic picture in many ways. For example, while the best established case involves finite\,/\,discrete symmetries, there have been recent proposals for how to extend this to the case of continuous symmetries \cite{Brennan:2024fgj, Antinucci:2024zjp, Bonetti:2024cjk, Apruzzi:2024htg}. Notably, in this broader setting, there can be subtleties concerning the dynamics of the $(d+1)$-dimensional bulk system. This includes the appearance, for example, of non-compact gauge groups, as well as a choice of a metric dependent regulator to make sense of the boundary conditions and partition function.\footnote{Consider, for example, the case of classical 3D gravity formulated as a Chern-Simons theory with non-compact gauge group \cite{Witten:1988hc}.} A related issue is that even if one demands that the bulk is gapped or a collection of free fields, the structure of the boundary condition $\mathcal{B}_d$ which dictates the global form of the theory also need not be gapped, i.e., it might also have free fields. Along these lines, we comment that in many top down constructions the ``bulk'' often includes interacting degrees of freedom. These additional bulk degrees of freedom can sometimes be decoupled, but as far as we are aware, this need not be the case in general.

To illustrate some of these issues, consider a QFT$_d$ with a continuous flavor symmetry given by a Lie group $G$. To track the effects of gauging, it is convenient to first introduce a gauge theory but one in which the gauge coupling might have some dependence on the interval direction $r$ of the bulk system:
\begin{equation}\label{eq:bulkLag}
\mathcal{L}_{\mathrm{bulk}} \supset -\frac{1}{4g(r)^2}\mathrm{Tr}\, F \wedge \ast F\,.
\end{equation}
Different choices for the position dependent profile of $g(r)$ both in the bulk and the boundary $\mathcal{B}_d$ (at $r = \infty$) lead to different possible bulk symmetry theories with boundary conditions. One can view these as different choices for how to regulate the bulk symmetry theory. One canonical choice is to take a limit where one tunes $g^{2} \rightarrow 0$ so that one is only left with free fields / possible topological terms. Another natural choice is to allow a non-trivial value at $\mathcal{B}_d$, at the expense of having some explicit metric dependence in the boundary conditions. To a certain extent, taking such a limit is rather natural when the gauge theory is in $(d+1) > 4$ dimensions (since it always flows to weak coupling) but when the bulk gauge theory is in $(d +1) \leq 4$, the bulk might itself experience strong coupling dynamics. One must then tune the various scales to reach the desired bulk system with trivial local dynamics.

Rather than go this route, we shall instead opt for a different strategy to make sense of the topological structure of continuous symmetries.
Our aim will be to view $\mathcal{S}_{d+1}$ as obtained from a limit of QFTs $\mathcal{S}_{d+1}(g)$ via:
\begin{equation}
\mathcal{S}_{d+1} \equiv \underset{g \rightarrow 0}{\lim} \, \mathcal{S}_{d+1}(g).
\end{equation}
We then will aim to instead construct the SymTh$_{d+2}$ for $\mathcal{S}_{d+1}(g)$. Iterating this procedure multiple times, we expect to eventually filter the whole system to a bulk which is fully gapped. From a top down perspective, this appears to be a more canonical way to proceed. For example, in the explicit examples we introduce later, we will encounter a bulk gauge theory which naturally has a position dependent coupling in the radial direction. A related comment is that similar structures typically appear in holographic setups.

Focusing then on the bulk symmetry theory $\mathcal{S}_{d+1}(g)$, we now treat this as a QFT$_{d+1}$ in its own right. With this in mind,
suppose that we did not include any ``edge modes'' at all, i.e., we place the theory on a $(d+1)$-dimensional space with no boundaries.
In this case one can again take this absolute theory and decompress it:
\be\label{eq:decompressS}
\scalebox{0.9}{
\begin{tikzpicture}
	\begin{pgfonlayer}{nodelayer}
		\node [style=none] (0) at (0, 0) {};
		\node [style=none] (1) at (1, 0) {};
		\node [style=none] (5) at (-2, 0.5) {$\mathcal{S}_{d+1}(g)$};
		\node [style=none] (9) at (1, 1.25) {};
		\node [style=none] (10) at (-2.75, 0) {};
		\node [style=none] (11) at (-1.25, 0) {};
		\node [style=none] (12) at (-1, 0) {};
		\node [style=none] (13) at (-3, 0) {};
		\node [style=none] (14) at (2.25, 0.5) {};
		\node [style=none] (15) at (3.75, 0.5) {};
		\node [style=none] (16) at (4, 0.5) {};
		\node [style=none] (17) at (2, 0.5) {};
		\node [style=none] (18) at (2.25, -0.5) {};
		\node [style=none] (19) at (3.75, -0.5) {};
		\node [style=none] (20) at (4, -0.5) {};
		\node [style=none] (21) at (2, -0.5) {};
		\node [style=none] (22) at (3, -1) {$\mathcal{T}_{d+1}$};
		\node [style=none] (23) at (3, 1) {$\mathcal{B}_{d+1}$};
		\node [style=none] (24) at (3, 0) {$\mathcal{S}_{d+2}$};
	\end{pgfonlayer}
	\begin{pgfonlayer}{edgelayer}
		\filldraw[fill=gray!50, draw=gray!50]  (2.25, 0.5) -- (2.25, -0.5) -- (3.75, -0.5) -- (3.75, 0.5) -- cycle;
		\draw [style=ArrowLineRight] (0.center) to (1.center);
		\draw [style=ThickLine] (10.center) to (11.center);
		\draw [style=DottedLine] (11.center) to (12.center);
		\draw [style=DottedLine] (10.center) to (13.center);
		\draw [style=PurpleLine] (14.center) to (15.center);
		\draw [style=RedLine] (18.center) to (19.center);
		\draw [style=DottedRed] (20.center) to (19.center);
		\draw [style=DottedRed] (21.center) to (18.center);
		\draw [style=DottedPurple] (16.center) to (15.center);
		\draw [style=DottedPurple] (17.center) to (14.center);
	\end{pgfonlayer}
\end{tikzpicture}}
\ee
Here $ \mathcal{S}_{d+2}$ is the symmetry theory describing the symmetries of $\mathcal{S}_{d+1}(g)$ and the dots indicate the omission of edge modes. It is worth noting that this decompression step is not unique since we can in principle distribute the gapped and free contributions to each boundary system in different ways. From this perspective, one might simply wish to refer to both $d+1$ theories as ``relative theories.'' Suppose, then, that we make a different choice of decompression of the form:
\be \label{eq:primed}
\scalebox{0.9}{
\begin{tikzpicture}
	\begin{pgfonlayer}{nodelayer}
		\node [style=none] (0) at (0, 0) {};
		\node [style=none] (1) at (1, 0) {};
		\node [style=none] (5) at (-2, 0.5) {$\mathcal{S}_{d+1}(g)$};
		\node [style=none] (9) at (1, 1.25) {};
		\node [style=none] (10) at (-2.75, 0) {};
		\node [style=none] (11) at (-1.25, 0) {};
		\node [style=none] (12) at (-1, 0) {};
		\node [style=none] (13) at (-3, 0) {};
		\node [style=none] (14) at (2.25, 0.5) {};
		\node [style=none] (15) at (3.75, 0.5) {};
		\node [style=none] (16) at (4, 0.5) {};
		\node [style=none] (17) at (2, 0.5) {};
		\node [style=none] (18) at (2.25, -0.5) {};
		\node [style=none] (19) at (3.75, -0.5) {};
		\node [style=none] (20) at (4, -0.5) {};
		\node [style=none] (21) at (2, -0.5) {};
		\node [style=none] (22) at (3, -1) {$\mathcal{T}_{d+1}^{\:\!\prime}$};
		\node [style=none] (23) at (3, 1) {$\mathcal{B}_{d+1}^{\:\!\prime}$};
		\node [style=none] (24) at (3, 0) {$\mathcal{S}_{d+2}^{\prime}$};
	\end{pgfonlayer}
	\begin{pgfonlayer}{edgelayer}
		\filldraw[fill=gray!50, draw=gray!50]  (2.25, 0.5) -- (2.25, -0.5) -- (3.75, -0.5) -- (3.75, 0.5) -- cycle;
		\draw [style=ArrowLineRight] (0.center) to (1.center);
		\draw [style=ThickLine] (10.center) to (11.center);
		\draw [style=DottedLine] (11.center) to (12.center);
		\draw [style=DottedLine] (10.center) to (13.center);
		\draw [style=PurpleLine] (14.center) to (15.center);
		\draw [style=RedLine] (18.center) to (19.center);
		\draw [style=DottedRed] (20.center) to (19.center);
		\draw [style=DottedRed] (21.center) to (18.center);
		\draw [style=DottedPurple] (16.center) to (15.center);
		\draw [style=DottedPurple] (17.center) to (14.center);
	\end{pgfonlayer}
\end{tikzpicture}}
\ee
For example, it could happen that a gapped TFT initially localized on $\mathcal{T}_{d+1}$ has now been moved over to become a part of $\mathcal{B}_{d+1}$. In this case, the two bulk theories $\mathcal{S}_{d+2}$ and $\mathcal{S}^{\prime}_{d+2}$ may also differ. That being said, top down considerations typically lead to a canonical split between these pieces.

Even more generally it can happen that $\mathcal{T}_{d+1}^{}$ is a relative QFT with locally decoupled sectors in the sense of \cite{Baume:2023kkf}. Then the edge modes and the connecting symmetry theories are disjoint sums of theories, and the latter interact only through the boundary condition $\mathcal{B}_{d+1}^{}$. In this case it is more accurate to speak of decompressing $\mathcal{S}_{d+1}$ into a SymTree with junction $\mathcal{J}_{d+1}^{}$.  This will be the case when we have multiple flavor branes in string constructions. We will not dwell on this distinction and universally represent the decompression as in line \eqref{eq:decompressS}.

We now ask how decompression of $\mathcal{S}_{d+1}(g)$ works when we consider it in the presence of the original boundary theories $\mathcal{B}_d$ and $\mathcal{T}_d$. To this end, suppose that we only decompress $\mathcal{S}_{d+1}(g)$ in
the interior of our system. Doing so, we obtain:
\be\label{eq:junctionlangles}
\scalebox{0.95}{
\begin{tikzpicture}
	\begin{pgfonlayer}{nodelayer}
		\node [style=SmallCircle] (0) at (-5, 0) {};
		\node [style=SmallCircleRed] (1) at (-2, 0) {};
		\node [style=none] (2) at (-2, 0.5) {$\mathcal{T}_d$};
		\node [style=none] (3) at (-5, 0.5) {$\mathcal{B}_d$};
		\node [style=none] (4) at (3.25, 0.5) {$\mathcal{S}_{d+1}(g)$};
		\node [style=none] (5) at (-0.5, 0) {};
		\node [style=none] (6) at (0.5, 0) {};
		\node [style=SmallCircle] (7) at (2, 0) {};
		\node [style=SmallCircleRed] (8) at (9, 0) {};
		\node [style=none] (9) at (9, 0.5) {$\mathcal{T}_d$};
		\node [style=none] (10) at (2, 0.5) {$\mathcal{B}_d$};
		\node [style=none] (11) at (4.75, 0) {};
		\node [style=none] (12) at (6.25, 0) {};
		\node [style=none] (13) at (5.5, 0.75) {};
		\node [style=none] (14) at (5.5, -0.75) {};
		\node [style=none] (15) at (5.5, -1.25) {$\mathcal{T}_{d+1}$};
		\node [style=none] (16) at (7.75, 0.5) {$\mathcal{S}_{d+1}(g)$};
		\node [style=none] (17) at (5.5, 1.25) {$\mathcal{B}_{d+1}$};
		\node [style=none] (22) at (4.25, -0.45) {$\overline{\mathcal{J}}_d$};
		\node [style=none] (23) at (6.75, -0.5) {$\mathcal{J}_d$};
		\node [style=none] (24) at (-3.5, 0.5) {$\mathcal{S}_{d+1}(g)$};
		\node [style=none] (25) at (5.5, 0) {$\mathcal{S}_{d+2}$};
	\end{pgfonlayer}
	\begin{pgfonlayer}{edgelayer}
		\draw[fill=gray!50, draw=gray!50] (5.5,0) circle (4ex);
		\draw [style=ThickLine] (0) to (1);
		\draw [style=ArrowLineRight] (5.center) to (6.center);
		\draw [style=ThickLine] (7) to (11.center);
		\draw [style=ThickLine] (12.center) to (8);
		\draw [style=RedLine, bend right=45] (11.center) to (14.center);
		\draw [style=RedLine, bend right=45] (14.center) to (12.center);
		\draw [style=PurpleLine, bend left=45] (11.center) to (13.center);
		\draw [style=PurpleLine, bend left=45] (13.center) to (12.center);
	\end{pgfonlayer}
\end{tikzpicture}}
\ee
In more detail, we now have trivalent junctions at the left and righthand sides, which we refer to as $\mathcal{J}_d$ and its orientation reversed counterpart $\overline{\mathcal{J}}_d$. These junctions fuse $\mathcal{S}_{d+1}(g)$ with $\mathcal{B}_{d+1}$ and $\mathcal{T}_{d+1}$. As explained in \cite{Baume:2023kkf}, these junction theories are typically non-topological, but support free fields which serve to match the boundary conditions of other symmetry theories / SymTFTs. In the explicit examples we consider later, it will also turn out that these junction theories are trivial. As such, we shall not dwell on them further.

Now, with the decompression (\ref{eq:junctionlangles}) in place, we can then proceed to compress back the finite segments supporting $\mathcal{S}_{d+1}(g)$ which separate $\mathcal{B}_d$ and $\mathcal{T}_d$ from their respective junctions:
\be
\scalebox{0.9}{
\begin{tikzpicture}
	\begin{pgfonlayer}{nodelayer}
		\node [style=none] (0) at (-7.75, 0.5) {$\mathcal{S}_{d+1}(g)$};
		\node [style=SmallCircle] (1) at (-9, 0) {};
		\node [style=SmallCircleRed] (2) at (-2, 0) {};
		\node [style=none] (3) at (-2, 0.5) {$\mathcal{T}_d$};
		\node [style=none] (4) at (-9, 0.5) {$\mathcal{B}_d$};
		\node [style=none] (5) at (-6.25, 0) {};
		\node [style=none] (6) at (-4.75, 0) {};
		\node [style=none] (7) at (-5.5, 0.75) {};
		\node [style=none] (8) at (-5.5, -0.75) {};
		\node [style=none] (9) at (-5.5, -1.25) {$\mathcal{T}_{d+1}$};
		\node [style=none] (10) at (-3.25, 0.5) {$\mathcal{S}_{d+1}(g)$};
		\node [style=none] (11) at (-5.5, 1.25) {$\mathcal{B}_{d+1}$};
		\node [style=none] (12) at (-6.75, -0.45) {$\overline{\mathcal{J}}_d$};
		\node [style=none] (13) at (-4.25, -0.5) {$\mathcal{J}_d$};
		\node [style=none] (14) at (-5.5, 0) {$\mathcal{S}_{d+2}$};
		\node [style=none] (15) at (-0.5, 0) {};
		\node [style=none] (16) at (0.5, 0) {};
		\node [style=SmallCircle] (18) at (2.25, 0) {};
		\node [style=SmallCircleRed] (19) at (5.25, 0) {};
		\node [style=none] (22) at (3, 0.75) {};
		\node [style=none] (23) at (3, -0.75) {};
		\node [style=none] (24) at (4.5, -0.75) {};
		\node [style=none] (25) at (4.5, 0.75) {};
		\node [style=none] (28) at (3.75, 0) {$\mathcal{S}_{d+2}$};
		\node [style=none] (29) at (3.75, -1.5) {$\mathcal{T}_{d+1}$};
		\node [style=none] (30) at (5.625, 0.375) {$\mathcal{T}_d$};
		\node [style=none] (31) at (1.875, 0.375) {$\mathcal{B}_d$};
		\node [style=none] (32) at (3.75, 1.5) {$\mathcal{B}_{d+1}$};
	\end{pgfonlayer}
	\begin{pgfonlayer}{edgelayer}
		\draw[fill=gray!50, draw=gray!50] (-5.5,0) circle (4ex);
		\draw[fill=gray!50, draw=gray!50] (3.75,0) circle (6.04ex);
		\filldraw[fill=gray!50, draw=gray!50]  (2.25, 0) -- (3, 0.75) -- (3, -0.75)  -- cycle;
		\filldraw[fill=gray!50, draw=gray!50]  (5.25, 0) -- (4.5, 0.75) -- (4.5, -0.75) -- cycle;
		\draw [style=ThickLine] (1) to (5.center);
		\draw [style=ThickLine] (6.center) to (2);
		\draw [style=RedLine, bend right=45] (5.center) to (8.center);
		\draw [style=RedLine, bend right=45] (8.center) to (6.center);
		\draw [style=PurpleLine, bend left=45] (5.center) to (7.center);
		\draw [style=PurpleLine, bend left=45] (7.center) to (6.center);
		\draw [style=ArrowLineRight] (15.center) to (16.center);
		\draw [style=RedLine] (24.center) to (19);
		\draw [style=RedLine, bend right=45, looseness=1.25] (23.center) to (24.center);
		\draw [style=RedLine] (23.center) to (18);
		\draw [style=PurpleLine] (18) to (22.center);
		\draw [style=PurpleLine, bend left=45, looseness=1.25] (22.center) to (25.center);
		\draw [style=PurpleLine] (25.center) to (19);
	\end{pgfonlayer}
\end{tikzpicture}}
\ee
Here we have used the triviality of the junctions in identifying the corners with $\mathcal{T}_d$ and $\mathcal{B}_d$.


There are again some different choices we could have made in the treatment of these corner and edge mode theories. For example, once we are dealing with distinct relative theories, we might instead have a decompression to:
\be\label{eq:primed2}
\scalebox{0.9}{
\begin{tikzpicture}
	\begin{pgfonlayer}{nodelayer}
		\node [style=SmallCircle] (18) at (2.25, 0) {};
		\node [style=SmallCircleRed] (19) at (5.25, 0) {};
		\node [style=none] (22) at (3, 0.75) {};
		\node [style=none] (23) at (3, -0.75) {};
		\node [style=none] (24) at (4.5, -0.75) {};
		\node [style=none] (25) at (4.5, 0.75) {};
		\node [style=none] (28) at (3.75, 0) {$\mathcal{S}_{d+2}^{\prime}$};
		\node [style=none] (29) at (3.75, -1.5) {$\mathcal{T}_{d+1}^{\:\!\prime}$};
		\node [style=none] (30) at (5.625, 0.375) {$\mathcal{T}_d^{\:\!\prime}$};
		\node [style=none] (31) at (1.875, 0.375) {$\mathcal{B}_d^{\:\!\prime}$};
		\node [style=none] (32) at (3.75, 1.5) {$\mathcal{B}_{d+1}^{\:\!\prime}$};
	\end{pgfonlayer}
	\begin{pgfonlayer}{edgelayer}
		\draw[fill=gray!50, draw=gray!50] (3.75,0) circle (6.04ex);
		\filldraw[fill=gray!50, draw=gray!50]  (2.25, 0) -- (3, 0.75) -- (3, -0.75)  -- cycle;
		\filldraw[fill=gray!50, draw=gray!50]  (5.25, 0) -- (4.5, 0.75) -- (4.5, -0.75) -- cycle;
		\draw [style=RedLine] (24.center) to (19);
		\draw [style=RedLine, bend right=45, looseness=1.25] (23.center) to (24.center);
		\draw [style=RedLine] (23.center) to (18);
		\draw [style=PurpleLine] (18) to (22.center);
		\draw [style=PurpleLine, bend left=45, looseness=1.25] (22.center) to (25.center);
		\draw [style=PurpleLine] (25.center) to (19);
	\end{pgfonlayer}
\end{tikzpicture}}
\ee
Namely, we redistribute the locations of some of the gapped and gapless degrees of freedom for the different systems. This is primarily a matter of perspective, and hinges on whether we are interested in the symmetry theory specified by QFT$_d$ directly, or whether we instead want to view the corner theory $\mathcal{T}_{d}^{\:\!\prime}$ as a defect\,/\,corner mode in some bigger bulk system. Equivalently, in the latter perspective the bigger bulk system $\mathcal{T}_{d+1}^{\:\!\prime}$ is specified as initial data, and in asking how its symmetries relate to the defect insertion $\mathcal{T}_d^{\:\!\prime}$ we should consider decompressions which manifest $\mathcal{T}_{d+1}^{\:\!\prime}$ as an edgemode, although other decompressions may be available from a purely $d$-dimensional perspective.

Further, observe that in the treatment where $\mathcal{T}_d^{\:\!\prime}$ is considered as a defect theory (or end of the world theory) in a given ambient bulk  $\mathcal{T}_{d+1}^{\:\!\prime}$, one might wish to view $\mathcal{B}_{d+1}^{\:\!\prime}$ as a symmetry theory $\mathcal{S}^{\prime}_{d+1}$ (in the sense of line (\ref{eq:primed})) which carries only ``partial information'' of the original symmetry theory $\mathcal{S}_{d+1}$.

Geometrically we will find a collection of decompressions satisfying
\be\label{eq:NiceProperty}
\mathcal{T}_d'=\mathcal{T}_d\,,
\ee
however with distinct data in dimensions $(d+1)$ and $(d+2)$ as in line \eqref{eq:primed2}. As such, some of these distinctions will not play much of a role. Finally, observe that we can in principle keep decompressing the bulk symmetry theory until we eventually reach a fully gapped bulk. In the case just discussed, we have essentially assumed that this procedure terminates after two steps of decompression, and we have handpicked our string examples such that this will be the case.

\subsection{Boundary Conditions}
\label{ssec:BCs}

Having introduced a bulk theory $\mathcal{S}_{d+2}$, we now turn to a discussion of boundary conditions. We will need to deal with the boundary conditions for bulk fields near $\mathcal{B}_{d+1}$, $\mathcal{T}_{d+1}$, as well as the further specialization to the corners $\mathcal{T}_d, \mathcal{B}_d$. We will reserve the main analysis of the corners to cases with a top down construction, as there $\mathcal{B}_d$ manifestly decompresses to the tuples $\mathbb{B}$ or $\widetilde{\mathbb{B}}$ which we will characterize instead of $ \mathcal{B}_d$ (see figure \ref{fig:Summary}). For ease of exposition, we specialize to the case where $\mathcal{B}_{d+1}$ is gapped and $\mathcal{T}_{d+1}$ is gapless.\footnote{One might be tempted to always split up $\mathcal{B}_{d+1}$ and $\mathcal{T}_{d+1}$ in this way. In the context of string constructions, however, it sometimes happens that free $\mathrm{U}(1)$ factors also geometrically localize near $\mathcal{B}_{d+1}$. That being said, the considerations we present here naturally extend to this more general situation as well.}

\begin{figure}
 \centering
 \scalebox{0.8}{
\begin{tikzpicture}
	\begin{pgfonlayer}{nodelayer}
		\node [style=SmallCircle] (0) at (-5, 0) {};
		\node [style=SmallCircleRed] (1) at (-2, 0) {};
		\node [style=none] (2) at (-4.25, 0.75) {};
		\node [style=none] (3) at (-4.25, -0.75) {};
		\node [style=none] (4) at (-2.75, -0.75) {};
		\node [style=none] (5) at (-2.75, 0.75) {};
		\node [style=none] (6) at (-3.5, 0) {$\mathcal{S}_{d+2}$};
		\node [style=none] (7) at (-3.5, -1.5) {$\mathcal{T}_{d+1}$};
		\node [style=none] (8) at (-1.5, 0.5) {$\mathcal{T}_d$};
		\node [style=none] (9) at (-5.5, 0.5) {$\mathcal{B}_d$};
		\node [style=none] (10) at (-3.5, 1.5) {$\mathcal{B}_{d+1}$};
		\node [style=SmallCircle] (11) at (2, 0) {};
		\node [style=SmallCircleRed] (12) at (5, 0) {};
		\node [style=none] (13) at (2.75, 0.75) {};
		\node [style=none] (14) at (2.75, -0.75) {};
		\node [style=none] (15) at (4.25, -0.75) {};
		\node [style=none] (16) at (4.25, 0.75) {};
		\node [style=none] (17) at (3.5, 0) {$Q$};
		\node [style=none] (18) at (3.5, -1.5) {$\partial_{(\mathcal{T}_{d+1})}Q$};
		\node [style=none] (21) at (3.5, 1.5) {$\partial_{(\mathcal{B}_{d+1})}Q$};
		\node [style=none] (22) at (1.5, 0.5) {$\partial^2_{(\mathcal{B}_{d})} Q$};
		\node [style=none] (23) at (-3.5, -2.5) {(i)};
		\node [style=none] (24) at (3.5, -2.5) {(ii)};
		\node [style=none] (25) at (0, -3) {};
		\node [style=none] (26) at (4.5, 0) {};
		\node [style=none] (27) at (5.5, 0) {};
		\node [style=none] (28) at (5, 0.5) {};
		\node [style=none] (29) at (5, -0.5) {};
		\node [style=none] (30) at (5.5, 0.25) {};
		\node [style=none] (31) at (5.5, -0.25) {};
		\node [style=none] (32) at (8, 1) {};
		\node [style=none] (33) at (8, -1) {};
		\node [style=none] (34) at (9.75, -0.25) {};
		\node [style=none] (35) at (9.75, 0.875) {};
		\node [style=none] (36) at (8.625, -0.25) {};
		\node [style=none] (37) at (9.75, 1.25) {};
		\node [style=none] (38) at (8.25, -0.25) {};
		\node [style=none] (39) at (10.25, -0.75) {$\partial^2_{(\mathcal{T}_{d})} Q$};
		\node [style=none] (40) at (7.75, 0) {};
		\node [style=none] (41) at (11.25, 0) {};
		\node [style=none] (42) at (9.5, 1.75) {};
		\node [style=none] (43) at (9.5, -1.75) {};
		\node [style=SmallCircleRed] (44) at (9.75, -0.25) {};
		\node [style=none] (45) at (9, -0.75) {$\mathbb{R}_{\:\!r \:\!\geq \:\!0}$};
		\node [style=none] (46) at (10.55, 0.375) {$\mathbb{R}_{\:\!x_\perp \geq \:\!0}$};
		\node [style=none] (47) at (9.5, -2.5) {(iii)};
		\node [style=none] (48) at (12.5, 0) {};
	\end{pgfonlayer}
	\begin{pgfonlayer}{edgelayer}
		\draw[fill=gray!50, draw=gray!50] (-3.5,0) circle (6.04ex);
		\filldraw[fill=gray!50, draw=gray!50]  (-2, 0) -- (-2.75, 0.75) -- (-2.75, -0.75)  -- cycle;
		\filldraw[fill=gray!50, draw=gray!50]  (-5, 0) -- (-4.25, 0.75) -- (-4.25, -0.75) -- cycle;
		\draw[fill=gray!50, draw=gray!50] (3.5,0) circle (6.04ex);
		\filldraw[fill=gray!50, draw=gray!50]  (2, 0) -- (2.75, 0.75) -- (2.75, -0.75)  -- cycle;
		\filldraw[fill=gray!50, draw=gray!50]  (5, 0) -- (4.25, 0.75) -- (4.25, -0.75) -- cycle;
		\filldraw[fill=gray!50, draw=gray!50]  (9.75, -0.25) -- (9.75, 0.875) -- (8.625, 0.875) --  (8.625, -0.25) -- cycle;
		\draw [style=RedLine] (4.center) to (1);
		\draw [style=RedLine, bend right=45, looseness=1.25] (3.center) to (4.center);
		\draw [style=RedLine] (3.center) to (0);
		\draw [style=PurpleLine] (0) to (2.center);
		\draw [style=PurpleLine, bend left=45, looseness=1.25] (2.center) to (5.center);
		\draw [style=PurpleLine] (5.center) to (1);
		\draw [style=RedLine] (15.center) to (12);
		\draw [style=RedLine, bend right=45, looseness=1.25] (14.center) to (15.center);
		\draw [style=RedLine] (14.center) to (11);
		\draw [style=PurpleLine] (11) to (13.center);
		\draw [style=PurpleLine, bend left=45, looseness=1.25] (13.center) to (16.center);
		\draw [style=PurpleLine] (16.center) to (12);
		\draw [style=DottedLine, bend left=45] (26.center) to (28.center);
		\draw [style=DottedLine, bend left=45] (28.center) to (27.center);
		\draw [style=DottedLine, bend left=45] (27.center) to (29.center);
		\draw [style=DottedLine, bend right=315] (29.center) to (26.center);
		\draw [style=DottedLine] (30.center) to (32.center);
		\draw [style=DottedLine] (33.center) to (31.center);
		\draw [style=RedLine] (34.center) to (36.center);
		\draw [style=DottedRed] (36.center) to (38.center);
		\draw [style=PurpleLine] (34.center) to (35.center);
		\draw [style=DottedPurple] (35.center) to (37.center);
		\draw [style=DottedLine, bend left=45] (40.center) to (42.center);
		\draw [style=DottedLine, bend left=45] (42.center) to (41.center);
		\draw [style=DottedLine, bend left=45] (41.center) to (43.center);
		\draw [style=DottedLine, bend right=315] (43.center) to (40.center);
	\end{pgfonlayer}
\end{tikzpicture}

}
\caption{Notational conventions and parametrization of the corner patch centered on $\mathcal{T}_d$.}
\label{fig:ConventionsLabelling}
\end{figure}
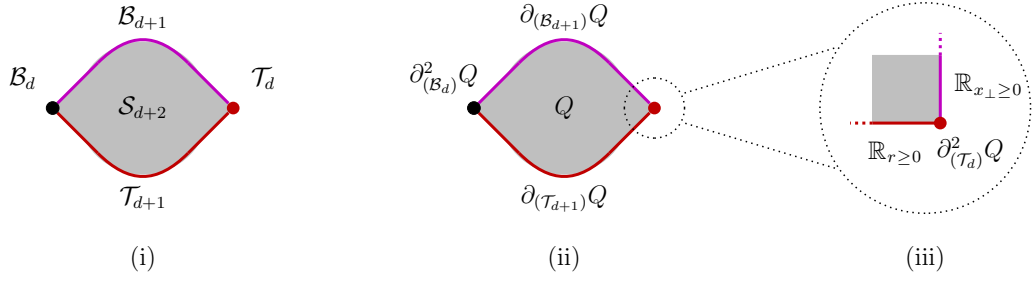

To set conventions (see figure  \ref{fig:ConventionsLabelling} for a summary),
let $Q$ be the $(d+2)$-dimensional manifold with corners supporting $\mathcal{S}_{d+2}$. It has two boundary components
\be
\partial Q=\partial_{(\mathcal{B}_{d+1})}Q\cup\partial_{(\mathcal{T}_{d+1})}Q
\ee
 where $\partial_{(\mathcal{B}_{d+1})}Q,\partial_{(\mathcal{T}_{d+1})}Q$ support $\mathcal{B}_{d+1}, \mathcal{T}_{d+1}$ respectively. We abbreviate $ \partial_{(\mathcal{B}_{d+1})}Q\equiv \partial_{(\mathcal{B}_{d+1})}$ and $\partial_{(\mathcal{T}_{d+1})}Q=\partial_{(\mathcal{T}_{d+1})}$ and denote the restrictions to these as $|_{\mathcal{B}_{d+1}}$ and $|_{\mathcal{T}_{d+1}}$ respectively.
  The boundaries $\partial_{(\mathcal{B}_{d+1})}Q,\partial_{(\mathcal{T}_{d+1})}Q$ themselves have boundaries that are oppositely oriented
 \be\label{eq:Opporient}
 \partial(\partial_{(\mathcal{B}_{d+1})}Q)=-\partial(\partial_{(\mathcal{T}_{d+1})}Q)\,,
 \ee
 and support $\mathcal{B}_{d}$, $\mathcal{T}_{d}$. We denote their supports as $\partial^2_{(\mathcal{B}_{d})} Q$ as $\partial^2_{(\mathcal{T}_{d})} Q$ where we realize the notation `$\partial^2$' to manifestly and compactly emphasize the notion of ``boundaries of boundaries'' (which is non-trivial when we have a space with corners).

Next, consider the local corner patch centered on $\mathcal{T}_d$ and modeled on $\R_{\geq 0}\times \R_{\geq 0}\times M_d$ with coordinates $r, x_\perp,y$. The edge $\partial_{(\mathcal{B}_{d+1})}Q$ is parametrized by $x_\perp,y$ and sits at $r=0$,  the edge $\partial_{(\mathcal{T}_{d+1})}Q$ is parametrized by $r,y$ and sits at $x_\perp=0$, and $y$ denotes collectively the coordinates of $M_d$ which is a copy of spacetime (see subfigure (iii) of figure \ref{fig:ConventionsLabelling}).

\subsubsection{Boundary Conditions: $\mathcal{S}_{d+2} |_{ \mathcal{T}_{d+1}}$}

Let us now turn to the boundary conditions involving restriction of the bulk modes of $\mathcal{S}_{d+2}$ onto $\mathcal{T}_{d+1}$.
Recall that in our simplified exposition, the gapped modes are on $\mathcal{B}_{d+1}$ and the gapless modes are on $\mathcal{T}_{d+1}$.
The relative theory $\mathcal{T}_{d+1}$ is associated to the higher-dimensional QFT $\mathcal{S}_{d+1}(g)$ (and its subsequent limit to $\mathcal{S}_{d+1}$, as obtained from the decompression:
\be\label{eq:BC1}
\scalebox{0.9}{
\begin{tikzpicture}
	\begin{pgfonlayer}{nodelayer}
		\node [style=none] (0) at (0, 0) {};
		\node [style=none] (1) at (1, 0) {};
		\node [style=none] (5) at (-2, 0.5) {$\mathcal{S}_{d+1}(g)$};
		\node [style=none] (9) at (1, 1.25) {};
		\node [style=none] (10) at (-2.75, 0) {};
		\node [style=none] (11) at (-1.25, 0) {};
		\node [style=none] (12) at (-1, 0) {};
		\node [style=none] (13) at (-3, 0) {};
		\node [style=none] (14) at (2.25, 0.5) {};
		\node [style=none] (15) at (3.75, 0.5) {};
		\node [style=none] (16) at (4, 0.5) {};
		\node [style=none] (17) at (2, 0.5) {};
		\node [style=none] (18) at (2.25, -0.5) {};
		\node [style=none] (19) at (3.75, -0.5) {};
		\node [style=none] (20) at (4, -0.5) {};
		\node [style=none] (21) at (2, -0.5) {};
		\node [style=none] (22) at (3, -1) {$\mathcal{T}_{d+1}$};
		\node [style=none] (23) at (3, 1) {$\mathcal{B}_{d+1}$};
		\node [style=none] (24) at (3, 0) {$\mathcal{S}_{d+2}$};
	\end{pgfonlayer}
	\begin{pgfonlayer}{edgelayer}
		\filldraw[fill=gray!50, draw=gray!50]  (2.25, 0.5) -- (2.25, -0.5) -- (3.75, -0.5) -- (3.75, 0.5) -- cycle;
		\draw [style=ArrowLineRight] (0.center) to (1.center);
		\draw [style=ThickLine] (10.center) to (11.center);
		\draw [style=DottedLine] (11.center) to (12.center);
		\draw [style=DottedLine] (10.center) to (13.center);
		\draw [style=PurpleLine] (14.center) to (15.center);
		\draw [style=RedLine] (18.center) to (19.center);
		\draw [style=DottedRed] (20.center) to (19.center);
		\draw [style=DottedRed] (21.center) to (18.center);
		\draw [style=DottedPurple] (16.center) to (15.center);
		\draw [style=DottedPurple] (17.center) to (14.center);
	\end{pgfonlayer}
\end{tikzpicture}}
\ee
As such, we impose Neumann boundary conditions of the bulk $\mathcal{S}_{d+2}$ fields near $\mathcal{S}_{d+2} |_{ \mathcal{T}_{d+1}}$. Given the additional gapless degrees of freedom on $\mathcal{T}_{d+1}$ this is also known as enriched Neumann boundary conditions in the literature \cite{Kaidi:2022cpf}. In the string construction we will consider, $\mathcal{T}_{d+1}$ is determined by a collection (possibly just one) of (KK-reduced) super-Yang-Mills theories.

\subsubsection{Boundary Conditions: $\mathcal{S}_{d+2}|_{ \mathcal{B}_{d+1}}$}\label{section:BC}

We now turn to boundary conditions for $\mathcal{S}_{d+2}$ near the gapped edge mode $\mathcal{B}_{d+1}$. We ask how bulk operators push onto $\mathcal{B}_{d+1}$, the top edge in \eqref{eq:BC1}.

To set notation, we denote the electric bulk fields by $a^{(i)}$ and their magnetic duals by $\widetilde{a}_{(i)}$, where $i$ denotes an indexing of our fields. Denote similarly the electric fields of $\mathcal{B}_{d+1}$ by $b^{(j)}$.
We also allow for a non-trivial background value in this boundary which we write as ${B}^{(j)}$.
A  general comment here is that while the fields of the theory $\mathcal{S}_{d+2}$ must admit a canonical electric-magnetic pairing, in the case of the relative theory $\mathcal{B}_{d+1}$ this need not hold.
Restricting any bulk field to the boundary
 supporting  $\mathcal{B}_{d+1}$ yields the boundary conditions:
\be\ba \label{eq:generaldiscretegluing}
\alpha^s_i a_n^{(i)} \Big|_{\mathcal{B}_{d+1} }&=\beta^s_j b^{(j)}_{\;\!n}+\gamma^s_j B^{(j)}_{n}&&\quad\qquad \textnormal{(electric)}\\[0.5em]
 \lb  \partial/\partial r  \,\lrcorner\, \lambda^s_i a_{n+1}^{(i)} \rb \Big|_{\mathcal{B}_{d+1} }&=\mu^s_j b^{(j)}_{\;\!n}+\nu^s_jB^{(j)}_{n}&&\quad\qquad \textnormal{(magnetic)}
\ea \ee
where here we have allowed for a non-trivial background value $B$ on the boundary, as a generalization of the Dirichlet boundary conditions one often imposes in the standard SymTFT formalism. In the above, the subscripts $n$ and $n+1$ on the fields denote the cocycle degree, and
$\alpha, \beta, \gamma$ and $\lambda,\mu,\nu$ are integer matrices. The notation $\partial/\partial r\,\lrcorner$ denotes the interior product with the unit vector field $\partial/\partial r$ associated to the coordinate $r$ orthogonal to the boundary. Both the bulk and boundary fields can be torsional and, in these cases, the matrices (and equations) are only defined with entries in $\Z_N$ for some integers $N$ depending on the labels $i,j,s,n$ above.

The boundary conditions \eqref{eq:generaldiscretegluing} give (affine) linear mappings:
\be\label{eq:BCmaps} \ba
E_n\,:&\quad \{a_n^{(i)}\}&&\rightarrow~\quad\{b_n^{(i)}\}\\
M_{n}\,:&\quad \{a_{n+1}^{(i)}\}&&\rightarrow~\quad\{b_n^{(i)}\}\,,
\ea \ee
and compatibility of the boundary conditions amounts to $E_n$ and $M_n$ respectively descending to well-defined mappings with codomain $\textnormal{coker}\,M_n$ and $\textnormal{coker}\,E_n$ respectively so their images do not overlap. Note further, that the cokernel $\{b_n^{(i)}\}/(\textnormal{Im}\,E_n\oplus \textnormal{Im}\,M_{n})$ need not be trivial; it characterizes the degrees of freedom of $\mathcal{B}_{d+1}$ of the boundary system which are not fixed by boundary conditions.

Finally, we note that boundary conditions for $\mathcal{S}_{d+2}$ need to be complete. This amounts to imposing a maximal set of compatible mixed Dirichlet and Neumann boundary conditions for $a_n^{(i)}$ and their magnetic duals $\widetilde{a}_m^{(i)}$ (with $n+m+1 = d+2$ for discrete symmetries). Dirichlet boundary conditions for $a_n^{(i)}$ are dually described and equivalent to Neumann boundary conditions for $\widetilde{a}_m^{(i)}$ and, conversely, Neumann boundary conditions dualize to Dirichlet boundary conditions. A compatible maximal set of boundary conditions can therefore be specified as mixed Neumann / Dirichlet boundary conditions of the pair $a_n^{(i)},\widetilde{a}_m^{(i)}$. As such the boundary conditions \eqref{eq:generaldiscretegluing} are equivalently given as\footnote{These considtions are written for discrete valued fields, if we consider abelian continous valued fields then $n+m+2=d+2$ and the second line would instead be $\lambda^s_{i\;} *_{d+1}d\widetilde{a}_{m}^{(i)}\Big|_{\mathcal{B}_{d+1} }=\mu^s_j b^{(j)}_{\;\!n}+\nu^s_j B^{(j)}_{n}$.}
\be\ba \label{eq:generaldiscretegluing2}
\alpha^s_i a_n^{(i)}  \Big|_{\mathcal{B}_{d+1} }&=\beta^s_j b^{(j)}_{\;\!n}+\gamma^s_j B^{(j)}_{n}\\[0.5em]
\lambda^s_{i\;} *_{d+1}\widetilde{a}_{m}^{(i)}\Big|_{\mathcal{B}_{d+1} }&=\mu^s_j b^{(j)}_{\;\!n}+\nu^s_j B^{(j)}_{n}\,,
\ea \ee
The second line of the boundary conditions \eqref{eq:generaldiscretegluing}
expresses the boundary conditions for $\widetilde{a}_m^{(i)}$ in terms of $a_n^{(i)}$ and $*_{d+1}$ is the Hodge star along $\mathcal{B}_{d+1}$. We will work with the boundary conditions  \eqref{eq:generaldiscretegluing} as these will be manifestly read from geometry in our top down examples.

Note again that we have not introduced fields $\widetilde{b}$ conjugate to $b$. This is because the fields $b$ do not characterize the degrees of freedom of $\mathcal{B}_{d+1}$. Rather, the $\mathcal{B}_{d+1}$ degrees of freedom are characterized by the cokernel $\{b_n^{(i)}\}/(\textnormal{Im}\,E_n\oplus \textnormal{Im}\,M_{n})$ and we will encounter examples where we can introduce conjugate variables corresponding to the dual of such an equivalence class. In the above parametrization these dual modes are subsumed into the $\{b_n^{(i)}\}$. Equivalently, the fields $\{a_n^{(i)}\},\{\widetilde{a}_m^{(i)}\},\{b_n^{(i)}\}$ over-parametrize the degrees of freedom of the system, and we only identify electromagnetic pairs once identifications between modes have been realized.

With this we can also clarify some aspects of our mixed Dirichlet\,/\,Neumann boundary conditions. In considering the pair $\mathcal{S}_{d+2}$ and ${\mathcal{B}_{d+1}}$, the path integrated fields include $\{b_n^{(i)} \}$ modulo the constraints of line \eqref{eq:generaldiscretegluing}. For example, turning all backgrounds off, bulk modes not in the kernel of the maps $E_n,M_n$ of \eqref{eq:BCmaps} remain fluctuating and as such $\mathcal{B}_{d+1}$ results in Neumann boundary conditions for these. The bulk and boundary fields are identified but the resulting mode remains fluctuating. True Dirichlet boundary conditions are imposed only for bulk modes in the kernel of $E_n,M_n$ (again with backgrounds turned off). This clarifies the sense in which line \eqref{eq:generaldiscretegluing} formulates mixed Dirichlet\,/\,Neumann boundary conditions.

\subsubsection{Corners}

In geometric string constructions of $d$-dimensional QFTs the $d$-dimensional degrees of freedom are usually localized in codimension larger than one. This will allows us to determine a bulk symmetry theory $\mathcal{S}_{d+2}$ such that $\mathcal{T}_d$ simultaneously specifies boundary mode for $\mathcal{S}_{d+1}$ and a corner mode for $\mathcal{S}_{d+2}$. In general, the initial bulk-boundary systems given by $\mathcal{S}_{d+1},\mathcal{T}_d$ unfold to corners supporting some different theory $\mathcal{T}_d'\neq \mathcal{T}_d$ and the geometric prescription we give via string theory should be understood as a toolset to avoid this general case. In the therefore special setups we consider the corner mode is fully specified by the boundary mode $\mathcal{S}_{d+1}|_{\mathcal{T}_d}$, i.e., from this perspective by data in one dimension higher.
An interesting issue concerns the structure of anomaly inflow from our bulk system(s) to the boundary\,/\,corner. There is clearly an inflow where we descend by one dimension at a time:
\begin{equation}
(d+2) \rightarrow (d+1) \rightarrow d\,.
\end{equation}
Based on the way we have constructed the bulk theory, observe also that there is no inflow directly from the $(d+2)$-dimensional bulk to the $d$-dimensional corner; instead, we can always factor through the intermediate edges of dimension $d+1$. Any inflow from dimension $(d+2)$ descending to dimension $(d+1)$ then cancels when pushed further to dimension $d$ by \eqref{eq:Opporient}. For this reason the main approach to characterizing corners we will take is to view them simply as interfaces between edges.

\subsection{Defect and Symmetry Operators}\label{ssec:defectandsymops}

\begin{figure}
\centering
\scalebox{0.8}{
\begin{tikzpicture}
	\begin{pgfonlayer}{nodelayer}
		\node [style=SmallCircle] (0) at (2, 6) {};
		\node [style=SmallCircleRed] (1) at (5, 6) {};
		\node [style=none] (2) at (2.75, 6.75) {};
		\node [style=none] (3) at (2.75, 5.25) {};
		\node [style=none] (4) at (4.25, 5.25) {};
		\node [style=none] (5) at (4.25, 6.75) {};
		\node [style=none] (7) at (3.5, 4.5) {$\mathcal{T}_{d+1}$};
		\node [style=none] (8) at (5.5, 6.5) {$\mathcal{T}_d$};
		\node [style=none] (9) at (1.5, 6.5) {$\mathcal{B}_d$};
		\node [style=none] (10) at (3.5, 7.5) {$\mathcal{B}_{d+1}$};
		\node [style=SmallCircle] (11) at (2, 2) {};
		\node [style=SmallCircleRed] (12) at (5, 2) {};
		\node [style=none] (13) at (2.75, 2.75) {};
		\node [style=none] (14) at (2.75, 1.25) {};
		\node [style=none] (15) at (4.25, 1.25) {};
		\node [style=none] (16) at (4.25, 2.75) {};
		\node [style=none] (18) at (3.5, 0.5) {$\mathcal{T}_{d+1}$};
		\node [style=none] (19) at (5.5, 2.5) {$\mathcal{T}_d$};
		\node [style=none] (20) at (1.5, 2.5) {$\mathcal{B}_d$};
		\node [style=none] (21) at (3.5, 3.5) {$\mathcal{B}_{d+1}$};
		\node [style=SmallCircle] (22) at (2, -2) {};
		\node [style=SmallCircleRed] (23) at (5, -2) {};
		\node [style=none] (24) at (2.75, -1.25) {};
		\node [style=none] (25) at (2.75, -2.75) {};
		\node [style=none] (26) at (4.25, -2.75) {};
		\node [style=none] (27) at (4.25, -1.25) {};
		\node [style=none] (29) at (3.5, -3.5) {$\mathcal{T}_{d+1}$};
		\node [style=none] (30) at (5.5, -1.5) {$\mathcal{T}_d$};
		\node [style=none] (31) at (1.5, -1.5) {$\mathcal{B}_d$};
		\node [style=none] (32) at (3.5, -0.5) {$\mathcal{B}_{d+1}$};
		\node [style=SmallCircle] (33) at (2, -6) {};
		\node [style=SmallCircleRed] (34) at (5, -6) {};
		\node [style=none] (35) at (2.75, -5.25) {};
		\node [style=none] (36) at (2.75, -6.75) {};
		\node [style=none] (37) at (4.25, -6.75) {};
		\node [style=none] (38) at (4.25, -5.25) {};
		\node [style=none] (40) at (3.5, -7.5) {$\mathcal{T}_{d+1}$};
		\node [style=none] (41) at (5.5, -5.5) {$\mathcal{T}_d$};
		\node [style=none] (42) at (1.5, -5.5) {$\mathcal{B}_d$};
		\node [style=none] (43) at (3.5, -4.5) {$\mathcal{B}_{d+1}$};
		\node [style=SmallCircle] (44) at (-6, 0) {};
		\node [style=SmallCircleRed] (45) at (-3, 0) {};
		\node [style=none] (46) at (-3, 0.5) {$\mathcal{T}_d$};
		\node [style=none] (47) at (-6, 0.5) {$\mathcal{B}_d$};
		\node [style=none] (48) at (-1.5, 0.25) {};
		\node [style=none] (49) at (-1.5, -0.25) {};
		\node [style=none] (50) at (-1.5, -0.75) {};
		\node [style=none] (51) at (-1.5, 0.75) {};
		\node [style=none] (52) at (0.5, 4) {};
		\node [style=none] (53) at (0.5, 1.25) {};
		\node [style=none] (54) at (0.5, -1.25) {};
		\node [style=none] (55) at (0.5, -4) {};
		\node [style=none] (62) at (7, 6) {(i)};
		\node [style=none] (63) at (7, 2) {(ii)};
		\node [style=none] (64) at (7, -2) {(iii)};
		\node [style=none] (65) at (7, -6) {(iv)};
	\end{pgfonlayer}
	\begin{pgfonlayer}{edgelayer}
		\draw[fill=gray!50, draw=gray!50] (3.5,2) circle (6.04ex);
		\filldraw[fill=gray!50, draw=gray!50]  (2, 2) -- (2.75, 2.75) -- (2.75, 1.25)  -- cycle;
		\filldraw[fill=gray!50, draw=gray!50]  (5, 2) -- (4.25, 2.75) -- (4.25, 1.25) -- cycle;
		
		\draw[fill=gray!50, draw=gray!50] (3.5,-2) circle (6.04ex);
		\filldraw[fill=gray!50, draw=gray!50]  (2, -2) -- (2.75, -2.75) -- (2.75, -1.25)  -- cycle;
		\filldraw[fill=gray!50, draw=gray!50]  (5, -2) -- (4.25, -2.75) -- (4.25, -1.25) -- cycle;
		
		\draw[fill=gray!50, draw=gray!50] (3.5,6) circle (6.04ex);
		\filldraw[fill=gray!50, draw=gray!50]  (2, 6) -- (2.75, 6.75) -- (2.75, 5.25)  -- cycle;
		\filldraw[fill=gray!50, draw=gray!50]  (5, 6) -- (4.25, 6.75) -- (4.25, 5.25) -- cycle;
		
		\draw[fill=brown!50, draw=brown!50] (3.5,-6) circle (6.04ex);
		\filldraw[fill=brown!50, draw=brown!50]  (2, -6) -- (2.75, -6.75) -- (2.75, -5.25)  -- cycle;
		\filldraw[fill=brown!50, draw=brown!50]  (5, -6) -- (4.25, -6.75) -- (4.25, -5.25) -- cycle;
		
		\draw [style=RedLine] (4.center) to (1);
		\draw [style=RedLine, bend right=45, looseness=1.25] (3.center) to (4.center);
		\draw [style=RedLine] (3.center) to (0);
		\draw [style=PurpleLine] (0) to (2.center);
		\draw [style=PurpleLine, bend left=45, looseness=1.25] (2.center) to (5.center);
		\draw [style=PurpleLine] (5.center) to (1);
		\draw [style=RedLine] (15.center) to (12);
		\draw [style=RedLine, bend right=45, looseness=1.25] (14.center) to (15.center);
		\draw [style=RedLine] (14.center) to (11);
		\draw [style=PurpleLine] (11) to (13.center);
		\draw [style=PurpleLine, bend left=45, looseness=1.25] (13.center) to (16.center);
		\draw [style=PurpleLine] (16.center) to (12);
		\draw [style=RedLine] (26.center) to (23);
		\draw [style=RedLine, bend right=45, looseness=1.25] (25.center) to (26.center);
		\draw [style=RedLine] (25.center) to (22);
		\draw [style=PurpleLine] (22) to (24.center);
		\draw [style=PurpleLine, bend left=45, looseness=1.25] (24.center) to (27.center);
		\draw [style=PurpleLine] (27.center) to (23);
		\draw [style=RedLine] (37.center) to (34);
		\draw [style=RedLine, bend right=45, looseness=1.25] (36.center) to (37.center);
		\draw [style=RedLine] (36.center) to (33);
		\draw [style=PurpleLine] (33) to (35.center);
		\draw [style=PurpleLine, bend left=45, looseness=1.25] (35.center) to (38.center);
		\draw [style=PurpleLine] (38.center) to (34);
		\draw [style=ThickLine] (44) to (45);
		\draw [style=ArrowLineRight] (50.center) to (55.center);
		\draw [style=ArrowLineRight] (49.center) to (54.center);
		\draw [style=ArrowLineRight] (48.center) to (53.center);
		\draw [style=ArrowLineRight] (51.center) to (52.center);
		\draw [style=BrownLine, snake it] (44) to (45);
		\draw [style=BrownLine, snake it] (0) to (2.center);
		\draw [style=BrownLine, snake it, bend left=41.98625, looseness=0.9] (2.center) to (5.center);
		\draw [style=BrownLine, snake it] (5.center) to (1);
		\draw [style=BrownLine, snake it] (11) to (14.center);
		\draw [style=BrownLine, snake it, bend right=41.98625, looseness=0.9] (14.center) to (15.center);
		\draw [style=BrownLine, snake it] (15.center) to (12);
		\draw [style=BrownLine, snake it] (22) to (23);
	\end{pgfonlayer}
\end{tikzpicture}}
\caption{We sketch the possible fate of defect operators under decompression.}
\label{fig:fateofdefects}
\end{figure}
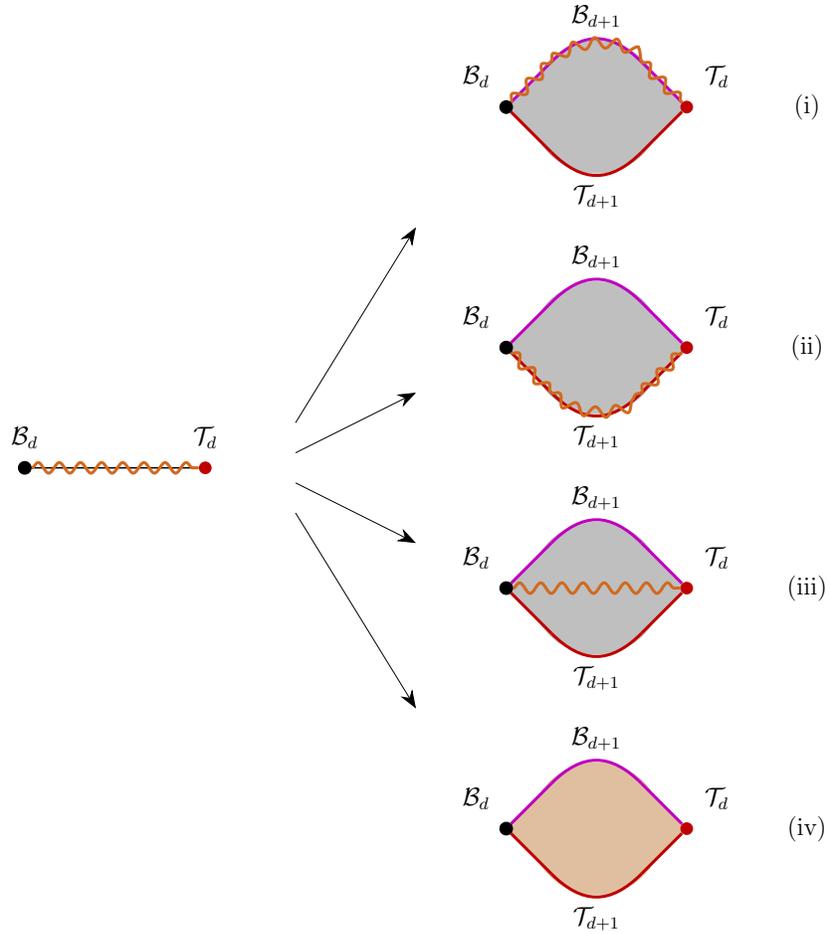

\begin{figure}
\centering
\scalebox{0.8}{
\begin{tikzpicture}
	\begin{pgfonlayer}{nodelayer}
		\node [style=SmallCircle] (0) at (2, 6) {};
		\node [style=SmallCircleRed] (1) at (5, 6) {};
		\node [style=none] (2) at (2.75, 6.75) {};
		\node [style=none] (3) at (2.75, 5.25) {};
		\node [style=none] (4) at (4.25, 5.25) {};
		\node [style=none] (5) at (4.25, 6.75) {};
		\node [style=none] (7) at (3.5, 4.5) {$\mathcal{T}_{d+1}$};
		\node [style=none] (8) at (5.5, 6.5) {$\mathcal{T}_d$};
		\node [style=none] (9) at (1.5, 6.5) {$\mathcal{B}_d$};
		\node [style=none] (10) at (3.5, 7.5) {$\mathcal{B}_{d+1}$};
		\node [style=SmallCircle] (11) at (2, 2) {};
		\node [style=SmallCircleRed] (12) at (5, 2) {};
		\node [style=none] (13) at (2.75, 2.75) {};
		\node [style=none] (14) at (2.75, 1.25) {};
		\node [style=none] (15) at (4.25, 1.25) {};
		\node [style=none] (16) at (4.25, 2.75) {};
		\node [style=none] (18) at (3.5, 0.5) {$\mathcal{T}_{d+1}$};
		\node [style=none] (19) at (5.5, 2.5) {$\mathcal{T}_d$};
		\node [style=none] (20) at (1.5, 2.5) {$\mathcal{B}_d$};
		\node [style=none] (21) at (3.5, 3.5) {$\mathcal{B}_{d+1}$};
		\node [style=SmallCircle] (22) at (2, -2) {};
		\node [style=SmallCircleRed] (23) at (5, -2) {};
		\node [style=none] (24) at (2.75, -1.25) {};
		\node [style=none] (25) at (2.75, -2.75) {};
		\node [style=none] (26) at (4.25, -2.75) {};
		\node [style=none] (27) at (4.25, -1.25) {};
		\node [style=none] (29) at (3.5, -3.5) {$\mathcal{T}_{d+1}$};
		\node [style=none] (30) at (5.5, -1.5) {$\mathcal{T}_d$};
		\node [style=none] (31) at (1.5, -1.5) {$\mathcal{B}_d$};
		\node [style=none] (32) at (3.5, -0.5) {$\mathcal{B}_{d+1}$};
		\node [style=SmallCircle] (33) at (2, -6) {};
		\node [style=SmallCircleRed] (34) at (5, -6) {};
		\node [style=none] (35) at (2.75, -5.25) {};
		\node [style=none] (36) at (2.75, -6.75) {};
		\node [style=none] (37) at (4.25, -6.75) {};
		\node [style=none] (38) at (4.25, -5.25) {};
		\node [style=none] (40) at (3.5, -7.5) {$\mathcal{T}_{d+1}$};
		\node [style=none] (41) at (5.5, -5.5) {$\mathcal{T}_d$};
		\node [style=none] (42) at (1.5, -5.5) {$\mathcal{B}_d$};
		\node [style=none] (43) at (3.5, -4.5) {$\mathcal{B}_{d+1}$};
		\node [style=SmallCircle] (44) at (-6, 0) {};
		\node [style=SmallCircleRed] (45) at (-3, 0) {};
		\node [style=none] (46) at (-3, 0.5) {$\mathcal{T}_d$};
		\node [style=none] (47) at (-6, 0.5) {$\mathcal{B}_d$};
		\node [style=none] (48) at (-1.5, 0.25) {};
		\node [style=none] (49) at (-1.5, -0.25) {};
		\node [style=none] (50) at (-1.5, -0.75) {};
		\node [style=none] (51) at (-1.5, 0.75) {};
		\node [style=none] (52) at (0.5, 4) {};
		\node [style=none] (53) at (0.5, 1.25) {};
		\node [style=none] (54) at (0.5, -1.25) {};
		\node [style=none] (55) at (0.5, -4) {};
		\node [style=SmallCircleBrown] (56) at (-4.5, 0) {};
		\node [style=SmallCircleBrown] (57) at (3.5, 7.13) {};
		\node [style=SmallCircleBrown] (58) at (3.5, 0.87) {};
		\node [style=SmallCircleBrown] (59) at (3.5, -2) {};
		\node [style=SmallCircleBrown] (60) at (3.5, -4.87) {};
		\node [style=SmallCircleBrown] (61) at (3.5, -7.13) {};
		\node [style=none] (62) at (7, 6) {(i)};
		\node [style=none] (63) at (7, 2) {(ii)};
		\node [style=none] (64) at (7, -2) {(iii)};
		\node [style=none] (65) at (7, -6) {(iv)};
	\end{pgfonlayer}
	\begin{pgfonlayer}{edgelayer}
		\draw[fill=gray!50, draw=gray!50] (3.5,2) circle (6.04ex);
		\filldraw[fill=gray!50, draw=gray!50]  (2, 2) -- (2.75, 2.75) -- (2.75, 1.25)  -- cycle;
		\filldraw[fill=gray!50, draw=gray!50]  (5, 2) -- (4.25, 2.75) -- (4.25, 1.25) -- cycle;
		
		\draw[fill=gray!50, draw=gray!50] (3.5,-2) circle (6.04ex);
		\filldraw[fill=gray!50, draw=gray!50]  (2, -2) -- (2.75, -2.75) -- (2.75, -1.25)  -- cycle;
		\filldraw[fill=gray!50, draw=gray!50]  (5, -2) -- (4.25, -2.75) -- (4.25, -1.25) -- cycle;
		
		\draw[fill=gray!50, draw=gray!50] (3.5,6) circle (6.04ex);
		\filldraw[fill=gray!50, draw=gray!50]  (2, 6) -- (2.75, 6.75) -- (2.75, 5.25)  -- cycle;
		\filldraw[fill=gray!50, draw=gray!50]  (5, 6) -- (4.25, 6.75) -- (4.25, 5.25) -- cycle;
		
		\draw[fill=gray!50, draw=gray!50] (3.5,-6) circle (6.04ex);
		\filldraw[fill=gray!50, draw=gray!50]  (2, -6) -- (2.75, -6.75) -- (2.75, -5.25)  -- cycle;
		\filldraw[fill=gray!50, draw=gray!50]  (5, -6) -- (4.25, -6.75) -- (4.25, -5.25) -- cycle;
		
		\draw [style=RedLine] (4.center) to (1);
		\draw [style=RedLine, bend right=45, looseness=1.25] (3.center) to (4.center);
		\draw [style=RedLine] (3.center) to (0);
		\draw [style=PurpleLine] (0) to (2.center);
		\draw [style=PurpleLine, bend left=45, looseness=1.25] (2.center) to (5.center);
		\draw [style=PurpleLine] (5.center) to (1);
		\draw [style=RedLine] (15.center) to (12);
		\draw [style=RedLine, bend right=45, looseness=1.25] (14.center) to (15.center);
		\draw [style=RedLine] (14.center) to (11);
		\draw [style=PurpleLine] (11) to (13.center);
		\draw [style=PurpleLine, bend left=45, looseness=1.25] (13.center) to (16.center);
		\draw [style=PurpleLine] (16.center) to (12);
		\draw [style=RedLine] (26.center) to (23);
		\draw [style=RedLine, bend right=45, looseness=1.25] (25.center) to (26.center);
		\draw [style=RedLine] (25.center) to (22);
		\draw [style=PurpleLine] (22) to (24.center);
		\draw [style=PurpleLine, bend left=45, looseness=1.25] (24.center) to (27.center);
		\draw [style=PurpleLine] (27.center) to (23);
		\draw [style=RedLine] (37.center) to (34);
		\draw [style=RedLine, bend right=45, looseness=1.25] (36.center) to (37.center);
		\draw [style=RedLine] (36.center) to (33);
		\draw [style=PurpleLine] (33) to (35.center);
		\draw [style=PurpleLine, bend left=45, looseness=1.25] (35.center) to (38.center);
		\draw [style=PurpleLine] (38.center) to (34);
		\draw [style=ThickLine] (44) to (45);
		\draw [style=ArrowLineRight] (50.center) to (55.center);
		\draw [style=ArrowLineRight] (49.center) to (54.center);
		\draw [style=ArrowLineRight] (48.center) to (53.center);
		\draw [style=ArrowLineRight] (51.center) to (52.center);
		\draw [style=BrownLine] (60) to (61);
	\end{pgfonlayer}
\end{tikzpicture}}
\caption{We sketch the possible fate of symmetry operators under decompression.}
\label{fig:fateofsyms}
\end{figure}
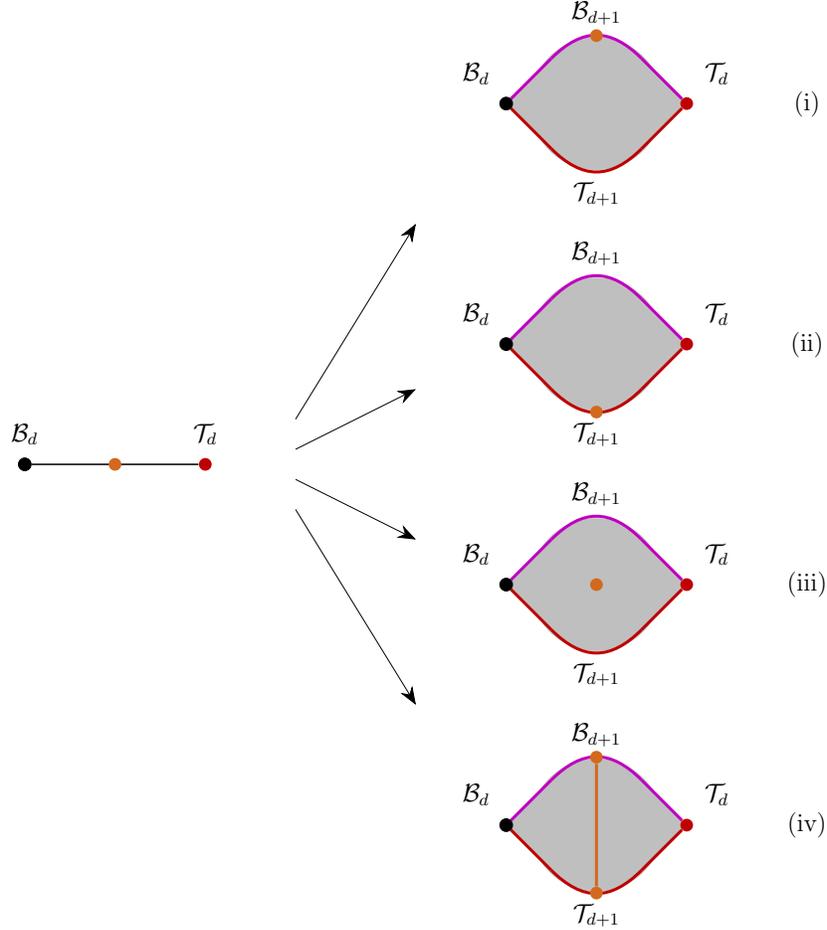

We now use the above considerations to address how one constructs symmetry operators and defect operators for the $\mathcal{T}_d$ theory. As from the usual sandwich construction, applied to the tuple $\mathcal{S}_{d+2}^{}, \mathcal{B}_{d+1}, \mathcal{T}_{d+1}, \mathcal{T}_d^{}, \mathcal{B}_d^{}$, we have simply:
\be\ba
   &\{\textnormal{Symmetry operators of $\mathcal{T}_d$}\} ~ \leftrightarrow ~  \\
   &\{\textnormal{Operators of $\mathcal{S}_{d+2}$, its edges $ \mathcal{B}_{d+1}, \mathcal{T}_{d+1}$, or $\mathcal{B}_d$ that do not terminate at $\mathcal{T}_d$}\} \\[0.75em]
   &\{\textnormal{Defect operators of $\mathcal{T}_d$}\}~ \leftrightarrow ~ \\  &\{\textnormal{Operators of $\mathcal{S}_{d+2}$, its edges $ \mathcal{B}_{d+1}, \mathcal{T}_{d+1}$, or $\mathcal{B}_d$ that terminate at $\mathcal{T}_d$}\}
\ea\ee
However, there are some novelties that arise for this more general setup. To illustrate, we begin with the standard decompression of an absolute QFT$_d$ involving its regulated SymTFT / SymTh $\mathcal{S}_{d+1}(g)$. In this setting, the topological operators of a SymTFT can in principle lift to a more general class of defects which have some dependent on local metric deformations. Even so, the passage back to the topological limit $\mathcal{S}_{d+1}(g) \rightarrow \mathcal{S}_{d+1}$ clearly still makes sense, in which case we again arrive at topological operators. With this caveat stated, we now ask how heavy defects and topological operators descend into the relative theory $\mathcal{T}_d$:
\be\label{eq:decompression}
\scalebox{0.9}{
\begin{tikzpicture}
	\begin{pgfonlayer}{nodelayer}
		\node [style=none] (0) at (-3.5, 0.5) {$\mathcal{S}_{d+1}(g)$};
		\node [style=SmallCircle] (1) at (-5, 0) {};
		\node [style=SmallCircleRed] (2) at (-2, 0) {};
		\node [style=none] (3) at (-2, 0.5) {$\mathcal{T}_d$};
		\node [style=none] (4) at (-5, 0.5) {$\mathcal{B}_d$};
		\node [style=none] (15) at (-0.5, 0) {};
		\node [style=none] (16) at (0.5, 0) {};
		\node [style=SmallCircle] (18) at (2.25, 0) {};
		\node [style=SmallCircleRed] (19) at (5.25, 0) {};
		\node [style=none] (22) at (3, 0.75) {};
		\node [style=none] (23) at (3, -0.75) {};
		\node [style=none] (24) at (4.5, -0.75) {};
		\node [style=none] (25) at (4.5, 0.75) {};
		\node [style=none] (28) at (3.75, 0) {$\mathcal{S}_{d+2}$};
		\node [style=none] (29) at (3.75, -1.5) {$\mathcal{T}_{d+1}$};
		\node [style=none] (30) at (5.625, 0.375) {$\mathcal{T}_d$};
		\node [style=none] (31) at (1.875, 0.375) {$\mathcal{B}_d$};
		\node [style=none] (32) at (3.75, 1.5) {$\mathcal{B}_{d+1}$};
	\end{pgfonlayer}
	\begin{pgfonlayer}{edgelayer}
		\draw[fill=gray!50, draw=gray!50] (3.75,0) circle (6.04ex);
		\filldraw[fill=gray!50, draw=gray!50]  (2.25, 0) -- (3, 0.75) -- (3, -0.75)  -- cycle;
		\filldraw[fill=gray!50, draw=gray!50]  (5.25, 0) -- (4.5, 0.75) -- (4.5, -0.75) -- cycle;
		\draw [style=ThickLine] (1) to (2.center);
		\draw [style=ArrowLineRight] (15.center) to (16.center);
		\draw [style=RedLine] (24.center) to (19);
		\draw [style=RedLine, bend right=45, looseness=1.25] (23.center) to (24.center);
		\draw [style=RedLine] (23.center) to (18);
		\draw [style=PurpleLine] (18) to (22.center);
		\draw [style=PurpleLine, bend left=45, looseness=1.25] (22.center) to (25.center);
		\draw [style=PurpleLine] (25.center) to (19);
	\end{pgfonlayer}
\end{tikzpicture}}
\ee
On the left we can depict the support of defect and symmetry operators\footnote{There are of course also more general cases where symmetry operator take the form of non-genuine bulk operators, attaching back to $\mathcal{B}_d$, see, e.g., \cite{Argurio:2024oym,WIPHHYZ}, which we will not consider here, however similar considerations also hold there.} respectively as:
\be\label{eq:defsym}\scalebox{0.95}{
\begin{tikzpicture}
	\begin{pgfonlayer}{nodelayer}
		\node [style=SmallCircle] (0) at (-4, 0) {};
		\node [style=SmallCircleRed] (1) at (-1, 0) {};
		\node [style=none] (2) at (-1, 0.5) {$\mathcal{T}_d$};
		\node [style=none] (3) at (-4, 0.5) {$\mathcal{B}_d$};
		\node [style=none] (4) at (-2.5, 0.5) {$\mathcal{S}_{d+1}$};
		\node [style=SmallCircle] (5) at (1, 0) {};
		\node [style=SmallCircleRed] (6) at (4, 0) {};
		\node [style=none] (7) at (4, 0.5) {$\mathcal{T}_d$};
		\node [style=none] (8) at (1, 0.5) {$\mathcal{B}_d$};
		\node [style=none] (9) at (2.5, 0.5) {$\mathcal{S}_{d+1}$};
		\node [style=SmallCircleBrown] (10) at (2.5, 0) {};
		\node [style=none] (11) at (4.75, 0) {};
	\end{pgfonlayer}
	\begin{pgfonlayer}{edgelayer}
		\draw [style=ThickLine] (0) to (1);
		\draw [style=ThickLine] (5) to (6);
		\draw [style=BrownLine, snake it] (0) to (1);
	\end{pgfonlayer}
\end{tikzpicture}}
\ee
Here we have presented a generic defect stretching from $\mathcal{T}_d$ to $\mathcal{B}_d$ and a generic symmetry operator localized in the bulk $\mathcal{S}_{d+1}$. We now apply \eqref{eq:decompression} to \eqref{eq:defsym} and note the various possibilities, see figure \ref{fig:fateofdefects} for defect operators and figure \ref{fig:fateofsyms} for symmetry operators. Notice that while some operators may appear as defects in the gapless $\mathcal{T}_{d+1}$ theory they can appear as topological operators from the point of view of a $\mathcal{T}_d$ observer.

How does a symmetry operator linking / acting on a defect operator deform under the decompression \eqref{eq:decompression}? We use the labelling in figures \ref{fig:fateofdefects} and \ref{fig:fateofsyms} to list pairs. The symmetry operator with support (i) can act on defects with support (i), (iv). Similarly (ii) can act on (ii), (iv). The symmetry operator with support (iii) can act on defects with support (iii), (iv). The symmetry operator with support (iv) can act on defects with support (i), (ii), (iii).


\subsection{Example: 2-group Symmetry / Symmetry Fractionalization}

We make the above considerations concrete in the context of two discrete fields $b_n,a_n$  and we find that our discussion naturally incorporates the effects of symmetry fractionalizations \cite{Barkeshli:2014cna,Chen:2014wse, Delmastro:2022pfo, Brennan:2022tyl} and extension properties of 2-group symmetries. Let us first set the background field $B_n$ for $b_n$ to zero, then the possible electric boundary conditions are:
\be \label{eq:exampleBC}
\alpha a_n|_{\mathcal{B}_{d+1}}=\beta b_{n}\,.
\ee
We take $b_n,a_n$ to be discrete of order $k_n,l_n$, respectively, i.e., they are torsional cocycles taking values in $\Z_{k_n},\Z_{l_n}$. For \eqref{eq:exampleBC} to be well-defined we require $\alpha,\beta$ to be multiples $k_n/g_n, l_n/g_n$ respectively with $g_n=\textnormal{gcd}(k_n,l_n)$. These factors multiply both sides to take values in $\Z_{g_n}$ which is the overall coefficient system of the equation. Bulk modes not taking values in $\Z_{g_n}\subset \Z_{k_n}$, characterized by the quotient $\Z_{k_n}/\Z_{g_n}\cong \Z_{k_n/g_n}$, are therefore not fixed by these boundary conditions and we need to supplement \eqref{eq:exampleBC} with boundary conditions for the dual fields $\widetilde{a}_m$.

We make the above constraints manifest and rewrite \eqref{eq:exampleBC} as a condition in $\mathbb{Z}_{g_n}$.
Working over the integers, we write:
\begin{equation}
\alpha = \alpha^{\prime} \frac{k_n}{g_n} \,, \qquad  \beta = \beta^{\prime} \frac{k_n}{g_n}.
\end{equation}
By abuse of notation, we also refer to the mod $\mathbb{Z}_{g_n}$ reduction by the same variables.
Then, we have:
\be \label{eq:exampleBC2}
\alpha' a_n'|_{\mathcal{B}_{d+1}}=\beta'b_{n}'\,,
\ee
now valued in $\Z_{g_n}$, i.e., $\alpha', \beta'\in \Z_{g_n}$ and $a_n'= ({k_n}/{g_n}) a_n$ and $b_{n}'=({l_n}/{g_n}) b_{n}$. Next, introduce $g_n''=\textnormal{gcd}(g_n,\alpha')$. Then the above multiplication by $\alpha'$ has a kernel isomorphic to $\Z_{g_n''}$. We consider the case with trivial kernel and set $\alpha'=1$ which implies $g_n''=1$. Further introduce $g_n'=\textnormal{gcd}(g_n,\beta')$. Then the above realizes Dirichlet boundary conditions on a $\Z_{g'_n}$ subgroup of $a_n'$ and outside of this subgroup specifies an identification of $a_n'$ profiles with boundary profiles. Boundary modes taking values in $\Z_{g_n/g_n'}$ are not mapped onto by the bulk, they remain fluctuating.

Let us discuss the physics of the case $\beta'=1$. The main idea will be to introduce a ``naive'' and ``true'' symmetry, which we refer to as $\mathcal{A}$ and $\widetilde{\mathcal{A}}$ for both the edge mode theory as well as the bulk theory. The boundary conditions connecting $\mathcal{S}_{d+2}$ with the gapped boundary $\mathcal{B}_{d+1}$ will result in a reduction in the naive symmetry to the true symmetry. With this in mind, we define:
\be
\mathcal{A}^{(\textnormal{edge})}_n\cong \Z_{\beta=k_n/g_n} \quad \widetilde{\mathcal{A}}^{(\textnormal{edge)}}_n\cong \Z_{k_n}\,, \quad  \widetilde{\mathcal{A}}^{\textnormal{(bulk)}}_n\cong \Z_{l_n}\,, \quad  \mathcal{A}^{\textnormal{(bulk)}}_n\cong\Z_{\alpha=l_n/g_n}\,,
\ee
which we can assemble into the long exact sequence familiar in the study of 2-groups
(see e.g., \cite{Kapustin:2013uxa, Benini:2018reh, Bhardwaj:2021wif, Lee:2021crt}):
\be\label{eq:2gpSeq}
0~\rightarrow~ \mathcal{A}_n^{\textnormal{(edge)}}~\xrightarrow[]{~\times g_n ~}~  \widetilde{\mathcal{A}}^{(\textnormal{edge})}_n ~\xrightarrow[]{~ \times\:\! l_n/g_n~}~ \widetilde{\mathcal{A}}^{\textnormal{(bulk)}}_n ~\xrightarrow[]{~\textnormal{mod\,}l_n/g_n   ~}~ \mathcal{A}^{\textnormal{(bulk)}}_n~\rightarrow~0\,.
\ee
To give a physical interpretation of this long exact sequence, introduce the topological operators
\be
U_{a_n}=\exp\lb \frac{2\pi i}{k_n}\int_{\Sigma_n} a_n \rb \,, \qquad U_{b_n}=\exp\lb \frac{2\pi i}{l_n}\int_{\Sigma_n} b_n\rb\,,
\ee
where $U_{a_n}, U_{b_n}$ are naive bulk and boundary topological operators built from periods of $a_n, b_n$ respectively and $\Sigma_n$ is some $n$-cycle. As we discussed in section \ref{ssec:defectandsymops}, these topological operators can be interpreted as symmetry operators for the $\mathcal{T}_d$ theory assuming the support of $\Sigma_n$ is separated from that of $\mathcal{T}_d$.

Exponentiating the periods of the boundary conditions appearing in equation \eqref{eq:exampleBC}, we get a relation between these naive symmetry operators:
\be\label{eq:alphagamma}
U_{a_n}^\alpha|_{\mathcal{B}_{d+1}}= U_{b_n}^\gamma.
\ee
Equivalently,  pushing $\alpha$ copies of $U_{a_n}$ into the boundary results in $\gamma$ copies of $ U_{b_n}$. The boundary symmetries described by $U_{b_n}$ are extended by the bulk symmetries described by $U_{a_n}$. We then see that the group $\mathcal{A}^{(\textnormal{edge})}_n$ describes symmetry operators which cannot be deformed off the boundary, and are thus intrinsic to the relative theory $\mathcal{B}_{d+1}$. Meanwhile the group $\mathcal{A}^{\textnormal{(bulk)}}_n$ describes the bulk symmetry operators which do not admit an interpretation as a symmetry operator of $\mathcal{B}_{d+1}$ when deformed into the boundary, i.e., they are transparent with respect to the boundary theory. The long exact sequence \eqref{eq:2gpSeq} connecting these groups defines a class
\be
\mathcal{P}\in H^3(B\mathcal{A}^{\textnormal{(bulk)}}_n, \mathcal{A}^{(\textnormal{edge})}_n)\,,
\ee which precisely describes the Postnikov class of a 2-group which modifies the associativity relation of $\mathcal{A}^{\textnormal{(bulk)}}_n$ symmetry operators \cite{Kapustin:2013uxa, Benini:2018reh}.

We can dualize the exact sequence \eqref{eq:2gpSeq} to:
\be
0~\rightarrow~ \mathcal{A}_n^{\textnormal{(bulk)},\vee}~\xrightarrow[]{}~  \widetilde{\mathcal{A}}^{(\textnormal{bulk}),\vee}_n ~\xrightarrow[]{}~ \widetilde{\mathcal{A}}^{(\textnormal{edge)},\vee}_n ~\xrightarrow[]{}~ \mathcal{A}^{\textnormal{(edge),}\vee}_n~\rightarrow~0\,.
\ee
where the image of the mapping
\be
\widetilde{\mathcal{A}}^{(\textnormal{bulk}),\vee}_n ~\xrightarrow[]{}~ \widetilde{\mathcal{A}}^{(\textnormal{edge)},\vee}_n
\ee
specifies the bulk defect operators which can end on the boundary. Indeed as emphasized in\footnote{Technically \cite{Lee:2021crt} mentions the case of a 2-group involving the mixture of a $0$-form symmetry and a $1$-form symmetry but their consideration generalize straightforwardly.} \cite{Lee:2021crt}, the group $\widetilde{\mathcal{A}}^{(\textnormal{edge)},\vee}_n$ can be interpreted as an equivalence class of $k$-dimensional defect operators modulo $n$-dimensional defect operators which can end on $(n-1)$-dimensional defect operators which transform faithfully under $\mathcal{A}_n^{\textnormal{(bulk)}}$. This in particular means that there are elements in $\widetilde{\mathcal{A}}^{(\textnormal{edge)},\vee}_n$ which are endable, albeit on $(n-1)$-dimensional defect operators that transform projectively under $\mathcal{A}_n^{\textnormal{(bulk)}}$.

Note that even if the Postnikov class of the 2-group vanishes, there can still be symmetry fractionalization effects between the bulk and boundaries. In particular, if we have that $\beta=1$ and $\alpha>1$ in \eqref{eq:alphagamma} then $\mathcal{P}=0$ but we still have the non-trivial effect that symmetry operators from the boundary can fractionate when pulled into the bulk. Conversely, if $\alpha=1$ and $\beta>1$, then we have that again $\mathcal{P}=0$ but we have that bulk symmetry operators can fractionate when taken to the boundary.\footnote{This is also why 2-groups with non-trivial Postnikov classes were once phrased as ``obstructions to symmetry fractionalization'' in the condensed matter literature \cite{Barkeshli:2014cna}. }

\section{Top Down Approach}
\label{sec:TopDown}

In the preceding sections we gave a bottom up discussion of symmetry theories on manifolds with corners. We now proceed with a top down perspective, showing how for QFTs engineered in string theory, the nested structures descend from the extra-dimensional geometry.

Throughout, we work on spacetimes of the form $\R^{d-1,1}\times X$. Here, $X$ is taken to be a non-compact background, preserving some amount of supersymmetry in the $d$-dimensional spacetime.\footnote{This assumption can be relaxed permitting even non-supersymmetric backgrounds, see for example \cite{Vafa:2001ra, Adams:2001sv, Morrison:2004fr, Braeger:2024jcj} and references therein.} Further, we require $X$ to be asymptotically conical\footnote{Examples include Calabi-Yau orbifolds $\mathbb{C}^n/\Gamma$ and $G_2$-holonomy orbifolds $X_7/\Gamma$ where $X_7$ is a Bryant-Salamon space \cite{bryant1989} (see also \cite{Cvetic:2001zx}), non-Higgsable clusters \cite{Morrison:2012np} and many more.} with radial coordinate $r$. This assumption is a technical simplification and we expect our considerations to hold more broadly. Concretely, we are interested in starting points in 10D (i.e., type IIA and IIB) or 11D (i.e., M-theory). The QFT will be localized along singularities
\be
 \R^{d-1,1}\times \mathscr{S}_0\subset \R^{d-1,1}\times X\,,
 \ee
where $\mathscr{S}_0\subset X$ is compact. In examples $\mathscr{S}_0$ will often simply be a point of maximal codimension in $X$. The $d$-dimensional QFT then follows from compactification on $\mathscr{S}_0$. In the extra-dimensional geometry $X$ the singularities $\mathscr{S}_0$ can either arise from a singular metric profile (as in geometric engineering) or from branes probing a local geometry, which itself might already have metric curvature singularities.

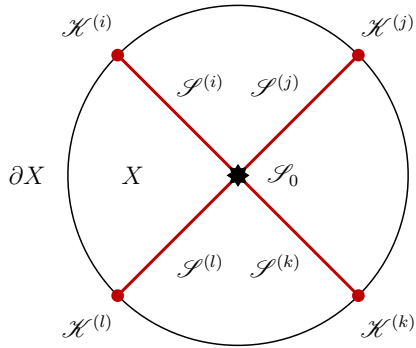
\begin{figure}
\centering
\scalebox{0.8}{
\begin{tikzpicture}
\begin{pgfonlayer}{nodelayer}
		\node [style=none] (0) at (-2, 2) {};
		\node [style=none] (1) at (2, 2) {};
		\node [style=none] (2) at (2, -2) {};
		\node [style=none] (3) at (-2, -2) {};
		\node [style=SmallCircleRed] (4) at (-2, 2) {};
		\node [style=SmallCircleRed] (5) at (2, 2) {};
		\node [style=SmallCircleRed] (6) at (2, -2) {};
		\node [style=SmallCircleRed] (7) at (-2, -2) {};
		\node [style=Star] (8) at (0, 0) {};
		\node [style=none] (9) at (0.75, 0) {$\mathscr{S}_0$};
		\node [style=none] (10) at (0.625, 1.5) {$\mathscr{S}^{(j)}$};
		\node [style=none] (11) at (-0.625, -1.5) {$\mathscr{S}^{(l)}$};
		\node [style=none] (12) at (0.625, -1.5) {$\mathscr{S}^{(k)}$};
		\node [style=none] (13) at (-0.625, 1.5) {$\mathscr{S}^{(i)}$};
		\node [style=none] (14) at (2.5, 2.5) {$\mathscr{K}^{(j)}$};
		\node [style=none] (15) at (2.5, -2.5) {$\mathscr{K}^{(k)}$};
		\node [style=none] (16) at (-2.5, -2.5) {$\mathscr{K}^{(l)}$};
		\node [style=none] (17) at (-2.5, 2.5) {$\mathscr{K}^{(i)}$};
		\node [style=none] (18) at (-1.75, 0) {$X$};
		\node [style=none] (19) at (-3.5, 0) {$\partial X$};
	\end{pgfonlayer}
	\begin{pgfonlayer}{edgelayer}
		\draw [style=ThickLine, bend right=45] (0.center) to (3.center);
		\draw [style=ThickLine, bend right=45] (3.center) to (2.center);
		\draw [style=ThickLine, bend right=45] (2.center) to (1.center);
		\draw [style=ThickLine, bend left=315] (1.center) to (0.center);
		\draw [style=RedLine] (3.center) to (1.center);
		\draw [style=RedLine] (0.center) to (2.center);
	\end{pgfonlayer}
\end{tikzpicture}}
\caption{Sketch of a typical geometry $X$. The generic singularity $\mathscr{S}$ consists of multiple irreducible components $\mathscr{S}^{(i)}$ these meet at the enhancement locus $\mathscr{S}_0$ and stretch to the asymptotic boundary resulting in $\mathscr{K}^{(i)}=\mathscr{S}^{(i)}\cap \partial X$. We refer to the $\mathscr{S}^{(i)}$ as ``flavor branes". }
\label{fig:GeometricData}
\end{figure}

Going forward, we specialize to purely geometric backgrounds with exclusively metric singularities. We consider cases in which $\mathscr{S}_0$ is of constant singularity type, i.e., there are no singularity enhancements along $\mathscr{S}_0$. However, conversely, $\mathscr{S}_0$ itself will be an enhancement within a generically less singular stratum $ \mathscr{S}$:
\be
\mathscr{S}_0\subset \mathscr{S}\subset X\,.
\ee
These additional singularities are taken to be supported on non-compact subspaces of $X$ and we will refer to their irreducible components as flavor branes. Whenever $\mathscr{S}$ has multiple irreducible components we say that $\mathscr{S}_0$ arises at the intersection of flavor branes, but more generally $\mathscr{S}_0$ is simply the locus along which the generic singularity of $\mathscr{S}$ worsens.

A representative example we will return to several times is the Calabi-Yau orbifold $\mathbb{C}^3/\Z_{2n}$ acted on according to $(z_1,z_2,z_3)\sim(\omega z_1, \omega z_2, \omega^{2n-2}z_3)$ with weight vector $(1,1,2n-2)$ and root of unity $\omega=\exp(2\pi i/N)$. Here $\mathscr{S}_0$ is the point $z_1=z_2=z_3=0$ and $\mathscr{S}$ is the locus $z_1=z_2=0$ which cuts out the flavor brane $\mathbb{C}/\Z_{n}\subset \mathbb{C}^3/\Z_{2n}$ supporting a singularity modeled on $\mathbb{C}^2/\Z_2$. In M-theory the resolution of this geometry engineers, in an electric frame, an SU$(n)_n$ 5D gauge theory with an SO$(3)$ flavor symmetry and a $\Z_n$ 1-form symmetry which combine into a 2-group symmetry.\footnote{In asserting that the $0$-form symmetry is SO$(3)$ we are neglecting contributions from both the R-symmetry and the tangent bundle directions of the 5D SCFT. For example, including the R-symmetry and tangent bundle structures, one could in principle have a global form for the symmetries such as $(\mathrm{SU}(2)_F \times \mathrm{SU}(2)_{R} \times \mathrm{Spin}(4,1)) / \mathbb{Z}_2$. This sort of correlated structure in the $0$-form symmetries has been observed in the related context of 6D SCFTs (see Appendix A of \cite{Heckman:2022suy}). One reason to suspect such a correlated structure for the global symmetries in this case is that a 5D $\mathcal{N} = 1$ hypermultiplet transforms in the $(\mathbf{2},\mathbf{2},\mathbf{4})$ of the corresponding Lie algebra. As such, the centers are expected to be non-trivially correlated. This will not impact the statements we make here, which exclusively focus on the $\mathfrak{su}(2)_F$ part of the $0$-form symmetry.} The singular geometry engineers a 5D SCFT. The flavor brane supports a 7D super-Yang-Mills theory twisted by the insertion of the 5D SCFT. The twist is reflected in geometry via a monodromy transformation of the normal geometry $\mathbb{C}^2/\Z_2$ when traversing a path linking the defect insertion in the flavor brane locus $\mathbb{C}/\Z_n$.

This example already allows us to highlight important features which will be universal in our top down analysis. Recall that in section \ref{sec:cheesesteak} we started our analysis in $d$ dimensions and then, via iterated decompression, added dimensions one at a time. Each decompression step was not unique and in decompressing a given QFT we in general had choices in how to distribute its degrees of freedom across various edges. In contrast, in our top down discussion the geometry $X$ will specify a relative higher dimensional theory $\mathcal{T}_D$ within which the $\mathcal{T}_d$ is realized as a defect theory where $D>d$.

This difference in starting point will have various natural consequences. First, in general we will have $D> d+1$ and therefore we will consider an additional compactification of $\mathcal{T}_D$ to a KK-theory $\mathcal{T}_{d+1}$ to map onto our previous considerations in adjacent dimensions. Second, the extra-dimensional geometry $X$ will always specify a preferred decompression of the type:
\be
\scalebox{0.9}{
\begin{tikzpicture}
	\begin{pgfonlayer}{nodelayer}
		\node [style=none] (0) at (0, 0) {};
		\node [style=none] (1) at (1, 0) {};
		\node [style=none] (5) at (-2, 0.5) {$\mathcal{S}_{d+1}(g)$};
		\node [style=none] (9) at (1, 1.25) {};
		\node [style=none] (10) at (-2.75, 0) {};
		\node [style=none] (11) at (-1.25, 0) {};
		\node [style=none] (12) at (-1, 0) {};
		\node [style=none] (13) at (-3, 0) {};
		\node [style=none] (14) at (2.25, 0.5) {};
		\node [style=none] (15) at (3.75, 0.5) {};
		\node [style=none] (16) at (4, 0.5) {};
		\node [style=none] (17) at (2, 0.5) {};
		\node [style=none] (18) at (2.25, -0.5) {};
		\node [style=none] (19) at (3.75, -0.5) {};
		\node [style=none] (20) at (4, -0.5) {};
		\node [style=none] (21) at (2, -0.5) {};
		\node [style=none] (22) at (3, -1) {$\mathcal{T}_{d+1}$};
		\node [style=none] (23) at (3, 1) {$\mathcal{B}_{d+1}$};
		\node [style=none] (24) at (3, 0) {$\mathcal{S}_{d+2}$};
	\end{pgfonlayer}
	\begin{pgfonlayer}{edgelayer}
		\filldraw[fill=gray!50, draw=gray!50]  (2.25, 0.5) -- (2.25, -0.5) -- (3.75, -0.5) -- (3.75, 0.5) -- cycle;
		\draw [style=ArrowLineRight] (0.center) to (1.center);
		\draw [style=ThickLine] (10.center) to (11.center);
		\draw [style=DottedLine] (11.center) to (12.center);
		\draw [style=DottedLine] (10.center) to (13.center);
		\draw [style=PurpleLine] (14.center) to (15.center);
		\draw [style=RedLine] (18.center) to (19.center);
		\draw [style=DottedRed] (20.center) to (19.center);
		\draw [style=DottedRed] (21.center) to (18.center);
		\draw [style=DottedPurple] (16.center) to (15.center);
		\draw [style=DottedPurple] (17.center) to (14.center);
	\end{pgfonlayer}
\end{tikzpicture}}
\ee
where the stringy construction automatically comes with a regulated $\mathcal{S}_{d+1}(g)$. Here $\mathcal{T}_{d+1}$ is specified by the singularities $\mathscr{S}\subset X$. This will also immediately identify $\mathcal{B}_{d+1}$ as the relative theory associated with modes which are not localized in $X$, i.e., so-called bulk modes. Further, we find a family of decompressions with identical corner modes
\be\label{eq:primed3}
\scalebox{0.9}{
\begin{tikzpicture}
	\begin{pgfonlayer}{nodelayer}
		\node [style=SmallCircle] (18) at (2.25, 0) {};
		\node [style=SmallCircleRed] (19) at (5.25, 0) {};
		\node [style=none] (22) at (3, 0.75) {};
		\node [style=none] (23) at (3, -0.75) {};
		\node [style=none] (24) at (4.5, -0.75) {};
		\node [style=none] (25) at (4.5, 0.75) {};
		\node [style=none] (28) at (3.75, 0) {$\mathcal{S}_{d+2}^{\prime}$};
		\node [style=none] (29) at (3.75, -1.5) {$\mathcal{T}_{d+1}^{\:\!\prime}$};
		\node [style=none] (30) at (5.625, 0.375) {$\mathcal{T}_d^{}$};
		\node [style=none] (31) at (1.875, 0.375) {$\mathcal{B}_d^{}$};
		\node [style=none] (32) at (3.75, 1.5) {$\mathcal{B}_{d+1}^{\:\!\prime}$};
	\end{pgfonlayer}
	\begin{pgfonlayer}{edgelayer}
		\draw[fill=gray!50, draw=gray!50] (3.75,0) circle (6.04ex);
		\filldraw[fill=gray!50, draw=gray!50]  (2.25, 0) -- (3, 0.75) -- (3, -0.75)  -- cycle;
		\filldraw[fill=gray!50, draw=gray!50]  (5.25, 0) -- (4.5, 0.75) -- (4.5, -0.75) -- cycle;
		\draw [style=RedLine] (24.center) to (19);
		\draw [style=RedLine, bend right=45, looseness=1.25] (23.center) to (24.center);
		\draw [style=RedLine] (23.center) to (18);
		\draw [style=PurpleLine] (18) to (22.center);
		\draw [style=PurpleLine, bend left=45, looseness=1.25] (22.center) to (25.center);
		\draw [style=PurpleLine] (25.center) to (19);
	\end{pgfonlayer}
\end{tikzpicture}}
\ee
by varying how we cut up the geometry $\partial X$, which is associated to $\mathcal{S}_{d+1}$. This allows us to add some of the degrees of freedom localized to $\mathcal{B}_{d+1}$ to $\mathcal{T}_{d+1}$ resulting in alternative edge modes $\mathcal{T}_{d+1}^{\:\!\prime}$  and $\mathcal{B}_{d+1}^{\:\!\prime}$ while leaving the corners unaltered. This plethora of options traces back to $\mathcal{T}_d$ being localized in $X$ in codimensions larger than one in all our examples.

Let us make some of the above explicit. In the example of M-theory on $\mathbb{C}^3/\Z_{2n}$ we have $d=5$ and the $\mathcal{T}_D$ is the $D=7$ super-Yang-Mills theory to the $\mathfrak{su}_2$ singular stratum. The 5D SCFT is a codimension-2 defect within it and compactifying over circles linking the defect in the 7D theory we find a 6D KK theory $\mathcal{T}_{d+1}$.

Returning to the general case, we also have that, in principle, $X$ can contain strata of singularities of depth larger than two
\be\label{eq:depth}
\mathscr{S}_0\subset \dots\subset \mathscr{S}_i\subset \dots\subset \mathscr{S}_I\equiv  \mathscr{S} \subset X\,,
\ee
where $0\leq i \leq I$ with $I>1$.
We immediately note that string theoretically engineered QFTs are necessarily of finite depth $(I +1) \leq  \dim X $. The main features we wish to highlight are already present when $I=1$, and we therefore focus on this case going forward. See figure \ref{fig:GeometricData} for such a space $X$ together with a typical singular locus $\mathscr{S}$.

Now, in the above example we utilized the fact that the bulk 7D super-Yang-Mills theory, playing the role of $\mathcal{T}_D$, is infrared free. This allowed us to take an IR decoupling limit such that ultimately only the 5D dynamics of the SCFT were retained. The existence of such a decoupling limit is by no means guaranteed, and is for example absent when considering defects theories in an ambient higher-dimensional SCFT \cite{Acharya:2023bth}. As such, in top down analyses it is much preferred to start with the ambient system and discuss $\mathcal{T}_d$, which might be as general as a quasi-SCFT \cite{Acharya:2023bth}, relative to it.

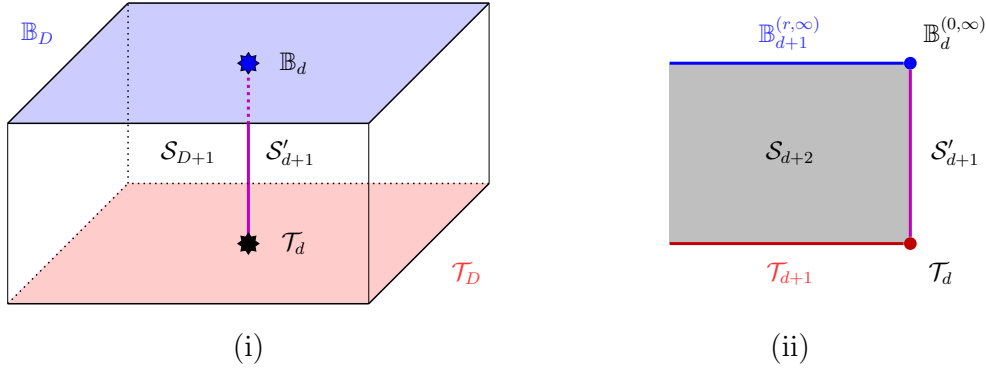
\begin{figure}
\centering
\scalebox{0.8}{
\begin{tikzpicture}
	\begin{pgfonlayer}{nodelayer}
		\node [style=none] (0) at (-8, -1) {};
		\node [style=none] (1) at (-6, 1) {};
		\node [style=none] (2) at (-2, -1) {};
		\node [style=none] (3) at (0, 1) {};
		\node [style=Star] (4) at (-4, 0) {};
		\node [style=none] (13) at (3, 0) {};
		\node [style=none] (14) at (7, 0) {};
		\node [style=SmallCircleRed] (15) at (7, 0) {};
		\node [style=none] (22) at (-4, -1.75) {\large (i)};
		\node [style=none] (23) at (5, -1.75) {\large (ii)};
		\node [style=none] (24) at (-8, 2) {};
		\node [style=none] (25) at (-6, 4) {};
		\node [style=none] (26) at (-2, 2) {};
		\node [style=none] (27) at (0, 4) {};
		\node [style=StarBlue] (28) at (-4, 3) {};
		\node [style=none] (29) at (3, 3) {};
		\node [style=none] (30) at (7, 3) {};
		\node [style=SmallCircleBlue] (31) at (7, 3) {};
		\node [style=none] (32) at (-4, 0) {};
		\node [style=none] (33) at (-4, 3) {};
		\node [style=none] (34) at (-4, 2) {};
		\node [style=none] (35) at (-0.375, -0.5) {{\color{red!80} $\mathcal{T}_{D}$}};
		\node [style=none] (36) at (-3.25, 0) {$\mathcal{T}_{d}$};
		\node [style=none] (37) at (7.5, -0.5) {$\mathcal{T}_{d}$};
		\node [style=none] (38) at (5, -0.5) {{\color{red!80} $\mathcal{T}_{d+1}$}};
		\node [style=none] (39) at (5, 3.5) {{\color{blue!80} $\mathbb{B}^{(r,\infty)}_{d+1}$}};
		\node [style=none] (40) at (7.75, 3.5) {$\mathbb{B}_{d}^{(0,\infty)}$};
		\node [style=none] (41) at (-7.5, 3.5) {{\color{blue!80} $\mathbb{B}_{D}$}};
		\node [style=none] (42) at (-3.25, 3) {$\mathbb{B}_{d}$};
		\node [style=none] (43) at (5, 1.5) {$\mathcal{S}_{d+2}$};
		\node [style=none] (44) at (7.75, 1.5) {$\mathcal{S}_{d+1}'$};
		\node [style=none] (45) at (-5, 1.5) {$\mathcal{S}_{D+1}$};
		\node [style=none] (46) at (-3.3, 1.5) {$\mathcal{S}_{d+1}'$};
	\end{pgfonlayer}
	\begin{pgfonlayer}{edgelayer}
	\filldraw[fill=blue!20, draw=blue!20]  (-8, 2) -- (-6, 4) -- (0, 4) -- (-2, 2) -- cycle;
	\filldraw[fill=red!20, draw=red!20]  (-6, 1) -- (0, 1) -- (-2, -1) -- (-8, -1) -- cycle;
	\filldraw[fill=gray!50, draw=gray!50]  (7, 3) -- (3, 3) -- (3, 0) -- (7, 0) -- cycle;
		
		\draw [style=ThickLine] (3.center) to (2.center);
		\draw [style=ThickLine] (2.center) to (0.center);
		\draw [style=RedLine] (13.center) to (15);
		\draw [style=ThickLine] (25.center) to (27.center);
		\draw [style=ThickLine] (27.center) to (26.center);
		\draw [style=ThickLine] (26.center) to (24.center);
		\draw [style=ThickLine] (24.center) to (25.center);
		\draw [style=BlueLine] (29.center) to (31);
		\draw [style=PurpleLine] (32.center) to (34.center);
		\draw [style=DottedPurple] (33.center) to (34.center);
		\draw [style=PurpleLine] (31) to (15);
		\draw (24.center) to (0.center);
		\draw (26.center) to (2.center);
		\draw (27.center) to (3.center);
		\draw [style=DottedLine] (0.center) to (1.center);
		\draw [style=DottedLine] (1.center) to (25.center);
		\draw [style=DottedLine] (3.center) to (1.center);
	\end{pgfonlayer}
\end{tikzpicture}}
\caption{In subfigure (i) we sketch the sandwich of the bulk QFT$_D$ with inserted defect QFT$_d$. In subfigure (ii) we KK-reduce subfigure (i) to an open cheesesteak.
}
\label{fig:SymTFTColorFlavor}
\end{figure}

With this, the primary question shifts to: What is the symmetry sandwich of an absolute QFT$_D$ with a $d$-dimensional defect theory inserted? To see how this question relates to our previous considerations, we naively extend such a coupled system by an additional dimension to a sandwich in overall dimension $(D+1)$ with interval coordinate $0\leq x_\perp \leq 1 $. For this, denote the relative theory associated with the bulk QFT$_D$ by $\mathcal{T}_D$ and that of the $d$-dimensional defect theory by $\mathcal{T}_d$. Similarly, introduce $\mathbb{B}_D,\mathbb{B}_d$ as the boundary conditions setting their global form. We sketch the associated sandwich as in subfigure (i) of figure \ref{fig:SymTFTColorFlavor}. The $(D+1)$-dimensional bulk is filled by the symmetry theory $\mathcal{S}_{D+1}$ and, as a defect within it, we have the symmetry theory $\mathcal{S}_{d+1}'$ (which we identify with a $\mathcal{B}_{d+1}$ in our bottom up discussion, however dropping the requirement that this boundary conditions is necessarily gapped\,/\,free).

From here, to take a step towards our bottom up considerations in adjacent dimension, we KK-reduce the bulk on $(D-d-1)$-dimensional concentric shells centered on and linking the defect $\mathcal{S}'_{d+1}$. This results in the $(D-d+1)$-dimensional bulk transverse to $\mathcal{S}'_{d+1}$ projecting down to a semi-infinite 2-dimensional strip which we parametrize with the radial coordinate $r\geq 0$ of the shells and the original decompression coordinate $x_\perp$.

The triple $(\mathcal{T}_D,\mathcal{S}_{D+1},\mathbb{B}_D)$ reduces to the triple $(\mathcal{T}_{d+1},\mathcal{S}_{d+2},\mathbb{B}_{d+1}^{(r,1)})$ where, in the last entry, we introduced an exponent recording $(r,x_\perp)$. Next, note that the defect triple $(\mathcal{T}_d,\mathcal{S}_{d+1}',\mathbb{B}_d)$ becomes an end of the world sandwich at $r=0$ which is relative to the former sandwich. In preparation of more unified notation we write $\mathbb{B}_d\equiv \mathbb{B}_d^{(0,1)}$. We sketch the two sandwiches in adjacent dimension in subfigure (ii) of figure \ref{fig:SymTFTColorFlavor}. Overall, we have a $(d+2)$-dimensional symmetry theory for an absolute $(d+1)$-dimensional KK-theory, with a $d$-dimensional end of the world defect, and we refer to this system as an ``open cheesesteak''.

So far we have described how $X$ may specify a $(d+1)$-dimensional absolute theory. In which situations, perhaps specifying yet more data can we hope to obtain a $d$-dimensional absolute theory? Recall the latter was our starting point in bottom up considerations.

Whenever the $(d+1)$-dimensional KK-theory is infrared free we can consider an IR limit localizing the ungapped dynamics to $d$-dimensions. However, this limit only results in an absolute $d$-dimensional theory if we specify an additional set of boundary conditions. This is due to the KK-theory containing extended defect operators, such as Wilson lines, which can end on the $d$-dimensional end of the world theory $\mathcal{T}_d$. The end points of such defects in $d$ dimensions are themselves defects of the end of the world theory, and to obtain an absolute theory in $d$ dimensions we must determine which maximally mutually local subset of these are to remain.

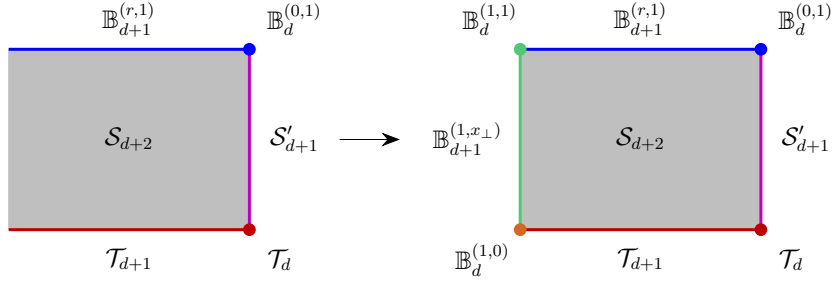
\begin{figure}
\centering
\scalebox{0.8}{\begin{tikzpicture}
	\begin{pgfonlayer}{nodelayer}
		\node [style=none] (0) at (-2.25, -1.5) {};
		\node [style=none] (1) at (-2.25, 1.5) {};
		\node [style=none] (2) at (-6.25, 1.5) {};
		\node [style=none] (3) at (-6.25, -1.5) {};
		\node [style=none] (4) at (-0.75, 0) {};
		\node [style=none] (5) at (0.25, 0) {};
		\node [style=none] (6) at (2.25, -1.5) {};
		\node [style=none] (7) at (2.25, 1.5) {};
		\node [style=none] (8) at (6.25, 1.5) {};
		\node [style=none] (9) at (6.25, -1.5) {};
		\node [style=SmallCircleBlue] (10) at (-2.25, 1.5) {};
		\node [style=SmallCircleBlue] (11) at (6.25, 1.5) {};
		\node [style=SmallCircleRed] (12) at (-2.25, -1.5) {};
		\node [style=SmallCircleRed] (13) at (6.25, -1.5) {};
		\node [style=SmallCircleGreen] (14) at (2.25, 1.5) {};
		\node [style=SmallCircleBrown] (15) at (2.25, -1.5) {};
		\node [style=none] (17) at (4.25, 2) {$\mathbb{B}_{d+1}^{(r,1)}$};
		\node [style=none] (18) at (7, 0) {$\mathcal{S}_{d+1}'$};
		\node [style=none] (19) at (4.25, 0) {$\mathcal{S}_{d+2}$};
		\node [style=none] (20) at (4.25, -2) {$\mathcal{T}_{d+1}$};
		\node [style=none] (21) at (-4.25, 2) {$\mathbb{B}_{d+1}^{(r,1)}$};
		\node [style=none] (22) at (-4.25, 0) {$\mathcal{S}_{d+2}$};
		\node [style=none] (23) at (-4.25, -2) {$\mathcal{T}_{d+1}$};
		\node [style=none] (24) at (-1.5, 0) {$\mathcal{S}_{d+1}'$};
		\node [style=none] (25) at (1.375, 0) {$\mathbb{B}_{d+1}^{(1,x_\perp)}$};
		\node [style=none] (26) at (1.75, 2) {$\mathbb{B}_{d}^{(1,1)}$};
		\node [style=none] (27) at (7, 2) {$\mathbb{B}_{d}^{(0,1)}$};
		\node [style=none] (28) at (6.75, -2) {$\mathcal{T}_{d}$};
		\node [style=none] (29) at (1.625, -2) {$\mathbb{B}_{d}^{(1,0)}$};
		\node [style=none] (30) at (-1.75, -2) {$\mathcal{T}_{d}$};
		\node [style=none] (31) at (-1.5, 2) {$\mathbb{B}_{d}^{(0,1)}$};
		\node [style=none] (32) at (-8, -3) {};
	\end{pgfonlayer}
	\begin{pgfonlayer}{edgelayer}
		\filldraw[fill=gray!50, draw=gray!50]  (-2.25, 1.5) -- (-2.25, -1.5) -- (-6.25, -1.50) -- (-6.25, 1.50)-- cycle;
		\filldraw[fill=gray!50, draw=gray!50]  (2.25, 1.5) -- (2.25, -1.5) -- (6.25, -1.50) -- (6.25, 1.50)-- cycle;
		\draw [style=GreenLine] (7.center) to (6.center);
		\draw [style=BlueLine] (2.center) to (1.center);
		\draw [style=BlueLine] (7.center) to (8.center);
		\draw [style=RedLine] (3.center) to (0.center);
		\draw [style=RedLine] (6.center) to (9.center);
		\draw [style=PurpleLine] (1.center) to (0.center);
		\draw [style=PurpleLine] (8.center) to (9.center);
		\draw [style=ArrowLineRight] (4.center) to (5.center);
	\end{pgfonlayer}
\end{tikzpicture}
}
\caption{To obtain an absolute $d$-dimensional theory, after taking a decoupling limit localizing dynamics to $d$-dimensions, we must impose additional boundary conditions (green).}
\label{fig:Closingcheesesteak}
\end{figure}

Such a choice is then realized by boundary conditions along the remaining internal non-compact dimension which are imposed at $r=\infty$. For now, let us simply note that this adds a triplet of boundary conditions which we denote by $\mathbb{B}_d^{(1,0)},\mathbb{B}_{d+1}^{(1,x_\perp)},\mathbb{B}_{d}^{(1,1)}$ respectively labelled by their location and dimension (see figure \ref{fig:Closingcheesesteak}).

\begin{figure}
\centering
\scalebox{0.8}{
\begin{tikzpicture}
	\begin{pgfonlayer}{nodelayer}
		\node [style=none] (0) at (0.125, 0) {};
		\node [style=none] (1) at (-0.875, 0) {};
		\node [style=none] (2) at (2.25, -1.5) {};
		\node [style=none] (3) at (2.25, 1.5) {};
		\node [style=none] (4) at (6.25, 1.5) {};
		\node [style=none] (5) at (6.25, -1.5) {};
		\node [style=SmallCircleBlue] (6) at (6.25, 1.5) {};
		\node [style=SmallCircleRed] (7) at (6.25, -1.5) {};
		\node [style=SmallCircleGreen] (8) at (2.25, 1.5) {};
		\node [style=SmallCircleBrown] (9) at (2.25, -1.5) {};
		\node [style=none] (10) at (4.25, 2) {$\mathbb{B}_{d+1}^{(r,1)}$};
		\node [style=none] (11) at (7, 0) {$\mathcal{B}_{d+1}$};
		\node [style=none] (12) at (4.25, 0) {$\mathcal{S}_{d+2}$};
		\node [style=none] (13) at (4.25, -2) {$\mathcal{T}_{d+1}$};
		\node [style=none] (14) at (1.375, 0) {$\mathbb{B}_{d+1}^{(1,x_\perp)}$};
		\node [style=none] (15) at (1.5, 2) {$\mathbb{B}_{d}^{(1,1)}$};
		\node [style=none] (16) at (7, 2) {$\mathbb{B}_{d}^{(0,1)}$};
		\node [style=none] (17) at (7, -2) {$\mathcal{T}_{d}$};
		\node [style=none] (18) at (1.5, -2) {$\mathbb{B}_{d}^{(1,0)}$};
		\node [style=none] (19) at (-4.25, 3.5) {};
		\node [style=none] (20) at (-4.25, 2.5) {};
		\node [style=none] (21) at (-6.25, -1.5) {};
		\node [style=none] (22) at (-6.25, 1.5) {};
		\node [style=none] (23) at (-6.25, 1.5) {};
		\node [style=none] (24) at (-2.25, -1.5) {};
		\node [style=SmallCircleRed] (26) at (-2.25, -1.5) {};
		\node [style=SmallCircleGrey] (27) at (-6.25, 1.5) {};
		\node [style=SmallCircleBrown] (28) at (-6.25, -1.5) {};
		\node [style=none] (30) at (-3.75, 0.5) {$\mathcal{B}_{d+1}$};
		\node [style=none] (31) at (-5, -0.5) {$\mathcal{S}_{d+2}$};
		\node [style=none] (32) at (-4.25, -2) {$\mathcal{T}_{d+1}$};
		\node [style=none] (33) at (-7.125, 0) {$\mathbb{B}_{d+1}^{(1,x_\perp)}$};
		\node [style=none] (34) at (-7, 2) {$\mathbb{B}_{d}^{(1,1),*}$};
		\node [style=none] (36) at (-1.5, -2) {$\mathcal{T}_{d}$};
		\node [style=none] (37) at (-7, -2) {$\mathbb{B}_{d}^{(1,0)}$};
		\node [style=SmallCircle] (38) at (-5.75, 6) {};
		\node [style=SmallCircleRed] (39) at (-2.75, 6) {};
		\node [style=none] (40) at (-5, 6.75) {};
		\node [style=none] (41) at (-5, 5.25) {};
		\node [style=none] (42) at (-3.5, 5.25) {};
		\node [style=none] (43) at (-3.5, 6.75) {};
		\node [style=none] (44) at (-4.25, 6) {$\mathcal{S}_{d+2}$};
		\node [style=none] (45) at (-4.25, 4.5) {$\mathcal{T}_{d+1}$};
		\node [style=none] (46) at (-2.25, 6.5) {$\mathcal{T}_d$};
		\node [style=none] (47) at (-6.25, 6.5) {$\mathcal{B}_d$};
		\node [style=none] (48) at (-4.25, 7.5) {$\mathcal{B}_{d+1}$};
		\node [style=none] (49) at (0, -2.5) {};
	\end{pgfonlayer}
	\begin{pgfonlayer}{edgelayer}
		\filldraw[fill=gray!50, draw=gray!50]  (-2.25, -1.5) -- (-6.25, -1.50) -- (-6.25, 1.50)-- cycle;
		\filldraw[fill=gray!50, draw=gray!50]  (2.25, 1.5) -- (2.25, -1.5) -- (6.25, -1.50) -- (6.25, 1.50)-- cycle;
		
		\draw[fill=gray!50, draw=gray!50] (-4.25,6) circle (6.04ex);
		\filldraw[fill=gray!50, draw=gray!50]  (-2.75, 6) -- (-3.5, 6.75) -- (-3.5, 5.25)  -- cycle;
		\filldraw[fill=gray!50, draw=gray!50]  (-5.75, 6) -- (-5, 6.75) -- (-5, 5.25) -- cycle;
		
		\draw [style=GreenLine] (3.center) to (2.center);
		\draw [style=BlueLine] (3.center) to (4.center);
		\draw [style=RedLine] (2.center) to (5.center);
		\draw [style=PurpleLine] (4.center) to (5.center);
		\draw [style=ArrowLineRight] (1.center) to (0.center);
		\draw [style=ArrowLineRight] (19.center) to (20.center);
		\draw [style=GreenLine] (22.center) to (21.center);
		\draw [style=RedLine] (21.center) to (24.center);
		\draw [style=PurpleLine] (23.center) to (24.center);
		\draw [style=RedLine] (42.center) to (39);
		\draw [style=RedLine, bend right=45, looseness=1.25] (41.center) to (42.center);
		\draw [style=RedLine] (41.center) to (38);
		\draw [style=PurpleLine] (38) to (40.center);
		\draw [style=PurpleLine, bend left=45, looseness=1.25] (40.center) to (43.center);
		\draw [style=PurpleLine] (43.center) to (39);
	\end{pgfonlayer}
\end{tikzpicture}
}
\caption{We depict the various $(d+2)$-dimensional symmetry theories we consider, and make explicit their relation: the tuple $\widetilde{\mathbb{B}}$ compresses to the tuple ${\mathbb{B}}$ which compresses to the corner $\mathcal{B}_d$. Here we have used $\mathcal{S}_{d+1}'\equiv \mathcal{B}_{d+1}$.}
\label{fig:Quesidilla}
\end{figure}
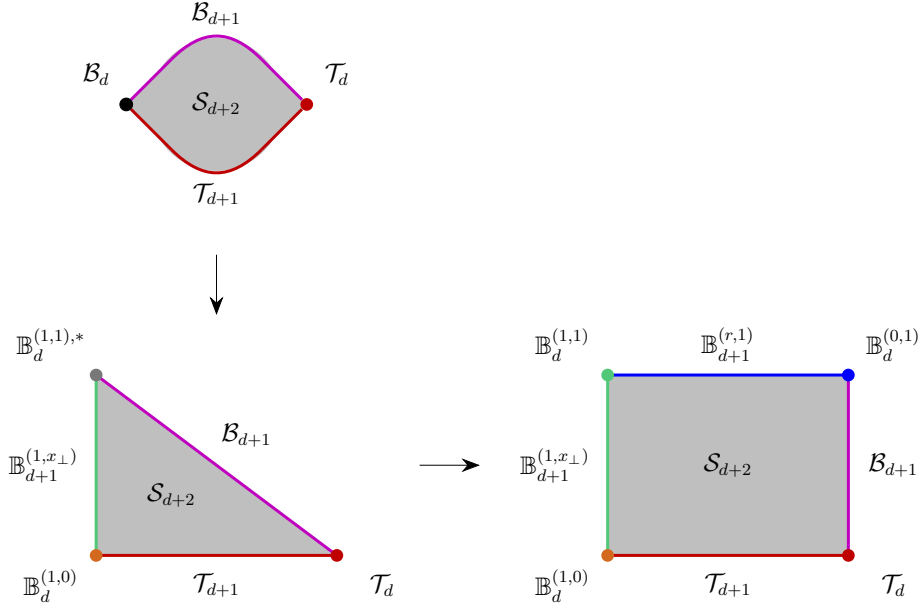

With this we can finally connect to our discussion in section \ref{sec:cheesesteak} by specifying where $\mathcal{B}_{d}$, and other data, is to be located in the above discussion. Our claim is that $X$ has naturally decompressed $\mathcal{B}_{d}$ into the tuple
\be
\widetilde{\mathbb{B}}=\lb \mathbb{B}_d^{(1,0)},\mathbb{B}_{d+1}^{(1,x_\perp)},\mathbb{B}_{d}^{(1,1)},\mathbb{B}^{(r,1)}_{d+1},\mathbb{B}_d^{(0,1)}\rb\,.
\ee
Conversely, contracting the $(d+1)$-dimensional edges $\mathbb{B}_{d+1}^{(1,x_\perp)},\mathbb{B}^{(r,1)}_{d+1}$ the three $d$-dimensional corners stack to $\mathcal{B}_{d}$. Further, we have the identification $\mathcal{S}'_{d+1}\equiv \mathcal{B}_{d+1}$ as a relative symmetry theory. The $(d+2)$-dimensional bulk and corner $\mathcal{T}_d$ are as in the bottom up discussion. In the notation of figure \ref{fig:Quesidilla} we have that $\mathcal{B}_{d}$ now results from compressing the edge of the tuple
\be
{\mathbb{B}}=\lb \mathbb{B}_d^{(1,0)},\mathbb{B}_{d+1}^{(1,x_\perp)},\mathbb{B}_{d}^{(1,1),*}\rb\,.
\ee
We give a geometric underpinning for $\widetilde{\mathbb{B}}$ and $\mathbb{B}$ in section \ref{ssec:GlobalForm}.

To round out this introduction we mention that whenever a limit isolating the $d$-dimensional dynamics engineered by $X$ exists we will contract only $\mathbb{B}^{(r,1)}_{d+1}$. In the geometry $X$ this will amount to a particularly natural partitioning of $X$ which pushes the flavor brane local model to the asymptotic boundary $\partial X$.

Further we comment that whenever $X_i$ are local patches covering some larger geometry $Y$ then the symmetry theory associated to $Y$ results from gluing the open cheesesteak of the patches $X_i$. Rather than lay out the general theory of this construction we will simply give an example with $Y=(T^2\times \mathbb{C}^2)/\Z_4$ and four copies of $X_i=\mathbb{C}^3/\Z_{4}$.

Finally, let us provide some geometric intuition for the ideas developed in this section. To begin, consider the very first decompression in our bottom up discussion where $\mathcal{S}_{d+1}$ is supported on $M_d\times I$ with interval $I=[0,1]$ with coordinate $r$. As in \cite{Apruzzi:2021nmk}, in top down constructions, each point $r\in I$ will be associated with a shell $\partial X$ at fixed radius $r$ of the internal geometry and scanning over $I$ we have that $X$ is swept out by radial shells. A point $r\in I$, assume for now $r\neq 0$ and $r\neq \infty$, can be decompressed into an interval $J$ with coordinate $x_\perp$. For this, pick a decomposition
\be \label{eq:decomposition}
\partial X= U\cup V\,,
\ee
where $U,V$ glue along their common boundary $\partial U=\partial V$ to $\partial X$. Note that formally $\partial X$ has infinite volume and M-theory or IIA\,/\,IIB describes a field theory with locally decoupled sectors as in \cite{Baume:2023kkf}. We can therefore open $\mathcal{S}_{d+1}|_{r=r_*}$ for a fixed value $r_*$ into a SymTree. This is achieved by deforming the gluing region into a cylinder
\be
\partial X= U'\cup Z\cup V'\,.
\ee
Here $U,U'$ and $V,V'$ are homotopic pairs. The cross-section of $Z$ is topologically $\partial U=\partial V\equiv W$ and $Z=J\times W$. The resulting SymTree has two ends corresponding to $U',V'$ connected by edges associated with $Z$.

With this any decomposition $\partial X= U\cup V$ gives a triple $\mathcal{T}_{d+1},\mathcal{S}^{}_{d+2},\mathcal{B}_{d+1}$. The triplets associated to two decompositions $\partial X= U_1\cup V_1$ and $\partial X= U_2\cup V_2$ agree if the $U_i,V_i$ are related by an ambient homotopy. Further, for geometries $X$ with flavor branes $\mathscr{S}$ there is a natural decomposition for $\partial X$. Flavor branes intersect $\partial X$ in a singular locus $\mathscr{K}$. A decomposition of $\partial X$ which isolates the flavor symmetries associated with $\mathscr{K}$ is therefore
\be \label{eq:PreferredDecomp}
\partial X= T_{\mathscr{K}}\cup \partial X^\circ\,,
\ee
where $\partial X^\circ$ is the complement $\partial X\setminus T_{\mathscr{K}}$ and $T_{\mathscr{K}}$ is the tubular neighbourhood of $\mathscr{K}\subset \partial X$. The $(d+2)$-dimensional decompressions described in this section are with respect to this natural decomposition. Overall, however there are as many decompression of type \eqref{eq:primed3} as decomposition \eqref{eq:decomposition} of $\partial X$.

We structure our top down discussion as follows. First, in section \ref{ssec:Geometry} we discuss in some detail how to favorably parametrize the geometry $X$. The main idea is to consider a local model $T_{\mathscr{S}}$ centered on the full singular locus. From this perspective we then discuss defect operators in section \ref{sec:DefectandSym}. In section \ref{sec:Gluing} we discuss the symmetry theories $\mathcal{S}_{d+2}$ and $\mathcal{B}_{d+1}$. This will involve a compactification on a manifold with boundary and an associated long exact sequence in relative homology will yield the boundary conditions \eqref{eq:generaldiscretegluing}. Next, making the local model $T_{\mathscr{S}}$ larger and larger we can fill the full geometry $X$, due to the homotopy relation $T_{\mathscr{S}}\sim X$, and we discuss how this deformation is realized in the cheesesteak (see figure \ref{fig:Quesidilla}). We then discuss how to close the open cheesesteak in section  \ref{sec:Gluing2}, imposing boundary conditions at asymptotic infinity $\partial X$, i.e., we specify the tuples $\mathbb{B},\widetilde{\mathbb{B}}$. Here, our main characterization of edge and corner conditions will be via their interaction with defect operators.
Finally, in section \ref{ssec:Comments}, we end with some general comments and a brief example demonstrating how open cheesesteaks enter as building blocks into more general considerations.

Where possible we make our discussion explicit using the running example of M-theory on $X=\mathbb{C}^3/\Z_{2n}$ acted on as $(z_1,z_2,z_3)\sim(\omega z_1, \omega z_2, \omega^{2n-2}z_3)$ with weight vector $(1,1,2n-2)$.

\subsection{Geometric Considerations}
\label{ssec:Geometry}

We begin by identifying the geometry corresponding to the bulk, edges and corners. We will naturally find a structure (see figure \ref{fig:Quesidilla}) in which additional decompression steps have taken place compared to the bottom up motivated approach.

\begin{figure}
\centering
\scalebox{0.8}{
\begin{tikzpicture}
\begin{pgfonlayer}{nodelayer}
		\node [style=none] (0) at (-2.75, 3) {};
		\node [style=none] (1) at (3.25, -3) {};
		\node [style=none] (2) at (3.25, 3) {};
		\node [style=none] (3) at (-2.75, -3) {};
		\node [style=none] (4) at (3.5, 2.75) {};
		\node [style=none] (5) at (3, 3.25) {};
		\node [style=none] (6) at (-2.5, 3.25) {};
		\node [style=none] (7) at (-3, 2.75) {};
		\node [style=none] (8) at (-3, -2.75) {};
		\node [style=none] (9) at (-2.5, -3.25) {};
		\node [style=none] (10) at (3, -3.25) {};
		\node [style=none] (11) at (3.5, -2.75) {};
		\node [style=none] (12) at (-1, -0.75) {};
		\node [style=none] (13) at (-0.5, -1.25) {};
		\node [style=none] (14) at (1, -1.25) {};
		\node [style=none] (15) at (1.5, -0.75) {};
		\node [style=none] (16) at (0.75, -1) {};
		\node [style=none] (17) at (1.25, -0.5) {};
		\node [style=none] (18) at (-0.25, -1) {};
		\node [style=none] (19) at (-0.75, -0.5) {};
		\node [style=none] (20) at (-1, 0.75) {};
		\node [style=none] (21) at (-0.5, 1.25) {};
		\node [style=none] (22) at (1, 1.25) {};
		\node [style=none] (23) at (1.5, 0.75) {};
		\node [style=none] (24) at (1.25, 0.5) {};
		\node [style=none] (25) at (0.75, 1) {};
		\node [style=none] (26) at (-0.25, 1) {};
		\node [style=none] (27) at (-0.75, 0.5) {};
		\node [style=SmallCircleBrown] (28) at (-2.75, 3) {};
		\node [style=SmallCircleBrown] (29) at (3.25, 3) {};
		\node [style=SmallCircleBrown] (30) at (3.25, -3) {};
		\node [style=SmallCircleBrown] (31) at (-2.75, -3) {};
		\node [style=none] (32) at (0.25, 5) {$\partial X^\circ$};
		\node [style=none] (41) at (0.25, 2.5) {$ X^\circ$};
		\node [style=SmallCircleBrown] (42) at (0, -7.5) {};
		\node [style=none] (43) at (-4.25, -7.5) {};
		\node [style=Star] (44) at (0.25, 0) {};
		\node [style=none] (45) at (-2.75, -7.5) {};
		\node [style=none] (46) at (-3.5, -8) {$ \mathscr{S}$};
		\node [style=none] (47) at (0, -8) {$\partial \mathscr{S}=\mathscr{K}$};
		\node [style=Star] (48) at (-1.75, -7.5) {};
		\node [style=none] (49) at (-1.75, -8) {$ \mathscr{S}_0$};
		\node [style=none] (50) at (1.75, -7.5) {};
		\node [style=none] (51) at (3.25, -7.5) {};
		\node [style=none] (52) at (2.5, -6.75) {$T_\mathscr{K}\subset \partial X$};
		\node [style=none] (53) at (2.6, -8.25) {$\partial T_\mathscr{K}= \partial^2 X^\circ$};
		\node [style=none] (54) at (7.75, 3) {};
		\node [style=none] (55) at (13.75, -3) {};
		\node [style=none] (56) at (13.75, 3) {};
		\node [style=none] (57) at (7.75, -3) {};
		\node [style=none] (58) at (14, 2.75) {};
		\node [style=none] (59) at (13.5, 3.25) {};
		\node [style=none] (60) at (8, 3.25) {};
		\node [style=none] (61) at (7.5, 2.75) {};
		\node [style=none] (62) at (7.5, -2.75) {};
		\node [style=none] (63) at (8, -3.25) {};
		\node [style=none] (64) at (13.5, -3.25) {};
		\node [style=none] (65) at (14, -2.75) {};
		\node [style=none] (66) at (9.5, -0.75) {};
		\node [style=none] (67) at (10, -1.25) {};
		\node [style=none] (68) at (11.5, -1.25) {};
		\node [style=none] (69) at (12, -0.75) {};
		\node [style=none] (74) at (9.5, 0.75) {};
		\node [style=none] (75) at (10, 1.25) {};
		\node [style=none] (76) at (11.5, 1.25) {};
		\node [style=none] (77) at (12, 0.75) {};
		\node [style=SmallCircleBrown] (82) at (7.75, 3) {};
		\node [style=SmallCircleBrown] (83) at (13.75, 3) {};
		\node [style=SmallCircleBrown] (84) at (13.75, -3) {};
		\node [style=SmallCircleBrown] (85) at (7.75, -3) {};
		\node [style=none] (86) at (10.75, 5) {$\partial X^\circ$};
		\node [style=none] (87) at (10.75, 2.5) {$ X^\circ$};
		\node [style=Star] (88) at (10.75, 0) {};
		\node [style=none] (89) at (10.75, 0.875) {};
		\node [style=none] (90) at (10.75, -0.875) {};
		\node [style=none] (91) at (11.625, 0) {};
		\node [style=none] (92) at (9.875, 0) {};
		\node [style=none] (93) at (10.75, 0) {};
		\node [style=none] (94) at (0.25, -5) {(i)};
		\node [style=none] (95) at (10.75, -5) {(ii)};
		\node [style=none] (96) at (6, -8.5) {};
		\node [style=none] (97) at (6.5, -8) {};
		\node [style=none] (98) at (5, -8.5) {};
		\node [style=none] (99) at (4.5, -8) {};
		\node [style=none] (100) at (6.5, -7) {};
		\node [style=none] (101) at (6, -6.5) {};
		\node [style=none] (102) at (5, -6.5) {};
		\node [style=none] (103) at (4.5, -7) {};
		\node [style=none] (104) at (5.5, -9.25) {$\partial T_{\mathscr{S}}=\partial X^\circ$};
		\node [style=none] (105) at (7.75, -7.5) {};
		\node [style=none] (106) at (9.25, -7.5) {};
		\node [style=none] (107) at (8.5, -6.75) {$T_{\mathscr{S}}|_{r>0}$};
		\node [style=none] (108) at (8.5, -8.25) {$\partial T_{\mathscr{S}}|_{r>0}$};
		\node [style=none] (109) at (11.5, -6.75) {};
		\node [style=none] (110) at (11.5, -8.25) {};
		\node [style=none] (111) at (12.25, -7.5) {};
		\node [style=none] (112) at (10.75, -7.5) {};
		\node [style=none] (113) at (11.5, -7.5) {};
		\node [style=none] (114) at (11.5, -8.75) {$\partial X^\circ_{\text{retract}}$};
		\node [style=none] (115) at (13.75, -7.5) {};
		\node [style=none] (116) at (15.25, -7.5) {};
		\node [style=none] (117) at (14.5, -8) {$\partial X^\circ$};
		\node [style=none] (118) at (5.5, -10) {};
	\end{pgfonlayer}
	\begin{pgfonlayer}{edgelayer}
	\filldraw[fill=green!10, draw=green!15] (2.5, -7.5) ellipse (0.7cm and 0.25cm);
	\filldraw[fill=blue!20, draw=blue!25] (8.5, -7.5) ellipse (0.7cm and 0.25cm);
		\draw [style=ThickLine, bend right=45] (11.center) to (4.center);
		\draw [style=ThickLine, bend right=45] (5.center) to (6.center);
		\draw [style=ThickLine, bend right=45] (7.center) to (8.center);
		\draw [style=ThickLine, bend right=45] (9.center) to (10.center);
		\draw [style=RedLine] (0.center) to (1.center);
		\draw [style=RedLine] (2.center) to (3.center);
		\draw [style=BlueLine] (13.center) to (9.center);
		\draw [style=BlueLine] (8.center) to (12.center);
		\draw [style=BlueLine] (20.center) to (7.center);
		\draw [style=BlueLine] (6.center) to (21.center);
		\draw [style=BlueLine] (22.center) to (5.center);
		\draw [style=BlueLine] (4.center) to (23.center);
		\draw [style=BlueLine] (15.center) to (11.center);
		\draw [style=BlueLine] (14.center) to (10.center);
		\draw [style=PurpleLine, bend left=45, looseness=1.50] (27.center) to (19.center);
		\draw [style=PurpleLine, bend left=45, looseness=1.50] (18.center) to (16.center);
		\draw [style=PurpleLine, bend left=45, looseness=1.50] (17.center) to (24.center);
		\draw [style=PurpleLine, bend left=45, looseness=1.50] (25.center) to (26.center);
		\draw [style=PurpleLine, bend right, looseness=0.75] (17.center) to (16.center);
		\draw [style=PurpleLine, bend right, looseness=0.75] (25.center) to (24.center);
		\draw [style=PurpleLine, bend left, looseness=0.75] (26.center) to (27.center);
		\draw [style=PurpleLine, bend left, looseness=0.75] (19.center) to (18.center);
		\draw [style=PurpleLine, bend left, looseness=0.75] (27.center) to (26.center);
		\draw [style=PurpleLine, bend left, looseness=0.75] (25.center) to (24.center);
		\draw [style=PurpleLine, bend right, looseness=0.75] (16.center) to (17.center);
		\draw [style=PurpleLine, bend right, looseness=0.75] (19.center) to (18.center);
		\draw [style=BlueLine, bend right, looseness=0.75] (22.center) to (23.center);
		\draw [style=BlueLine, bend left, looseness=0.75] (15.center) to (14.center);
		\draw [style=BlueLine, bend right, looseness=0.75] (13.center) to (12.center);
		\draw [style=BlueLine, bend right, looseness=0.75] (20.center) to (21.center);
		\draw [style=BlueLine, bend left, looseness=0.75] (22.center) to (23.center);
		\draw [style=BlueLine, bend right, looseness=0.75] (15.center) to (14.center);
		\draw [style=BlueLine, bend left, looseness=0.75] (13.center) to (12.center);
		\draw [style=BlueLine, bend left] (20.center) to (21.center);
		\draw [style=GreenLine, bend right, looseness=0.75] (5.center) to (4.center);
		\draw [style=GreenLine, bend left, looseness=0.75] (5.center) to (4.center);
		\draw [style=GreenLine, bend right, looseness=0.75] (11.center) to (10.center);
		\draw [style=GreenLine, bend right, looseness=0.75] (10.center) to (11.center);
		\draw [style=GreenLine, bend left, looseness=0.75] (8.center) to (9.center);
		\draw [style=GreenLine, bend left, looseness=0.75] (9.center) to (8.center);
		\draw [style=GreenLine, bend right] (7.center) to (6.center);
		\draw [style=GreenLine, bend right, looseness=0.75] (6.center) to (7.center);
		\draw [style=RedLine] (43.center) to (45.center);
		\draw [style=GreenLine, bend right=90, looseness=0.50] (51.center) to (50.center);
		\draw [style=GreenLine, bend left=90, looseness=0.50] (51.center) to (50.center);
		\draw [style=ThickLine, bend right=45] (65.center) to (58.center);
		\draw [style=ThickLine, bend right=45] (59.center) to (60.center);
		\draw [style=ThickLine, bend right=45] (61.center) to (62.center);
		\draw [style=ThickLine, bend right=45] (63.center) to (64.center);
		\draw [style=RedLine] (54.center) to (55.center);
		\draw [style=RedLine] (56.center) to (57.center);
		\draw [style=BlueLine] (67.center) to (63.center);
		\draw [style=BlueLine] (62.center) to (66.center);
		\draw [style=BlueLine] (74.center) to (61.center);
		\draw [style=BlueLine] (60.center) to (75.center);
		\draw [style=BlueLine] (76.center) to (59.center);
		\draw [style=BlueLine] (58.center) to (77.center);
		\draw [style=BlueLine] (69.center) to (65.center);
		\draw [style=BlueLine] (68.center) to (64.center);
		\draw [style=GreenLine, bend right, looseness=0.75] (59.center) to (58.center);
		\draw [style=GreenLine, bend left, looseness=0.75] (59.center) to (58.center);
		\draw [style=GreenLine, bend right, looseness=0.75] (65.center) to (64.center);
		\draw [style=GreenLine, bend right, looseness=0.75] (64.center) to (65.center);
		\draw [style=GreenLine, bend left, looseness=0.75] (62.center) to (63.center);
		\draw [style=GreenLine, bend left, looseness=0.75] (63.center) to (62.center);
		\draw [style=GreenLine, bend right] (61.center) to (60.center);
		\draw [style=GreenLine, bend right, looseness=0.75] (60.center) to (61.center);
		\draw [style=BlueLine, in=180, out=45] (67.center) to (90.center);
		\draw [style=BlueLine, in=135, out=0] (90.center) to (68.center);
		\draw [style=BlueLine, in=-90, out=135] (69.center) to (91.center);
		\draw [style=BlueLine, in=-135, out=90] (91.center) to (77.center);
		\draw [style=BlueLine, in=0, out=-135] (76.center) to (89.center);
		\draw [style=BlueLine, in=315, out=-180] (89.center) to (75.center);
		\draw [style=BlueLine, in=90, out=-45] (74.center) to (92.center);
		\draw [style=BlueLine, in=45, out=-90] (92.center) to (66.center);
		\draw [style=PurpleLine] (89.center) to (93.center);
		\draw [style=PurpleLine] (93.center) to (91.center);
		\draw [style=PurpleLine] (90.center) to (93.center);
		\draw [style=PurpleLine] (93.center) to (92.center);
		\draw [style=PurpleLine, bend left=45, looseness=1.50] (103.center) to (99.center);
		\draw [style=PurpleLine, bend left=45, looseness=1.50] (98.center) to (96.center);
		\draw [style=PurpleLine, bend left=45, looseness=1.50] (97.center) to (100.center);
		\draw [style=PurpleLine, bend left=45, looseness=1.50] (101.center) to (102.center);
		\draw [style=PurpleLine, bend right, looseness=0.75] (97.center) to (96.center);
		\draw [style=PurpleLine, bend right, looseness=0.75] (101.center) to (100.center);
		\draw [style=PurpleLine, bend left, looseness=0.75] (102.center) to (103.center);
		\draw [style=PurpleLine, bend left, looseness=0.75] (99.center) to (98.center);
		\draw [style=PurpleLine, bend left, looseness=0.75] (103.center) to (102.center);
		\draw [style=PurpleLine, bend left, looseness=0.75] (101.center) to (100.center);
		\draw [style=PurpleLine, bend right, looseness=0.75] (96.center) to (97.center);
		\draw [style=PurpleLine, bend right, looseness=0.75] (99.center) to (98.center);
		\draw [style=BlueLine, bend right=90, looseness=0.50] (105.center) to (106.center);
		\draw [style=BlueLine, bend left=90, looseness=0.50] (105.center) to (106.center);
		\draw [style=PurpleLine] (109.center) to (113.center);
		\draw [style=PurpleLine] (113.center) to (111.center);
		\draw [style=PurpleLine] (110.center) to (113.center);
		\draw [style=PurpleLine] (113.center) to (112.center);
		\draw [style=ThickLine] (115.center) to (116.center);
	\end{pgfonlayer}
\end{tikzpicture}}
\caption{We sketch the non-compact manifold $X$. In the figure two flavor branes (red) intersect at $\mathscr{S}_0$. The tubular neighborhoods of the flavor branes (blue) and their asymptotic boundaries (green) are depicted. We shade these tubes in the legend, in subfigures (i), (ii) we only mark their boundaries. These tubes glue along the neighborhood associated with the intersection $\mathscr{S}_0$ (purple) to the tubular neighborhood $T_{\mathscr{S}}$ of the full singular locus $\mathscr{S}$. The complement $X^\circ = X\setminus T_{\mathscr{S}}$ lies `behind' the local model boundary $\partial T_{\mathscr{S}}$ or ``at infinity'' with respect to $T_{\mathscr{S}}$.
The deformation retraction shrinking the neighborhood of $\mathscr{S}_0$, or equivalently, growing the neighborhoods of $\mathscr{S}\setminus \mathscr{S}_0$ deforms subfigure (i) to subfigure (ii). Further, we have the converse limit, growing the purple tube until it fills $X$.}
\label{fig:GeoEng}
\end{figure}
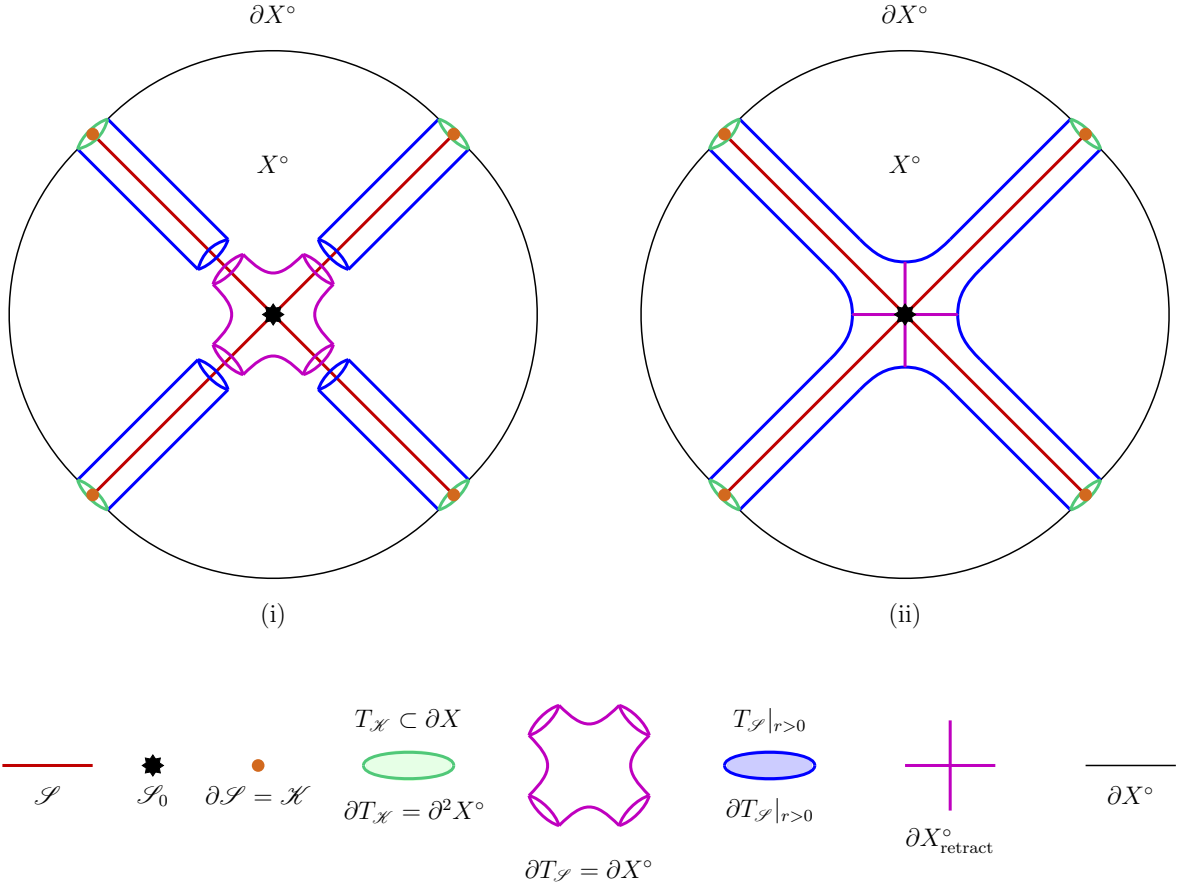

To begin, consider the non-compact manifold $X$ with singular locus $\mathscr{S}_0\subset \mathscr{S}\subset X$ and asymptotic boundary $\partial X$. The maximally singular sublocus $\mathscr{S}_0$ is compact and does not intersect $\partial X$. However, flavor branes may stretch to $\partial X$ and intersect it in $\mathscr{K}=\partial \mathscr{S}=\mathscr{S}\cap \partial X $. Further, we introduce the tubular neighborhoods $T_\mathscr{S}\subset X$  of $\mathscr{S}$ and $T_\mathscr{K}\subset \partial X$ of $\mathscr{K}$, and their complements
\be
X^\circ = X\setminus T_{\mathscr{S}}\,, \qquad (\partial X)^\circ =\partial X\setminus T_\mathscr{K}\,.
\ee
 We will often use the shorthand notation $(\partial X)^\circ\equiv \partial X^\circ$. The tubular neighborhood $T_{\mathscr{S}}$ is the local model for the bulk QFT$_D$. Both $X^\circ$ and $\partial X^\circ$ are manifolds with boundary. Their boundaries
\be
\partial (X^\circ)=D(\partial  \:\!   T_{\mathscr{S}})\,, \qquad \partial (\partial X)^\circ =\partial \:\! T_\mathscr{K}\,,
\ee
are of dimension $\dim X-1$ and $\dim X-2$, respectively, see figure \ref{fig:GeoEng}. Here $D(\partial  T_{\mathscr{S}})$ denotes the double of a manifold with boundary, given by two copies of $\partial  \:\!   T_{\mathscr{S}}$ glued along their boundary $\partial T_{\mathscr{K}}$. We will abbreviate\footnote{Here, `$\partial^2{}$' makes the nested structures and relevance of manifolds with corners explicit.} $\partial ( (\partial X)^\circ)\equiv \partial^2 X^\circ=\partial T_{\mathscr{K}}$.

Two very important, yet simple, relations are the homotopy equivalences
\be \label{eq:KeyRelations}
X  \sim  T_{\mathscr{S}}\,, \qquad \partial X^\circ  \sim \partial  \:\!   T_{\mathscr{S}}\,.
\ee
The excision operation $\circ$ and the boundary operation $\partial$ do not commute. Fundamentally, these relations are the reason why the construction we present in this section, which is based on $T_{\mathscr{S}}$, is applicable to the theory engineered by $X$.


In the setups we consider here, the asymptotic singular locus $\mathscr{K}$ is of fixed singularity type with no enhancements. We denote by $\mathscr{X}_f$ a model of the singularity, i.e., its normal geometry at some fixed point of $\mathscr{K}$. The tube $T_{\mathscr{K}}$ is modelled on a family of singularities $\{ \mathscr{X}_{f,k}\}_{k\:\!\in \mathscr{K}}$ and $T_{\mathscr{K}}$ is generically a fibration of $\mathscr{X}_f$ over $\mathscr{K}$ twisted by symmetries of $\mathscr{X}_f$. We write\footnote{ Here, $ \mathscr{X}_f \rtimes  \mathscr{K}$ denotes a fibration (twisted fiber product) of the singularity model $ \mathscr{X}_f $ over $  \mathscr{K}$. More generally we write $\textnormal{Fiber}\rtimes \textnormal{Base}$.}  $T_{\mathscr{K}}=  \mathscr{X}_f \rtimes \mathscr{K}$.
Further, note that the normal geometry $\mathscr{X}_f$ extends radially inwards and sets the singularity type along the full flavor locus. The coordinate, normal to the singular locus, i.e., parametrizing the radial coordinate of  $\mathscr{X}_f$, is denoted $x_\perp$.

Next we describe (non-flat) fibrations which will be the starting point for the supergravity compactification. We briefly summarize their structure here and then discuss them in greater detail. Underlying our constructions will be the projection
\be
\pi_{IJ}\,: \quad T_{\mathscr{S}} ~\rightarrow~ I\times J\,,
\ee
with intervals $I,J=[0,1]$ parametrized by $r,x_\perp$ respectively. From here, fixing $x_\perp$ we obtain the projection of a boundary slice
\be
\pi_{I}\,: \quad \partial T_{\mathscr{S}} ~\rightarrow~ I\,.
\ee
The generic fiber of both projections is topologically a copy of
\be
\partial T_{\mathscr{K}}=\partial^2 X^\circ=\partial \mathscr{X}_f \rtimes \partial \mathscr{S}=\partial \mathscr{X}_f \rtimes \mathscr{K}\,.
\ee
At the boundaries\,/\,corners of $I\times J$ the fiber can change, we refer to these fibers as exceptional. The supergravity compactification will be with respect to these generic and exceptional fibers and result in a theory supported on the $d$-dimensional spacetime of $\mathcal{T}_d$ times the two decompression dimensions $I\times J$.

\begin{figure}
\centering
\scalebox{0.9}{
\begin{tikzpicture}
	\begin{pgfonlayer}{nodelayer}
		\node [style=none] (0) at (-4, -1) {};
		\node [style=none] (1) at (-2, 1) {};
		\node [style=none] (2) at (2, -1) {};
		\node [style=none] (3) at (4, 1) {};
		\node [style=Star] (4) at (0, 0) {};
		\node [style=NodeCross] (5) at (-1.5, 0) {};
		\node [style=none] (6) at (-1.5, 0.25) {};
		\node [style=none] (7) at (-1.5, 2.25) {};
		\node [style=none] (8) at (0, 0.25) {};
		\node [style=none] (9) at (0, 2.25) {};
		\node [style=none] (10) at (-1.125, -0.325) {};
		\node [style=none] (11) at (-1.15, 0.25) {};
		\node [style=none] (13) at (1, 0) {};
		\node [style=none] (14) at (3.5, -0.5) {$\mathscr{S}$};
		\node [style=none] (15) at (-1.5, 2.75) {$\partial \mathscr{X}_f$};
		\node [style=none] (16) at (0, 2.75) {$\partial \mathscr{E}$};
		\node [style=none] (17) at (-2, 1.625) {$\pi_{\mathscr{S}}$};
		\node [style=none] (18) at (2, 0.25) {Sym$(\mathscr{X}_f)$};
		\node [style=none] (19) at (7, 0) {};
		\node [style=none] (20) at (11, 0) {};
		\node [style=Star] (21) at (11, 0) {};
		\node [style=NodeCross] (23) at (8, 0) {};
		\node [style=none] (24) at (8, 0.25) {};
		\node [style=none] (25) at (8, 2.25) {};
		\node [style=none] (26) at (11, 0.25) {};
		\node [style=none] (27) at (11, 2.25) {};
		\node [style=none] (31) at (8, 2.75) {$\partial \mathscr{X}_f \rtimes \partial \mathscr{S}$};
		\node [style=none] (32) at (11, 2.75) {$\partial \mathscr{E} \rtimes  \mathscr{S}_0$};
		\node [style=none] (33) at (7.5, 1.125) {$\pi_I$};
		\node [style=none] (34) at (9, -0.5) {$I$};
		\node [style=none] (35) at (11, -0.5) {$0$};
		\node [style=none] (36) at (0, -1.75) {(i)};
		\node [style=none] (37) at (9, -1.75) {(ii)};
		\node [style=none] (38) at (13, 0) {};
	\end{pgfonlayer}
	\begin{pgfonlayer}{edgelayer}
		\filldraw[fill=red!20, draw=red!20]  (-4, -1) -- (-2, 1) -- (4, 1) -- (2, -1) -- cycle;
		\draw [style=ThickLine] (1.center) to (3.center);
		\draw [style=ThickLine] (3.center) to (2.center);
		\draw [style=ThickLine] (2.center) to (0.center);
		\draw [style=ThickLine] (0.center) to (1.center);
		\draw [style=DottedLine] (7.center) to (6.center);
		\draw [style=DottedLine] (9.center) to (8.center);
		\draw [style=ArrowLineRight, in=45, out=90] (13.center) to (11.center);
		\draw [style=ThickLine, in=-90, out=-45] (10.center) to (13.center);
		\draw [style=ThickLine] (1.center) to (3.center);
		\draw [style=ThickLine] (3.center) to (2.center);
		\draw [style=ThickLine] (2.center) to (0.center);
		\draw [style=ThickLine] (0.center) to (1.center);
		\draw [style=DottedLine] (7.center) to (6.center);
		\draw [style=DottedLine] (9.center) to (8.center);
		\draw [style=ArrowLineRight, in=45, out=90] (13.center) to (11.center);
		\draw [style=ThickLine, in=-90, out=-45] (10.center) to (13.center);
		\draw [style=DottedLine] (25.center) to (24.center);
		\draw [style=DottedLine] (27.center) to (26.center);
		\draw [style=RedLine] (19.center) to (21);
	\end{pgfonlayer}
\end{tikzpicture}}
\caption{We sketch the pair of fibrations $\pi_{\mathscr{S}}: \partial \:\!  T_{\mathscr{S}} \rightarrow \mathscr{S}$ and $\pi_{I}: \partial \:\!  T_{\mathscr{S}} \rightarrow I$. The generic fibers are $\partial \mathscr{X}_f$ and $\partial \mathscr{X}_f \rtimes \partial \mathscr{S}$ while the exceptional fibers are $\partial \mathscr{E}$ and $\partial \mathscr{E} \rtimes  \mathscr{S}_0$,  respectively. The fibration $\pi_{\mathscr{S}}$ is twisted by the symmetry group $\textnormal{Sym}(\mathscr{X}_f)$.
}
\label{fig:Fibrations}
\end{figure}
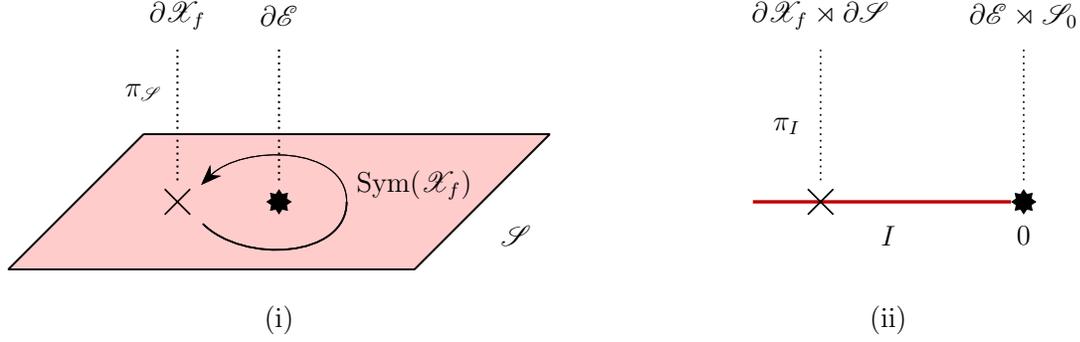

To discuss the above fibrations we begin by considering the related fibration $T_{\mathscr{S}}\rightarrow \mathscr{S}$ with generic fiber $\mathscr{X}_f$. Away from $\mathscr{S}_0$ the fibration is simply determined from the normal geometry to $\mathscr{S}\setminus \mathscr{S}_0$ which is modeled by assumption on $\mathscr{X}_f$. From here, fiberwise restricting to the boundary, we obtain the fibration
\be
\pi_{\mathscr{S}}\,:~ \partial \:\!  T_{\mathscr{S}}~\rightarrow~ \mathscr{S}\,.
\ee
 This fibration has exactly one type of exceptional fiber $\partial \mathscr{E}$ projecting to $\mathscr{S}_0$. For example, when $\mathscr{S}_0$ is a point we have
\be\label{eq:Retract}
\partial \mathscr{E} = \partial \:\! T_{\mathscr{S}}|_{\textnormal{retract}} = \partial X^\circ |_{\textnormal{retract}} \,,
\ee
where $|_{\textnormal{retract}}$ abbreviates a deformation retraction onto a lower dimensional skeleton. 
See subfigure (i) of figure \ref{fig:Fibrations} for a sketch of $\pi_{\mathscr{S}}$. Further, immediately compare subfigures (i) of figure \ref{fig:SymTFTColorFlavor} and \ref{fig:Fibrations} making the internal geometry related to the former clear.

We now produce a second, very related, fibration assuming that $\mathscr{S}$ respects the conical features of $X$. We assume that $\mathscr{S}$ is itself conical with link $\mathscr{K}=\partial \mathscr{S}$ and radial coordinate $0\leq r\leq 1$ such that at $r=0$ we find $\partial \mathscr{S}$ to collapse to $\mathscr{S}_0$. In that case, we may lift the link $\partial \mathscr{S}$ into the fiber and fiber the space exclusively over the radial direction $I=[0,1]$. This results in the fibration
\be
\pi_{I}\,:~ \partial \:\!  T_{\mathscr{S}}~\rightarrow~ I\,,
\ee
with generic fiber $\partial \mathscr{X}_f \rtimes \partial \mathscr{S}$. At $r=0$ the generic fiber degenerates to $\partial \mathcal{E} \rtimes  \mathscr{S}_0$. See subfigure (ii) of figure \ref{fig:Fibrations} for a sketch of $\pi_{I}$. Filling in the normal direction to the singular locus we have the fibration
\be \label{eq:SquareFibration}
\pi_{IJ}\,:\quad T_{\mathscr{S}}~\rightarrow~ I\times J\,,
\ee
mapping the locus $\mathscr{S}_0$ to $r,x_\perp=0$ and $\mathscr{S}$ at $x_\perp=0$. Copies of $ \partial X^\circ |_{\textnormal{retract}}$ are mapped to $r=0$ at fixed $x_\perp\neq 0$. The generic point with $r,x_\perp\neq 0$ is mapped onto by a copy of $\partial T_{\mathscr{K}}=\partial \mathscr{X}_f \rtimes \partial \mathscr{S}$. \medskip

{\bf Example} We make the above explicit for the Calabi-Yau orbifold $\mathbb{C}^3/\Z_{2n}$ with quotient $(z_1,z_2,z_3)\sim(\omega z_1, \omega z_2, \omega^{2n-2}z_3)$ and weights $(1,1,2n-2)$ and root of unity $\omega=\exp(2\pi i /2n)$. Here $T_{\mathscr{S}}$ is a tubular neighbourhood of the singularities $\mathscr{S}=\{z_1,z_2=0\}$ and a 6-dimensional manifold with boundary. With coordinates $r=|z_3|^2$ and $x_\perp=|z_1|^2+|z_2|^2$, we have
\be
T_{\mathscr{S}}=\{(z_1,z_2,z_3)\in X\,|\, x_\perp =|z_1|^2+|z_2|^2\leq \epsilon\}\,,
\ee
for some finite $\epsilon>0$. The first homotopy in \eqref{eq:KeyRelations} is realized by the deformation induced by growing $\epsilon $. To see the second homotopy take a model of the asymptotic boundary to be the copy of $S^5/\Z_{2n}$ cut out by $|z_1|^2+|z_2|^2+|z_3|^2=1$, similarly restricting $T_\mathscr{S}$ to the unit ball, and consider a scaling of the third coordinate until we realize the projection $(z_1,z_2,z_3)\mapsto (z_1,z_2,\kappa z_3)$ with $\kappa^2=(1-\epsilon)/|z_3|^2$ which maps the boundary of the cylinder $T_{\mathscr{S}}$ onto asymptotic infinity.

The model of the flavor singularity is $\mathscr{X}_f=\mathbb{C}^2/\Z_2$ and consequently $\partial T_{\mathscr{K}}=S^3/\Z_2 \rtimes S^1$ where the $S^1$ is the angular circle in $\mathbb{C}/\Z_n$ linking the codimension-6 locus $\mathscr{S}_0=\{z_1=z_2=z_3=0\}$ within the flavor brane. The coordinate $r$ parametrizes the radius of $\mathbb{C}/\Z_n$ while $x_\perp$ parametrizes the radius in $\mathscr{X}_f$. We have have the projections
\be \ba
\pi_{\mathscr{S}}\,&:~ T_{\mathscr{S}}\rightarrow  \mathscr{S}\,, && (z_1,z_2,z_3)~\mapsto~z_3\\
\pi_{I}\,&:~ T_{\mathscr{S}}\rightarrow  I\,, && (z_1,z_2,z_3)~\mapsto~|z_3|^2=r \\
\pi_{IJ}\,&:~ T_{\mathscr{S}}\rightarrow  I\times J\,, \qquad && (z_1,z_2,z_3)~\mapsto~(|z_3|^2,|z_1|^2+|z_2|^2)=(r,x_\perp)\,.
\ea \ee
Consider $\pi_{\mathscr{S}}$, at $z_3=0$ the normal geometry is $\mathbb{C}^2/\Z_{2n}$ and therefore $\partial \mathscr{E}=S^3/\Z_{2n}$ for this example, otherwise we have the generic fiber $S^3/\Z_2$ projecting onto $z_3\neq 0$. All other fibers follow from this observation.

\begin{figure}
\centering
\scalebox{0.8}{
\begin{tikzpicture}
	\begin{pgfonlayer}{nodelayer}
		\node [style=none] (0) at (0, 0) {};
		\node [style=none] (1) at (0, 3) {};
		\node [style=none] (2) at (3, 0) {};
		\node [style=SmallCircleRed] (3) at (0, 0) {};
		\node [style=SmallCircleBrown] (4) at (2, 0) {};
		\node [style=SmallCircleGrey] (5) at (0, 2) {};
		\node [style=none] (6) at (-0.65, 2.8) {$x_\perp$};
		\node [style=none] (7) at (2.8, -0.5) {$r$};
		\node [style=none] (8) at (0.625, 0.75) {};
		\node [style=none] (9) at (-1.25, 0.75) {};
		\node [style=none] (10) at (1.25, 1.25) {};
		\node [style=none] (11) at (2.25, 2.25) {};
		\node [style=none] (12) at (-2, 0.75) {$\mathbb{C}^3/\Z_{2n}$};
		\node [style=none] (13) at (2.75, 2.75) {asymptotic $S^5/\Z_{2n}$};
	\end{pgfonlayer}
	\begin{pgfonlayer}{edgelayer}
		\filldraw[fill=gray!50, draw=gray!50]  (0, 0) -- (0, 2) -- (2, 0) -- cycle;
		\draw [style=ArrowLineRight] (0.center) to (2.center);
		\draw [style=ArrowLineRight] (0.center) to (1.center);
		\draw [style=RedLine] (3) to (4);
		\draw [style=ArrowLineRight] (11.center) to (10.center);
		\draw [style=ArrowLineRight] (9.center) to (8.center);
		\draw [style=GreenLine] (5) to (4);
		\draw [style=PurpleLine] (5) to (3);
	\end{pgfonlayer}
\end{tikzpicture}
}
\caption{We sketch a projection of $X=\mathbb{C}^3/\Z_{2n}$ onto a 2-simplex with local coordinates $r,x_\perp$. The asymptotic boundary $\partial X=S^5/\Z_{2n}$ is projected on to the green line. }
\label{fig:Projection}
\end{figure}
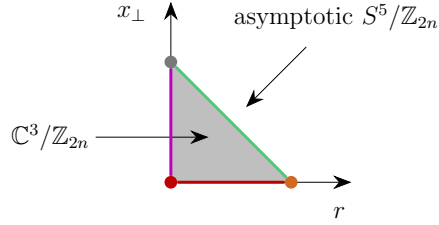

Further, we make the bottom line of figure \ref{fig:Quesidilla} explicit and explain how $I\times J$ can deform to a 2-simplex with the same local coordinates. For this note that the asymptotic boundary $S^5/\Z_{2n}$ can be viewed as cut out by
\be\label{eq:cutout}
1=|z_1|^2+|z_2|^2+|z_3|^2=r+x_\perp\,.
\ee
Now, in the quadrant $\mathbb{R}_{\geq 0}\times\mathbb{R}_{\geq 0} $ parametrized by $r,x_\perp$, the asymptotic $S^5/\Z_{2n}$ maps onto an interval. The space $\mathbb{C}^3/\Z_{2n}$, viewed as $r+x_\perp\leq 1$, maps onto a 2-simplex with 1-faces $r=0$ and $x_\perp \in [0,1]$, and  $x_\perp=0$ and $r \in [0,1]$, and \eqref{eq:cutout}. See figure \ref{fig:Projection}. \medskip

\begin{figure}
\centering
\scalebox{0.95}{
\begin{tikzpicture}
	\begin{pgfonlayer}{nodelayer}
		\node [style=none] (0) at (3, -1) {};
		\node [style=none] (1) at (3, -5) {};
		\node [style=none] (2) at (-3, -5) {};
		\node [style=none] (3) at (-3, -1) {};
		\node [style=SmallCircleRed] (4) at (3, -5) {};
		\node [style=SmallCircleBlue] (5) at (3, -1) {};
		\node [style=none] (6) at (3.25, 6.25) {};
		\node [style=none] (7) at (3.25, 5.5) {};
		\node [style=none] (8) at (3, 6.25) {};
		\node [style=none] (9) at (3, 5.5) {};
		\node [style=none] (10) at (3.25, 5) {};
		\node [style=none] (11) at (3.25, 4.25) {};
		\node [style=none] (12) at (3, 5) {};
		\node [style=none] (13) at (3, 4.25) {};
		\node [style=none] (14) at (3.25, 3.75) {};
		\node [style=none] (15) at (3.25, 3) {};
		\node [style=none] (16) at (3, 3.75) {};
		\node [style=none] (17) at (3, 3) {};
		\node [style=none] (18) at (3.25, 2.5) {};
		\node [style=none] (19) at (3.25, 1.75) {};
		\node [style=none] (20) at (3, 2.5) {};
		\node [style=none] (21) at (3, 1.75) {};
		\node [style=none] (23) at (4, 5.25) {};
		\node [style=none] (24) at (4, 4) {};
		\node [style=none] (25) at (4, 2.75) {};
		\node [style=none] (26) at (4, 1.75) {};
		\node [style=none] (27) at (4, 6.25) {};
		\node [style=none] (32) at (5.25, 3.9325) {};
		\node [style=none] (33) at (6, 3.9325) {};
		\node [style=none] (34) at (6.25, 4.0625) {};
		\node [style=none] (35) at (5, 4.0625) {};
		\node [style=none] (40) at (-3, 6.25) {};
		\node [style=none] (41) at (-3, 5.5) {};
		\node [style=none] (42) at (-3, 5) {};
		\node [style=none] (43) at (-3, 4.25) {};
		\node [style=none] (44) at (-3, 3.75) {};
		\node [style=none] (45) at (-3, 3) {};
		\node [style=none] (46) at (-3, 2.5) {};
		\node [style=none] (47) at (-3, 1.75) {};
		\node [style=none] (49) at (0, -5.5) {$\mathcal{T}_{d+1}$};
		\node [style=none] (50) at (3.625, -5.5) {$\mathcal{T}_{d}$};
		\node [style=none] (51) at (0, -0.5) {$\mathbb{B}^{(r,1)}_{d+1}$};
		\node [style=none] (52) at (3.75, -0.5) {$\mathbb{B}^{(0,1)}_{d}$};
		\node [style=none] (53) at (0, -2.75) {$\mathcal{S}_{d+2}$};
		\node [style=none] (54) at (4, -2.75) {$\mathcal{B}_{d+1}$};
		\node [style=none] (56) at (7.5, 4) {$\partial X^\circ$};
		\node [style=none] (57) at (-7.5, -1) {};
		\node [style=none] (58) at (0, -3.375) {Determined by $ \partial^2 X^\circ$};
		\node [style=none] (59) at (5.375, -3.375) {Determined by $\partial X^\circ$};
		\node [style=none] (61) at (0, 1.375) {};
		\node [style=none] (62) at (0, 0.25) {};
		\node [style=none] (64) at (-2.75, 5.875) {};
		\node [style=none] (65) at (2.75, 5.875) {};
		\node [style=none] (66) at (-2.75, 4.625) {};
		\node [style=none] (67) at (2.75, 4.625) {};
		\node [style=none] (68) at (-2.75, 3.375) {};
		\node [style=none] (69) at (2.75, 3.375) {};
		\node [style=none] (70) at (-2.75, 2.125) {};
		\node [style=none] (71) at (2.75, 2.125) {};
		\node [style=none] (72) at (3.5, 2.125) {};
		\node [style=none] (73) at (3.5, 3.375) {};
		\node [style=none] (74) at (4.5, 4) {};
		\node [style=none] (75) at (3.5, 4.625) {};
		\node [style=none] (76) at (3.5, 5.875) {};
		\node [style=Star] (77) at (4.5, 4) {};
		\node [style=none] (80) at (1.75, 9) {};
		\node [style=none] (81) at (2.25, 9.5) {};
		\node [style=none] (82) at (2.25, 13) {};
		\node [style=none] (83) at (1.75, 13.5) {};
		\node [style=none] (84) at (-1.75, 13.5) {};
		\node [style=none] (85) at (-2.25, 13) {};
		\node [style=none] (86) at (-2.25, 9.5) {};
		\node [style=none] (87) at (-1.75, 9) {};
		\node [style=Star] (88) at (0, 11.25) {};
		\node [style=none] (89) at (2.05, 9.2) {};
		\node [style=none] (90) at (2.05, 13.3) {};
		\node [style=none] (91) at (-2.05, 13.3) {};
		\node [style=none] (92) at (-2.05, 9.2) {};
		\node [style=none] (93) at (0.75, 12.5) {};
		\node [style=none] (94) at (1.25, 12) {};
		\node [style=none] (95) at (1.25, 10.5) {};
		\node [style=none] (96) at (0.75, 10) {};
		\node [style=none] (97) at (-0.75, 10) {};
		\node [style=none] (98) at (-1.25, 10.5) {};
		\node [style=none] (99) at (-1.25, 12) {};
		\node [style=none] (100) at (-0.75, 12.5) {};
		\node [style=none] (101) at (0.5, 12.25) {};
		\node [style=none] (102) at (1, 11.75) {};
		\node [style=none] (103) at (1, 10.75) {};
		\node [style=none] (104) at (0.5, 10.25) {};
		\node [style=none] (105) at (-0.5, 10.25) {};
		\node [style=none] (106) at (-1, 10.75) {};
		\node [style=none] (107) at (-1, 11.75) {};
		\node [style=none] (108) at (-0.5, 12.25) {};
		\node [style=none] (109) at (-2.25, 13) {};
		\node [style=none] (110) at (-1.75, 13.5) {};
		\node [style=none] (111) at (1.75, 9) {};
		\node [style=none] (112) at (2.25, 9.5) {};
		\node [style=none] (113) at (2.25, 13) {};
		\node [style=none] (114) at (1.75, 13.5) {};
		\node [style=none] (115) at (-1.75, 9) {};
		\node [style=none] (116) at (-2.25, 9.5) {};
		\node [style=none] (117) at (0, 8.25) {};
		\node [style=none] (118) at (0, 6.75) {};
		\node [style=none] (119) at (0, 7) {};
		\node [style=none] (120) at (0.5, 0.9) {$\pi_I$};
		\node [style=SmallCircleGreen] (121) at (-3, -1) {};
		\node [style=SmallCircleBrown] (122) at (-3, -5) {};
		\node [style=none] (123) at (-3.75, 6.5) {};
		\node [style=none] (124) at (-4.25, 4) {};
		\node [style=none] (125) at (-3.75, 1.5) {};
		\node [style=none] (126) at (-4.875, 4) {$T_{\mathscr{S}}$};
	\end{pgfonlayer}
	\begin{pgfonlayer}{edgelayer}
		\filldraw[fill=gray!50, draw=gray!50]  (-3, -1) -- (3, -1) -- (3, -5) -- (-3, -5) -- cycle;
		\draw [style=RedLine] (2.center) to (1.center);
		\draw [style=BlueLine] (3.center) to (0.center);
		\draw [style=PurpleLine] (0.center) to (1.center);
		\draw [style=PurpleLine, bend left=15, looseness=0.75] (6.center) to (7.center);
		\draw [style=PurpleLine, bend right=15, looseness=0.75] (6.center) to (7.center);
		\draw [style=BlueLine, bend right=15, looseness=0.75] (8.center) to (9.center);
		\draw [style=BlueLine, bend left=15, looseness=0.75] (8.center) to (9.center);
		\draw [style=PurpleLine, bend left=15, looseness=0.75] (10.center) to (11.center);
		\draw [style=PurpleLine, bend right=15, looseness=0.75] (10.center) to (11.center);
		\draw [style=BlueLine, bend right=15, looseness=0.75] (12.center) to (13.center);
		\draw [style=BlueLine, bend left=15, looseness=0.75] (12.center) to (13.center);
		\draw [style=PurpleLine, bend left=15, looseness=0.75] (14.center) to (15.center);
		\draw [style=PurpleLine, bend right=15, looseness=0.75] (14.center) to (15.center);
		\draw [style=BlueLine, bend right=15, looseness=0.75] (16.center) to (17.center);
		\draw [style=BlueLine, bend left=15, looseness=0.75] (16.center) to (17.center);
		\draw [style=PurpleLine, bend left=15, looseness=0.75] (18.center) to (19.center);
		\draw [style=PurpleLine, bend right=15, looseness=0.75] (18.center) to (19.center);
		\draw [style=BlueLine, bend right=15, looseness=0.75] (20.center) to (21.center);
		\draw [style=BlueLine, bend left=15, looseness=0.75] (20.center) to (21.center);
		\draw [style=PurpleLine, in=90, out=0, looseness=0.75] (7.center) to (23.center);
		\draw [style=PurpleLine, in=0, out=-90, looseness=0.75] (23.center) to (10.center);
		\draw [style=PurpleLine, in=90, out=0, looseness=0.75] (11.center) to (24.center);
		\draw [style=PurpleLine, in=0, out=-90, looseness=0.75] (24.center) to (14.center);
		\draw [style=PurpleLine, in=90, out=0, looseness=0.75] (15.center) to (25.center);
		\draw [style=PurpleLine, in=0, out=-90, looseness=0.75] (25.center) to (18.center);
		\draw [style=PurpleLine] (19.center) to (26.center);
		\draw [style=PurpleLine] (6.center) to (27.center);
		\draw [style=PurpleLine, bend left=90, looseness=2.00] (27.center) to (26.center);
		\draw [style=GreenLine, bend right=15, looseness=0.75] (40.center) to (41.center);
		\draw [style=GreenLine, bend left=15, looseness=0.75] (40.center) to (41.center);
		\draw [style=GreenLine, bend right=15, looseness=0.75] (42.center) to (43.center);
		\draw [style=GreenLine, bend left=15, looseness=0.75] (42.center) to (43.center);
		\draw [style=GreenLine, bend right=15, looseness=0.75] (44.center) to (45.center);
		\draw [style=GreenLine, bend left=15, looseness=0.75] (44.center) to (45.center);
		\draw [style=GreenLine, bend right=15, looseness=0.75] (46.center) to (47.center);
		\draw [style=GreenLine, bend left=15, looseness=0.75] (46.center) to (47.center);
		\draw [style=BlueLine] (40.center) to (8.center);
		\draw [style=BlueLine] (9.center) to (41.center);
		\draw [style=BlueLine] (42.center) to (12.center);
		\draw [style=BlueLine] (13.center) to (43.center);
		\draw [style=BlueLine] (44.center) to (16.center);
		\draw [style=BlueLine] (17.center) to (45.center);
		\draw [style=BlueLine] (46.center) to (20.center);
		\draw [style=BlueLine] (21.center) to (47.center);
		\draw [style=ArrowLineRight] (61.center) to (62.center);
		\draw [style=RedLine] (70.center) to (71.center);
		\draw [style=RedLine] (68.center) to (69.center);
		\draw [style=RedLine] (66.center) to (67.center);
		\draw [style=RedLine] (64.center) to (65.center);
		\draw [style=RedLine, in=60, out=0] (76.center) to (74.center);
		\draw [style=RedLine, in=120, out=0, looseness=0.75] (75.center) to (74.center);
		\draw [style=RedLine, in=0, out=-120] (74.center) to (73.center);
		\draw [style=RedLine, in=-60, out=0] (72.center) to (74.center);
		\draw [style=ThickLine, bend right=45, looseness=0.75] (87.center) to (80.center);
		\draw [style=ThickLine, bend left=45, looseness=0.75] (84.center) to (83.center);
		\draw [style=ThickLine, bend left=45, looseness=0.75] (82.center) to (81.center);
		\draw [style=ThickLine, bend left=45, looseness=0.75] (86.center) to (85.center);
		\draw [style=RedLine] (92.center) to (90.center);
		\draw [style=RedLine] (91.center) to (89.center);
		\draw [style=PurpleLine, bend right=45, looseness=1.55] (108.center) to (101.center);
		\draw [style=PurpleLine, bend right=45, looseness=1.55] (102.center) to (103.center);
		\draw [style=PurpleLine, bend left=315,  looseness=1.55] (104.center) to (105.center);
		\draw [style=PurpleLine, bend right=45, looseness=1.55] (106.center) to (107.center);
		\draw [style=BlueLine] (85.center) to (99.center);
		\draw [style=BlueLine] (100.center) to (84.center);
		\draw [style=BlueLine] (93.center) to (83.center);
		\draw [style=BlueLine] (94.center) to (82.center);
		\draw [style=BlueLine] (95.center) to (81.center);
		\draw [style=BlueLine] (80.center) to (96.center);
		\draw [style=BlueLine] (97.center) to (87.center);
		\draw [style=BlueLine] (86.center) to (98.center);
		\draw [style=PurpleLine, bend right=15] (101.center) to (102.center);
		\draw [style=PurpleLine, bend right=15] (103.center) to (104.center);
		\draw [style=PurpleLine, bend right=15] (105.center) to (106.center);
		\draw [style=PurpleLine, bend right=15] (107.center) to (108.center);
		\draw [style=PurpleLine, bend left=15] (107.center) to (108.center);
		\draw [style=PurpleLine, bend left=15] (101.center) to (102.center);
		\draw [style=PurpleLine, bend left=15] (103.center) to (104.center);
		\draw [style=PurpleLine, bend left=15] (105.center) to (106.center);
		\draw [style=BlueLine, bend left=15] (100.center) to (99.center);
		\draw [style=BlueLine, bend left=15] (99.center) to (100.center);
		\draw [style=BlueLine, bend right=15] (93.center) to (94.center);
		\draw [style=BlueLine, bend left=15] (95.center) to (96.center);
		\draw [style=BlueLine, bend left=15] (97.center) to (98.center);
		\draw [style=BlueLine, bend left=15] (98.center) to (97.center);
		\draw [style=BlueLine, bend left=15] (96.center) to (95.center);
		\draw [style=BlueLine, bend right=15] (94.center) to (93.center);
		\draw [style=GreenLine, bend left=15] (110.center) to (109.center);
		\draw [style=GreenLine, bend left=15] (109.center) to (110.center);
		\draw [style=GreenLine, bend left=15] (112.center) to (111.center);
		\draw [style=GreenLine, bend left=15] (111.center) to (112.center);
		\draw [style=GreenLine, bend left=15] (113.center) to (114.center);
		\draw [style=GreenLine, bend left=15] (114.center) to (113.center);
		\draw [style=GreenLine, bend left=15] (115.center) to (116.center);
		\draw [style=GreenLine, bend left=15] (116.center) to (115.center);
		\draw [style=ArrowLineRight] (119.center) to (118.center);
		\draw [style=ThickLine, snake it] (117.center) to (119.center);
		\draw [style=GreenLine] (121.center) to (122.center);
		\draw [style=ThickLine, in=0, out=-180, looseness=0.75] (123.center) to (124.center);
		\draw [style=ThickLine, in=180, out=0, looseness=0.75] (124.center) to (125.center);
	\end{pgfonlayer}
\end{tikzpicture}}
\caption{We sketch how the geometric structures map project onto $(r,x_\perp)$. We show the projection for the tubular neighbourhood $T_\mathscr{S}$. The homotopy $X\sim T_{\mathscr{S}}$ realized by growing the purple region / shrinking the blue tubes in the top figure gives a similar projection for $X$. The arrow denotes fiberwise compactification with respect to the fibration $\pi_I$. The boundary conditions (green) set at metric infinity of $X$ are unspecified in an open cheesesteak.}
\label{fig:CheeseGeometry}
\end{figure}
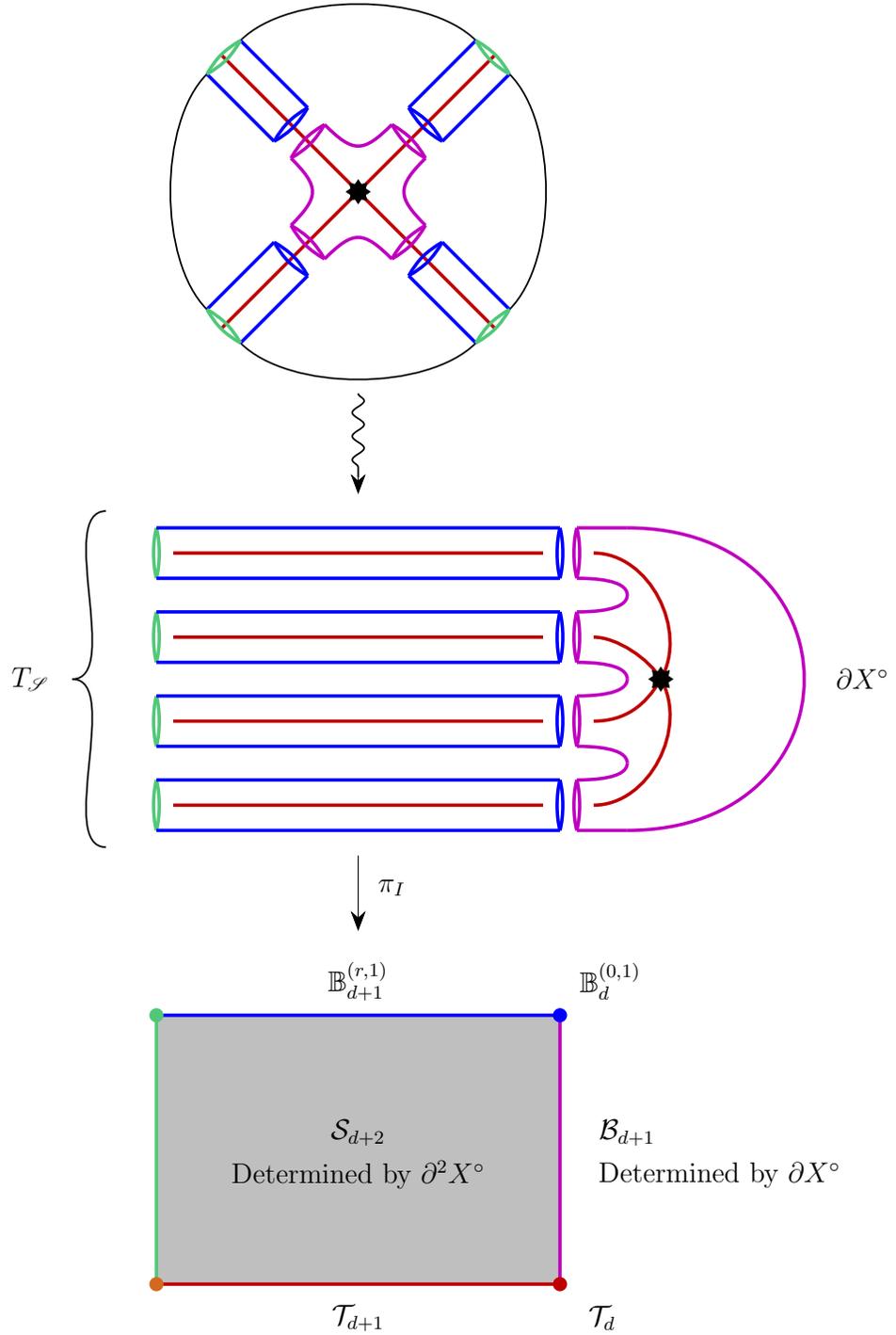

Let us now return to our general discussion. We identify the internal geometry corresponding to bulk, edges and corner of the open cheesesteak as already sketched in figure \ref{fig:GeoEng}. The bulk corresponds to a two-parameter family of copies of $\partial^2 X^\circ$. There are three edges, two of which are parametrized by $r$ with $x_\perp=0,1$ respectively and one parametrized by $x_\perp$ with $r=0$. The former pair corresponds to a 1-parameter family of copies of $\mathscr{K}=\partial{\mathscr{S}}$ and $\partial^2 X^\circ$ respectively. The latter corresponds to a 1-parameter family of copies of $\partial X^\circ$ (a fattening of $\partial \mathscr{E}$). The corner at $(r,x_\perp)=(0,1)$ corresponds to a copy of $\partial X^\circ$ and the corner at $(r,x_\perp)=(1,0)$ corresponds to $\mathscr{S}_0$. See figure \ref{fig:CheeseGeometry}.

From the perspective of the relative bulk QFT only edges and corners with $x_\perp=1$ are ``at infinity.'' Here, boundary conditions must be imposed to realize an absolute theory. From the perspective of the corner $\mathcal{T}_d$ both edges and corner with $x_\perp=1$ and\,/\,or $r=1$ are ``at infinity'' and any limit localizing modes in $d$ dimensions needs to be supplemented with boundary conditions at $r=1$ if it is to produce and absolute theory in $d$ dimensions.

\subsection{Defect Groups}
\label{sec:DefectandSym}

In the preceeding section we described a fibration of the geometry $ T_{\mathscr{S}}\sim X$ by fibers labelled by the coordinates $r,x_\perp$. The two symmetry theories $\mathcal{S}_{d+2}$ and $\mathcal{S}_{d+1}'=\mathcal{B}_{d+1}$ will derive from compactification of some supergravity theory on these fibers to $d+2$ or $d+1$ dimensions. They are in part characterized by their operator content. Favorably, the stringy construction of many such operators can be realized without much of this field theory analysis: they result from branes wrapped on cycles of the geometry following \cite{Heckman:2022muc}, see also \cite{GarciaEtxebarria:2022vzq,Apruzzi:2022rei}.

Here, we therefore take the perspective of studying $\mathcal{S}_{d+2}$ and $\mathcal{S}_{d+1}'=\mathcal{B}_{d+1}$ via the top down construction of their operators which are characterized by the kind of $p$-brane wrapped and its wrapping locus, a relative homology cycle. With this, we now extend the notion of defect group \cite{DelZotto:2015isa} to the QFT system engineered by $\mathscr{S}_0\subset \mathscr{S}$, i.e., the flavor brane theory $\mathcal{T}_{d+1}$ with end of the world theory $\mathcal{T}_{d}$. We further discuss similar generalizations in the constructions of symmetry operators.

To motivate the necessity of these considerations, consider again the example of M-theory on $\mathbb{C}^3/\Z_{2n}$. The boundary homology groups\footnote{Here and throughout this paper, unless stated otherwise, we denote singular homology and cohomology groups with integer coefficient of some space $X$ by $H_n(X)\equiv H_n(X;\Z)$ and $H^n(X)\equiv H^n(X; \Z)$ respectively.}  we focus on are
\be\label{eq:seedexample}
H_1(S^5/\Z_{2n})\cong \Z_{n}\,, \qquad H_3(S^5/\Z_{2n})\cong \Z_{2n}\,.
\ee
In an electric frame, wrapping M2-branes over cones on the 1-cycles produces defect line operators charged under a $\Z_n$ 1-form symmetry. Naively, wrapping M5-branes on asymptotic 3-cycles constructs the corresponding symmetry operators. We expect a 1:1 correspondence between these objects,  however, the count is off, $\Z_n\neq \Z_{2n}$. Clearly we ought to restrict the 3-cycles which link non-trivially with the 1-cycles, but even so, it remains to settle how the unaccounted for wrappings are to be interpreted and specify ``linking'' in singular spaces.

\begin{figure}
\centering
\scalebox{0.9}{
\begin{tikzpicture}
	\begin{pgfonlayer}{nodelayer}
		\node [style=none] (0) at (-1.5, 0) {};
		\node [style=none] (1) at (-1.5, 2.5) {};
		\node [style=none] (3) at (-5.5, 0) {};
		\node [style=SmallCircleRed] (5) at (-1.5, 0) {};
		\node [style=SmallCircleBrown] (6) at (-3.5, 0) {};
		\node [style=none] (7) at (-3.5, 1.5) {};
		\node [style=none] (8) at (-3.5, 2) {};
		\node [style=none] (9) at (-3.5, -1.5) {(i)};
		\node [style=none] (10) at (-1.5, 3) {};
		\node [style=none] (11) at (5.5, 0) {};
		\node [style=none] (12) at (5.5, 2.5) {};
		\node [style=none] (13) at (1.5, 0) {};
		\node [style=SmallCircleRed] (14) at (5.5, 0) {};
		\node [style=SmallCircleBrown] (15) at (5.5, 0) {};
		\node [style=none] (16) at (5.5, 1.5) {};
		\node [style=none] (17) at (5.5, 2) {};
		\node [style=none] (18) at (3.5, -1.5) {(ii)};
		\node [style=none] (19) at (5.5, 3) {};
		\node [style=none] (20) at (-6, 0) {};
		\node [style=none] (21) at (1, 0) {};
		\node [style=none] (22) at (3.5, -0.5) {$\mathcal{T}_{d+1}$};
		\node [style=none] (23) at (-3.5, -0.5) {$\mathcal{T}_{d+1}$};
		\node [style=none] (24) at (6, -0.5) {$\mathcal{T}_{d}$};
		\node [style=none] (25) at (-1, -0.5) {$\mathcal{T}_{d}$};
		\node [style=none] (24) at (6.25, 1.5) {$\mathcal{B}_{d+1}$};
		\node [style=none] (25) at (-0.75, 1.5) {$\mathcal{B}_{d+1}$};
	\end{pgfonlayer}
	\begin{pgfonlayer}{edgelayer}
	\filldraw[fill=gray!50, draw=gray!50]  (1.5, 0) -- (5.5, 0) -- (5.5, 2.5) -- (1.5, 2.5) -- cycle;
	\filldraw[fill=gray!50, draw=gray!50]   (-1.5, 0) -- (-5.5, 0) -- (-5.5, 2.5) -- (-1.5, 2.5) -- cycle;
		\draw [style=RedLine] (3.center) to (0.center);
		\draw [style=PurpleLine] (1.center) to (0.center);
		\draw [style=DottedBrown] (7.center) to (8.center);
		\draw [style=BrownLine, snake it] (7.center) to (6);
		\draw [style=DottedPurple] (10.center) to (1.center);
		\draw [style=RedLine] (13.center) to (11.center);
		\draw [style=PurpleLine] (12.center) to (11.center);
		\draw [style=DottedBrown] (16.center) to (17.center);
		\draw [style=BrownLine, snake it] (16.center) to (15);
		\draw [style=DottedPurple] (19.center) to (12.center);
		\draw [style=DottedRed] (3.center) to (20.center);
		\draw [style=DottedRed] (13.center) to (21.center);
	\end{pgfonlayer}
\end{tikzpicture}}
\caption{We sketch defects (snaked brown line) contributing to the defect group of genuine non-endable defects of the coupled edge corner system $\mathcal{T}_{d+1},\mathcal{T}_d$. Defects depicted in (i) are defects of $\mathcal{T}_{d+1}$ and can be deformed into the boundary supporting $\mathcal{B}_{d+1}$ resulting in defects of the corner mode $\mathcal{T}_{d}$ as depicted in (ii).}
\label{fig:DefecGp}
\end{figure}
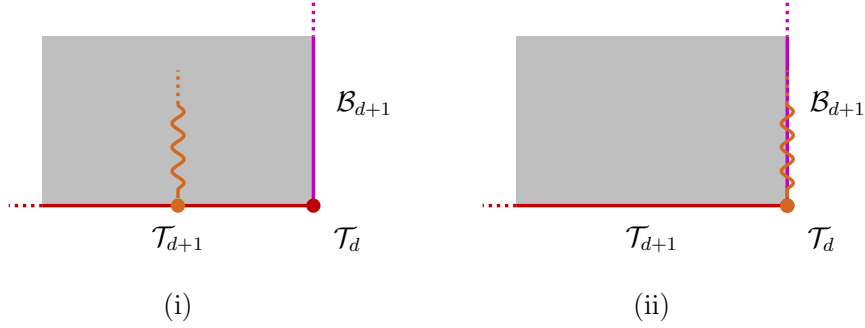

To proceed in generality, and fix such mismatches, first consider the defects of the flavor brane theory $\mathcal{T}_{d+1}$. Wrapping branes on cones over classes in $H_n(\partial \mathscr{X}_f)$ construct defect operators of $\mathcal{T}_{D}$ extending in the singular locus $(\mathscr{S}\setminus \mathscr{S}_0)\times \R^{d-1,1}$. Now, after KK reducing further on $\partial \mathscr{S}$, see figure \ref{fig:SymTFTColorFlavor}, these descend to defects of $\mathcal{T}_{d+1}$ extending in $ (I\setminus \{0\})\times  \R^{d-1,1}$. The relevant cycles for brane wrappings are now cones over classes in:
\be\label{eq:DefectGpFlavor}
H_n(\partial \mathscr{X}_f\rtimes \partial \mathscr{S} )\cong H_n(\partial^2 X^\circ )\,.
\ee
In the cheesesteak these wrappings result in defects sketched in subfigure (i) of figure \ref{fig:DefecGp}.

In a similar fashion, the loci naively relevant for constructing defects of the corner theory $\mathcal{T}_d$ are cones over classes in $\partial \mathscr{E}\rtimes \mathscr{S}_0 \sim \partial X^\circ|_{\textnormal{retract}}$ and for $\mathcal{T}_d$  the groups \eqref{eq:DefectGpFlavor} are replaced with
\be \label{eq:naive}
H_n(\partial X^\circ)\cong H_n(\partial X^\circ|_{\textnormal{retract}} )\,.
\ee
In the cheesesteak these wrappings result in defects sketched in subfigure (ii) of figure \ref{fig:DefecGp}.

To relate the two sets of defects above, consider sliding a cycle in $\partial^2 X^\circ$, characterizing defects of flavor brane theory, into the geometry $\partial X^\circ$ which we associated with the corner theory. This is formalized by the mapping
\be
D_n\,:\quad H_n(\partial^2 X^\circ ) ~\rightarrow ~ H_n(\partial X^\circ ) \,,
\ee
which is the lift of the embedding $\partial^2 X^\circ \hookrightarrow\partial X^\circ $ to homology. Defects of $\mathcal{T}_d$ which do not result from flavor brane defects pushed into $\mathcal{B}_{d+1}$ are therefore characterized by the cokernel
\be
 H_n(\partial X^\circ ) /H_n(\partial^2 X^\circ )\,.
\ee
However, to determine the defects intrinsic to the corner theory $\mathcal{T}_{d}$ we should ask the converse question: which of the naive defects, constructed by wrappings on \eqref{eq:naive}, can be moved off $\mathcal{T}_{d}$? That is, in figure \ref{fig:DefecGp}, can we deform subfigure (ii) to subfigure (i)? This is only possible for some maximal subgroup $F_n$ of $\textnormal{Im}\, D_n$ on which we can define
\be \label{eq:InverseD}
D_n^{-1}\,:\quad H_n(\partial^2 X^\circ ) ~\leftarrow ~ F_n\subset \textnormal{Im}\, D_n \subset H_n(\partial X^\circ ) \,.
\ee

Motivated by these considerations we define the defect group of genuine, non-endable defects of the $ \mathcal{T}_{d+1} , \mathcal{T}_d$ system derived from $p\;\!$-brane wrappings:
\be \label{eq:DefectGroup}\ba
\mathbb{D}[\mathcal{T}_d]&=\underset{m}{\mathbb{%
{\displaystyle\bigoplus}
}}\,\mathbb{D}_{m}[\mathcal{T}_d]&&\text{ \ \ with \ \ }\quad \mathbb{D}_{m}[\mathcal{T}_d]=\underset{p-n=m-2}{%
{\displaystyle\bigoplus}
}\lb H_n(\partial X^\circ)/F_n\rb |_{\textnormal{triv}}\\[0.5em]
\mathbb{D}[\mathcal{T}_{d+1}]&=\underset{m}{\mathbb{%
{\displaystyle\bigoplus}
}}\,\mathbb{D}_{m}[\mathcal{T}_{d+1}]&&\text{ \ \ with \ \ }\quad \mathbb{D}_{m}[\mathcal{T}_{d+1}]=\underset{p-n=m-2}{%
{\displaystyle\bigoplus}
}H_n(\partial^2 X^\circ) |_{\textnormal{triv}}
\ea \ee
where in giving the homology groups we have characterized wrapped cones by their cross-section. Here $m$ is the dimensionality of the defects in $\mathcal{T}_d$ and $\mathcal{T}_D$, and $|_{\textnormal{triv}}$ means the cycle trivializes when embedded into the corresponding fibers of $T_{\mathscr{S}}\rightarrow I$. Similarly, we can also introduce the naive defect group
\be\label{eq:naivedefects}\ba
\widetilde{\mathbb{D}}[\mathcal{T}_d]&=\underset{m}{\mathbb{%
{\displaystyle\bigoplus}
}}\,\widetilde{\mathbb{D}}_{m}[\mathcal{T}_d]&&\text{ \ \ with \ \ }\quad \widetilde{\mathbb{D}}_{m}[\mathcal{T}_d]=\underset{p-n=m-2}{%
{\displaystyle\bigoplus}
}H_n(\partial X^\circ) |_{\textnormal{triv}}
\ea\ee
built from cycles of $\partial X^\circ$. Symmetry operators acting on these defects are constructed by wrapping branes on the asymptotic cycles themselves. Choices of polarizations for defect operators of \eqref{eq:DefectGroup} can of course be obstructed by anomalies.\medskip


{\bf Example} Let us apply the above to the example of M-theory on $\mathbb{C}^3/\Z_{2n}$. There we compute (see appendix \ref{app:A}) the groups
\be \ba
H_3(\partial^2 X^\circ)&\cong \Z\,, && H_3(\partial X^\circ)\cong \Z\,, \\
H_1(\partial^2 X^\circ)&\cong \Z\oplus \Z_2\,, \quad &&H_1(\partial X^\circ)\cong \Z_{2n}\,.
\ea \ee
We compute the mapping $H_3(\partial^2 X^\circ)\rightarrow H_3(\partial X^\circ)$ to be multiplication by $n$. The mapping $H_1(\partial^2 X^\circ)\rightarrow H_1(\partial X^\circ)$ is computed to $(f,t)\mapsto f+nt$ where $(1,0)$ generates $\Z$ and $(0,1)$ generates $\Z_2$. It follows here that $F_3=\Z$ and $F_1=\Z_2$. Therefore the group of line operators in $\mathcal{T}_d$ from M2-brane wrappings on non-compact 2-cycles computes to
\be \ba
\mathbb{D}_{1}[\mathcal{T}_d]&\cong H_1(\partial X^\circ)/ F_1 \cong \Z_n\,, \ea \ee
while the symmetry operators acting on these are M5-branes wrapped over $H_3(\partial X^\circ)/ F_3 \cong \Z_n$. With respect to the orbifolding group $\Z_{2n}$ we find the symmetry operators here to be a $\Z_n\subset \Z_{2n}$ subgroup while the defect operators are a $\Z_n \cong \Z_{2n}/\Z_2$ quotient. Overall, we have correctly determined M2- (resp. M5-brane) wrappings associated with defects (resp. symmetry operators), accurately capturing the $\Z_n$ 1-form symmetry of the 5D SCFT in an electric frame. It now remains to discuss the defect operators one would naively attribute to the subgroup $\Z_2\subset \Z_{2n}$ and the symmetry operators similarly attributed to $\Z_2 \cong \Z_{2n}/\Z_n$. The former are clear, they are flavor line operators, as constructed from wrapping M2-branes on non-compact 2-cycles, captured by the subgroup $\Z_2\subset \mathbb{D}_{1}[\mathcal{T}_{d+1}]$. For the latter we require more general considerations. \medskip

Returning to the general case, there are further defect operators which are genuine from the perspective of the corner theory $\mathcal{T}_{d}$ but non-genuine in the flavor brane theory $\mathcal{T}_{d+1}$. The kernel of the mapping $D_n$ corresponds to those defects which can end in the geometry $\partial X^\circ$, i.e., there exists a chain in $\partial X^\circ$ bounding the cycle, and wrapping branes on this chain constructs boundaries of flavor defects. Using these loci we can build further defects (see subfigure (ii) of figure \ref{fig:DefecGp2}). Motivated by these considerations we introduce the mixed defect group derived from $p\;\!$-brane wrappings
\be \label{eq:DefectGroup2}\ba
\mathbb{D}_{\textnormal{mix}}[\mathcal{T}_d,\mathcal{T}_{d+1}]&=\underset{m}{\mathbb{%
{\displaystyle\bigoplus}
}}\,\mathbb{D}^{(m)}_{\textnormal{mix}}[\mathcal{T}_d,\mathcal{T}_{d+1}]&&\text{ with  }\quad \mathbb{D}^{(m)}_{\text{mix}}[\mathcal{T}_d,\mathcal{T}_{d+1}]=\underset{p-n=m-2}{%
{\displaystyle\bigoplus}
}\lb \textnormal{Ker}\,D_n \rb |_{\textnormal{triv}}\,.
\ea \ee


 \begin{figure}
\centering
\scalebox{0.8}{
 \begin{tikzpicture}
	\begin{pgfonlayer}{nodelayer}
		\node [style=none] (0) at (-1.5, 0) {};
		\node [style=none] (1) at (-1.5, 2.5) {};
		\node [style=none] (2) at (-5.5, 0) {};
		\node [style=SmallCircleRed] (3) at (-1.5, 0) {};
		\node [style=SmallCircleBrown] (4) at (-1.5, 1.5) {};
		\node [style=none] (5) at (-3.5, 1.5) {};
		\node [style=none] (6) at (-4, 1.5) {};
		\node [style=none] (7) at (-3.5, -1.5) {(i)};
		\node [style=none] (8) at (-1.5, 3) {};
		\node [style=none] (9) at (5.5, 0) {};
		\node [style=none] (10) at (5.5, 2.5) {};
		\node [style=none] (11) at (1.5, 0) {};
		\node [style=SmallCircleRed] (12) at (5.5, 0) {};
		\node [style=SmallCircleBrown] (13) at (5.5, 0) {};
		\node [style=none] (14) at (4, 1.5) {};
		\node [style=none] (15) at (3.5, 2) {};
		\node [style=none] (16) at (3.5, -1.5) {(ii)};
		\node [style=none] (17) at (5.5, 3) {};
		\node [style=none] (18) at (-6, 0) {};
		\node [style=none] (19) at (1, 0) {};
		\node [style=none] (22) at (3.5, -0.5) {$\mathcal{T}_{d+1}$};
		\node [style=none] (23) at (-3.5, -0.5) {$\mathcal{T}_{d+1}$};
		\node [style=none] (24) at (6, -0.5) {$\mathcal{T}_{d}$};
		\node [style=none] (25) at (-1, -0.5) {$\mathcal{T}_{d}$};
		\node [style=none] (24) at (6.25, 1.5) {$\mathcal{B}_{d+1}$};
		\node [style=none] (25) at (-0.75, 1.5) {$\mathcal{B}_{d+1}$};
	\end{pgfonlayer}
	\begin{pgfonlayer}{edgelayer}
		\filldraw[fill=gray!50, draw=gray!50]  (1.5, 0) -- (5.5, 0) -- (5.5, 2.5) -- (1.5, 2.5) -- cycle;
		\filldraw[fill=gray!50, draw=gray!50]   (-1.5, 0) -- (-5.5, 0) -- (-5.5, 2.5) -- (-1.5, 2.5) -- cycle;
		\draw [style=RedLine] (2.center) to (0.center);
		\draw [style=PurpleLine] (1.center) to (0.center);
		\draw [style=DottedBrown] (5.center) to (6.center);
		\draw [style=BrownLine, snake it] (5.center) to (4);
		\draw [style=DottedPurple] (8.center) to (1.center);
		\draw [style=RedLine] (11.center) to (9.center);
		\draw [style=PurpleLine] (10.center) to (9.center);
		\draw [style=DottedBrown] (14.center) to (15.center);
		\draw [style=BrownLine, snake it] (14.center) to (13);
		\draw [style=DottedPurple] (17.center) to (10.center);
		\draw [style=DottedRed] (2.center) to (18.center);
		\draw [style=DottedRed] (11.center) to (19.center);
	\end{pgfonlayer}
\end{tikzpicture}}
\caption{In (i) we sketch a defect (snaked brown line) ending on the boundary supporting $\mathcal{B}_{d+1}$. We can deform it onto the corner, (i) $\rightarrow$ (ii), and this gives another class of defects which are genuine in $\mathcal{T}_{d}$. However, taking the perspective of $\mathcal{T}_{d+1}$, they are seen to live at the boundary of a higher-dimensional defect. }
\label{fig:DefecGp2}
\end{figure}
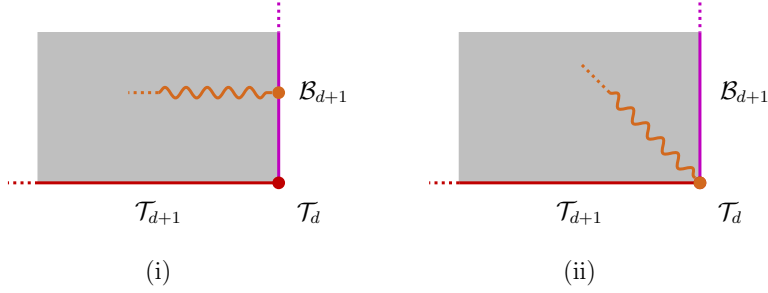 \medskip

{\bf Our Running Example} In our running example of M-theory on $\mathbb{C}^3/\Z_{2n}$ we indeed compute
\be
\mathbb{D}^{(2)}_{\text{mix}}[\mathcal{T}_d,\mathcal{T}_{d+1}]\cong \Z_2\,,
\ee
and have therefore now accounted for the missing $\Z_2$ (and the symmetry operators wrapped on the corresponding asymptotic cycles) which is geometrized as $H_3(\partial X)\cong \Z_{2n}$ mod $2$ as seen from the long exact sequence in relative homology of the pair $\partial X^\circ,\partial^2X^\circ$.\medskip

 Have we described the full set of non-endable defects? For the flavor brane theory $\mathcal{T}_{d+1}$ we have, as this theory is well-defined in isolation such that the methods of \cite{DelZotto:2015isa} directly apply. For the corner theory $\mathcal{T}_{d}$ we described the full set of defects constructed by wrapping non-compact cycles of $\partial X$. In appendix \ref{app:A} we show that
 we have the short exact sequence
 \be
 0~\rightarrow~    H_n(\partial X^\circ)/ F_n~\rightarrow~  H_n(\partial X) ~\rightarrow~  \textnormal{Ker}\,D_n   ~\rightarrow~  0\,,
 \ee
for all examples we will consider, i.e., all cycles in $H_n(\partial X)$ are accounted for.

\subsection{Bulk $\mathcal{S}_{d+2}$ and Boundaries ${\mathcal{B}_{d+1}},\mathcal{T}_{d+1}$}
\label{sec:Gluing}

We now discuss the symmetry theory $\mathcal{S}_{d+2}$ with boundary theory $\mathcal{B}_{d+1}$ explicitly by determining their field contents, actions and boundary conditions. This in particular improves on the defect group discussion by giving a Dirac pairing for the groups in \eqref{eq:DefectGroup}, which determines the set of possible polarizations, and anomalies obstructing choices of polarizations. Both of these will be characterized by interaction terms of the action. Once this bulk-boundary system is discussed we turn to a similar analysis for $\mathcal{T}_{d+1}$.

\subsubsection{Bulk $\mathcal{S}_{d+2}$ and Boundary ${\mathcal{B}_{d+1}}$}

To begin, note that the symmetry theory $\mathcal{S}_{d+2}$ with boundary theory $\mathcal{B}_{d+1}$ is computed by compactification of some supergravity theory on the fibration $\pi_{IJ}:\partial X^\circ \rightarrow I\times J$.
Here, compactification on the generic fibers $\partial \mathscr{X}_f\rtimes \partial \mathscr{S}=\partial T_{\mathscr{K}}$ which are copies of $\partial^2 X^\circ$ computes the symmetry theory of the flavor brane theory, at $r=0$ and along the edge $J$, this fiber degenerates to $\partial \mathscr{E}\rtimes \partial \mathscr{S}_0$ which is topologically (after fattening) a copy of $\partial X^\circ$ determining the boundary theory $\mathcal{B}_{d+1}$. See figure \ref{fig:CheeseGeometry}. As such we will be concerned with compactification on $\partial^2 X^\circ$ and $\partial X^\circ$.

Immediately we comment that the flavor symmetry theory $\mathcal{S}_{d+2}$ is a standard absolute symmetry theory as the relevant space $\partial^2 X^\circ=\partial T_{\mathscr{K}}$ is a smooth compact manifold. Therefore, $\mathcal{S}_{d+2}$ follows by considering the geometry $T_{\mathscr{K}}$ normal to the flavor brane and applying the prescription laid out in \cite{Apruzzi:2021nmk, Baume:2023kkf, Apruzzi:2024htg}. On the other hand, the space  $\partial X^\circ$, associated with the boundary theory $\mathcal{B}_{d+1}$, is a manifold with boundary and the compactification, which results in a relative symmetry theory, requires more care.

In both cases we give our analysis at the level of singular (co)homology.

\paragraph{Field Content}\mbox{}\medskip

First, we determine the field content of $\mathcal{B}_{d+1}, \mathcal{S}_{d+2}$, then we discuss the boundary condition $\mathcal{S}_{d+2}|_{\mathcal{B}_{d+1}}$, and from here turn to discuss the action of $\mathcal{B}_{d+1}, \mathcal{S}_{d+2}$. We begin our analysis by clarifying the cocycles we expand fields in.

The geometry determining the field content of $\mathcal{S}_{d+2}$ is the smooth compact space $\partial^2X^\circ$ and fields follow by expanding the relevant supergravity fields in the cocycles $\mathscr{F}^n=H^n(\partial^2 X^\circ)$ in standard fashion.

The geometry relevant for determining the field content of $\mathcal{B}_{d+1}$ is the manifold with boundary $\partial X^\circ$, or equivalently, a fattening of the exceptional fiber. Therefore, there are two natural options for internal cocycles when expanding supergravity profiles, which are $H^n(\partial X^\circ)$ and $H^n(\partial X^\circ, \partial^2 X^\circ)$. We will employ both simultaneously, and correct for an ``overcounting". We say we reduce supergravity fields on
\be\label{eq:forms}
{H}^n(\partial X^\circ,\partial^2 X^\circ)~\xrightarrow[]{}~ {H}^n(\partial X^\circ)\,,
\ee
or equivalently
\be
\mathscr{C}^n= {H}^n(\partial X^\circ,\partial^2 X^\circ) \oplus {H}^n(\partial X^\circ) /\sim\,,
\ee
where we identify cocycles in \eqref{eq:forms} with their image. The mapping in \eqref{eq:forms} is taken from the long exact sequence in relative homology for the pair $\partial X^\circ, \partial^2 X^\circ$.


\paragraph{Boundary Conditions}\mbox{}\medskip

Now, with both the field content of $\mathcal{S}_{d+2}$ and ${\mathcal{B}_{d+1}}$ in hand we can formulate boundary conditions between these. Boundary conditions are determined by the mappings
\be\label{eq:BCGeo}
{H}^n(\partial X^\circ)~\xrightarrow[]{}~ {H}^n(\partial^2 X^\circ)~\xrightarrow[]{}~{H}^{n+1}(\partial X^\circ,\partial^2 X^\circ)\,,
\ee
which relate the internal cocycles of the geometry resulting in relations of their spacetime coefficients, the symmetry theory fields. Here, the cycles associated with fields of $\mathcal{S}_{d+2}$ are the central entry and they are mapped into\,/\,onto to cycles associated with fields of $\mathcal{B}_{d+1}$.

Consider, for example, a supergravity field strength $G_{n+m}$ of degree $n+m$. Let $\omega_n^{(\partial X^\circ)}\in {H}^n(\partial X^\circ) $ and $\omega_n^{(\partial^2 X^\circ)}\in {H}^n(\partial^2 X^\circ)$ be two cocycles such that the former is mapped onto the latter. We make the expansions
\be \ba
G_{n+m}&= B_{m}^{(\partial X^\circ)}\cup\omega_n^{(\partial X^\circ)}+\dots \\
G_{n+m}&= B_{m}^{(\partial^2 X^\circ)}\cup\omega_n^{(\partial^2 X^\circ)}+\dots
\ea \ee
with $m$-cocycle coefficients. The fields $B_{m}^{(\partial^2 X^\circ)}$ and $B_{m}^{(\partial X^\circ)}$ propagate in the bulk and boundary theory $\mathcal{S}_{d+2}$ and $\mathcal{B}_{d+1}$ respectively. The first mapping in \eqref{eq:BCGeo} implies
\be
\omega_n^{(\partial X^\circ)}|_{\partial^2X^\circ}=s\:\! \omega_n^{(\partial^2 X^\circ)} \qquad \textnormal{mod}\,t\,,
\ee
with some integers $s,t$ depending on the geometry, also possibly without the modulo condition. Therefore boundary conditions between the symmetry theory  fields are
\be \label{eq:BCs}
B_{m}^{(\partial^2 X^\circ)}\Big|_{\mathcal{B}_{d+1}} = s B_{m}^{(\partial X^\circ)} \qquad \textnormal{mod}\,t\,,
\ee
in the natural normalization with respect to \eqref{eq:BCGeo}. Similarly, we can consider the pair
\be \label{eq:Expansions}\ba
G_{n+m}&= B_{m-1}^{(\partial X^\circ,\;\!\partial^2 X^\circ)}\cup\omega_{n+1}^{(\partial X^\circ,\:\!\partial^2 X^\circ)}+\dots \\
G_{n+m}&= B_{m}^{(\partial^2 X^\circ)}\cup\omega_n^{(\partial^2 X^\circ)}+\dots
\ea \ee
which, following analogous steps as described above, now for the second map in \eqref{eq:BCGeo}, gives the relation
\be
\omega_n^{(\partial^2 X^\circ)}=p  \lbb \partial /\partial r \,\lrcorner\, \omega_{n+1}^{(\partial X^\circ,\partial^2 X^\circ)}\rbb  \Bigg|_{\mathcal{B}_{d+1}}  \qquad \textnormal{mod}\,q\,,
\ee
for some integers $p,q$ again depending on the geometry and possibly without the modulo condition. This relation on the internal cocycles translates into following boundary condition for the symmetry theory fields
\be \label{eq:dualBCs}
  \lbb \partial /\partial r \,\lrcorner\, B^{\:\!m}_{(\partial^2 X^\circ)} \rbb\! \Bigg|_{\mathcal{B}_{d+1}}  =  p  B^{\:\!m}_{(\partial X^\circ, \partial^2 X^\circ)}  \qquad \textnormal{mod}\,q\,,
\ee
for some integers $p,q$ depending on the geometry. Here $r$ is the coordinate of the half-space $I$ which is the coordinate normal to the boundary theory $\mathcal{B}_{d+1}$.

Exactness in the central entry in \eqref{eq:BCGeo} implies that \eqref{eq:BCs} and \eqref{eq:dualBCs} lead to a complete set of boundary conditions in the sense of section \ref{sec:cheesesteak}.

\paragraph{Boundary Degrees of Freedom}\mbox{}\medskip

Let us now discuss the intrinsic degrees of freedom of the boundary theory $\mathcal{B}_{d+1}$, i.e., the degrees of freedom not fixed by the boundary conditions of lines \eqref{eq:BCs} and \eqref{eq:dualBCs}. Recall that the field content of $\mathcal{B}_{d+1}$ derives via reduction on cocycles
\be\label{eq:Identification}
{H}^n(\partial X^\circ,\partial^2 X^\circ)~\xrightarrow[]{}~ {H}^n(\partial X^\circ)\,,
\ee
boundary conditions eat up degrees of freedom according to the flanking terms
\be
{H}^{n-1}(\partial^2 X^\circ)~\xrightarrow[]{}~ {H}^n(\partial X^\circ,\partial^2 X^\circ)~\xrightarrow[]{}~ {H}^n(\partial X^\circ)~\xrightarrow[]{}~ {H}^{n}(\partial^2 X^\circ)\,,
\ee
by exactness the unconstrained\,/\,intrinsic fields descend precisely from the reduction on the image of the mapping \eqref{eq:forms}, i.e., the cocycles
\be \label{eq:trapped}
{H}^n(\partial X^\circ,\partial^2 X^\circ)\subset {H}^n(\partial X^\circ)\,,
\ee
which are the cocycles vanishing along $\partial^2 X^\circ$ and are not related to ${H}^{n-1}(\partial^2 X^\circ)$ associated with flavor field content.
In the symmetry theory $\mathcal{B}_{d+1}$ the associated degrees of freedom are not fixed by the bulk symmetry theory $\mathcal{S}_{d+2}$, remaining free to fluctuate, and are path integrated.

Overall the path integral of the symmetry theory pair $\mathcal{S}_{d+2},\mathcal{B}_{d+1}$ is given by a path integral over all fields of $\mathcal{S}_{d+2}$, as result from reduction along $H^n(\partial^2 X^\circ)$, and all fields of $\mathcal{B}_{d+1}$ as result from the reduction along $H^n(\partial X^\circ,\partial^2 X^\circ)\rightarrow H^n(\partial X^\circ)$. The field space integrated over is subject to identifications induced by $ H^n(\partial X^\circ) \rightarrow H^n(\partial^2 X^\circ) \rightarrow H^{n+1}(\partial X^\circ,\partial^2 X^\circ)$ which set the boundary conditions. The path integral, after identifications are made, is over the full set of flavor brane fields and those intrinsic to $\mathcal{B}_{d+1}$ determined by ${H}^n(\partial X^\circ,\partial^2 X^\circ)\subset {H}^n(\partial X^\circ)$.\medskip

{\bf Our Running Example}  Consider M-theory on $\mathbb{C}^3/\Z_{2n}(1,1,2n-2)$. In an electric frame this theory as a $\Z_n$ 1-form symmetry. However, the (co)homology groups $\partial^2 X^\circ, \partial X^\circ$ only feature instances of $\Z_2$ and $\Z_{2n}$ subgroups, and naively the background field for the $\Z_n$ 1-form symmetry, obtained via expansions of the type \eqref{eq:Expansions}, seems absent. However, we compute
\be
\text{Im}\lb {H}^2(\partial X^\circ,\partial^2 X^\circ)\rightarrow {H}^2(\partial X^\circ)\rb =\Z_n
\ee
which we interpret to mean that the boundary modes of $\mathcal{B}_{d+1}$, which are left to fluctuate after the boundary conditions to the bulk $\mathcal{S}_{d+2}$ have been imposed, precisely capture the background fields for the $\Z_n$ 1-form symmetry. In this sense these degrees of freedom are only manifest after imposing boundary conditions and are ultimately localized to $\mathcal{B}_{d+1}$.

\paragraph{Interactions}\mbox{}\medskip

Interaction terms of the symmetry theories $\mathcal{S}_{d+2},\mathcal{B}_{d+1}$ reflect anomalies of the theories $\mathcal{T}_d,\mathcal{T}_{d+1}$ respectively. In our setup $\mathcal{S}_{d+2}$ is an absolute symmetry theory while $\mathcal{B}_{d+1}$ is relative thereto and as such the interaction terms for $\mathcal{S}_{d+2}$ are computed via standard reduction using the differential cohomology ring of $\partial^2 X^\circ$ much as in \cite{Apruzzi:2021nmk, Apruzzi:2024htg, GarciaEtxebarria:2024fuk}.

However, as $\mathcal{B}_{d+1}$ is relative to $\mathcal{S}_{d+2}$ and the geometry playing the role of the closed space $\partial^2X^\circ$ for the latter is now replaced with the manifold with boundary  $\partial X^\circ$, we need to revisit the computation of the interaction terms for $\mathcal{B}_{d+1}$. We describe an approximation to this problem in integral cohomology.

For this recall that following \eqref{eq:BCGeo} we discussed derivation of the field content and boundary conditions for the bulk-boundary system  $\mathcal{S}_{d+2},\,\mathcal{B}_{d+1}$. Key was the insight that supergravity fields are expanded both in cohomology groups $H^n(\partial X^\circ)$ and $H^n(\partial X^\circ,\partial^2 X^\circ)$ or, more precisely, in differential cocycles projecting to these groups. As such, reducing the terms of a supergravity theory the internal integrals are determined from a cup products and linkings of these two groups of cocycles.

With this let us first focus on the case of discrete symmetries. In this case, in a cohomology approximation, the final pairing is a linking. The relevant linking form in (co)homology is
\be \ba
\ell^\vee \,: \quad&\textnormal{Tor}\,H_n(\partial X^\circ) \times \textnormal{Tor}\,H_{M-n-1}(\partial X^\circ,\partial^2X^\circ) ~\rightarrow~ \Q/\Z\,,\\
\ell \,: \quad &\textnormal{Tor}\,H^{n+1}(\partial X^\circ) \times \textnormal{Tor}\,H^{M-n}(\partial X^\circ,\partial^2X^\circ) ~\rightarrow~ \Q/\Z\,,
\ea \ee
where $M=\textnormal{dim}\,\partial X^\circ=\textnormal{dim}\,\partial X=\textnormal{dim}\,X-1$. These forms follow from Poincar\'e-Lefschetz duality and the universal coefficient theorem which respectively assert
\be
 \textnormal{Tor}\,H_{M-n-1}(\partial X^\circ,\partial^2X^\circ) \cong  \textnormal{Tor}\,H^{n+1}(\partial X^\circ) \cong  \textnormal{Tor}\,H_{n}(\partial X^\circ)^\vee\,,
\ee
giving a non-degenerate pairing.

Such quadratic pairings give rise to discrete $BF$-like couplings when reducing $p$-form fields of the supergravity theory on $H^*(\partial X^\circ)$ and $H^*(\partial X^\circ, \partial^2 X^\circ)$. See Appendix B of \cite{Baume:2023kkf} for how the discrete BF terms reduce from the kinetic terms of the $p$-form fields. More generally, the coefficients in the Lagrangian of $\mathcal{B}_{d+1}$ are proportional to
\be \label{eq:linking}
\ell(\omega_1\cup \dots\cup \omega_{k},\omega_1'\cup \dots\cup \omega_{k'}' )\in \Q/\Z\,,
\ee
where $\omega_{i}$ with $i=1,\dots,k$ and  $\omega_{i'}'$ with $i'=1,\dots,k'$ where are cocycles such that their cup product is an element of $H^{n+1}(\partial X^\circ)$ and $H^{M-n}(\partial X^\circ,\partial^2 X^\circ)$ respectively. The linking \eqref{eq:linking} is thus determined from various cohomology rings and their linking forms and generally straightforwardly computable. Generally only an integer multiple of the coefficient appearing in the Lagrangian of $\mathcal{B}_{d+1}$ can be computed in this manner and \eqref{eq:linking}, for example in the context of M-theory compactifications, this often needs to be divided by a combinatorial factor of $2, 3$ or $6$. In principle, such a refinement requires the use of index theory or differential cohomology which we defer to future work, however in cases we will present there is a work around based on self-consistency considerations.

We can make similar remarks for continuous abelian symmetries, where now the relevant pairings are the intersection pairings
\be \ba
i^\vee \,: \quad&\textnormal{Free}\,H_n(\partial X^\circ) \times \textnormal{Free}\,H_{M-n}(\partial X^\circ,\partial^2X^\circ) ~\rightarrow~ \Z\,,\\
i \,: \quad &\textnormal{Free}\,H^{n}(\partial X^\circ) \times \textnormal{Free}\,H^{M-n}(\partial X^\circ,\partial^2X^\circ) ~\rightarrow~ \Z\,,
\ea \ee
with $\textnormal{Free}\; H_*\equiv H_*/\textnormal{Tor}H_*$ and analogously for cohomology. We again obtain $BF$-like couplings from reducing the kinetic terms for the $p$-form fields as well as higher order terms analogous to \eqref{eq:linking}. For instance, let us consider reducing M-theory on some $\partial X^\circ$ such that $H^n(\partial X^\circ)\simeq H^{M-n}(\partial X^\circ,\partial^2X^\circ)$ is non-zero. Then the relevant 11D kinetic term for the $C_3$ potential is proportional to
\begin{equation}
    \frac{1}{G_N} \int_{11D} G_4 \wedge * G_4
\end{equation}
where we have included the 11D Newton's constant $G_N$. We can rewrite this action as
\begin{equation}\label{eq:lagrangemultiplier}
    \int_{11D} \left( G_4 \wedge h_{7} + G_N \; h_7 \wedge * h_7\right)
\end{equation}
after using the Lagrange multiplier $h_7$. After reducing $G_4$ and $h_7$ on some $\omega_n\in H^n(\partial X^\circ)$ and $\omega'_{M-n}\in H^{M-n}(\partial X^\circ,\partial^2X^\circ)$ respectively, we obtain a term in the action of $\mathcal{B}_{d+1}$
\begin{equation}\label{eq:lagrangemult2}
    K\int \left( G_{4-n}\wedge h_{7-M+n} + G_N \; h_{7-M+n} \wedge *h_{7-M+n}\right)
\end{equation}
where $K\equiv i(\omega_n, \omega'_{M-n})$ is the intersection number. After taking a decoupling limit from the gravitational degrees of freedom, $G_N\rightarrow 0$, we see that we reduce to the same $BF$-like terms that have appeared in recent proposals for describing SymTFT for $U(1)$ symmetries \cite{Brennan:2024fgj, Antinucci:2024zjp, Bonetti:2024cjk, Apruzzi:2024htg}. The conceptual difference is that there are no gauge fields valued in $\mathbb{R}$ in sight in our construction. This apparent conceptual mismatch is mended by simply noting that all of the observables in $\mathcal{B}_{d+1}$ are invariant under $G_{4-n}\rightarrow G_{4-n}+dc_{3-n}$ and $h_{7-M+n}\rightarrow h_{7-M+n}+dc_{6-M+n}$, i.e. only dependence of the cohomology class of the fields are observable, not the particular representatives. Given such a redundancy, it would appear that one could regard $G_{4-n}$ as an $\mathbb{R}$-valued gauge potential and $h_7$ as a $U(1)$ curvature, or vice-versa although no honest $\mathbb{R}$-valued gauging has appeared at any step.

\subsubsection{Bulk $\mathcal{S}_{d+2}$ and Boundary ${\mathcal{T}_{d+1}}$}

We now turn to discuss the boundary conditions imposed on $\mathcal{S}_{d+2}$ by the relative theory ${\mathcal{T}_{d+1}}$ associated to the singular locus $\mathscr{S}\setminus \mathscr{S}_0$. Generically, these will be enriched Neumann boundary conditions in the sense of \cite{Kaidi:2022cpf}.

To say more, let us specialize to the case in which ${\mathcal{T}_{d+1}}$ is an infrared free gauge theory as will be the case whenever $\mathscr{S}\setminus \mathscr{S}_0$ is an ADE singularity in M-theory. In taking the IR limit, localizing the dynamics of our system to $d$ dimensions, the flavor brane gauge coupling decreases $g\rightarrow 0$. As a result we find a copy of Horowitz's non-Abelian BF theory \cite{Horowitz:1989ng} localized to ${\mathcal{T}_{d+1}}$ in the limit, as previously noted in \cite{Bonetti:2024cjk}. In addition, the twisted normal geometry to $\mathscr{S}\setminus \mathscr{S}_0$ leads to additional topological terms in the flavor brane fields localized to ${\mathcal{T}_{d+1}}$.\medskip

{\bf Our Running Example}  Consider M-theory on $\mathbb{C}^3/\Z_{2n}(1,1,2n-2)$. The orbifold contains an $\mathfrak{su}_2$ singularity supported on $\mathbb{C}/\Z_{n}$ and the normal geometry $\mathbb{C}^2/\Z_2$ thereof is acted on by a $\Z_n$ monodromy along paths linking the tip of the cone $\mathbb{C}/\Z_{n}$. The 7D super-Yang-Mills theory supported on $\mathbb{C}/\Z_{n}$ has a corresponding background profile turned on. To obtain the 6D KK theory, for which $\mathcal{T}_d$ functions as an end of the world brane, we compactify on the circle parametrized by the angular argument of  $\mathbb{C}/\Z_{n}$ (this results in a radius dependent gauge coupling, see our discussion around \eqref{eq:bulkLag}).  The KK theory contains massless fields organizing into a 6D $\mathfrak{su}(2)$ super-Yang-Mills theory and the non-trivial normal geometry gives rise to an additional topological term. In the limit $g\rightarrow 0$ we have the overall contribution
\be \label{eq:thetatermspecific}
2\pi \int_{\R_{r\geq 0} \times \R^{1,4}} \tr\lb \frac{f_2}{2\pi} \cup \frac{h_4}{2\pi} \rb -\frac{2\pi}{6n}\int_{\R_{r\geq 0} \times \R^{1,4}} \tr\lb \frac{f_2}{2\pi}\cup  \frac{f_2}{2\pi} \cup  \frac{f_2}{2\pi}\rb \,,
\ee
to the 6-dimensional edgemode theory $\mathcal{T}_{d+1}$. Here $f_2$ is the non-abelian field strength of the 6D $\mathfrak{su}(2)$ SYM theory, $h_4$ is a Lagrange multiplier and $r$ is the radial coordinate of $\mathbb{C}/\Z_{n}$. We discuss the geometry and the derivation of \eqref{eq:thetatermspecific}, using a IIA dual frame and the Wess-Zumino couplings on the world volume of a D6-brane stack in section \ref{sec:Illustrative} which is similar to the steps leading to \eqref{eq:lagrangemult2}.

\subsection{Boundary and Corner Conditions $\mathbb{B}, \widetilde{\mathbb{B}}$}\label{sec:Gluing2}

So far we have discussed our various $(d+2)$-dimensional constructions in a local patch centered on the theory $\mathcal{T}_d$ which realizes an interface theory between the edges $\mathcal{T}_{d+1},\mathcal{B}_{d+1}$ and, simultaneously, a corner theory to $\mathcal{S}_{d+2}$. With this, we now discuss the remaining edges and corners to $\mathcal{S}_{d+2}$.

We begin with the open cheesesteak
\be\label{eq:open}
\scalebox{0.9}{
\begin{tikzpicture}
	\begin{pgfonlayer}{nodelayer}
		\node [style=none] (0) at (1.75, -1.25) {};
		\node [style=none] (1) at (1.75, 1.25) {};
		\node [style=none] (2) at (-2.25, 1.25) {};
		\node [style=none] (3) at (-2.25, -1.25) {};
		\node [style=SmallCircleBlue] (4) at (1.75, 1.25) {};
		\node [style=SmallCircleRed] (5) at (1.75, -1.25) {};
		\node [style=none] (6) at (-0.25, 1.75) {$\mathbb{B}_{d+1}^{(r,1)}$};
		\node [style=none] (7) at (-0.25, 0) {$\mathcal{S}_{d+2}$};
		\node [style=none] (8) at (-0.25, -1.75) {$\mathcal{T}_{d+1}$};
		\node [style=none] (9) at (2.5, 0) {$\mathcal{B}_{d+1}$};
		\node [style=none] (10) at (2.25, -1.75) {$\mathcal{T}_{d}$};
		\node [style=none] (11) at (2.25, 1.75) {$\mathbb{B}_{d}^{(0,1)}$};
		\node [style=none] (12) at (-2.75, 1.25) {};
		\node [style=none] (13) at (-2.75, -1.25) {};
	\end{pgfonlayer}
	\begin{pgfonlayer}{edgelayer}
		\filldraw[fill=gray!50, draw=gray!50]  (1.75, -1.25) -- (1.75, 1.25) -- (-2.25, 1.25) -- (-2.25, -1.25)-- cycle;
		\draw [style=BlueLine] (2.center) to (1.center);
		\draw [style=RedLine] (3.center) to (0.center);
		\draw [style=PurpleLine] (1.center) to (0.center);
		\draw [style=DottedBlue] (12.center) to (2.center);
		\draw [style=DottedRed] (13.center) to (3.center);
	\end{pgfonlayer}
\end{tikzpicture}}
\ee
which contains two more pieces of data, the boundary $\mathbb{B}_{d+1}^{(r,1)}$ and the corner $\mathbb{B}_{d}^{(0,1)}$ serving as boundary conditions to $\mathcal{B}_{d+1},\mathcal{S}_{d+2}$ respectively. Upon specifying these we achieve an absolute theory in $d+1$ dimensions (KK theories in our top down constructions), i.e., collapsing \eqref{eq:open} along the vertical $x_\perp$-\:\!direction we achieve the system:
\be \label{eq:EOW}
\scalebox{0.9}{
\begin{tikzpicture}
	\begin{pgfonlayer}{nodelayer}
		\node [style=none] (0) at (1, 0) {};
		\node [style=none] (1) at (-1.5, 0) {};
		\node [style=none] (2) at (-2, 0) {};
		\node [style=Star] (3) at (1, 0) {};
		\node [style=none] (4) at (-0.25, 0.5) {QFT$_{d+1}$};
		\node [style=none] (5) at (2, 0) {EOW$_{d}$};
	\end{pgfonlayer}
	\begin{pgfonlayer}{edgelayer}
		\draw [style=ThickLine] (1.center) to (0.center);
		\draw [style=DottedLine] (2.center) to (1.center);
	\end{pgfonlayer}
\end{tikzpicture}}
\ee
Here EOW$_{d}$ is the end of the world theory. Clearly $\mathbb{B}_{d+1}^{(r,1)}$ sets the global form of the absolute theory QFT$_{d+1}$ and similarly $\mathbb{B}_{d+1}^{(0,1)}$ will select the spectrum of defects localized to EOW$_{d}$. Whenever we have a limit isolating the $d$-dimensional dynamics we see that $\mathbb{B}_{d+1}^{(r,1)}$ will set the global form of the flavor symmetry associated with $\mathcal{T}_{d+1}$.

Let us now consider setups where such a limit exists and has been taken. Upon imposing additional boundary conditions at $r=1$ we obtain an absolute QFT$_d$ from EOW$_{d}$ and the overall symmetry data is organized as:
\be\label{eq:closed}
\scalebox{0.9}{
\begin{tikzpicture}
	\begin{pgfonlayer}{nodelayer}
		\node [style=none] (0) at (1.75, -1.25) {};
		\node [style=none] (1) at (1.75, 1.25) {};
		\node [style=none] (2) at (-2.25, 1.25) {};
		\node [style=none] (3) at (-2.25, -1.25) {};
		\node [style=SmallCircleBlue] (4) at (1.75, 1.25) {};
		\node [style=SmallCircleRed] (5) at (1.75, -1.25) {};
				\node [style=none] (6) at (-0.25, 1.75) {$\mathbb{B}_{d+1}^{(r,1)}$};
		\node [style=none] (7) at (-0.25, 0) {$\mathcal{S}_{d+2}$};
		\node [style=none] (8) at (-0.25, -1.75) {$\mathcal{T}_{d+1}$};
		\node [style=none] (9) at (2.5, 0) {$\mathcal{B}_{d+1}$};
		\node [style=none] (10) at (2.25, -1.75) {$\mathcal{T}_{d}$};
		\node [style=none] (11) at (2.25, 1.75) {$\mathbb{B}_{d}^{(0,1)}$};
		\node [style=SmallCircleGreen] (12) at (-2.25, 1.25) {};
		\node [style=SmallCircleBrown] (13) at (-2.25, -1.25) {};
		\node [style=none] (14) at (-3, 0) {$\mathbb{B}_{d+1}^{(1,x_\perp)}$};
		\node [style=none] (15) at (-3, -1.75) {$\mathbb{B}_{d}^{(1,0)}$};
		\node [style=none] (16) at (-3, 1.75) {$\mathbb{B}_{d}^{(1,1)}$};
	\end{pgfonlayer}
	\begin{pgfonlayer}{edgelayer}
		\filldraw[fill=gray!50, draw=gray!50]  (1.75, -1.25) -- (1.75, 1.25) -- (-2.25, 1.25) -- (-2.25, -1.25)-- cycle;
		\draw [style=BlueLine] (2.center) to (1.center);
		\draw [style=RedLine] (3.center) to (0.center);
		\draw [style=PurpleLine] (1.center) to (0.center);
		\draw [style=GreenLine] (12) to (13);
	\end{pgfonlayer}
\end{tikzpicture}}
\ee
We group the gapped\,/\,free edges and corners into the tuple
\be
\widetilde{\mathbb{B}}=\lb \mathbb{B}_d^{(1,0)},\mathbb{B}_{d+1}^{(1,x_\perp)},\mathbb{B}_{d}^{(1,1)},\mathbb{B}^{(r,1)}_{d+1},\mathbb{B}_d^{(0,1)}\rb\,.
\ee

Let us focus on the boundary condition $\mathbb{B}_{d+1}^{(1,x_\perp)}$ first and give an argument and a consistency check that it is Fourier dual to the boundary condition $\mathbb{B}_{d+1}^{(r,1)}$. This also fully fixes the corner $\mathbb{B}_{d}^{(1,1)}$ as the interface realizing this Fourier transformation.

The argument is based on considering a small neighborhood of the corner $\mathbb{B}_d^{(1,1)}$ where the edges $\mathbb{B}_{d+1}^{(1,x_\perp)}$ and  $\mathbb{B}_{d+1}^{(r,1)}$ meet. There, consider a field $B$ of $\mathcal{S}_{d+2}$ with, for example, Dirichlet boundary conditions imposed by $\mathbb{B}_{d+1}^{(r,1)}$, i.e., the field $B$ restricted to this edge vanishes. The field $B$ is a cochain and then, near the boundary  $\mathbb{B}_{d+1}^{(r,1)}$, we can expand it as $B=B' \cup dx_\perp$, with some cochain $B'$ in one degree lower. Next, assuming continuity in a neighborhood of the corner, approach the boundary condition supporting $\mathbb{B}_{d+1}^{(1,x_\perp)}$, here $B=B'\cup dx_\perp$ is now compatible with Neumann boundary conditions, the Fourier dual boundary conditions.

For the consistency check start with the sandwich $(\mathbb{B}_{d+1}^{(r,1)};\mathcal{S}_{d+2};\mathcal{T}_{d+1})$ which describes an absolute QFT in dimension $d+1$. Then, the boundary conditions $\mathbb{B}_{d+1}^{(r,1)}$ picks out a subset of genuine defects, the remaining defects are instead realized as non-genuine defects. Start by constructing the non-genuine defect via an extended operator of $\mathcal{S}_{d+2}$ which runs at $r=\textnormal{const}$. This defect can not terminate at $\mathbb{B}_{d+1}^{(r,1)}$, otherwise it would be genuine, hence we need to run it to a different boundary to realize the desired non-genuine defect. This is achieved by taking a turn at some value of $x_\perp$ and then continue extending the defect along the direction parametrized by $r$:
\be
\scalebox{0.8}{
\begin{tikzpicture}
	\begin{pgfonlayer}{nodelayer}
		\node [style=none] (0) at (-2, 1.5) {};
		\node [style=none] (1) at (2, 1.5) {};
		\node [style=none] (2) at (2, -1.5) {};
		\node [style=none] (3) at (-2, -1.5) {};
		\node [style=SmallCircleRed] (4) at (2, -1.5) {};
		\node [style=SmallCircleBlue] (5) at (2, 1.5) {};
		\node [style=SmallCircleGreen] (6) at (-2, 1.5) {};
		\node [style=SmallCircleBrown] (7) at (-2, -1.5) {};
		\node [style=none] (8) at (-2.75, 0) {$\mathbb{B}_{d+1}^{(1,x_\perp)}$};
		\node [style=none] (10) at (-2.5, 2.125) {$\mathbb{B}_{d}^{(1,1)}$};
		\node [style=none] (11) at (0, 2.125) {$\mathbb{B}_{d+1}^{(r,1)}$};
		\node [style=none] (12) at (-0.1, -1.5) {};
		\node [style=none] (13) at (-0.1, -0.1) {};
		\node [style=none] (14) at (-2, -0.1) {};
		\node [style=none] (15) at (2.5, -0.75) {};
		\node [style=none] (16) at (2.5, 0.75) {};
		\node [style=none] (17) at (1.25, -2) {};
		\node [style=none] (18) at (-1.25, -2) {};
		\node [style=none] (19) at (3, 0) {$x_\perp$};
		\node [style=none] (20) at (0, -2.5) {$r$};
	\end{pgfonlayer}
	\begin{pgfonlayer}{edgelayer}
		\filldraw[fill=gray!50, draw=gray!50]  (-2, 1.5) -- (2, 1.5) -- (2, -1.5) -- (-2, -1.5) -- cycle;
		\draw [style=RedLine] (3.center) to (2.center);
		\draw [style=BlueLine] (0.center) to (1.center);
		\draw [style=PurpleLine] (1.center) to (2.center);
		\draw [style=GreenLine] (0.center) to (3.center);
		\draw [style=ArrowLineRight] (15.center) to (16.center);
		\draw [style=ArrowLineRight] (17.center) to (18.center);
		\draw [style=BrownLine, snake it] (12.center) to (13.center);
		\draw [style=BrownLine,  snake it] (14.center) to (13.center);
	\end{pgfonlayer}
\end{tikzpicture}}
\ee
In order to realize the non-genuine defect in the QFT we must be able to terminate the defect at $\mathbb{B}_{d+1}^{(1,x_\perp)}$, i.e., Neumann boundary conditions along $\mathbb{B}_{d+1}^{(r,1)}$ are consistent with Dirichlet boundary conditions at $\mathbb{B}_{d+1}^{(1,x_\perp)}$.

Overall, the triple $(\mathbb{B}_{d+1}^{(r,1)}; \mathbb{B}_d^{(1,1)}; \mathbb{B}_{d+1}^{(1,x_\perp)})$ is seen to be associated to the global form of the flavor brane theory $\mathcal{T}_{d+1}$ and fixed by either of the two edges. For this reason we can focus on $\mathbb{B}_{d+1}^{(r,1)}$ and in subsection \ref{ssec:GlobalForm} we discuss how the geometry $X$ determines $\mathbb{B}_{d+1}^{(r,1)}$.

With this, we turn to discuss the corners $\mathbb{B}^{(0,1)}_d,\mathbb{B}^{(1,0)}_d$ which are respectively interfaces between edges $\mathcal{B}_{d+1},\mathcal{T}_{d+1}$ and $\mathbb{B}_{d+1}^{(r,1)},\mathbb{B}_{d+1}^{(1,x_\perp)}$, where the latter pair is gapped or free. We characterize the corners via the boundary conditions they impose on these edges.

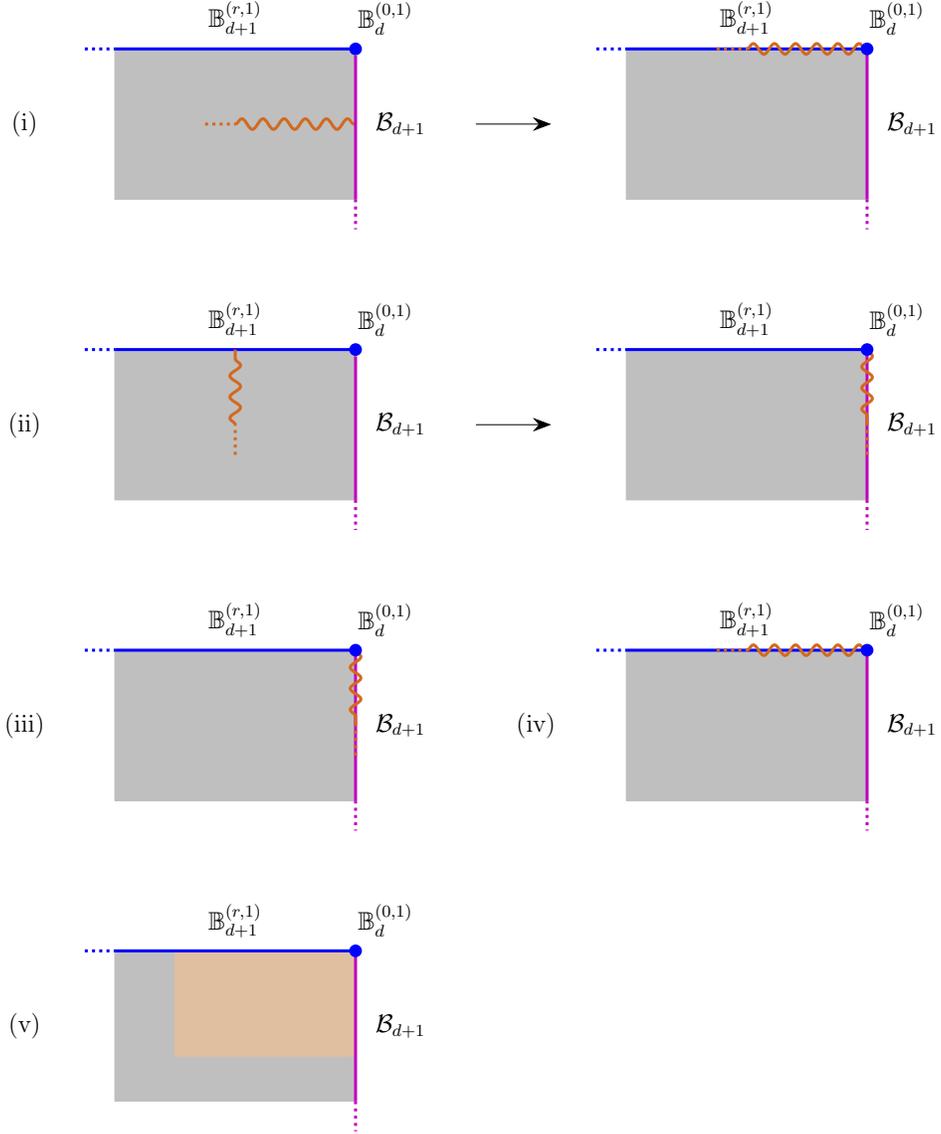
\begin{figure}
\centering
\scalebox{0.8}{\begin{tikzpicture}
	\begin{pgfonlayer}{nodelayer}
		\node [style=none] (1) at (2, 1.25) {};
		\node [style=none] (2) at (-2, 1.25) {};
		\node [style=SmallCircleBlue] (3) at (2, 1.25) {};
		\node [style=none] (5) at (0, 1.75) {$\mathbb{B}^{(r,1)}_{d+1}$};
		\node [style=none] (6) at (2.75, 0) {$\mathcal{B}_{d+1}$};
		\node [style=none] (8) at (2.5, 1.75) {$\mathbb{B}^{(0,1)}_d$};
		\node [style=none] (11) at (2, 0) {};
		\node [style=none] (12) at (0, 0) {};
		\node [style=none] (13) at (-0.5, 0) {};
		\node [style=none] (15) at (10.5, 1.25) {};
		\node [style=none] (16) at (6.5, 1.25) {};
		\node [style=SmallCircleBlue] (17) at (10.5, 1.25) {};
		\node [style=none] (19) at (8.5, 1.75) {$\mathbb{B}^{(r,1)}_{d+1}$};
		\node [style=none] (20) at (11.25, 0) {$\mathcal{B}_{d+1}$};
		\node [style=none] (22) at (11, 1.75) {$\mathbb{B}^{(0,1)}_d$};
		\node [style=none] (25) at (10.5, 1.25) {};
		\node [style=none] (26) at (8.5, 1.25) {};
		\node [style=none] (27) at (8, 1.25) {};
		\node [style=none] (28) at (4, 0) {};
		\node [style=none] (29) at (5.25, 0) {};
		\node [style=none] (30) at (2, -6.25) {};
		\node [style=none] (31) at (2, -3.75) {};
		\node [style=none] (32) at (-2, -3.75) {};
		\node [style=SmallCircleBlue] (33) at (2, -3.75) {};
		\node [style=none] (35) at (0, -3.25) {$\mathbb{B}^{(r,1)}_{d+1}$};
		\node [style=none] (36) at (2.75, -5) {$\mathcal{B}_{d+1}$};
		\node [style=none] (38) at (2.5, -3.25) {$\mathbb{B}^{(0,1)}_d$};
		\node [style=none] (41) at (0, -3.75) {};
		\node [style=none] (42) at (0, -5) {};
		\node [style=none] (43) at (0, -5.5) {};
		\node [style=none] (44) at (10.5, -6.25) {};
		\node [style=none] (45) at (10.5, -3.75) {};
		\node [style=none] (46) at (6.5, -3.75) {};
		\node [style=SmallCircleBlue] (47) at (10.5, -3.75) {};
		\node [style=none] (49) at (8.5, -3.25) {$\mathbb{B}^{(r,1)}_{d+1}$};
		\node [style=none] (50) at (11.25, -5) {$\mathcal{B}_{d+1}$};
		\node [style=none] (52) at (11, -3.25) {$\mathbb{B}^{(0,1)}_d$};
		\node [style=none] (55) at (10.5, -3.75) {};
		\node [style=none] (56) at (10.5, -5) {};
		\node [style=none] (57) at (10.5, -5.5) {};
		\node [style=none] (58) at (4, -5) {};
		\node [style=none] (59) at (5.25, -5) {};
		\node [style=none] (60) at (2, -1.25) {};
		\node [style=none] (61) at (-2.5, 1.25) {};
		\node [style=none] (62) at (2, -1.75) {};
		\node [style=none] (63) at (10.5, -6.75) {};
		\node [style=none] (64) at (2, -6.75) {};
		\node [style=none] (65) at (-2.5, -3.75) {};
		\node [style=none] (66) at (10.5, -1.25) {};
		\node [style=none] (67) at (10.5, -1.75) {};
		\node [style=none] (68) at (6, 1.25) {};
		\node [style=none] (69) at (6, -3.75) {};
		\node [style=none] (70) at (2, -8.75) {};
		\node [style=none] (71) at (-2, -8.75) {};
		\node [style=SmallCircleBlue] (72) at (2, -8.75) {};
		\node [style=none] (73) at (0, -8.25) {$\mathbb{B}^{(r,1)}_{d+1}$};
		\node [style=none] (74) at (2.75, -10) {$\mathcal{B}_{d+1}$};
		\node [style=none] (75) at (2.5, -8.25) {$\mathbb{B}^{(0,1)}_d$};
		\node [style=none] (76) at (2, -8.75) {};
		\node [style=none] (79) at (10.5, -8.75) {};
		\node [style=none] (80) at (6.5, -8.75) {};
		\node [style=SmallCircleBlue] (81) at (10.5, -8.75) {};
		\node [style=none] (82) at (8.5, -8.25) {$\mathbb{B}^{(r,1)}_{d+1}$};
		\node [style=none] (83) at (11.25, -10) {$\mathcal{B}_{d+1}$};
		\node [style=none] (84) at (11, -8.25) {$\mathbb{B}^{(0,1)}_d$};
		\node [style=none] (85) at (10.5, -8.75) {};
		\node [style=none] (86) at (8.5, -8.75) {};
		\node [style=none] (87) at (8, -8.75) {};
		\node [style=none] (90) at (2, -11.25) {};
		\node [style=none] (91) at (-2.5, -8.75) {};
		\node [style=none] (92) at (2, -11.75) {};
		\node [style=none] (93) at (10.5, -11.25) {};
		\node [style=none] (94) at (10.5, -11.75) {};
		\node [style=none] (95) at (6, -8.75) {};
		\node [style=none] (96) at (-3.5, 0) {(i)};
		\node [style=none] (97) at (-3.5, -5) {(ii)};
		\node [style=none] (98) at (-3.5, -10) {(iii)};
		\node [style=none] (99) at (5, -10) {(iv)};
		\node [style=none] (100) at (2, -8.75) {};
		\node [style=none] (101) at (2, -10) {};
		\node [style=none] (102) at (2, -10.5) {};
		\node [style=none] (103) at (2, -13.75) {};
		\node [style=none] (104) at (-2, -13.75) {};
		\node [style=SmallCircleBlue] (105) at (2, -13.75) {};
		\node [style=none] (106) at (0, -13.25) {$\mathbb{B}^{(r,1)}_{d+1}$};
		\node [style=none] (107) at (2.75, -15) {$\mathcal{B}_{d+1}$};
		\node [style=none] (108) at (2.5, -13.25) {$\mathbb{B}^{(0,1)}_d$};
		\node [style=none] (109) at (2, -13.75) {};
		\node [style=none] (112) at (2, -16.25) {};
		\node [style=none] (113) at (2, -16.75) {};
		\node [style=none] (114) at (-2.5, -13.75) {};
		\node [style=none] (115) at (-3.5, -15) {(v)};
		\node [style=none] (116) at (2, -17.25) {};
	\end{pgfonlayer}
	\begin{pgfonlayer}{edgelayer}
		\filldraw[fill=gray!50, draw=gray!50]  (-2, 1.25) -- (2, 1.25) -- (2, -1.25) -- (-2, -1.25) -- cycle;
		
		\filldraw[fill=gray!50, draw=gray!50]  (-2, -6.25) -- (2, -6.25) -- (2, -3.75) -- (-2, -3.75) -- cycle;
		
		\filldraw[fill=gray!50, draw=gray!50]  (-2, -11.25) -- (2, -11.25) -- (2, -8.75) -- (-2, -8.75) -- cycle;
		
		\filldraw[fill=gray!50, draw=gray!50]  (-2, -16.25) -- (2, -16.25) -- (2, -13.75) -- (-2, -13.75) -- cycle;
		
		\filldraw[fill=brown!50, draw=brown!50]  (-1, -15.5) -- (2, -15.5) -- (2, -13.75) -- (-1, -13.75) -- cycle;
		
		\filldraw[fill=gray!50, draw=gray!50]  (10.5, 1.25) -- (6.5, 1.25) -- (6.5, -1.25) -- (10.5, -1.25) -- cycle;
		
		\filldraw[fill=gray!50, draw=gray!50]  (10.5, -6.25) -- (6.5, -6.25) -- (6.5, -3.75) -- (10.5, -3.75) -- cycle;
		
		\filldraw[fill=gray!50, draw=gray!50]  (10.5, -11.25) -- (6.5, -11.25) -- (6.5, -8.75) -- (10.5, -8.75) -- cycle;
		\draw [style=BlueLine] (2.center) to (1.center);
		\draw [style=BrownLine, snake it] (12.center) to (11.center);
		\draw [style=DottedBrown] (13.center) to (12.center);
		\draw [style=BlueLine] (16.center) to (15.center);
		\draw [style=BrownLine, snake it] (26.center) to (25.center);
		\draw [style=DottedBrown] (27.center) to (26.center);
		\draw [style=ArrowLineRight] (28.center) to (29.center);
		\draw [style=BlueLine] (32.center) to (31.center);
		\draw [style=BrownLine, snake it] (42.center) to (41.center);
		\draw [style=DottedBrown] (43.center) to (42.center);
		\draw [style=BlueLine] (46.center) to (45.center);
		\draw [style=ArrowLineRight] (58.center) to (59.center);
		\draw [style=DottedBlue] (68.center) to (16.center);
		\draw [style=BlueLine] (71.center) to (70.center);
		\draw [style=BlueLine] (80.center) to (79.center);
		\draw [style=BrownLine, snake it] (86.center) to (85.center);
		\draw [style=DottedBrown] (87.center) to (86.center);
		\draw [style=DottedPurple] (30.center) to (64.center);
		\draw [style=DottedPurple] (44.center) to (63.center);
		\draw [style=DottedPurple] (93.center) to (94.center);
		\draw [style=DottedPurple] (90.center) to (92.center);
		\draw [style=PurpleLine] (85.center) to (93.center);
		\draw [style=PurpleLine] (76.center) to (90.center);
		\draw [style=DottedBlue] (65.center) to (32.center);
		\draw [style=DottedBlue] (69.center) to (46.center);
		\draw [style=DottedBlue] (95.center) to (80.center);
		\draw [style=DottedBlue] (91.center) to (71.center);
		\draw [style=PurpleLine] (33) to (30.center);
		\draw [style=PurpleLine] (55.center) to (44.center);
		\draw [style=PurpleLine] (3) to (60.center);
		\draw [style=PurpleLine] (25.center) to (66.center);
		\draw [style=DottedPurple] (60.center) to (62.center);
		\draw [style=DottedPurple] (66.center) to (67.center);
		\draw [style=DottedBlue] (61.center) to (2.center);
		\draw [style=BrownLine, snake it] (55.center) to (56.center);
		\draw [style=DottedBrown] (57.center) to (56.center);
		\draw [style=BrownLine, snake it] (100.center) to (101.center);
		\draw [style=DottedBrown] (102.center) to (101.center);
		\draw [style=BlueLine] (104.center) to (103.center);
		\draw [style=DottedPurple] (112.center) to (113.center);
		\draw [style=PurpleLine] (109.center) to (112.center);
		\draw [style=DottedBlue] (114.center) to (104.center);
	\end{pgfonlayer}
\end{tikzpicture}
}
\caption{Defect deformations and configurations determining properties of the corner $\mathbb{B}_{d}^{(0,1)}$. The snaked lines indicate defect\,/\,symmetry operators, shaded brown square in (v) also indicates a defect operator. Defects can be trivial in the $(d+2)$-dimensional bulk, e.g., subfigure (i) includes the case in which a genuine operator of $\mathcal{B}_{d+1}$ is deformed into $\mathbb{B}_{d}^{(0,1)}$. }
\label{fig:CornerCond}
\end{figure}

We give this characterization in terms of a defect analysis in subsection \ref{ssec:Corners}. For instance we determine the fate of bulk operators of the edges (which possibly extend into $\mathcal{S}_{d+2}$) when pushed into $\mathbb{B}^{(0,1)}_d$ (see subfigure (i), (ii) of figure \ref{fig:CornerCond}). We also determine which of the defects, that can not be deformed away from the edges, can be terminated at $\mathbb{B}^{(0,1)}_d$ (see subfigure (iii), (iv), (v) of figure \ref{fig:CornerCond}). These two considerations cover how the defect\,/\,symmetry operators of the edges $\mathbb{B}_{d+1}^{(r,1)},\mathcal{B}^{}_{d+1}$ interact with $\mathbb{B}^{(0,1)}_d$. Identically we study the other corners.

Finally, we note that in our geometric constructions it is extremely natural to deform the closed cheesesteak \eqref{eq:closed} by contracting the edge $\mathbb{B}_{d+1}^{(r,1)}$ and stacking the corner $\mathbb{B}^{(0,1)}_d,\mathbb{B}^{(1,1)}_d$ resulting in $\mathbb{B}^{(1,1),*}_d$. Overall this contraction leads to:
\be\scalebox{0.9}{
\begin{tikzpicture}
	\begin{pgfonlayer}{nodelayer}
		\node [style=none] (0) at (2, -1.25) {};
		\node [style=none] (1) at (-2, 1.25) {};
		\node [style=none] (2) at (-2, 1.25) {};
		\node [style=none] (3) at (-2, -1.25) {};
		\node [style=SmallCircleGrey] (4) at (-2, 1.25) {};
		\node [style=SmallCircleRed] (5) at (2, -1.25) {};
		\node [style=none] (7) at (-0.75, -0.5) {$\mathcal{S}_{d+2}$};
		\node [style=none] (8) at (0, -1.75) {$\mathcal{T}_{d+1}$};
		\node [style=none] (9) at (0.75, 0.5) {$\mathcal{B}_{d+1}$};
		\node [style=none] (10) at (2.5, -1.75) {$\mathcal{T}_{d}$};
		\node [style=SmallCircleBrown] (13) at (-2, -1.25) {};
		\node [style=none] (14) at (-2.75, 0) {$\mathbb{B}_{d+1}^{(1,x_\perp)}$};
		\node [style=none] (15) at (-2.625, -1.75) {$\mathbb{B}_{d}^{(1,0)}$};
		\node [style=none] (15) at (-2.5, 1.75) {$\mathbb{B}_{d}^{(1,1),*}$};
	\end{pgfonlayer}
	\begin{pgfonlayer}{edgelayer}
		\filldraw[fill=gray!50, draw=gray!50]  (2, -1.25) -- (-2, 1.25) -- (-2, -1.25)-- cycle;
		\draw [style=RedLine] (3.center) to (0.center);
		\draw [style=PurpleLine] (1.center) to (0.center);
		\draw [style=GreenLine] (2.center) to (13);
	\end{pgfonlayer}
\end{tikzpicture}}
\ee
Let us motivate this contraction by consideing figure \ref{fig:CheeseGeometry} where we execute the homotopy $T_{\mathscr{S}}\sim X$. In the figure the contraction of the edge lifts to contraction of the blue neighborhoods, simultaneously growing the purple neighborhood. The new corner $\mathbb{B}^{(1,1),*}_d$ can then be analyzed as sketched above. We group the gapped\,/\,free edges and corners into the tuple
\be
{\mathbb{B}}=\lb \mathbb{B}_d^{(1,0)},\mathbb{B}_{d+1}^{(1,x_\perp)},\mathbb{B}_{d}^{(1,1),*}\rb\,.
\ee

\subsubsection{Global Form of Flavor Symmetries and $(\mathbb{B}_{d+1}^{(r,1)}; \mathbb{B}_d^{(1,1)}; \mathbb{B}_{d+1}^{(1,x_\perp)})$}
\label{ssec:GlobalForm}

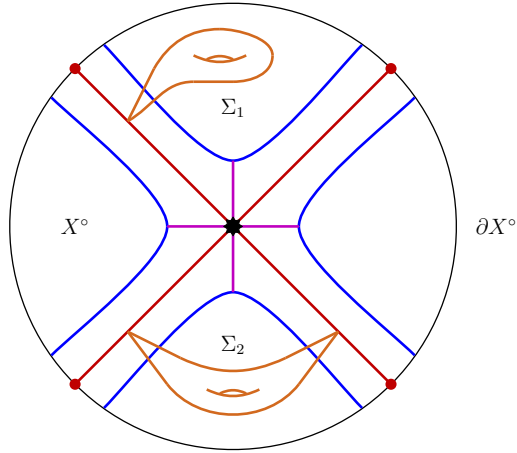
\begin{figure}
\centering
\scalebox{0.7}{
\begin{tikzpicture}
	\begin{pgfonlayer}{nodelayer}
		\node [style=none] (0) at (-3, 3) {};
		\node [style=none] (1) at (3, 3) {};
		\node [style=none] (2) at (3, -3) {};
		\node [style=none] (3) at (-3, -3) {};
		\node [style=Star] (4) at (0, 0) {};
		\node [style=SmallCircleRed] (5) at (-3, 3) {};
		\node [style=SmallCircleRed] (6) at (3, 3) {};
		\node [style=SmallCircleRed] (7) at (3, -3) {};
		\node [style=SmallCircleRed] (8) at (-3, -3) {};
		\node [style=none] (9) at (0, 1.25) {};
		\node [style=none] (10) at (0, -1.25) {};
		\node [style=none] (11) at (1.25, 0) {};
		\node [style=none] (12) at (-1.25, 0) {};
		\node [style=none] (13) at (-2.45, 3.45) {};
		\node [style=none] (14) at (-3.45, 2.45) {};
		\node [style=none] (15) at (2.45, 3.45) {};
		\node [style=none] (16) at (3.45, 2.45) {};
		\node [style=none] (17) at (3.45, -2.45) {};
		\node [style=none] (18) at (2.45, -3.45) {};
		\node [style=none] (19) at (-2.45, -3.45) {};
		\node [style=none] (20) at (-3.45, -2.45) {};
		\node [style=none] (21) at (-1.5, 3) {};
		\node [style=none] (22) at (-0.75, 2.75) {};
		\node [style=none] (23) at (-0.5, 3.175) {};
		\node [style=none] (24) at (0, 3.175) {};
		\node [style=none] (25) at (0.25, 3.25) {};
		\node [style=none] (26) at (-0.75, 3.25) {};
		\node [style=none] (27) at (-2, 2) {};
		\node [style=none] (28) at (-0.25, 3.75) {};
		\node [style=none] (29) at (0, 2.75) {};
		\node [style=none] (30) at (0.75, 3.25) {};
		\node [style=none] (31) at (-2, -2) {};
		\node [style=none] (32) at (2, -2) {};
		\node [style=none] (33) at (1, -2.5) {};
		\node [style=none] (34) at (1.25, -3) {};
		\node [style=none] (35) at (-1, -2.5) {};
		\node [style=none] (36) at (-1.25, -3) {};
		\node [style=none] (37) at (-0.25, -3.175) {};
		\node [style=none] (38) at (0.25, -3.175) {};
		\node [style=none] (39) at (0.5, -3.1) {};
		\node [style=none] (40) at (-0.5, -3.1) {};
		\node [style=none] (41) at (-3, 0) {$X^\circ$};
		\node [style=none] (42) at (5, 0) {$\partial X^\circ$};
		\node [style=none] (43) at (0, 2.25) {$\Sigma_1$};
		\node [style=none] (44) at (0, -2.25) {$\Sigma_2$};
	\end{pgfonlayer}
	\begin{pgfonlayer}{edgelayer}
		\draw [style=ThickLine, bend left=45] (0.center) to (1.center);
		\draw [style=ThickLine, bend left=45] (1.center) to (2.center);
		\draw [style=ThickLine, bend left=45] (2.center) to (3.center);
		\draw [style=ThickLine, bend left=45] (3.center) to (0.center);
		\draw [style=RedLine] (0.center) to (2.center);
		\draw [style=RedLine] (1.center) to (3.center);
		\draw [style=BlueLine, in=-180, out=-45, looseness=0.50] (13.center) to (9.center);
		\draw [style=BlueLine, in=225, out=0, looseness=0.50] (9.center) to (15.center);
		\draw [style=BlueLine, in=90, out=-45, looseness=0.50] (14.center) to (12.center);
		\draw [style=BlueLine, in=45, out=-90, looseness=0.50] (12.center) to (20.center);
		\draw [style=BlueLine, in=45, out=180, looseness=0.50] (10.center) to (19.center);
		\draw [style=BlueLine, in=135, out=0, looseness=0.50] (10.center) to (18.center);
		\draw [style=BlueLine, in=-90, out=135, looseness=0.50] (17.center) to (11.center);
		\draw [style=BlueLine, in=225, out=90, looseness=0.50] (11.center) to (16.center);
		\draw [style=PurpleLine] (9.center) to (10.center);
		\draw [style=PurpleLine] (12.center) to (11.center);
		\draw [style=BrownLine] (21.center) to (27.center);
		\draw [style=BrownLine, in=180, out=30] (27.center) to (22.center);
		\draw [style=BrownLine, bend left=25] (23.center) to (24.center);
		\draw [style=BrownLine, bend right=25] (26.center) to (25.center);
		\draw [style=BrownLine] (22.center) to (29.center);
		\draw [style=BrownLine, in=-180, out=60] (21.center) to (28.center);
		\draw [style=BrownLine, in=90, out=0, looseness=0.75] (28.center) to (30.center);
		\draw [style=BrownLine, in=0, out=-90] (30.center) to (29.center);
		\draw [style=BrownLine] (31.center) to (35.center);
		\draw [style=BrownLine] (31.center) to (36.center);
		\draw [style=BrownLine] (33.center) to (32.center);
		\draw [style=BrownLine] (32.center) to (34.center);
		\draw [style=BrownLine, bend left=52] (34.center) to (36.center);
		\draw [style=BrownLine, bend right=25] (35.center) to (33.center);
		\draw [style=BrownLine, bend left=25] (37.center) to (38.center);
		\draw [style=BrownLine, bend right=25] (40.center) to (39.center);
	\end{pgfonlayer}
\end{tikzpicture}
}
\caption{Resketch of subfigure (ii) of figure \ref{fig:GeoEng}. The space $X^\circ=X\setminus T_{\mathscr{S}}$, delineated by the blue lines, is ``at infinity" with respect to the tubular neighbourhood $T_{\mathscr{S}}$ centered on the singular locus $\mathscr{S}$. Topological boundary condition are imposed at $\partial T_{\mathscr{S}} $. They are determined by the topology of $X^\circ$. Flavor defects in the flavor brane local model are given by brane wrappings of $\Sigma_i\cap T_{\mathscr{S}}$. They can only occur in combinations permitted by $X^\circ$, i.e., they must extend to a cycle (brown) beyond $\partial T_{\mathscr{S}}$. Boundary conditions which imply other collection of defects are obstructed by the geometry.   }
\label{fig:FlavorDefects}
\end{figure}

We now give a geometric characterization of $\mathbb{B}_{d+1}^{(r,1)}$ setting the global form of $\mathcal{T}_{d+1}$. Recall, the vertical direction in \eqref{eq:open} is parametrized by $x_\perp$ which is the direction normal to both the flavor brane and normal to the radial direction $r$ in the geometry $X$. As such defects which run along $x_\perp$ at constant $r$, see subfigure (i) of figure \ref{fig:DefecGp}, are engineered by brane wrappings on cycles at constant radius. The wrapping loci are characterized by elements of $H_n(\partial X)$. Now, given a class in $H_n(\partial X)$ we have the mapping
\be \label{eq:condition}
R_n\,:\quad H_n(\partial X) \rightarrow H_n(T_{\mathscr{K}},\partial T_{\mathscr{K}})
\ee
defined by intersection with the tube $T_{\mathscr{K}}$. The image under this map corresponds, via all possible brane wrappings, to the  defect operators which can end on $\mathbb{B}_{d+1}^{(r,1)}$. In particular, unlike in the standard SymTFT discussion, $\mathbb{B}_{d+1}^{(r,1)}$ is fixed by the geometry and not subject to a choice. We sketch in figure \ref{fig:FlavorDefects} how flavor defects lift to $X$, beyond the lcoal model $T_{\mathscr{K}}$.

\medskip

{\bf Our Running Example} Consider the 5D SCFT engineered by M-theory on $\mathbb{C}^3/\Z_{2n}$. This SCFT has an $\mathfrak{su}_2$ flavor brane corresponding to an $A_1$ singularity at $z_1,z_2=0$. In \cite{Apruzzi:2021vcu}, see also \cite{Cvetic:2022imb, DelZotto:2022joo}, the flavor symmetry of the 5D SCFT was computed to be SO$(3)$, which we interpret to be the global form of the flavor brane theory. We now show that the geometry fixes the boundary condition $\mathbb{B}_{d+1}^{(r,1)}$ such that no genuine non-endable flavor Wilson lines can be constructed. For this note that, in the local model $T_{\mathscr{K}}$, such lines would be constructed from M2-branes wrappings of cones over 1-cycles in $H_1(\partial T_{\mathscr{K}})$. These cones are elements of $H_2(T_{\mathscr{K}},\partial T_{\mathscr{K}})$. The discussion around \eqref{eq:condition} can then be understood as the condition that these cones result as a restriction of some 2-cycle in $\partial X$ to $T_{\mathscr{K}}$. However, we have
\be
H_2(S^5/\Z_{2n})=0\,,
\ee
and therefore no such 2-cycles exist. Therefore, in the 6D KK flavor brane theory we must therefore instead have the dual 3-form $\Z_2$ symmetry acting on defects constructed via M5-brane wrappings. Indeed, we can wrap M5-branes on classes $H_3(S^5/\Z_{2n})$ which restrict in $T_{\mathscr{K}}$ to the correct defect. Equivalently, from the perspective of the flavor brane, we have determined which of the cones
\be \ba
\textnormal{Tor}\,H_2(T_{\mathscr{K}},\partial T_{\mathscr{K}})&\cong \textnormal{Tor}\, H_1(\partial T_{\mathscr{K}})\cong \Z_2\,, \\ \textnormal{Tor}\, H_3(T_{\mathscr{K}},\partial T_{\mathscr{K}})&\cong \textnormal{Tor}\, H_2(\partial T_{\mathscr{K}})\cong \Z_2\,,
\ea \ee
can be wrapped by M2-, M5-branes. The cycles in $ \textnormal{Tor}\, H_1(\partial T_{\mathscr{K}})$ and $ \textnormal{Tor}\, H_2(\partial T_{\mathscr{K}})$ link in $\partial T_{\mathscr{K}}$ and are therefore the cycles from which mutually non-local defects are constructed. \medskip

In summary, the boundary condition $\mathbb{B}_{d+1}^{(r,x_\perp)}$ is such that the defects realized by $p$-brane wrappings which can end on the boundary are isomorphic to
\be
\mathcal{P}=\oplus_n\mathcal{P}_{n}\,,\qquad \mathcal{P}_n=\underset{p-n=m-1}{%
{\displaystyle\bigoplus}
} \textnormal{Im}\, R_n \,,
\ee
in the obvious notation.

\subsubsection{Corners $\mathbb{B}^{(0,1)}_d,\mathbb{B}^{(1,0)}_d$}
\label{ssec:Corners}

We now characterize the corners $\mathbb{B}^{(0,1)}_d,\mathbb{B}^{(1,0)}_d$ through their interplay with defect\,/\,symmetry operators. First, we consider the corner $\mathbb{B}^{(0,1)}_d$ which is present in both the open \eqref{eq:open} and closed \eqref{eq:closed} cheesesteak. The presented analysis will also immediately generalizes to the very related corner $\mathbb{B}^{(1,1),*}_d$. We will find $\mathbb{B}^{(0,1)}_d$ to determine global properties of the end of the world theory \eqref{eq:EOW}. Then, whenever a limit isolating the $d$-dimensional dynamics exists the corner $\mathbb{B}^{(1,0)}_d$ is found to provide the missing information in setting the global form of the resulting $d$-dimensional theory.

\paragraph{Corner $\mathbb{B}^{(0,1)}_d$}\mbox{}\medskip

We now argue that there is a choice of polarization to be made in specifying the corner condition $\mathbb{B}^{(0,1)}_d$.

First, recall that the projection $\pi_{IJ}$, given in \eqref{eq:SquareFibration}, maps copies of $\partial X^\circ$ onto both $\mathbb{B}^{(0,1)}_d$ and points of the edge $\mathcal{B}_{d+1}$. Viewing the coordinate $x_\perp$ as parametrizing the decompression dimension of a symmetry sandwich, it follows that $\mathbb{B}^{(0,1)}_d$ specifies which of the defects naive defects $\widetilde{\mathbb{D}}[\mathcal{T}_d]$, given in \eqref{eq:naivedefects}, can terminate at the corner. This choice of such defects is not without constraints. Note, we have that $\mathbb{B}^{(0,1)}_d$ also is a boundary to the edge $\mathbb{B}^{(r,1)}_{d+1}$ and therefore $\mathbb{B}^{(0,1)}_d$ must permit the termination of all defects which can be deformed off $\mathcal{B}_{d+1}$ into $\mathcal{S}_{d+2}$ and which are simultaneously permitted to terminate on $\mathbb{B}^{(r,1)}_d$ (see subfigure (ii) of figure \ref{fig:CornerCond}). Equivalently, the endability of these defects at the corner $\mathbb{B}^{(0,1)}_d$  is inherited from the connecting edge $\mathbb{B}^{(r,1)}_{d+1}$.  This determines the endability of a subgroup of the naive defects $\widetilde{\mathbb{D}}[\mathcal{T}_d]$ and the quotient, which remains unconstrained by this condition, is precisely the defect group $\mathbb{D}[\mathcal{T}_d]$ defined in \eqref{eq:DefectGroup}.

We conclude, the data specifying the corner condition $\mathbb{B}^{(0,1)}_d$ includes a choice of maximally mutually local subgroup of the defect group $\mathbb{D}[\mathcal{T}_d]$, referred to as a polarization. See subfigure (iii) of figure \ref{fig:CornerCond} for a sketch of the type of defects selected in this choice. There is also a symmetry operator version of this analysis centered on considerations such as subfigure (i), (iv) of figure \ref{fig:CornerCond}.

We now discuss an extension of this notion of polarization, mixing structures of $\mathcal{T}_{d+1}$ and $\mathcal{T}_d$, which geometrically will be a consequence of $\partial^2 X^\circ$ being a smooth compact manifold and boundary to the smooth space $\partial X^\circ$. For convenience we restrict ourselves to discrete data although we expect our considerations to hold more broadly. Our considerations will also result in a Dirac pairing, filling a gap in the previous discussion.

We begin by studying extension properties of various defect operators. Consider the long exact sequence in relative homology of the pair $\partial^2 X^\circ, \partial X^\circ$:
\be
\dots  ~\rightarrow~ H_{n+1}(\partial X^\circ) ~\xrightarrow[]{~\iota_{n+1} ~}~ H_{n+1}(\partial X^\circ, \partial^2 X^\circ) ~\xrightarrow[]{~\partial_n ~}~ H_n(\partial^2 X^\circ) ~\xrightarrow[]{~\jmath_{n} ~}~ H_n(\partial X^\circ)~\rightarrow~\dots\,.
\ee
The image of the mapping $\partial_n$ is associated with defects of the type:
\be\label{eq:EXT1}
\scalebox{0.9}{
\begin{tikzpicture}
	\begin{pgfonlayer}{nodelayer}
		\node [style=SmallCircleBlue] (0) at (1, 1) {};
		\node [style=none] (1) at (-1, 1) {};
		\node [style=none] (2) at (-1.5, 1) {};
		\node [style=none] (3) at (1, -1) {};
		\node [style=none] (4) at (1, -1.5) {};
		\node [style=none] (5) at (1, -0.5) {};
		\node [style=none] (6) at (-0.5, 1) {};
		\node [style=none] (7) at (0, 1.5) {$\mathbb{B}_{d+1}^{(r,1)}$};
		\node [style=none] (8) at (1.75, 0) { $\mathcal{B}_{d+1}$};
		\node [style=none] (9) at (1.75, 1.5) {$\mathbb{B}_{d}^{(0,1)}$};
			
	\end{pgfonlayer}
	\begin{pgfonlayer}{edgelayer}
		\filldraw[fill=gray!50, draw=gray!50]  (1, -1) -- (1, 1) -- (-1, 1) -- (-1, -1) -- cycle;
		\filldraw[fill=brown!50, draw=brown!50]  (1, -0.75) -- (1, 1) -- (-0.75, 1) -- (-0.75, -0.75) -- cycle;
		\draw [style=BlueLine] (1.center) to (0);
		\draw [style=PurpleLine] (0) to (3.center);
		\draw [style=DottedPurple] (3.center) to (4.center);
		\draw [style=DottedBlue] (2.center) to (1.center);
	\end{pgfonlayer}
\end{tikzpicture}}
\ee
which when constructed via a $p$-brane wrapping have dimension $p+1-(k+1+1)=p-k-1$. A multiple of such defects can lie in the kernel of the map $\partial_n$ (their bulk part trivializes), and by exactness in the image of $\iota_{n+1}$, and such defects are then of the same dimension and localized to the edge:
\be\label{eq:EXT2}
\scalebox{0.9}{
\begin{tikzpicture}
	\begin{pgfonlayer}{nodelayer}
		\node [style=SmallCircleBlue] (0) at (1, 1) {};
		\node [style=none] (1) at (-1, 1) {};
		\node [style=none] (2) at (-1.5, 1) {};
		\node [style=none] (3) at (1, -1) {};
		\node [style=none] (4) at (1, -1.5) {};
		\node [style=none] (5) at (1, -0.5) {};
		\node [style=none] (6) at (-0.5, 1) {};
		\node [style=none] (7) at (0, 1.5) {$\mathbb{B}_{d+1}^{(r,1)}$};
		\node [style=none] (8) at (1.75, 0) {$\mathcal{B}_{d+1}$};
		\node [style=none] (9) at (1.75, 1.5) {$\mathbb{B}_{d}^{(0,1)}$};
	\end{pgfonlayer}
	\begin{pgfonlayer}{edgelayer}
		\filldraw[fill=gray!50, draw=gray!50]  (1, -1) -- (1, 1) -- (-1, 1) -- (-1, -1) -- cycle;
		\draw [style=BlueLine] (1.center) to (0);
		\draw [style=PurpleLine] (0) to (3.center);
		\draw [style=DottedPurple] (3.center) to (4.center);
		\draw [style=DottedBlue] (2.center) to (1.center);
		\draw [style=BrownLine, snake it] (0) to (5.center);
		\draw [style=DottedBrown] (5.center) to (3.center);
	\end{pgfonlayer}
\end{tikzpicture}}
\ee
In turn, multiples of such defects can then lie in the kernel of $\iota_{n+1}$ and result from bulk defects of the following type pushed onto the edge $\mathcal{B}_{d+1}$:
\be \label{eq:EXT3}
\scalebox{0.9}{
\begin{tikzpicture}
	\begin{pgfonlayer}{nodelayer}
		\node [style=SmallCircleBlue] (0) at (1, 1) {};
		\node [style=none] (1) at (-1, 1) {};
		\node [style=none] (2) at (-1.5, 1) {};
		\node [style=none] (3) at (1, -1) {};
		\node [style=none] (4) at (1, -1.5) {};
		\node [style=none] (5) at (1, -0.5) {};
		\node [style=none] (6) at (-0.5, 1) {};
		\node [style=none] (7) at (0, 1.5) {$\mathbb{B}_{d+1}^{(r,1)}$};
		\node [style=none] (8) at (1.75, 0) {$\mathcal{B}_{d+1}$};
		\node [style=none] (9) at (1.75, 1.5) {$\mathbb{B}_{d}^{(0,1)}$};
		\node [style=none] (10) at (0, 1) {};
		\node [style=none] (11) at (0, -0.5) {};
		\node [style=none] (12) at (0, -1) {};
	\end{pgfonlayer}
	\begin{pgfonlayer}{edgelayer}
		\filldraw[fill=gray!50, draw=gray!50]  (1, -1) -- (1, 1) -- (-1, 1) -- (-1, -1) -- cycle;
		\draw [style=BlueLine] (1.center) to (0);
		\draw [style=PurpleLine] (0) to (3.center);
		\draw [style=DottedPurple] (3.center) to (4.center);
		\draw [style=DottedBlue] (2.center) to (1.center);
		\draw [style=BrownLine, snake it] (10.center) to (11.center);
		\draw [style=DottedBrown] (11.center) to (12.center);
	\end{pgfonlayer}
\end{tikzpicture}}
\ee
The natural link pairings between all these defects then follow from dualities in algebraic topology. We have, by Poincar\'e duality and the universal coefficient theorem, the isomorphism
\be\label{eq:Dirac1}
\textnormal{Tor}\,H_{k}(\partial^2X^\circ)\cong \textnormal{Tor}\,H_{n-k-2}(\partial^2X^\circ)^\vee\,,
\ee
where $n=\dim \partial X^\circ=\dim \partial^2 X^\circ+1$ and $G^\vee=\textnormal{Hom}(G,\mathrm{U}(1))$ denotes the Pontryagin dual of a group $G$. Similarly we also have, by Poincar\'e-Lefschetz duality and the universal coefficient theorem, the isomorphism
\be \label{eq:Dirac2}
\textnormal{Tor}\,H_{k}(\partial X^\circ)\cong \textnormal{Tor}\,H_{n-k-1}(\partial X^\circ, \partial^2 X^\circ)^\vee\,.
\ee
These dualities extend to mappings in the sequence, i.e., the mappings
\be\label{eq:Dualmaps}\ba
\textnormal{Tor}\,H_{k}(\partial^2 X^\circ) ~&\rightarrow~\textnormal{Tor}\,H_{k}(\partial X^\circ)  \\
\textnormal{Tor}\,H_{n-k-1}(\partial X^\circ, \partial^2 X^\circ) ~&\rightarrow~\textnormal{Tor}\,H_{n-k-2}(\partial^2 X^\circ) \,,
\ea \ee
are related by duality. One consequence of this is that every extension relation of type \eqref{eq:EXT1} $\rightarrow$ \eqref{eq:EXT2}, for some $p$-brane construction of a defect, in accompanied by the extension relation of type \eqref{eq:EXT2} $\rightarrow$ \eqref{eq:EXT3} for the electromagnetically dual $q$-brane. The linkings \eqref{eq:Dirac1} and  \eqref{eq:Dirac2} then compute the Dirac pairings for such pairs compatible with these extension properties.

{\bf Our Running Example} Consider the above for the 5D SCFT engineered by M-theory on  $\mathbb{C}^3/\Z_{2n}$. For this geometry the long exact sequence above decomposes into four exact subsequences (see appendix \ref{app:A}). They are
\be \ba
0 ~\rightarrow~ H_{2}(\partial X^\circ, \partial^2 X^\circ) ~\rightarrow~  H_1(\partial^2 X^\circ)~\rightarrow~ H_1(\partial X^\circ)~\rightarrow~&0 \\
0 ~\rightarrow~ \Z  ~\xrightarrow[]{~1\,\mapsto\, (n,1) ~}~ \Z\oplus \Z_2 ~\xrightarrow[]{~(a,b)\,\mapsto\, a+nb ~}~ \Z_{2n}~\rightarrow~&0\,,
\ea
\ee
and
\be \ba
0 ~\rightarrow~  H_3(\partial^2 X^\circ)~\rightarrow~ H_3(\partial X^\circ)~\rightarrow~ H_3(\partial X^\circ, \partial^2 X^\circ)~\rightarrow~H_2(\partial^2 X^\circ)~\rightarrow~&0 \\
0 ~\rightarrow~ \Z ~\xrightarrow[]{~1\,\mapsto\, n ~}~ \Z~\xrightarrow[]{~1\,\mapsto\, 2 ~}~ \Z_{2n}~\xrightarrow[]{~1\,\mapsto\, 1 ~}~\Z_2~\rightarrow~&0\,,
\ea
\ee
together with a trivial sequence in degree 4, 5 and 0 which we omit here. Now, consider the defect group of electric line operators constructed by wrapping M2-branes on a cone over elements in $H_1(\partial X^\circ)\cong \Z_{2n}$ modulo $\Z_2$, i.e., $\Z_n$. We see that $n$ copies of any defect line is zero in $\Z_n$ but possibly non-zero in $\Z_{2n}$ and if non-trivial can be deformed to a genuine line operator of type \eqref{eq:EXT3} . Similarly, consider a defect of type \eqref{eq:EXT3} which projects to fill the full 2-plane and is constructed by wrapping an M5-brane on a cone over a 3-cycle given by a family of cycles $H_2(\partial^2 X^\circ)\cong \Z_2$ capped off by relative cycle in $H_3(\partial X^\circ,\partial^2 X^\circ)$. Two copies of this defect trivialize in the bulk of the 2-plane but remain non-trivial on the edge.

Overall, we see that the defect group of electric lines and magnetic surfaces, individually isomorphic to $\Z_n$, are extended both to $\Z_{2n}$. Physically, this can be interpreted as the defect group of the electric lines and magnetic surfaces if we were to gauge the $\mathfrak{su}_2$ flavor symmetry. Equivalently, the above reflects aspects of the 2-group symmetry of the 5D SCFT which informs on how to combine the defect groups of the SCFT and the flavor brane theory.

Let us push the example further, given the interpretation in the previous paragraph we expect a Dirac pairing on $\Z_{2n}\times \Z_{2n}$. Indeed, geometrically we will argue it to be determined via linking of the groups $H_1(\partial X^\circ)$ and $H_3(\partial X^\circ, \partial^2 X^\circ)\cong H_3(\partial X)$. We compute this pairing:
\be\label{eq:pairing} \ba
H_1(\partial X^\circ) \times H_3(\partial X^\circ, \partial^2 X^\circ)~&\rightarrow ~\Q/\Z  \\
\Z_{2n}\times\Z_{2n}~&\rightarrow ~\Q/\Z  \\[0.25em]
(a,b)~&\mapsto\;-\frac{ab}{2n}\,.
\ea \ee
This reduced to the standard Dirac pairing on the lines and surfaces of $\mathbb{D}[\mathcal{T}_d]$ by restriction. From the above we see that the lines are to be identified as the $\Z_n$ obtained from $\Z_{2n}$ after modding by $n$ and the surfaces result as the $\Z_n$ subgroup of $\Z_{2n}$. The latter gives a factor of 2, we thus have
\be \ba
\Z_{n}\times\Z_{n}~&\rightarrow ~\Q/\Z  \\[0.25em]
(a,b)~&\mapsto\;-\frac{ab}{n}\,,
\ea \ee
for the 5D defect group of electric lines and magnetic surfaces.

\paragraph{Corner $\mathbb{B}^{(1,0)}_d$}\mbox{}\medskip

We now argue that there is no second choice of polarization to be made in specifying the corner condition $\mathbb{B}^{(1,0)}_d$ of the closed cheesesteak, however we need to impose boundary conditions for fields localized to $\mathcal{T}_{d+1}$.

First, recall that the projection $\pi_{IJ}$, given in \eqref{eq:SquareFibration}, maps copies of the locus $\mathscr{K}=\partial \mathscr{S}$ onto both $\mathbb{B}^{(0,1)}_d$ and points of the edge $\mathcal{T}_{d+1}$. Points of the other edge $\mathbb{B}^{(1,x_\perp)}_d$ connecting to $\mathbb{B}^{(0,1)}_d$ are mapped onto by copies of $\partial^2 X^\circ$ which collapses to $\mathscr{K}$ when $x_\perp\rightarrow0$. In principle, the previous analysis now simply repeats with this altered starting point, however, due to our particular choice of decomposition $\partial X=\partial X^\circ \cup T_{\mathscr{K}}$, see \eqref{eq:PreferredDecomp}, we have that the mappings
\be \label{eq:mappingIn}
H_n(\partial^2 X^\circ)=H_n(\partial T_{\mathscr{K}})\rightarrow H_n(T_{\mathscr{K}})\cong H_n(\mathscr{K})\,,
\ee
are surjective. Consequently, the geometry projecting to the corner supports no additional cycles and no additional defects are localized to the edge $\mathcal{T}_{d+1}$ for which we would have to specify a polarization. More precisely, lifting $\mathcal{T}_{d+1}$ we have less cycles\footnote{Such statements of course depend on the (co)homology theory employed. Here, we are working with singular (co)homology, which however is expected to be too coarse in non-smooth settings. A next better approximation would be the orbifold cohoomology of Chen and Ruan \cite{Chen:2000cy}, we defer this to future work.} for our constructions as the above mapping can have a kernel, and overall the set of defects admissible to terminate at  $\mathbb{B}^{(1,0)}_d$ is inherited from the boundary conditions imposed along $\mathbb{B}^{(1,x_\perp)}_{d+1}$.

Recall next that $\mathcal{T}_{d+1}$ constitutes an enriched Neumann boundary condition for the bulk $\mathcal{S}_{d+2}$, as such we also need to impose boundary conditions for the degrees of freedom related to this ``enrichment''. This problem is fairly situational, depending the degrees of freedom along $\mathcal{T}_d$ and we defer a careful treatment of this question to future work. However, we note, by way of example, that these boundary conditions of $\mathcal{T}_{d+1}$ at  $\mathbb{B}^{(1,0)}_d$ are not independent from our defect discussion.\medskip

{\bf Our Running Example} Consider the 5D SCFT engineered by M-theory on $X=\mathbb{C}^3/\Z_{2n}$ in an electric polarization. The support of a defect constructed from a M2-brane wrapping over a cone on a free generator of $H_1(\partial^2 X^\circ)\cong \Z\oplus \Z_2$, away from the singular locus $\mathscr{S}$, is visualized as:
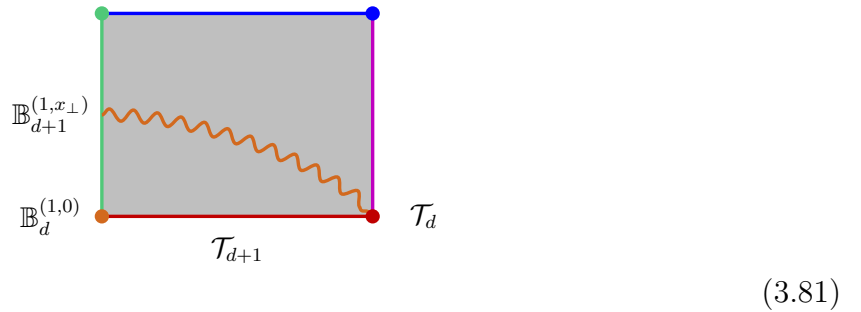
\be
\scalebox{0.9}{
\begin{tikzpicture}
	\begin{pgfonlayer}{nodelayer}
		\node [style=none] (0) at (2, -1.5) {};
		\node [style=SmallCircleRed] (1) at (2, -1.5) {};
		\node [style=none] (2) at (2, -1.5) {};
		\node [style=none] (5) at (0, -2) {$\mathcal{T}_{d+1}$};
		\node [style=none] (6) at (2.75, -1.5) {$\mathcal{T}_d$};
		\node [style=none] (7) at (-2, 1.5) {};
		\node [style=none] (8) at (-2, -1.5) {};
		\node [style=none] (9) at (2, 1.5) {};
		\node [style=SmallCircleBrown] (10) at (-2, -1.5) {};
		\node [style=SmallCircleGreen] (11) at (-2, 1.5) {};
		\node [style=SmallCircleBlue] (12) at (2, 1.5) {};
		\node [style=none] (13) at (-2.75, 0) {$\mathbb{B}^{(1,x_\perp)}_{d+1}$};
		\node [style=none] (14) at (0, -2.75) {};
		\node [style=none] (15) at (-2.75, -1.5) {$\mathbb{B}_d^{(1,0)}$};
		\node [style=none] (16) at (-2, 0) {};
	\end{pgfonlayer}
	\begin{pgfonlayer}{edgelayer}
		\filldraw[fill=gray!50, draw=gray!50]  (2, -1.5) -- (2, 1.5) -- (-2, 1.5) -- (-2, -1.5) -- cycle;
		\draw [style=ThickLine] (2.center) to (1);
		\draw [style=GreenLine] (7.center) to (8.center);
		\draw [style=RedLine] (8.center) to (2.center);
		\draw [style=BlueLine] (7.center) to (9.center);
		\draw [style=PurpleLine] (9.center) to (2.center);
		\draw [style=BrownLine, snake it, in=135, out=0, looseness=0.75] (16.center) to (2.center);
	\end{pgfonlayer}
\end{tikzpicture}}
\ee
Under the mapping \eqref{eq:mappingIn} the cross section of the wrapped cone does not trivialize and we obtain the non-trivial defect:
\be\label{eq:InstantonString}
\scalebox{0.9}{
\begin{tikzpicture}
	\begin{pgfonlayer}{nodelayer}
		\node [style=none] (0) at (2, -1.5) {};
		\node [style=SmallCircleRed] (1) at (2, -1.5) {};
		\node [style=none] (2) at (2, -1.5) {};
		\node [style=none] (5) at (0, -2) {$\mathcal{T}_{d+1}$};
		\node [style=none] (6) at (2.75, -1.5) {$\mathcal{T}_d$};
		\node [style=none] (7) at (-2, 1.5) {};
		\node [style=none] (8) at (-2, -1.5) {};
		\node [style=none] (9) at (2, 1.5) {};
		\node [style=SmallCircleBrown] (10) at (-2, -1.5) {};
		\node [style=SmallCircleGreen] (11) at (-2, 1.5) {};
		\node [style=SmallCircleBlue] (12) at (2, 1.5) {};
		\node [style=none] (13) at (-2.75, 0) {$\mathbb{B}^{(1,x_\perp)}_{d+1}$};
		\node [style=none] (14) at (0, -2.75) {};
		\node [style=none] (15) at (-2.75, -1.5) {$\mathbb{B}_d^{(1,0)}$};
		\node [style=none] (16) at (-2, -1.5) {};
	\end{pgfonlayer}
	\begin{pgfonlayer}{edgelayer}
		\filldraw[fill=gray!50, draw=gray!50]  (2, -1.5) -- (2, 1.5) -- (-2, 1.5) -- (-2, -1.5) -- cycle;
		\draw [style=ThickLine] (2.center) to (1);
		\draw [style=GreenLine] (7.center) to (8.center);
		\draw [style=RedLine] (8.center) to (2.center);
		\draw [style=BlueLine] (7.center) to (9.center);
		\draw [style=PurpleLine] (9.center) to (2.center);
		\draw [style=BrownLine, snake it] (16.center) to (2.center);
	\end{pgfonlayer}
\end{tikzpicture}}
\ee
In the original geometry we have fully submerged the M2-brane in the ADE locus associated with the 7D super-Yang-Mills flavor brane theory. Locally, the IIA dual frame of this configuration is a D2-brane inside of a D6-brane, i.e., in the infrared limit of the KK theory \eqref{eq:InstantonString} is an instanton string involving the non-abelian world-volume gauge field of $\mathcal{T}_{d+1}$ with field strength $f_2$. See \eqref{eq:thetatermspecific} for a discussion of some of the worldvolume dynamics in the infrared limit. Therefore, we have a consistency condition, the boundary condition for the world volume gauge field at $\mathbb{B}_d^{(1,0)}$ must be such that instanton strings can terminate there.

\subsection{Further Comments}
\label{ssec:Comments}
Finally, let us make some comments on how our considerations generalize. In general settings the singular locus $\mathscr{S}$ splits into a disjoint union of connected loci. Each of these can contain subloci along which the generic singularity worsens. When we have a nesting as in \eqref{eq:depth} this happens multiple times and our constructions generalizes to yield a symmetry theory $\mathcal{S}_{d+m}$ for each connected component where $m=I+1$ is the length of the respective chain \eqref{eq:depth}. In the total geometry these then glue together to a SymTree of cheesesteaks extrapolating the considerations in \cite{Baume:2023kkf}.

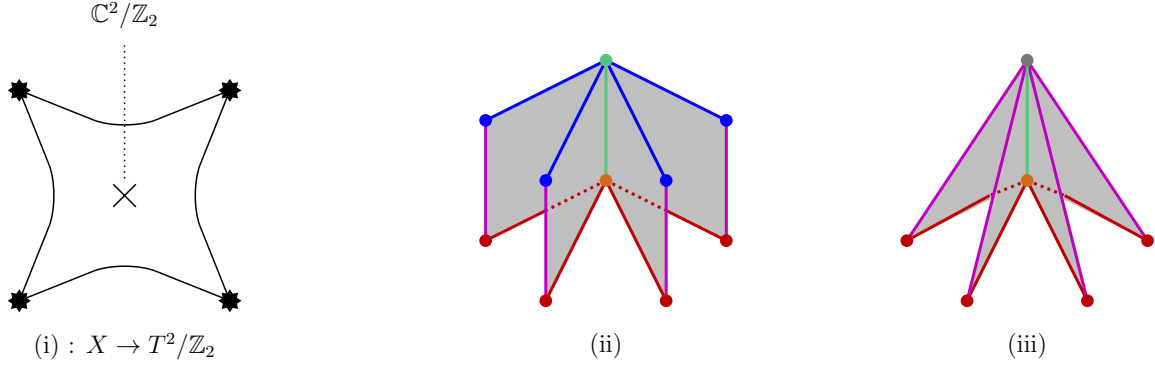
\begin{figure}
\centering
\scalebox{0.8}{
\begin{tikzpicture}
	\begin{pgfonlayer}{nodelayer}
		\node [style=none] (0) at (-1.75, 1.75) {};
		\node [style=none] (1) at (-0.5, 1.25) {};
		\node [style=none] (2) at (-1.25, 0.5) {};
		\node [style=none] (3) at (0.5, 1.25) {};
		\node [style=none] (4) at (1.25, 0.5) {};
		\node [style=none] (5) at (1.25, -0.5) {};
		\node [style=none] (6) at (0.5, -1.25) {};
		\node [style=none] (7) at (-0.5, -1.25) {};
		\node [style=none] (8) at (-1.25, -0.5) {};
		\node [style=none] (9) at (1.75, -1.75) {};
		\node [style=none] (10) at (1.75, 1.75) {};
		\node [style=none] (11) at (-1.75, -1.75) {};
		\node [style=none] (12) at (0, -2.5) {(i) : $X \rightarrow T^2/\Z_2$};
		\node [style=NodeCross] (13) at (0, 0) {};
		\node [style=none] (14) at (0, 0.25) {};
		\node [style=none] (15) at (0, 2.5) {};
		\node [style=none] (16) at (0, 3) {$\mathbb{C}^2/\Z_2$};
		\node [style=none] (17) at (7, -1.75) {};
		\node [style=none] (18) at (7, 0.25) {};
		\node [style=none] (19) at (8, 2.25) {};
		\node [style=none] (20) at (8, 0.25) {};
		\node [style=none] (21) at (9, -1.75) {};
		\node [style=none] (22) at (9, 0.25) {};
		\node [style=none] (23) at (10, 1.25) {};
		\node [style=none] (24) at (10, -0.75) {};
		\node [style=none] (25) at (6, -0.75) {};
		\node [style=none] (26) at (6, 1.25) {};
		\node [style=none] (27) at (7, -0.25) {};
		\node [style=none] (28) at (9, -0.25) {};
		\node [style=SmallCircleRed] (29) at (7, -1.75) {};
		\node [style=SmallCircleRed] (30) at (9, -1.75) {};
		\node [style=SmallCircleRed] (31) at (10, -0.75) {};
		\node [style=SmallCircleRed] (32) at (6, -0.75) {};
		\node [style=SmallCircleBlue] (33) at (6, 1.25) {};
		\node [style=SmallCircleBlue] (34) at (7, 0.25) {};
		\node [style=SmallCircleBlue] (35) at (9, 0.25) {};
		\node [style=SmallCircleBlue] (36) at (10, 1.25) {};
		\node [style=SmallCircleBrown] (37) at (8, 0.25) {};
		\node [style=SmallCircleGreen] (38) at (8, 2.25) {};
		\node [style=none] (39) at (8, -2.5) {(ii)};
		\node [style=none] (40) at (3.75, -3) {};
		\node [style=Star] (41) at (-1.75, -1.75) {};
		\node [style=Star] (42) at (1.75, -1.75) {};
		\node [style=Star] (43) at (1.75, 1.75) {};
		\node [style=Star] (44) at (-1.75, 1.75) {};
		\node [style=none] (45) at (14, -1.75) {};
		\node [style=none] (46) at (15, 2.25) {};
		\node [style=none] (47) at (15, 2.25) {};
		\node [style=none] (48) at (15, 0.25) {};
		\node [style=none] (49) at (16, -1.75) {};
		\node [style=none] (50) at (15, 2.25) {};
		\node [style=none] (51) at (15, 2.25) {};
		\node [style=none] (52) at (17, -0.75) {};
		\node [style=none] (53) at (13, -0.75) {};
		\node [style=none] (54) at (15, 2.25) {};
		\node [style=none] (55) at (14.375, 0) {};
		\node [style=none] (56) at (15.625, 0) {};
		\node [style=SmallCircleRed] (57) at (14, -1.75) {};
		\node [style=SmallCircleRed] (58) at (16, -1.75) {};
		\node [style=SmallCircleRed] (59) at (17, -0.75) {};
		\node [style=SmallCircleRed] (60) at (13, -0.75) {};
		\node [style=SmallCircleBrown] (65) at (15, 0.25) {};
		\node [style=none] (67) at (15, -2.5) {(iii)};
		\node [style=SmallCircleGrey] (68) at (15, 2.25) {};
	\end{pgfonlayer}
	\begin{pgfonlayer}{edgelayer}
		\filldraw[fill=gray!50, draw=gray!50]  (6, 1.25) -- (6, -0.75) -- (8, 0.25) -- (8, 2.25) -- cycle;
		\filldraw[fill=gray!50, draw=gray!50]  (7, 0.25) -- (7, -1.75) -- (8, 0.25) -- (8, 2.25) -- cycle;
		\filldraw[fill=gray!50, draw=gray!50]  (9, 0.25) -- (9, -1.75) -- (8, 0.25) -- (8, 2.25) -- cycle;
		\filldraw[fill=gray!50, draw=gray!50]  (10, 1.25) -- (10, -0.75) -- (8, 0.25) -- (8, 2.25) -- cycle;
		\filldraw[fill=gray!50, draw=gray!50]  (13, -0.75) -- (15, 0.25) -- (15, 2.25) -- cycle;
		\filldraw[fill=gray!50, draw=gray!50]  (14, -1.75) -- (15, 0.25) -- (15, 2.25) -- cycle;
		\filldraw[fill=gray!50, draw=gray!50]  (16, -1.75) -- (15, 0.25) -- (15, 2.25) -- cycle;
		\filldraw[fill=gray!50, draw=gray!50]  (17, -0.75) -- (15, 0.25) -- (15, 2.25) -- cycle;
		\draw [style=ThickLine] (0.center) to (1.center);
		\draw [style=ThickLine, bend right=-340, looseness=0.75] (1.center) to (3.center);
		\draw [style=ThickLine] (3.center) to (10.center);
		\draw [style=ThickLine] (10.center) to (4.center);
		\draw [style=ThickLine, bend right=-340, looseness=0.75] (4.center) to (5.center);
		\draw [style=ThickLine] (5.center) to (9.center);
		\draw [style=ThickLine] (6.center) to (9.center);
		\draw [style=ThickLine, bend left=340, looseness=0.75] (6.center) to (7.center);
		\draw [style=ThickLine] (7.center) to (11.center);
		\draw [style=ThickLine] (11.center) to (8.center);
		\draw [style=ThickLine, bend right=-340, looseness=0.75] (8.center) to (2.center);
		\draw [style=ThickLine] (2.center) to (0.center);
		\draw [style=DottedLine] (15.center) to (14.center);
		\draw [style=RedLine] (24.center) to (28.center);
		\draw [style=RedLine] (20.center) to (21.center);
		\draw [style=RedLine] (20.center) to (17.center);
		\draw [style=RedLine] (27.center) to (25.center);
		\draw [style=DottedRed] (20.center) to (27.center);
		\draw [style=DottedRed] (20.center) to (28.center);
		\draw [style=PurpleLine] (26.center) to (25.center);
		\draw [style=PurpleLine] (18.center) to (17.center);
		\draw [style=PurpleLine] (22.center) to (21.center);
		\draw [style=PurpleLine] (23.center) to (24.center);
		\draw [style=BlueLine] (26.center) to (19.center);
		\draw [style=BlueLine] (19.center) to (18.center);
		\draw [style=BlueLine] (19.center) to (22.center);
		\draw [style=BlueLine] (19.center) to (23.center);
		\draw [style=GreenLine] (19.center) to (20.center);
		\draw [style=RedLine] (52.center) to (56.center);
		\draw [style=RedLine] (48.center) to (49.center);
		\draw [style=RedLine] (48.center) to (45.center);
		\draw [style=RedLine] (55.center) to (53.center);
		\draw [style=DottedRed] (48.center) to (55.center);
		\draw [style=DottedRed] (48.center) to (56.center);
		\draw [style=PurpleLine] (54.center) to (53.center);
		\draw [style=PurpleLine] (46.center) to (45.center);
		\draw [style=PurpleLine] (50.center) to (49.center);
		\draw [style=PurpleLine] (51.center) to (52.center);
		\draw [style=BlueLine] (54.center) to (47.center);
		\draw [style=BlueLine] (47.center) to (46.center);
		\draw [style=BlueLine] (47.center) to (50.center);
		\draw [style=BlueLine] (47.center) to (51.center);
		\draw [style=GreenLine] (47.center) to (48.center);
	\end{pgfonlayer}
\end{tikzpicture}}
\caption{In subfigure (i) we sketch the space $(\mathbb{C}^2\times T^2)/\Z_4$ as projection onto a $T^2/\Z_2$ base with generic fiber $\mathbb{C}^2/\Z_{2}$. The $T^2/\Z_2$ base has four orbifold points modeled on $\mathbb{C}/\Z_2$ which locally, in the total space, are associated with patches modeled on $\mathbb{C}^3/\Z_4$. In subfigure (ii) we show the support of the resulting symmetry theory, consisting of four open cheesesteaks glued together along a common spine. In subfigure (iii) we collapse the four blue edges, resulting in a collection of four 2-simplices glued along a common 1-face.}
\label{fig:BookOfCheesesteaks}
\end{figure}

For instance, when the generic singularity worsens at multiple disjoint loci in $\mathscr{S}$ once then the resulting structure can be viewed as glued from the open cheesesteak we have discussed here. It constitutes a building block.

We demonstrate this in terms of an example. Consider the geometry $X= (\mathbb{C}^2\times T^2 )/\Z_4$ (also see the discussion in \cite{Cvetic:2023pgm}) where the torus has complex structure $\tau=i$. Denote the coordinates of $\mathbb{C}^2$ by $z_1,z_2$ and that of $T^2$ by $z_3$. We consider the weight vector $(1,1,2)$ and can rewrite the quotient as
\be
X= (\mathbb{C}^2\times T^2) /\Z_4= (\mathbb{C}^2/\Z_2\times T^2) /\Z_2\,,
\ee
to see that the space consists of an $\mathfrak{su}_2$ singularity along $T^2/\Z_2$. The latter is topologically a sphere with four points modeled on $\mathbb{C}/\Z_2$. There, in the total geometry, the singularity is modeled on $\mathbb{C}^3/\Z_4$ with weight vector $(1,1,2)$ which is exactly our running example $\mathbb{C}^3/\Z_{2n}$ for the case $n=2$ (see subfigure (i) figure \ref{fig:BookOfCheesesteaks}). Conversely, we can view $X$ as glued together from four patches modeled on $\mathbb{C}^3/\Z_4$. To each of these we can associate an open cheesesteak which we then glue along a junction, see subfigure (ii) figure \ref{fig:BookOfCheesesteaks}. We can also degenerate subfigure (ii) to subfigure (iii) resulting in a structure given by four 2-simplices glued along a 1-face.

\section{Illustrative Examples in $d=5,4$}
\label{sec:Illustrative}

We now give some illustrative examples. These will be minimally supersymmetric theories in 5D and 4D engineered via an exceptional holonomy cone $X=\textnormal{Cone}(\partial X)$ in M-theory. Much of the extra dimensional analysis will be deferred to appendix \ref{app:A} and we focus here mostly on specifying the two symmetry theories $\mathcal{S}_{d+2}$ and relative to it $\mathcal{B}_{d+1}$, and the flavor brane theory $\mathcal{T}_{d+1}$, i.e., we specify the full neighborhood of the corner mode $\mathcal{T}_{d}$.

The extra-dimensional input we start with will be the long exact sequence in relative cohomology for the pair $\partial X^\circ,\partial^2X^\circ$ which reads
\be
\dots  ~\rightarrow~  H^{n}(\partial X^\circ, \partial^2 X^\circ)  ~\rightarrow~H^{n}(\partial X^\circ) ~\rightarrow~H^{n}(\partial^2 X^\circ) ~\rightarrow~\dots
\ee
and the Chern-Simons term\footnote{Although the differential cohomology uplift of these terms are a more appropriate starting point, see for example \cite{Apruzzi:2021nmk, GarciaEtxebarria:2024fuk}, we restrict our considerations to ordinary cohomology which we will find to be sufficient in most instances. We defer the more careful treatment to future work.} of 11D supergravity
\be \label{eq:11DCS}
2\pi \int_{M_{11}}\lb -\frac{1}{6} \frac{C_3}{2\pi}\cup \frac{G_4}{2\pi}\cup \frac{G_4}{2\pi}+\frac{1}{48} \frac{C_3}{2\pi}\cup \lbb  p_2(M_{11})-p_1(M_{11})^2/4 \rbb\rb \,,
\ee
where $G_4=dC_3$ is the field strength of the M-theory 3-form $C_3$ and $p_i(M_{11})$ are the Pontryagin classes of $M_{11}$. We have normalized $G_4/2\pi$ to have integral periods.

We structure our analysis by first describing various geometric features of the internal geometry $X$ and its singularities $\mathscr{S}$, and discuss the flavor brane theory $\mathcal{T}_{d+1}$ via its relation to the world volume theory of a D6-brane in a IIA dual frame. Then we address the field content of the pair  $\mathcal{S}_{d+2},\mathcal{B}_{d+1}$ and the boundary condition imposed by the latter on the former. From here we turn to discuss the Lagrangian governing the interactions and anomalies.

Once the neighborhood of the corner mode $\mathcal{T}_{d}$ is discussed we turn to specify the boundary conditions $\mathbb{B},\widetilde{\mathbb{B}}$. Overall we will focus in the various examples on different features. For the 5D examples we focus on discrete symmetries and 2-group symmetries. For the 4D examples we focus on continuous symmetries.

\subsection{Example: 5D SCFTs}

M-theory on the Calabi-Yau orbifold cone $X=\mathbb{C}^3/\Z_N$ engineers a large class of 5D superconformal field theories, see e.g., \cite{Intriligator:1997pq} as well as the more recent \cite{DelZotto:2017pti, Xie:2017pfl, Jefferson:2018irk, Apruzzi:2019vpe, Tian:2021cif, Acharya:2021jsp, Argyres:2022mnu}. Here we are considering a faithful $\Gamma \cong \Z_N$ group action
\be
(z_1,z_2,z_3)~\mapsto~(\omega^{m_1}z_1,\omega^{m_2}z_2,\omega^{m_3}z_3)\,,
\ee
on $\mathbb{C}^3$ parametrized by $z_1,z_2,z_3$ with $m_1+m_2+m_3=0$ mod $N$ and $\omega=\exp(2\pi i/N)$.

To frame our discussion to follow, we being by considering the singularities $\mathscr{S}\subset X$ and describe the pair $(\partial X^\circ,\partial^2X^\circ)$. First, from the group action, we read that the orbifold cones $\mathbb{C}^3/\Z_N=\textnormal{Cone}(S^5/\Z_N)$ have a codimension 6 singularity at their tip and up to three codimension 4 singularities at $z_i,z_j=0$. The latter are supported on a copy of $\C/\Z_N$ parametrized by the coordinate which does not vanish. As such the asymptotic singular locus $\mathscr{K}\subset \partial X=S^5/\Z_N$ consists of up to three disjoint circles and the tubular neighborhood $T_{\mathscr{K}}\subset \partial X$ is a disjoint union of up to three 4-disk bundles with circle base. The 4-disks are singular and modeled on a neighborhood of the origin of an A-type ADE singularity. Excision of the singularities then gives the smooth 5-dimensional manifold with boundary $\partial X^\circ=\partial X\setminus T_{\mathscr{K}}$ and therefore $\partial^2 X^\circ$ is a smooth closed 4-dimensional manifold which projects onto $\mathscr{K}$ with 3-dimensional lens space fibers.

From here, compute the long exact sequence in relative cohomology for the pair $(\partial X^\circ,\partial^2X^\circ)$ (see Appendix \ref{app:A} for details). The sequence reads
\begin{equation}\label{eq:RCHcirc}
    \begin{array}{c||cccccc}
& H^k(\partial X^\circ,\partial^2X^\circ) &  & H^k(\partial X^\circ) &  &   H^k(\partial^2X^\circ)  &      \\[0.4em] \hline \hline \\[-0.9em]
       k=0 ~&   0  & \rightarrow & \Z & \rightarrow &  \mathbb{Z}^{| \mathscr{K}|} & \rightarrow  \\
      k=1 ~&   \Z^{{| \mathscr{K}|}-1}  & \rightarrow & 0 & \rightarrow &  \mathbb{Z}^{| \mathscr{K}|} & \xrightarrow[]{}  \\
      k=2 ~&   \Z^{| \mathscr{K}|} & \rightarrow & \Gamma^\vee & \rightarrow &\Gamma_{\textnormal{fix}}^\vee& \rightarrow  \\
         k=3~ &  0  &\rightarrow & \Z^{{| \mathscr{K}|}} & \rightarrow& \Z^{| \mathscr{K}|}\oplus \Gamma_{\textnormal{fix}} &\rightarrow  \\
       k=4~ &    \Gamma  &\rightarrow & \Z^{{| \mathscr{K}|}-1} &\rightarrow &\Z^{| \mathscr{K}|} &\rightarrow  \\
        k=5 ~&   \mathbb{Z}  &\rightarrow & 0 &\rightarrow & 0  & \rightarrow
    \end{array}
\end{equation}
where $\Gamma=\Z_N$ and $\Gamma_{\textnormal{fix}}\subset \Gamma$ is the subgroup generated by all group elements which have fixed points on $S^5$. We denote the Pontryagin dual of an abelian group $G$ by $G^\vee=\textnormal{Hom}(G,\textnormal{U}(1))$. Concretely we have,
\be\label{eq:H}
\Gamma_{\textnormal{fix}}\cong \Z_{g}\cong \Z_{g_1}\times \Z_{g_2}\times \Z_{g_3}
\ee
with coprime product $g=g_1g_2g_3$ and $g_i=\textnormal{gcd}(m_i,N)$. We denote the number of connected components of $\mathscr{K}$ by $|\mathscr{K}|=1,2,3$ where the precise value depends on the weights $m_i$ which we assume to have chosen such that $|\mathscr{K}|\geq 1$.

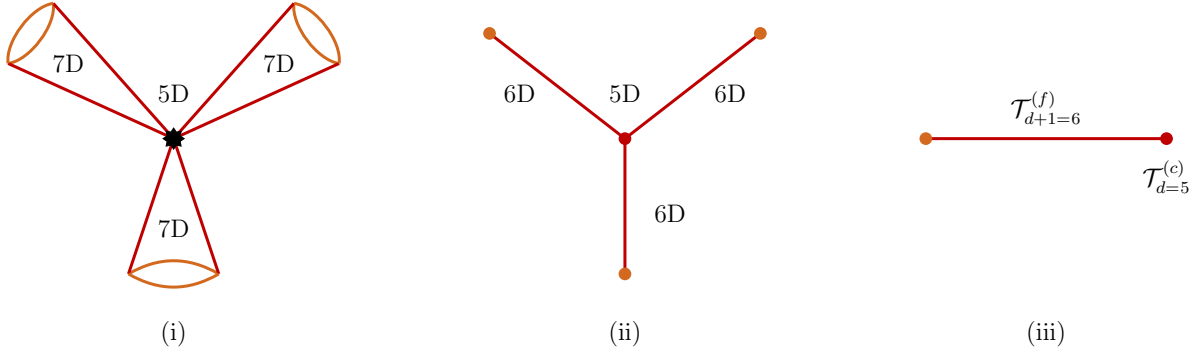
\begin{figure}
\centering
\scalebox{0.8}{
\begin{tikzpicture}
	\begin{pgfonlayer}{nodelayer}
		\node [style=none] (0) at (-7.25, 1.75) {};
		\node [style=none] (1) at (-6.5, 2.75) {};
		\node [style=none] (2) at (-1.75, 1.75) {};
		\node [style=none] (3) at (-2.5, 2.75) {};
		\node [style=none] (4) at (-3.75, -1.75) {};
		\node [style=none] (5) at (-5.25, -1.75) {};
		\node [style=none] (6) at (-4.5, 0.5) {};
		\node [style=Star] (7) at (-4.5, 0.5) {};
		\node [style=none] (8) at (-2.75, 1.75) {7D};
		\node [style=none] (9) at (-6.25, 1.75) {7D};
		\node [style=none] (10) at (-4.5, -1) {7D};
		\node [style=none] (11) at (-4.5, 1.25) {5D};
		\node [style=none] (12) at (-4.5, -2.75) {(i)};
		\node [style=none] (13) at (0.75, 2.25) {};
		\node [style=none] (15) at (5.25, 2.25) {};
		\node [style=none] (18) at (3, -1.75) {};
		\node [style=none] (19) at (3, 0.5) {};
		\node [style=none] (21) at (4.75, 1.25) {6D};
		\node [style=none] (22) at (1.25, 1.25) {6D};
		\node [style=none] (23) at (3.75, -0.75) {6D};
		\node [style=none] (24) at (3, 1.25) {5D};
		\node [style=SmallCircleBrown] (25) at (3, -1.75) {};
		\node [style=SmallCircleBrown] (26) at (0.75, 2.25) {};
		\node [style=SmallCircleBrown] (27) at (5.25, 2.25) {};
		\node [style=SmallCircleRed] (28) at (3, 0.5) {};
		\node [style=none] (29) at (3, -2.75) {(ii)};
		\node [style=none] (30) at (8, 0.5) {};
		\node [style=none] (31) at (12, 0.5) {};
		\node [style=SmallCircleBrown] (32) at (8, 0.5) {};
		\node [style=SmallCircleRed] (33) at (12, 0.5) {};
		\node [style=none] (34) at (10, 1) {$\mathcal{T}_{d+1=\:\!6}^{(f)}$};
		\node [style=none] (35) at (12, -0.125) {$\mathcal{T}_{d=\:\!5}^{(c)}$};
		\node [style=none] (36) at (10, -2.75) {(iii)};
		\node [style=none] (37) at (3, -3.5) {};
	\end{pgfonlayer}
	\begin{pgfonlayer}{edgelayer}
		\draw [style=BrownLine, bend left] (5.center) to (4.center);
		\draw [style=BrownLine, bend left=45, looseness=0.75] (2.center) to (3.center);
		\draw [style=BrownLine, bend left=45, looseness=0.75] (1.center) to (0.center);
		\draw [style=RedLine] (1.center) to (6.center);
		\draw [style=RedLine] (6.center) to (3.center);
		\draw [style=RedLine] (2.center) to (6.center);
		\draw [style=RedLine] (6.center) to (4.center);
		\draw [style=RedLine] (6.center) to (5.center);
		\draw [style=RedLine] (6.center) to (0.center);
		\draw [style=BrownLine, bend left=45, looseness=0.75] (0.center) to (1.center);
		\draw [style=BrownLine, bend left=45, looseness=0.75] (3.center) to (2.center);
		\draw [style=BrownLine, bend right] (5.center) to (4.center);
		\draw [style=RedLine] (15.center) to (19.center);
		\draw [style=RedLine] (19.center) to (18.center);
		\draw [style=RedLine] (19.center) to (13.center);
		\draw [style=RedLine] (30.center) to (31.center);
	\end{pgfonlayer}
\end{tikzpicture}}
\caption{Sketch of the relative theories in 5D, 6D, 7D for $\mathbb{C}^3/\Z_N$ in M-theory. In (i) we sketch the 5D SCFT as a defect in three 7D super-Yang-Mills theories supported on the 5D spacetime times three cones meeting at their apex. In (ii) we KK reduce these cones to their radial half lines. Folding (iii) we arrive at (ii). The relative flavor brane theory is a disjoint union of the three sets of 6D KK theories.}
\label{fig:RelativeTheories65D}
\end{figure}

In M-theory the codimension 4 singularities of $\mathscr{S}$ engineer 7D super-Yang-Mills theories which intersect at the codimension 6 singularity where a 5D SCFT is supported. In our examples each 7D super-Yang-Mills theory can be analyzed locally in the geometry, where it is associated with an A-type ADE singularity, and be described dually by a D6-brane in IIA. With respect to each such 7D sector the 5D SCFT is a codimension two defect and its insertion into the 7D theory turns on a non-trivial background field profile in the 7D bulk. We can analyze this profile by noting that in the geometry the 7D super-Yang-Mills theory is supported on $\mathbb{R}^{1,4}\times \lb \C/\Z_K\setminus \{ 0\} \rb$, for some $K$, with normal geometry $\C^2/(\Z_N/\Z_K)$. Along a path linking the puncture in $ \C/\Z_K\setminus \{ 0\} $ this normal geometry is acted on by a monodromy. In the IIA dual frame this monodromy is captured by a holonomy for the RR 1-form field $C_1$ which interacts with the D6-brane world volume theory via the Wess-Zumino coupling
\be
\frac{2\pi}{6(2\pi)^3}\int C_1\cup \tr \lb (f_2-B_2)\cup (f_2-B_2)\cup (f_2-B_2) \rb\,,
\ee
where $f_2$ is the non-abelian field strength on the D6-brane stack and $B_2$ is the NSNS 2-form. We will consider backgrounds with the NSNS 2-form turned off. Upon KK reducing on concentric circles $S^1$ linking the puncture in $ \C/\Z_K\setminus \{ 0\} $ the above setup reduces to a 6D KK theory with a massless 6D super-Yang-Mills sector containing the coupling
\be \label{eq:Theta}
\frac{2\pi q}{6} \int  \tr \lb \frac{f_2}{2\pi}\cup \frac{f_2}{2\pi}\cup \frac{f_2}{2\pi} \rb\,,
\ee
with period $q=  \int_{S^1} C_1$ which is some rational number constrained by $Kq\in \Z$.
The disjoint union over these 6D theories is then the relative flavor brane theory $\mathcal{T}_{d+1}$ and the corner theory $\mathcal{T}_{d}$ realizes an end of the world brane thereof, with $d=5$, here the 5D SCFT.

The KK reduction is not strictly necessary, however, it is favorable to collapse the system into a bulk and boundary in adjacent dimensions when discussing their symmetries. Further, let us already mention here that in taking the IR limit which localizes the dyanmics of the system to $d$ dimensions we will need to carefully consider which bulk operators $\mathcal{S}_{d+2}$ remain as symmetry operators for the resulting $d$-dimensional QFT.

\subsubsection{Field Content of $\mathcal{S}_{d+2}, \mathcal{B}_{d+1}$}

We begin with deriving the field content of $\mathcal{S}_{d+2}$. We expand the supergravity field strength ${G}_4$ in classes of ${H}^n(\partial^2 X^\circ)$. We have, see \eqref{eq:RCHcirc} for the basis of cocycles of the expansion,
\be \ba
{G}_4=\sum_{i=1}^{|\mathscr{K}|}\Big(  {H}_4^{(i)} &\cup {\mathbf{1}}_{(i)} + {H}_3^{(i)} \cup {v}_{1,(i)} +{B}_2^{(i)}\cup {t}_{2,(i)}\,+ \\ {B}_1^{(i)}&\cup {t}_{3,(i)}+ {H}_1^{(i)} \cup {v}_{3,(i)} + {H}_0^{(i)} \cup \textnormal{vol}_{\:\!4,(i)}  \Big)\,,\\[0.2em]
\ea \ee
where $i$ runs over the flavor branes and ${H}_n,{B}_n$ denotes an abelian $\textnormal{U}(1)$ field strength of degree $n$ and abelian discrete background field in degree $n$ respectively, and ${v}_n,{t}_n$ denote free, torsional integral coycles of degree $n$ respectively. The field content of $\mathcal{S}_{d+2}$ is
\be
( {H}_0^{(i)},  {H}_1^{(i)})\,,  ( {H}_3^{(i)},  {H}_4^{(i)})\,,  \quad \textnormal{and}\quad ( {B}_1^{(i)},  {B}_2^{(i)})\,,
\ee
 where we indicated KK pairs and grouped continuous and discrete fields. In order to make the order of the discrete fields explicit we also write $ {B}_k^{(\Z_{g_i})}$ for a degree $k$ cocycle of order $g_i$ or group these into $B_2^{(\Gamma_{\textnormal{fix}}^\vee)}\equiv B_2^{(\Z_g)} $. Therefore, using the isomorphism \eqref{eq:H} we equivalently have the expansion
 \be \ba
{G}_4=\sum_{i=1}^{|\mathscr{K}|}\Big(  {H}_4^{(i)} &\cup {\mathbf{1}}_{(i)} + {H}_3^{(i)} \cup {v}_{1,(i)} +{B}_2^{(\Z_g)}\cup {t}_{2}\,+ \\ {B}_1^{(\Z_g)}&\cup {t}_{3}+ {H}_1^{(i)} \cup {v}_{3,(i)} + {H}_0^{(i)} \cup \textnormal{vol}_{\:\!4,(i)}  \Big)\,,\\[0.2em]
\ea \ee
replacing the triplet of doublets $ ( {B}_1^{(i)},  {B}_2^{(i)})$ with $ ( {B}_1^{(\Z_g)},  {B}_2^{(\Z_g)})$

 Next we discuss the field content of $\mathcal{B}_{d+1}$.  We expand the supergravity field strength ${G}_4$ in classes of ${H}^n(\partial X^\circ,\partial^2 X^\circ)$ and ${H}^n(\partial X^\circ)$. We have
\be\ba \label{eq:Expansion2}
{G}_4&={G}_4 \cup {\mathbf{1}} +{B}_2^{(\Z_N)}\cup {s}_{2}+ \sum_{i=1}^{|\mathscr{K}|}  {G}_1^{(i)} \cup {u}_{3,(i)} +\sum_{j=1}^{|\mathscr{K}|-1}  {G}_0^{(j)} \cup {u}_{4,(j)}  \\[0.2em]
&~~~\,+\sum_{j=1}^{|\mathscr{K}|-1}  {F}_3^{(j)} \cup {w}_{1,(j)} +\sum_{i=1}^{|\mathscr{K}|}  {F}_2^{(i)} \cup {w}_{2,(i)} + {B}_0^{(\Z_N)}\cup {r}_{4}\,,
\ea\ee
 where $i$ again runs over the flavor branes, $j$ over some of their differences, and ${G}_n,{F}_n$ denote abelian $\textnormal{U}(1)$ field strengths of degree $n$. Further, we have the discrete symmetry background fields $B_0^{(\Z_N)},B_2^{(\Z_N)} $. The cohomology classes ${u}_{n}$, $({w}_n)$ and ${s}_{n}$, $({r}_n)$ are free and torsional (relative) integral cohomology classes of $\partial X^\circ$ respectively. The field content is
 \be
 G_0^{(j)}, G_1^{(i)},F_2^{(i)},F_3^{(j)},  G_4,  \quad\textnormal{and}\quad {B}_0^{(\Z_N)},{B}_2^{(\Z_N)},
 \ee
 where we have grouped continuous and discrete fields respectively. These fields are subject to identifications as specified by the mapping $\iota^{n}:\,H^n(\partial X^\circ,\partial^2 X^\circ)\rightarrow H^n(\partial X^\circ)$.

Let us make these identifications explicit for the case $|\mathscr{K}|=1$. In this case we drop the index $i$ as it takes a single value and delete the fields labelled $j$ from the spectrum as this index runs over the empty set. Then we only have the single identification
 \be\label{eq:Identifications} \ba
 |\Gamma_{\textnormal{fix}}^\vee|{B}_2^{(\Gamma^\vee)}&={F}_2\\
g{B}_2^{(\Z_N)}&={F}_2\,,
 \ea \ee
 where in the second line, equivalent to the first, we have made the group orders explicit via $\Gamma^\vee\cong \Z_N$ and $\Gamma_{\textnormal{fix}}^\vee \cong \Z_g$. All fields take values in $\textnormal{U}(1)$; this is the
 natural normalization given we are relating discrete and continuous fields. The identification constrains  $F_2$ to take finitely many values, isomorphic to $(\Gamma/\Gamma_{\textnormal{fix}})^\vee$, and with this constraint in place the above simply correspond to the subgroup relation $(\Gamma/\Gamma_{\textnormal{fix}})^\vee\subset \Gamma^\vee$.

 Let us recall the derivation of \eqref{eq:Identifications}. Denote by $w_2$ and $s_2$ the 2-cocycles generating the groups $H^2(\partial X^\circ,\partial^2 X^\circ)\cong \Z$ and $H^2(\partial X^\circ)\cong \Gamma^\vee$ respectively. Applying $\iota^2$ we have
 \be
{F}_2 \cup w_2 ~\mapsto ~  {F}_2 \cup \iota^2(w_2)={F}_2 \cup gs_2
 \ee
which we compare to $ B_2^{(\Z_N)}\cup s_2$, multiplying through by $g$ we conclude that $F_2$ and $gB_2^{(\Z_N)}$ are to be identified. See the discussion following \eqref{eq:Identification} for more details.

In addition we have dual fields resulting from similar expansions for $G_7$. For example, the discrete fields resulting from this expansion are
\be
B_5^{(\Z_g)}, \quad \textnormal{and}\quad B^{(\Z_N)}_2, B^{(\Z_N)}_3\,.
\ee
These fields are analyzed identically to those derived from $G_4$.

\subsubsection{Boundary Condition of $\mathcal{S}_{d+2} |_{\mathcal{B}_{d+1}}$}

The relative symmetry theory $ \mathcal{B}_{d+1}$ realizes a boundary condition for $\mathcal{S}_{d+2}$. Restricting a field $\mathcal{S}_{d+2}$ to the support of $\mathcal{B}_{d+1}$ we can relate the corresponding internal cycles, given they now embed into the same geometry, a copy of $\partial X^\circ$, via the mappings $H^n(\partial X^\circ)\rightarrow H^n(\partial^2 X^\circ)$ and $H^n(\partial^2 X^\circ)\rightarrow {H}^{n+1}(\partial X^\circ,\partial^2 X^\circ)$. The former mapping is degree preserving and relates cocycles of $\mathcal{B}_{d+1}$ and $\mathcal{S}_{d+2}$ in identical degree. The latter mapping relates cocycles in adjacent degrees. See the discussion surrounding \eqref{eq:BCGeo} for more details.

We make these boundary conditions explicit for the case $|\mathscr{K}|=1$. We have the degree preserving boundary conditions for fields from expansions along  $H^n(\partial^2 X^\circ)$ and  $H^n(\partial X^\circ)$ given by
\be \ba  \label{eq:DBC}
 H_4\big|_{\mathcal{B}_{d+1}} &= G_4\\
  H_3\big|_{\mathcal{B}_{d+1}} &= 0\\
  B_2^{(\Z_g)}\big|_{\mathcal{B}_{d+1}}&= (N/g)B_2^{(\Z_N)}\\
 \lbb (N/g) H_1 +B_1^{(\Z_g)}\rbb\!\Big|_{\mathcal{B}_{d+1}}&=G_1 \\
   H_0\big|_{\mathcal{B}_{d+1}}&= 0\,,\\
 \ea \ee
 where we denote restriction to the support of $\mathcal{B}_{d+1}$ at $r=0$ by $|_{\mathcal{B}_{d+1}}$. Next, the degree lowering boundary conditions for fields from expansions along  $H^n(\partial^2 X^\circ)$ and  $H^{n+1}(\partial X^\circ,\partial^2X^\circ)$ are given by
\be \label{eq:NBC}\ba
\Big( {\partial}/{\partial r} \,\lrcorner\, H_4\Big)\!\; \Big|_{\mathcal{B}_{d+1}}&= 0\\[0.2em]
 \Big(  {\partial}/{\partial r}  \,\lrcorner\, H_3\Big)\!\;  \Big|_{\mathcal{B}_{d+1}}&= (N/g)F_2\\[0.2em]
\lb {\partial}/{\partial r}  \,\lrcorner\, B_2^{(\Z_g)}\rb\!\!\;  \Big|_{\mathcal{B}_{d+1}} &= 0\\[0.2em]
\lb {\partial}/{\partial r}  \,\lrcorner\, \lbb H_1+B_1^{(\Z_g)}\rbb\rb\!\!\;  \Big|_{\mathcal{B}_{d+1}} &=  B_0^{(\Z_N)} \\[0.2em]
\Big(  {\partial}/{\partial r}  \,\lrcorner\, H_0\Big)\!\;  \Big|_{\mathcal{B}_{d+1}} &= 0\,.\\
\ea \ee

Let us discuss some of these boundary conditions in greater detail. For example, how are the identification and boundary condition
\be \ba
 g{B}_2^{(\Z_N)}={F}_2\,,  \qquad \Big(  {\partial}/{\partial r}  \,\lrcorner\, H_3\Big)\!\!\;  \Big|_{\partial} = (N/g)\;\!F_2\,,
\ea \ee
consistent? The fields $F_2,H_3$ are valued in $\textnormal{U}(1)$, hence the latter boundary condition fixed $F_2$ up to a $\Z_{N/g}$ phase. This phase is then fixed by the former identification. The field $F_2$ is completely eaten up. Next, how are the pair of boundary conditions
\be\ba
 \lbb (N/g) H_1 +B_1^{(\Z_g)}\rbb \!\big|_\partial =G_1  \,, ~~\qquad \lb {\partial}/{\partial r}  \,\lrcorner\, \lbb H_1+B_1^{(\Z_g)}\rbb\rb\!\!\;  \Big|_{\partial} &=  B_0^{(\Z_N)} \,,\\[0.3em]
\ea \ee
compatible? The naively first condition fixed $H_1$ up to a $\Z_{N/g}$ phase. However, the first term in the first condition can be compensate by the second term extending the undermined  $\Z_{N/g}$ phase to a $\Z_{N}$ phase shared between the pair $H_1,B_1^{(\Z_g)}$. This phase then determines the profile for $B_0^{(\Z_N)}$ by the second condition.

Next we ask: what are the fields of $\mathcal{B}_{d+1}$ which are not constrained by the bulk fields $\mathcal{S}_{d+2}$, i.e., the degrees of freedom propagating as boundary modes in $\mathcal{S}_{d+1}$? On general grounds we argued that these result from the image of the mapping $H^n(\partial X^\circ, \partial^2 X^\circ)\rightarrow H^n(\partial X^\circ)$. In the case $|\mathscr{K}|=1$ we thus find a single field (and its magnetic dual) propagating in $\mathcal{B}_{d+1}$ which is:
\be \label{eq:1FS}
B_2^{(\Z_{N/g})}\equiv B_2^{(\Gamma/\Gamma_{\text{fix}})^\vee} \equiv gB_2^{(\Gamma^\vee)} \,.
\ee

It is well-known that the defect group of lines constructed from wrapped M2-branes of the 5D SCFTs engineered by $\mathbb{C}^3/\Gamma$ is isomorphic to $\textnormal{Ab}(\Gamma/\Gamma_{\text{fix}})$ \cite{DelZotto:2015isa, DelZotto:2022fnw}. We have found the corresponding background field to be localized to $\mathcal{B}_{d+1}$. This is compatible with our defect group analysis in section \ref{sec:TopDown} which here results in the defect group
$
\mathbb{D}_{1}[\mathcal{T}_d]\cong \textnormal{Ab}(\Gamma/\Gamma_{\text{fix}})
$.

We also immediately recover the results of the 2-group analysis of \cite{Cvetic:2022imb, DelZotto:2022joo} for the case of 5D SCFTs engineered by $\mathbb{C}^3/\Gamma$ with $\Gamma\cong \Z_N$. The fields $B_2^{(\Z_g)},B_2^{(\Z_N)},B_2^{(\Z_{N/g})}$ are related by the short exact sequence
\be \ba
&0~\rightarrow~(\Gamma/\Gamma_{\text{fix}})^\vee~\rightarrow~\Gamma^\vee~\rightarrow~\Gamma_{\text{fix}}^\vee~\rightarrow~0\,, \\
&0~\rightarrow~\Z_{N/g}~\rightarrow~\Z_N~\rightarrow~\Z_g~\rightarrow~0\,,
\ea \ee
which is here realized by the boundary conditions\footnote{We briefly review our normalization and notation conventions. In this section all fields, including discrete fields, are valued in $\textnormal{U}(1)$. The relation between the normalizations for a discrete field $B$ of order $K$ is
\be
B_{\:\!\textnormal{U}(1)}=\frac{2\pi }{K}B_{\;\!\Z_K}\,.
\ee
This has consequences for how fields associated with a short exact sequence
\be
1~\rightarrow~\Z_K^{(a)}~\rightarrow~\Z_M^{(b)}~\rightarrow~\Z_M/\Z_K\cong \Z_{M/K}^{(c)}~\rightarrow~1\,,
\ee
relate. Here $K$ divides $M$ and the exponents denote the fields we associate to these groups now. When normalized to take values in finite groups the sequence implies
\be
a_{\;\!\Z_K}= \frac{M}{K}b_{\;\!\Z_M}\,, \qquad  c_{\;\!\Z_{M/K}}=\lbb b_{\;\!\Z_M}~ \textnormal{mod}\,\frac{M}{K}\rbb\,.
\ee
When normalized with values in $\textnormal{U}(1)$ we have instead
\be \label{eq:SESNormalization}
a_{\:\!\textnormal{U}(1)}=b_{\:\!\textnormal{U}(1)}\,, \qquad K b_{\:\!\textnormal{U}(1)}=c_{\:\!\textnormal{U}(1)}\,.
\ee
We remind that our notation in this section will be $B_d^{(G)}$ for a degree $d$ cocycle of order $|G|$ valued in $\textnormal{U}(1)$. In particular to exact sequence we associate relations such as \eqref{eq:SESNormalization}. For instance, in \eqref{eq:DBC} and \eqref{eq:NBC} no `mod' appears. This is the preferred convention when discrete fields mix with continuous fields.
 }
\be
B_2^{(\Z_{N/g})}=B_2^{(\Z_N)}\,, \qquad B_2^{(\Z_g)}|_{\mathcal{B}_{d+1}}=(N/g)B_2^{(\Z_N)}\,,
\ee
and the relation \eqref{eq:1FS}. These determined the 2-group sequence (see \cite{Lee:2021crt})
\be
0 ~\rightarrow~ \mathcal{A}_{\mathcal{B}_{d+1}}\cong \Z_{N/g} ~\rightarrow~ \widetilde{\mathcal{A}}_{\mathcal{B}_{d+1}}\cong \Z_N  ~\rightarrow~ \widetilde{G}_{\mathcal{T}_{d+1}} \cong \mathrm{SU}(g)~\rightarrow~  {G}_{\mathcal{T}_{d+1}}  \cong \mathrm{PSU}(g)~\rightarrow~1
\ee
where $ \mathcal{A}_{\mathcal{B}_{d+1}}$ is the 1-form symmetry group,  $\widetilde{\mathcal{A}}_{\mathcal{B}_{d+1}}$ the naive 1-form symmetry group, $\widetilde{G}_{\mathcal{T}_{d+1}} $ the naive simply connected continuous 0-form flavor symmetry group and $  \widetilde{G}_{\mathcal{T}_{d+1}} $ the  0-form flavor symmetry group of the 5D SCFT. Making the split
\be \ba
&0 ~\rightarrow~ \mathcal{A}_{\mathcal{B}_{d+1}}\cong \Z_{N/g} ~\rightarrow~ \widetilde{\mathcal{A}}_{\mathcal{B}_{d+1}}\cong \Z_N   ~\rightarrow~  \widetilde{\mathcal{A}}_{\mathcal{B}_{d+1}}/\mathcal{A}_{\mathcal{B}_{d+1}} \cong \Z_{N/g}~\rightarrow~ 0 \\
&0 ~\rightarrow~Z( \widetilde{G}_{\mathcal{T}_{d+1}}) \cong \Z_{N/g} ~\rightarrow~ \widetilde{G}_{\mathcal{T}_{d+1}} \cong \mathrm{SU}(g)~\rightarrow~  {G}_{\mathcal{T}_{d+1}}  \cong \mathrm{PSU}(g)~\rightarrow~1\,,
\ea \ee
we decompose the 4-term sequence into two short exact sequences each associated with an edge of the cheesesteak. Here $Z( \widetilde{G}_{\mathcal{T}_{d+1}}) $ is the center of the simply connected Lie group $\widetilde{G}_{\mathcal{T}_{d+1}} $  and we have
\be
 \widetilde{\mathcal{A}}_{\mathcal{B}_{d+1}}/\mathcal{A}_{\mathcal{B}_{d+1}} \cong Z( \widetilde{G}_{\mathcal{T}_{d+1}})
\ee
mediated via the $d+2$ dimensional bulk. We summarize the discussion by indicating where the various piece of data are found within the cheesesteak:
\be
\scalebox{0.9}{\begin{tikzpicture}
	\begin{pgfonlayer}{nodelayer}
		\node [style=none] (0) at (2, -1.5) {};
		\node [style=SmallCircleRed] (1) at (2, -1.5) {};
		\node [style=none] (2) at (2, -1.5) {};
		\node [style=none] (4) at (2.75, -2) {2-group};
		\node [style=none] (5) at (-3.5, 1.5) {};
		\node [style=none] (6) at (-3.5, -1.5) {};
		\node [style=none] (7) at (2, 1.5) {};
		\node [style=SmallCircleBrown] (8) at (-3.5, -1.5) {};
		\node [style=SmallCircleGreen] (9) at (-3.5, 1.5) {};
		\node [style=SmallCircleBlue] (10) at (2, 1.5) {};
		\node [style=none] (11) at (2.75, 0) {$\widetilde{\mathcal{A}}_{\mathcal{B}_{d+1}}$};
		\node [style=none] (12) at (-0.75, -2.75) {};
		\node [style=none] (13) at (-0.75, 0) {$ Z( \widetilde{G}_{\mathcal{T}_{d+1}}) \cong \widetilde{\mathcal{A}}_{\mathcal{B}_{d+1}}/\mathcal{A}_{\mathcal{B}_{d+1}}$};
		\node [style=none] (14) at (-0.75, -2) {$\mathfrak{g}_{\mathcal{T}_{d+1}} $};
		\node [style=none] (29) at (-4.5, 0) {};
		\node [style=none] (30) at (-0.75, 2.25) {};
		\node [style=none] (31) at (-0.75, 2) {${G}_{\mathcal{T}_{d+1}} $};
		\node [style=none] (32) at (2.75, 2) {$ {\mathcal{A}}_{\mathcal{B}_{d+1}} $};
	\end{pgfonlayer}
	\begin{pgfonlayer}{edgelayer}
		\filldraw[fill=gray!50, draw=gray!50]  (2, -1.5) -- (2, 1.5) -- (-3.5, 1.5) -- (-3.5, -1.5) -- cycle;
		\draw [style=ThickLine] (2.center) to (1);
		\draw [style=GreenLine] (5.center) to (6.center);
		\draw [style=RedLine] (6.center) to (2.center);
		\draw [style=BlueLine] (5.center) to (7.center);
		\draw [style=PurpleLine] (7.center) to (2.center);
	\end{pgfonlayer}
\end{tikzpicture}}
\ee
Reading the figure bottom to top, we have first the boundary conditions realizing the specified symmetries, the naive superstructures, and at the bottom we labelled the edge $\mathcal{T}_{d+1}$ with the Lie algebra of $ \widetilde{G}_{\mathcal{T}_{d+1}} $ and indicated that the corner theory $\mathcal{T}_{d+1}$ has a 2-group symmetry. An isomorphic labelling is:
\be
\scalebox{0.9}{
\begin{tikzpicture}
	\begin{pgfonlayer}{nodelayer}
		\node [style=none] (0) at (2, -1.5) {};
		\node [style=SmallCircleRed] (1) at (2, -1.5) {};
		\node [style=none] (2) at (2, -1.5) {};
		\node [style=none] (3) at (2.75, -2) {2-group};
		\node [style=none] (4) at (-2, 1.5) {};
		\node [style=none] (5) at (-2, -1.5) {};
		\node [style=none] (6) at (2, 1.5) {};
		\node [style=SmallCircleBrown] (7) at (-2, -1.5) {};
		\node [style=SmallCircleGreen] (8) at (-2, 1.5) {};
		\node [style=SmallCircleBlue] (9) at (2, 1.5) {};
		\node [style=none] (10) at (2.75, 0) {$\Z_N$};
		\node [style=none] (11) at (0, -2.75) {};
		\node [style=none] (12) at (0, 0) {$\Z_g$};
		\node [style=none] (13) at (0, -2) {$\mathfrak{su}_g$};
		\node [style=none] (14) at (-3, 0) {};
		\node [style=none] (31) at (0, 2) {$\mathrm{PSU}(g) $};
		\node [style=none] (32) at (2.75, 2) {$ \Z_{N/g} $};
	\end{pgfonlayer}
	\begin{pgfonlayer}{edgelayer}
		\draw [style=ThickLine] (2.center) to (1);
		\filldraw[fill=gray!50, draw=gray!50]  (2, -1.5) -- (2, 1.5) -- (-2, 1.5) -- (-2, -1.5) -- cycle;
		\draw [style=GreenLine] (4.center) to (5.center);
		\draw [style=RedLine] (5.center) to (2.center);
		\draw [style=BlueLine] (4.center) to (6.center);
		\draw [style=PurpleLine] (6.center) to (2.center);
	\end{pgfonlayer}
\end{tikzpicture}
}
\ee

\subsubsection{Actions of $\mathcal{S}_{d+2}, \mathcal{B}_{d+1}$}

The action for $\mathcal{S}_{d+2}$ is determined from reduction of 11D supergravity on $\partial^2 X^\circ$. For illustrative purposes we again focus on the case $|\mathscr{K}|=1$. In the general cases we have $|\mathscr{K}|$ copies of the following discussion, one for each flavor brane. From the Chern-Simons term $C_3{G}_4G_4$ we find
\be\ba \label{eq:7D}
\mathcal{S}_{d+2}&\supset-\frac{1}{6(2\pi)^3}\Big[ 6\alpha \int  H_4 \cup B_1^{(\Z_g)}  \cup B_2^{(\Z_g)}+3\beta  \int  H_3 \cup B_2^{(\Z_g)}  \cup B_2^{(\Z_g)}\\[0.2em]
&~~~\,+6\gamma \int H_0 \cup H_4\cup C_3+6\delta \int H_1 \cup H_3\cup C_3\Big]\,,\\[0.35em]
\ea\ee
where $dC_3=H_4$. The flavor brane is an $\mathfrak{su}_g$ singularity and therefore, \cite{Apruzzi:2021nmk, Hubner:2022kxr},
\be
\alpha/2=\beta/2=-\frac{g-1}{2g}\qquad \textnormal{mod}\,1\,.
\ee
We also have $\gamma=\delta=1$ as these are computed to a trivial pairing and a base-fiber intersection respectively. We focus here on the discrete fields $B_1^{(\Z_g)}$ and $B_2^{(\Z_g)}$ which also come with the canonical single derivative terms
\be\label{eq:standardonedivterm}
\mathcal{S}_{d+2}\supset \frac{g}{2\pi}\int\lb  B_1^{(\Z_g)}\cup d B_5^{(\Z_g)}+B_2^{(\Z_g)}\cup d B_4^{(\Z_g)}\rb\,.
\ee
derived following \cite{Baume:2023kkf,GarciaEtxebarria:2024fuk}. The interpretation of these interactions are discussed in detail in \cite{Apruzzi:2021nmk, Apruzzi:2023uma, Baume:2023kkf}. The terms above are those of 7D super-Yang-Mills theory KK reduced on a circle. Geometrically, this is a non-trivial consequence of the cohomology ring of $\partial^2 X^\circ$, which is a twisted fiber product of a 3D lens space over a circle, being isomorphic to the cohomology ring of the direct product of this base and fiber. Physically, this is matched by the monodromy term \eqref{eq:Theta} having no consequence for the symmetries of the flavor theory. Indeed, it is a total derivative and can be expressed as a term exclusively supported on the support of the 5D SCFT locus living on the boundary to the 6D flavor theory.

The action for $\mathcal{B}_{d+1}$ is determined from reduction of 11D supergravity on $\partial X^\circ$. This reduction can be performed directly from geometry, and is further constrained to be compatible with the boundary conditions \eqref{eq:DBC} and \eqref{eq:NBC} and the bulk action \eqref{eq:7D}. For example, apply the boundary conditions
\be \ba
 B_2^{(\Z_g)}\big|_\partial = (N/g)B_2^{(\Z_N)}\,,\qquad  \lb {\partial}/{\partial r}  \,\lrcorner\, B_1^{(\Z_g)}\rb\!\!\;  \Big|_{\partial} =  B_0^{(\Z_N)}\,,\\
  B_5^{(\Z_g)}\big|_\partial = (N/g)B_5^{(\Z_N)}\,,\qquad  \lb {\partial}/{\partial r}  \,\lrcorner\, B_4^{(\Z_g)}\rb\!\!\;  \Big|_{\partial} =  B_3^{(\Z_N)}\,,\\
\ea \ee
to the term \eqref{eq:standardonedivterm} to find the single derivative boundary term
\be
\mathcal{B}_{d+1}\supset \frac{N}{2\pi}\int\lb  B_0^{(\Z_N)}\cup d B_5^{(\Z_N)}+B_2^{(\Z_N)}\cup d B_3^{(\Z_N)}\rb\,,
\ee
where the cocycles maintaining degree contribute a factor of $N/g$. As an example performed directly via geometry consider the coupling
\be\ba
6\;\!\mathcal{S}_{d+1=6}^{(c)}&\supset-  2\pi\epsilon \lb \frac{N/g}{2\pi} \rb^2 \lb \frac{N}{2\pi} \rb \int {B}_2^{(\Z_{N/g})}\cup {B}_2^{(\Z_{N/g})}\cup {B}_2^{(\Z_N)}\\[0.25em]
&=- 2\pi \epsilon \int \widetilde{B}_2^{(\Z_{N/g})}\cup \widetilde{B}_2^{(\Z_{N/g})}\cup \widetilde{B}_2^{(\Z_N)}\,,
\ea \ee
where the tilde now indicates a change in normalization to discrete fields taking values in $\Z_{N/g}$ and $\Z_{N}$. The change in normalization makes it clear that $\epsilon$ is computed by the triple product
\be \label{eq:Anomaly} \ba
H^2(\partial X^\circ, \partial^2 X^\circ)\times H^2(\partial X^\circ, \partial^2 X^\circ) \times H^2(\partial X^\circ) ~&\rightarrow~\Q/\Z \\
\Z \times\Z \times \Gamma^\vee ~&\rightarrow~\Q/\Z \\
(1,1,1)~&\mapsto ~\epsilon= \ell \lb \omega_2^{(\partial X^\circ\!,\, \partial^2 X^\circ) }\cup \omega_2^{(\partial X^\circ\!,\, \partial^2 X^\circ)}, \omega_2^{(\partial X^\circ)}\rb \\[0.5em]
\ea \ee
where the cup product $\cup \,: H^2(\partial X^\circ, \partial^2 X^\circ)\times H^2(\partial X^\circ, \partial^2 X^\circ) \rightarrow H^4(\partial X^\circ, \partial^2 X^\circ)\cong \Gamma$ maps into a finite group and $\ell$ denotes a linking form. Upon modding out the kernel, by using the isomorphism $\textnormal{Tor}\,H^2(\partial X)\cong H^2(\partial X^\circ,\partial^2 X^\circ)/H^1(\partial^2X^\circ)\cong (\Gamma/\Gamma_{\text{fix}})^\vee$, we can also consider
\be \ba
\textnormal{Tor}\,H^2(\partial X)\times \textnormal{Tor}\,H^2(\partial X) \times \textnormal{Tor}\,H^2(\partial X^\circ) ~&\rightarrow~\Q/\Z \\
(\Gamma/\Gamma_{\text{fix}})^\vee \times (\Gamma/\Gamma_{\text{fix}})^\vee\times \Gamma^\vee ~&\rightarrow~\Q/\Z \\
(1,1,1)~&\mapsto ~\epsilon= \ell \lb \omega_2^{(\partial X)}\cup \omega_2^{(\partial X)}, \omega_2^{(\partial X^\circ)}\rb
\ea \ee
where the cup product $\cup \,: H^2(\partial X)\times H^2(\partial X) \rightarrow H^4(\partial X)\cong \Gamma$ is determined from the previously considered cup product. Next, we compute the triple product,
\be
\epsilon=\frac{(L_m^1)^3}{gL_M^3}\,,
\ee
where we have $m=(m_1,m_2,m_3)$ and $M=(N,m_1,m_2,m_3)$ and $L^k_m,L^k_M$ are some combinatorial factors (see Appendix \ref{sec:AnomalyTriple5DSCFTs} for details).

Note that $(\Gamma/\Gamma_{\text{fix}})^\vee\cong \Z_{N/g}$ is the 1-form symmetry group and $\Gamma^\vee\cong \Z_N$ is the 1-form symmetry group extended by the 1-form symmetry of the flavor brane. This unrefined anomaly is therefore finer than the pure 5D 1-form symmetry self-anomaly, which is recovered by restricting to a subgroup $(\Gamma/\Gamma_{\text{fix}})^\vee\subset \Gamma^\vee$.

For example, in our illustrative example of M-theory on $\mathbb{C}^3/\Z_{2n}$, where we have $M=(2n,1,1,2n-2)$ and $m=(1,1,2n-2)$, we compute $\epsilon=1/n$ compatible with the pure 1-form anomaly $g\epsilon=2\epsilon=2/n$ compute in \cite{Apruzzi:2021nmk}.

\subsubsection{Summary for the Example of M-theory on $\mathbb{C}^3/\Z_{2n}(1,1,2n-2)$}

\begin{figure}
\centering
\scalebox{0.8}{
\begin{tikzpicture}
	\begin{pgfonlayer}{nodelayer}
		\node [style=none] (0) at (-3, -3) {};
		\node [style=none] (1) at (3, -3) {};
		\node [style=none] (2) at (-3, 3) {};
		\node [style=SmallCircleRed] (3) at (3, -3) {};
		\node [style=SmallCircleBrown] (4) at (-3, -3) {};
		\node [style=SmallCircleGrey] (5) at (-3, 3) {};
		\node [style=none] (6) at (4.625, -3.5) {Relative 5D SCFT};
		\node [style=none] (7) at (0, -3.5) {6D SYM $+$ $f_2^3$-term $+\dots$};
		\node [style=none] (9) at (1.375, 0.875) { \rotatebox{-45}{$\frac{2n}{2\pi}\int B_2^{(\Z_{2n})}\cup dB_3^{(\Z_{2n})}$+\dots} };
		\node [style=none] (11) at (-4, 3.5) {BC for $B_2^{(\Z_{2n})},B_3^{(\Z_{2n})}$};
		\node [style=none] (12) at (-5.5, 0.375) {Dirichlet BC for $B_2^{(\Z_2)}$};
		\node [style=none] (14) at (-5.5, -0.375) {Neumann BC for $B_4^{(\Z_2)}$};
		\node [style=none] (15) at (-1, -1.5) {$\frac{2}{2\pi}\int B_2^{(\Z_2)}\cup dB_4^{(\Z_2)}$};
		\node [style=none] (16) at (0.625, 0.25) {\footnotesize \rotatebox{-45}{$B_2^{(\Z_2)}|_{\partial}=nB_2^{(\Z_{2n})}\,,~ \lb {\partial}/{\partial r}  \,\lrcorner\, B_4^{(\Z_2)}\rb\!\!\;  \Big|_{\partial} =  B_3^{(\Z_{2n})}$}};
		\node [style=none] (17) at (0, -4.5) {};
		\node [style=none] (19) at (-1.5, -2.125) {$+\dots$};
		\node [style=none] (20) at (-5.5, -1.875) {$\dots$};
	\end{pgfonlayer}
	\begin{pgfonlayer}{edgelayer}
		\filldraw[fill=gray!50, draw=gray!50]  (-3, 3) -- (-3, -3) -- (3, -3)  -- cycle;
		\draw [style=GreenLine] (2.center) to (0.center);
		\draw [style=RedLine] (0.center) to (1.center);
		\draw [style=PurpleLine] (2.center) to (1.center);
	\end{pgfonlayer}
\end{tikzpicture}}
\caption{Sketch of some features of the cheesesteak for M-theory on $\mathbb{C}^3/\Z_{2n}(1,1,2n-2)$ related to the 5D defect groups of lines and surfaces. We have $\Gamma^\vee \cong \Z_{2n}$ and $(\Gamma/\Gamma_{\text{fix}})^\vee \cong \Z_n$ and $\Gamma_{\text{fix}}^\vee \cong \Z_2$.}
\label{Fig:SummaryFields}
\end{figure}
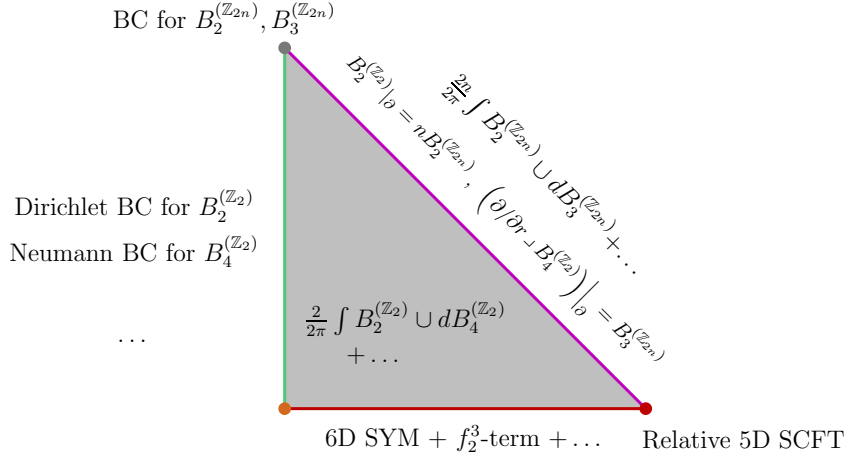

Let us summarize the overall cheesesteak for the example of M-theory on $\mathbb{C}^3/\Z_{2n}$ with weights $(1,1,2n-2)$. We focus on the discrete data. See figure \ref{Fig:SummaryFields} for a summary where we present the cheesesteak with the edge $\mathbb{B}^{(r,1)}_{d+1}$ collapsed.

The flavor symmetry\footnote{Here we ignore possible mixing of this 0-form symmetry with the R-symmetry and the structure group of the tangent bundle.} of the 5D SCFT is SO$(3)$ as determined by the boundary conditions $\mathbb{B}_{d+1}^{(1,x_\perp)}$ (green edge) which fixes the global form of the flavor brane theory $\mathcal{T}_{d+1}$ (red edge). The relative symmetry theory $\mathcal{B}_{d+1}$ (purple edge) with discrete fields of order $2n$ adds defects such that the lines and 3-surfaces present in the bulk $\mathcal{S}_{d+2}$ (grey face), individually isomorphic to $\Z_{2}$, are extended to $\Z_{2n}$ as characterized by the boundary conditions $\mathcal{S}_{d+2}|_{\mathcal{B}_{d+1}}$. The polarization for these additional defects is determined at $\mathbb{B}_{d}^{(1,1),*}$ (grey dot). This polarization is constrained by $\mathbb{B}_{d+1}^{(1,x_\perp)}$. Finally, the 5D SCFT is an end of the world theory to the 6D flavor brane (red dot). Overall we have the actions:
\be\ba
\mathcal{S}_{d+2}&=\frac{2}{2\pi}\int\lb  B_1^{(\Z_2)}\cup d B_5^{(\Z_2)}+B_2^{(\Z_2)}\cup d B_4^{(\Z_2)}\rb- \frac{1}{2(2\pi)^2} \int  H_4 \cup B_1^{(\Z_2)}  \cup B_2^{(\Z_2)} \\[0.25em]
&~~~\, + \frac{1}{2(2\pi)^2}  \int  H_3 \cup B_2^{(\Z_2)}  \cup B_2^{(\Z_2)}+\dots \\[0.5em]
\mathcal{B}_{d+1}&= \frac{2n}{2\pi}\int\lb  B_0^{(\Z_{2n})}\cup d B_5^{(\Z_{2n})}+B_2^{(\Z_{2n})}\cup d B_3^{(\Z_{2n})}\rb-  \frac{4\pi}{3n} \lb \frac{2n}{2\pi} \rb^3  \int {B}_2^{(\Z_{2n})}\cup {B}_2^{(\Z_{2n})}\cup {B}_2^{(\Z_{2n})} \\[0.25em]
&~~~\, + \dots\,,
\ea\ee
where we wrote out the data associated with discrete symmetries. Here made replacements using $2B_2^{(\Z_{2n})}=B_2^{(\Z_{n})}$ to achieve a cubic term in $\mathcal{B}_{d+1}$ which permitted us the give the refined anomaly, the coefficient of the $(B_2^{(\Z_{2n})})^3$ term. The latter follows from the 1-form self-anomaly being $2/6n$, hence we know that $2\times \epsilon/6=2/6n$ mod 1 and therefore either $\epsilon=1/6n$ or  $\epsilon=1/6n+1/2$. Additionally, from $\Z_n$ embedding into a larger $\Z_{2n}$ we know that $\epsilon/6$ must follow from a pure triple $\Z_{2n}$ anomaly term (as written above) by field redefinition, and reversing back to $B_2^{(\Z_{n})}$ we have $\epsilon=1/6n$.

\subsection{Example: 4D Chiral Matter}
\label{sec:BiFund}

We now turn to an example of a free 4D theory. From the associated extra-dimensional construction we will find many continuous symmetries, and one key question will be which of these remain, once the 4D dynamics have been isolated.

To begin, let $\mathbb{W}\P^3$ denoted the weighted projective space with projective coordinates $z_1,\bar z_2,z_3,\bar z_4$ carrying weights $N_1,N_1,N_2,N_2$ respectively with $N_1,N_2$ coprime.  In M-theory the cone
\be
X= \textnormal{Cone}(\mathbb{W}\P^3)\,,
\ee
which is conjectured to admit a $G_2$-holonomy metric, engineers a 4D $\mathcal{N}=1$ chiral superfield $\Phi$ in the bifundamental representation $({\bf N_1},{\bf \overline N_2})_{N_1+N_2}$ of a non-abelian flavor symmetry algebra $\mathfrak{su}_{N_1}\oplus \mathfrak{su}_{N_2}\oplus \mathfrak{u}_1$ \cite{Atiyah:2001qf, Acharya:2001gy, Witten:2001uq}. See also \cite{Cvetic:2001kk, Cvetic:2001nr, Cvetic:2001tj} for related discussions in the context of intersecting D6-branes and their lifts to $G_2$ spaces.


We consider the singularities of $X$. There are two codimension four A-type ADE loci (flavor branes) with Lie algebras $\mathfrak{su}_{N_1}$ and $\mathfrak{su}_{N_2}$, wrapping $z_3,z_4=0$ and $z_1,z_2=0$ respectively. These ADE loci extend radially and enhance to a codimension-7 singularity at the tip of the cone. The full singular locus is the union of these cones
\be
\mathscr{S}=\textnormal{Cone}(\P^1_{12}) \cup \textnormal{Cone}(\P^1_{34})=\R^3_{N_1}\!\cup \R^3_{N_2}\,.
\ee
Here $\P^1_{12}$ and $\P^1_{34}$ are parametrized by $z_1,z_2$ and $z_3,z_4$ respectively, and give $\R^3_{N_i}$ which support $\C^2/\Z_{N_i}$ ADE singularities. We have the disjoint union $\mathscr{K}=\P^1_{12}\cup \P^1_{34}$ and the chiral superfield $\Phi$ is supported at the tip of the cone.

The IIA dual of this geometry are two stacks of $N_1,N_2$ D6-branes intersecting supersymmetrically at an angle in $\mathbb{R}^6$. Their worldvolume lifts to the codimension four locus $\R^3_{N_1}\!\cup \R^3_{N_2}$ and their intersection point lifts to the tip of the cone.

Going forward we focus on the case $(N_1,N_2)=(N,1)$. The discussion generalizes straightforwardly to coprime pairs $(N_1,N_2)$. We can also consider pairs $(N_1,N_2)$ with $\gcd(N_1,N_2)\neq 1$, in these cases however the geometry $X$ changes, becoming a global quotient of an already singular cone \cite{Acharya:2001gy}, and our analysis will need to be slightly modified.

In any case, while the IIA dual still contains two sets of D6-branes the M-theory uplift only contains a single A-type ADE locus with Lie algebra $\mathfrak{su}_N$, the single D6-brane uplifts to a smooth patch of the geometry. We now have $\mathscr{K}=\P^1_{12} $ and $\mathscr{S}=\R^3_{N}$ where the singularity worsens from codimension four to codimension seven at the origin of $\R^3_{N}$.

From here we see that the 6-dimensional manifold with boundary $\partial X^\circ=\mathbb{W}\P^3\setminus \textnormal{Tube}(\P^1_{12})$ retracts onto the 2-sphere $\P^1_{34}$ and that the closed 5-dimensional manifold $\partial^2X^\circ$ is given by a bundle over $\P^1_{12}$ with lens space fibers $S^3/\Z_N$ (see Appendix \ref{app:A} for details). With this input we compute the long exact sequence of the pair $(\partial X^\circ,\partial^2X^\circ)$, which reads:
\begin{equation}\label{eq:RCHcircG2}
    \begin{array}{c||cccccc}
& H^n(\partial X^\circ, \partial^2 X^\circ) &  & H^n(\partial X^\circ) &  &   H^n(\partial^2 X^\circ)  &      \\[0.4em] \hline \hline \\[-0.9em]
       n=0 ~&   0  & \rightarrow & \Z & \rightarrow &  \mathbb{Z} & \rightarrow  \\
      n=1 ~& 0  & \rightarrow & 0 & \rightarrow &  0 & \xrightarrow[]{}  \\
      n=2 ~&   0  & \rightarrow &\Z & \rightarrow &\Z& \rightarrow  \\
         n=3~ &  0  &\rightarrow &0& \rightarrow& \Z &\rightarrow  \\
       n=4~ &    \Z  &\rightarrow &0&\rightarrow &0&\rightarrow  \\
        n=5 ~&  0 &\rightarrow & 0 &\rightarrow & \Z  & \rightarrow  \\
               n=6 ~&   \Z  & \rightarrow &0 & \rightarrow &  0 & \rightarrow  \\
    \end{array}
\end{equation}

The IIA dual already illuminated some of the physical aspects, however we will be interested in the 5D KK theory $\mathcal{T}_{d+1}$ obtained by reducing $\textnormal{Cone}(\mathscr{K})$ to a half-line by KK reducing the theory on the ADE flavor brane on the link $\mathscr{K}$.

To understand this 5D KK theory, recall that in M-theory the codimension four singularities of $\mathscr{S}$ engineers 7D $\mathfrak{su}_N$ super-Yang-Mills theory. The defect at the tip of the cone, corresponding to the intersection of the single otherwise disjoint D6-brane, sources a background field configuration for this super-Yang-Mills theory. This background profile is analyzed geometrically, noting that the M-theory circle restricted to $\mathbb{P}_{12}$ twists with unit monopole number. In the IIA dual frame this is captured by a flux for the RR 1-form field $C_1$, with field strength $F_2^{(\text{RR})}$ through $\mathbb{P}_{12}$, i.e., the 10D supergravity background we are considering is such that
\be
\int_{\mathbb{P}_{12}}F_2^{(\text{RR})}=2\pi\,.
\ee
There are now two D6-brane Wess-Zumino terms on the stack of D6-branes which give rise to interesting terms given such a background. They are
\be\ba \label{eq:TopTerms}
\int_{N\times\textnormal{D6}} F_2^{(\text{RR})} \wedge \frac{\textnormal{tr\,}(a_1 \wedge f_2\wedge f_2)}{24 \pi^2}&=\int_{N\times\textnormal{D6}} F_2^{(\text{RR})} \wedge \textnormal{CS}_5^{(1)}(a_1)\,, \\[0.5em] \int_{N\times \textnormal{D6}} C_3 \wedge \frac{\textnormal{tr\,}(f_2\wedge f_2)}{8\pi^2}&=\int_{N\times \textnormal{D6}} C_3 \wedge \Theta_4^{(1)}(f_2)
\ea \ee
where $f_2$ is an $\mathfrak{su}_N$ field strength with connection $a_1$. Here $\textnormal{CS}^{(1)}_5$ denotes a level 1 5D Chern-simons term and $\Theta^{(1)}_4$ denotes a unit monopole density. KK reducing the $C_3$ field contributes a $\mathfrak{u}_1$ gauge field $A_1$ and overall we obtain from the above interactions two terms:
\be\label{eq:WVterms}
\int \frac{\textnormal{tr}\,(a_1\wedge f_2\wedge f_2)}{24\pi^2}\,, \qquad
\int A_1 \wedge \frac{\textnormal{tr\,}(f_2\wedge f_2)}{8\pi^2}\,.
\ee
The gauge fields $A_1,a_1$ combine to a 5D $\mathfrak{u}_N$ gauge field.

One consequence of the 5D $\mathfrak{su}_N$ Chern-Simons term, i.e., the first term in \eqref{eq:WVterms}, is that the center symmetry of the $\mathfrak{su}_N$ sector of the 5D KK theory is completely broken. This clearly matches \eqref{eq:RCHcircG2} where the cohomology groups of $\partial^2 X^\circ$ were found to be torsion-free. One consequence of the 5D instanton density interaction, the second term \eqref{eq:WVterms}, is that we have anomaly inflow onto the 4D theory, matched by a chiral fermion \cite{Witten:2001uq}.

With this we have now described the 4D chiral superfield $\Phi$ as an end of the world theory $\mathcal{T}_d$ to a 5D KK theory $\mathcal{T}_{d+1}$. We now turn to discuss the symmetries of this bulk-edge system via their respective symmetry theories $\mathcal{S}_{d+2}, \mathcal{B}_{d+1}$.

  \subsubsection{Field Content of $\mathcal{S}_{d+2},\mathcal{B}_{d+1}$}

We begin with the field content of $\mathcal{S}_{d+2}$. We expand the supergravity field strength ${G}_4$ in classes of ${H}^n(\partial^2 X^\circ)$. We have
\be\label{eq:Expansion3} \ba
{G}_4={H}_4 \cup {\mathbf{1}} + {H}_2 \cup {v}_{2}+ {H}_1 \cup {v}_{3} \,,\\[0.2em]
\ea \ee
where $i$ runs over the flavor branes and ${H}_n$ denotes abelian $\textnormal{U}(1)$ field strength of degree $n$ and ${v}_k$ denote  cohomology classes of degree $k$. The field content of $\mathcal{S}_{d+2}$ consists purely of abelian field strengths
\be
H_1,H_2,H_4\,.
\ee

Next we discuss the field content of $\mathcal{B}_{d+1}$.  We expand the supergravity field strength ${G}_4$ in classes of ${H}^n(\partial X^\circ,\partial^2 X^\circ)$ and ${H}^n(\partial X^\circ)$. We have
\be\ba \label{eq:Expansion4}
{G}_4&={G}_0 \cup {u}_4  + {F}_4 \cup {\mathbf{1}} + {F}_2 \cup {w}_{2} \,.
\ea\ee
Here ${G}_n,{F}_n$ denote abelian $\textnormal{U}(1)$ field strengths of degree $n$. The classes ${u}_{n}$, $({w}_n)$ are free (relative) integral cohomology classes of $\partial X^\circ$ respectively. The field content of $ \mathcal{B}_{d+1}$ is
 \be
 G_0,F_2,F_4\,.
 \ee
There are no equivalence relations\,/\,identifications due to the mappings $\iota^{n}:\,H^n(\partial X^\circ,\partial^2 X^\circ)\rightarrow H^n(\partial X^\circ)$ as these are all trivial. Equivalently, the edge $ \mathcal{B}_{d+1}$ supports no degrees of freedom which are not inherited via restriction from the bulk.

Let us comment on a distinguishing feature of $H_2$. Geometrically, this field strength is associated to an internal 2-cocycle which is mapped onto via restriction of the 2-cocycle generating $H^2(T_{\mathscr{K}})$. Conversely, when attempting to build the defects acted on by this 0-form symmetry, we would wrap an M2-brane on a cone over the 2-cycle generating $H_2(\partial^2 X^\circ)=H_2(\partial T_{\mathscr{K}})$. However, this 2-cycle does not trivialize under the inclusion $\partial  T_{\mathscr{K}}\hookrightarrow T_{\mathscr{K}}$ obstructing this construction. Therefore, $H_2$ should be interpreted as the background profile to the $\mathfrak{u}_1$ field strength of $A_1$ \cite{Atiyah:2001qf}.

\subsubsection{Boundary Conditions $\mathcal{S}_{d+2}|_{\mathcal{B}_{d+1}}$}

The relative symmetry theory $ \mathcal{B}_{d+1}$ realizes a boundary condition for $\mathcal{S}_{d+2}$. Restricting a field $\mathcal{S}_{d+2}$ to the support of $\mathcal{B}_{d+1}$ we can relate the corresponding internal cycles, given they now embed into the same geometry, a copy of $\partial X^\circ$, via the mappings $H^n(\partial X^\circ)\rightarrow H^n(\partial^2 X^\circ)$ and $H^n(\partial^2 X^\circ)\rightarrow {H}^{n+1}(\partial X^\circ,\partial^2 X^\circ)$. The former mapping is degree preserving and relates cocycles of the $\mathcal{B}_{d+1}$ and $\mathcal{S}_{d+2}$ in identical degree. The latter mapping relates cocycles in adjacent degrees.

We make these boundary conditions explicit. We have degree preserving boundary conditions for fields from expansions along  $H^n(\partial^2 X^\circ)$ and  $H^n(\partial X^\circ)$ given by
\be \ba  \label{eq:DBC2}
 H_2\big|_{\mathcal{B}_{d+1}} &= F_2\\
  H_4\big|_{\mathcal{B}_{d+1}} &= F_4
 \ea \ee
 where we denote restriction to the support of $\mathcal{B}_{d+1}$ at $r=0$ by $|_{\mathcal{B}_{d+1}}$. Next, degree lowering boundary conditions for fields from expansions along  $H^n(\partial^2 X^\circ)$ and  $H^n(\partial X^\circ,\partial^2X^\circ)$ are given by
\be \label{eq:NBC2}\ba
\Big( {\partial}/{\partial r} \,\lrcorner\, H_1\Big)\!\; \Big|_{\mathcal{B}_{d+1}} &= G_0\,.
\ea \ee
See section \ref{sec:Gluing} for general discussion.

\subsubsection{Actions of $\mathcal{S}_{d+2}, \mathcal{B}_{d+1}$}

The action for $\mathcal{S}_{d+2}$  determined from reduction of 11D supergravity on $\partial^2 X^\circ$. From the Chern-Simons term $C_3{G}_4G_4$ we find
\be\ba \label{eq:6D2}
\mathcal{S}_{d+2}&\supset-2\pi\int \frac{C_3}{2\pi}\cup \frac{H_2}{2\pi}\cup \frac{H_1}{2\pi}\,,
\ea\ee
where locally $dC_3=H_4$. Similarly, reducing $C_3{G}_4G_4$ on $\partial X^\circ$, we find
\be\ba \label{eq:5D}
\mathcal{B}_{d+1}&\supset-2\pi\int \frac{C_3'}{2\pi}\cup \frac{F_2}{2\pi}\cup \frac{G_0}{2\pi}-\frac{2\pi }{6} \int \frac{A_1}{2\pi}\cup \frac{F_2}{2\pi}\cup \frac{F_2}{2\pi}\,,
\ea\ee
where locally $dC_3'=F_4$ (see Appendix \ref{app:A} for details). The first term contribution to $\mathcal{B}_{d+1}$ can be derived from the flavor symmetry theory and the boundary conditions \eqref{eq:DBC2} and \eqref{eq:NBC2}.

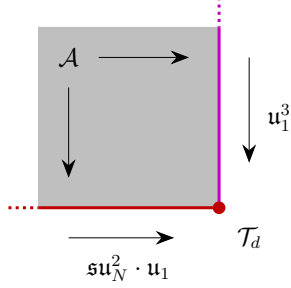
\begin{figure}[t!]
    \centering
    \scalebox{0.8}{\begin{tikzpicture}
	\begin{pgfonlayer}{nodelayer}
		\node [style=none] (0) at (0, 0) {};
		\node [style=none] (1) at (0, 3) {};
		\node [style=none] (2) at (-3, 0) {};
		\node [style=none] (3) at (0, 3.5) {};
		\node [style=none] (4) at (-3.5, 0) {};
		\node [style=SmallCircleRed] (5) at (0, 0) {};
		\node [style=none] (6) at (0.5, -0.5) {$\mathcal{T}_{d}$};
		\node [style=none] (7) at (-2.5, -0.5) {};
		\node [style=none] (8) at (-0.75, -0.5) {};
		\node [style=none] (9) at (0.5, 0.75) {};
		\node [style=none] (10) at (0.5, 2.5) {};
		\node [style=none] (11) at (1, 1.5) {$\mathfrak{u}_1^3$};
		\node [style=none] (12) at (-1.5, -1) {$\mathfrak{su}_N^2\cdot \mathfrak{u}_1$};
		\node [style=none] (13) at (-2.5, 2.5) {$\mathcal{A}$};
		\node [style=none] (14) at (-0.5, 2.5) {};
		\node [style=none] (15) at (-2.5, 0.5) {};
		\node [style=none] (16) at (-2.5, 2) {};
		\node [style=none] (17) at (-2, 2.5) {};
        \node [style=none] (18) at (-4, 1.5) {};
	\end{pgfonlayer}
	\begin{pgfonlayer}{edgelayer}
        \filldraw[fill=gray!50, draw=gray!50]  (-3, 0) -- (0, 0) -- (0, 3) --(-3,3)  -- cycle;
		\draw [style=RedLine] (2.center) to (0.center);
		\draw [style=PurpleLine] (1.center) to (0.center);
		\draw [style=DottedPurple] (3.center) to (1.center);
		\draw [style=DottedRed] (4.center) to (2.center);
		\draw [style=ArrowLineRight] (10.center) to (9.center);
		\draw [style=ArrowLineRight] (7.center) to (8.center);
		\draw [style=ArrowLineRight] (17.center) to (14.center);
		\draw [style=ArrowLineRight] (16.center) to (15.center);
	\end{pgfonlayer}
\end{tikzpicture}}
    \caption{Anomaly flow in the corner patch centered on the 4D corner mode theory. The gauge anomalies of a chiral fermion are determined from inflow along the edges connecting to the corner. A bulk anomaly $\mathcal{A}$ flows onto the edges.}
    \label{fig:anomalyflow}
\end{figure}

The above realizes various instances of anomaly inflow. Gauge transformations of $C_3$ are anomalous, due to the flavor symmetry theory term of line \eqref{eq:6D2}, and contribute a boundary term which is cancelled by the induced gauge transformation of $C_3'$ and the term \eqref{eq:5D}. Recall we simply have $C_3|_{\mathcal{B}_{d+1}}=C_3'$. Similarly, the 4D chiral fermion lives at the boundary of both $\mathcal{B}_{d+1}$ and the flavor brane $\mathcal{T}_{d+1}$. Gauge transformations of $A_1$ then give a $\mathfrak{u}_1^3$ anomaly of a 4D chiral fermion and the $\mathfrak{su}_N^2\cdot \mathfrak{u}_1$ anomaly which all must be cancelled by the corner mode.

We compute the above anomalies via Stokes' theorem. Consequently, anomalies flow in adjacent dimensions, i.e., they flow from the flavor symmetry theory $\mathcal{S}_{d+2}$ onto its boundaries, the relative symmetry theory $\mathcal{B}_{d+1}$ and the flavor brane $\mathcal{T}_{d+1}$, and they flow from $\mathcal{S}_{d+1}$ and $ \mathcal{T}_{d+1}$ onto their end of the world theory $\mathcal{T}_{d}$ (see figure \ref{fig:anomalyflow}).

\section{Conclusions} \label{sec:CONC}

In this paper we have studied a top down motivated generalization of SymTFTs / SymThs which unifies and extends earlier bottom up approaches.
We have proposed a filtration of the symmetries of a QFT to higher-dimensional systems which eventually terminates in a fully gapped bulk theory. In this approach, one views the SymTFT / SymTh of the original QFT as obtained from a regulated limit of some broader class of QFTs. Each such bulk QFT has its own SymTFT / SymTh, and filtering this structure repeatedly terminates. The original relative QFT now sits at a corner of a larger manifold with corners. Edges and faces of this higher-dimensional system can in principle support either gapped or gapless degrees of freedom.

While motivated by top down considerations, we have given a purely bottom up construction. From this perspective, the top down approach serves more as a guide in making various ``canonical'' choices in the construction. We have also presented some string / M-theory based examples which both motivate and illustrate this basic construction. In particular, in the case of 5D SCFTs with a continuous flavor symmetry, we have shown that there is a fully gapped bulk SymTFT, but one which lives in more than six dimensions. As another example, we showed that even in seemingly ``simple'' systems such as a collection of $N$ free chiral multiplets there is a surprisingly rich topological structure once the physics of the associated flavor symmetries is fully taken into account. In the remainder of this section we discuss some avenues of future investigation.

At the level of explicit computations, we have shown that there is a canonical way to read off triple product structures directly from geometry. An important subtlety here is that the original M-theory interaction term likely supports a choice of refinement in this product structure, analogous to the refinement of the link pairing observed in related field theoretic and geometric computations which enters in passing, for example, from Chern-Simons theory to Spin-Chern-Simons theory. It would be interesting to develop this refinement of link products since it will provide additional data on the structure of anomalies in the strongly coupled QFTs engineered via geometry.

The primary examples considered in this paper have involved a QFT with a continuous flavor symmetry. More broadly, however, one can consider a strongly coupled edge mode theory coupled to a bulk system which is itself strongly coupled. Explicit examples of this sort were constructed in
\cite{Acharya:2023bth}. It would be quite interesting to determine the associated relative symmetry theories with edges and corners which govern such systems.

A related comment is that especially in lower-dimensional systems, the dynamics of the bulk system is often a strongly interacting system in its own right. Determining the associated relative symmetry theories for this class of systems would also be quite interesting.

One of the general lessons from this work is that in top down realizations of generalized symmetries, the bulk symmetry theory is often more intricate than just a TFT in one higher dimension. Related considerations were also observed in the context of ``SymTree'' constructions where one has a collection of SymTFTs glued together along non-topological junctions \cite{Baume:2023kkf}. It would be natural to investigate broader generalizations of such treelike structures, but now where the branches themselves are also decompressed to a higher-dimensional gapped system of the sort considered in this paper.

Another well-motivated generalization involves including possible time dependent effects, as captured by the celestial topology of a string background (in the sense of \cite{Heckman:2024zdo}). This has mainly been developed in the case where the bulk consists of gapped bulk theories which interact across an interface theory. It would be natural to extend this to more general situations where the higher-dimensional bulk system is again a relative symmetry theory. Related considerations apply to non-supersymmetric intersecting brane configurations. In this setting notions such as ``edges and corners'' will likely also need to be revisited to take into account time dependent phenomena.

Finally, while these structures are best defined in limits where the effects of gravity are switched off, it is of course interesting to study the consequences of recoupling them to gravity, much as in \cite{Cvetic:2023pgm} (see also \cite{Gould:2023wgl}). The very fact that flavor branes of a local model are often shared across multiple throats suggests a generalization of the cutting and gluing of local models presented here.

\newpage

\section*{Acknowledgements}
We thank H.Y. Zhang for collaboration at an early stage of this work. We thank M. Del Zotto, I. Garcia Etxebarria, and X. Yu for helpful discussions. The work of MC and JJH is supported by DOE (HEP) Award DE-SC0013528. The work of MC is supported in part by
Slovenian Research Agency (ARRS No. P1-0306) and Fay R. and Eugene L. Langberg Endowed Chair funds. The work of MC and MH was supported by the Simons Foundation Collaboration grant \#724069 on ``Special Holonomy in Geometry, Analysis and Physics''. The work of MC, JJH, and MH is supported in part by a University Research Foundation grant at the University of Pennsylvania. The work of RD was supported in part by NSF grants DMS 2001673 and 2401422, by NSF FRG grant DMS 2244978, and by Simons HMS Collaboration grant 390287. The work of JJH and MH is supported in part by BSF grant 2022100. The work of ET is partly supported by the ERC Starting Grant QGuide-101042568 - StG 2021. MC, JJH, MH and ET thank the Simons Summer workshops in 2023 and 2024 for hospitality during part of this work. MC and ET would like to thank the Harvard Swampland Initiative for their hospitality during the completion of this work. RD and MH would like to thank the Mainz Institute of Theoretical Physics (MITP) for their hospitality during the completion of this work. JJH thanks the KITP for hospitality during part of this work; this research was supported in part by grant NSF PHY-2309135 to the Kavli Institute for Theoretical Physics (KITP). MH would like to thank the CERN Theoretical Physics Department for their hospitality during the completion of this work. ET thanks the UPenn Department of Physics and Astronomy for their hospitality during  completion of this work.

\appendix

\section{Fibrations, Sequences and Products}
\label{app:A}
In this Appendix we present the top down derivation of some results stated in section \ref{sec:Illustrative}. Our starting point will be a purely geometric M-theory background  $M_d\times X$ with $d$-dimensional spacetime $M_d$. The non-compact geometry  $X=\textnormal{Cone}(\partial X)$ will be a special holonomy cone engineering minimally supersymmetric theories. Concretely, we considered the two examples
\begin{itemize}
\item $X=\textnormal{Cone}(\mathbb{W}\P^3)$ with $\mathbb{W}\P^3$ a weighted projective space whose projective coordinates $z_1,\bar z_2,z_3,\bar z_4$ have weights $N,N,1,1$ respectively. This space engineers a 4d $\mathcal{N}=1$ chiral superfield $\Phi$ in the fundamental representation of an $\mathfrak{u}_N$ flavor algebra.
\item $X=\C^3/\Z_{2n}= \textnormal{Cone}(S^5/\Z_{2n})$ where the complex coordinates $z_1,z_2,z_3$ have weights $1,1,2n-2$ respectively. This space engineers a 5D SCFT with SU$(n)_n$ IR gauge theory phase.
\end{itemize}
and generalizations thereof.

\subsection{Families of ADE Singularities}

The geometries $X$ both exhibit singular loci consisting of a single irreducible component of non-compact codimension 4 ADE singularities of A-type which enhance at the tip of the cone $X$ to a maximal codimension singularity. In both cases the geometries will belong to larger classes of examples with multiple non-compact codimension 4 singularities of A-type which we touched on in section \ref{sec:Illustrative}.

As ADE singularities appear in both examples, let us begin here with their definition and recall their symmetries. An ADE singularity is a quotient singularity of the form $\mathscr{X}_\Gamma=\mathbb{R}^4/\Gamma$ where $\Gamma$ is a finite subgroup of $\textnormal{SU}(2)_L$ acting on $\mathbb{R}^4$ via rotations as induced by the identification $\textnormal{SO}(4)=\textnormal{SU}(2)_L\times \textnormal{SU}(2)_R/\Z_2$.  The isometry\,/\,symmetry group of an ADE singularity is therefore inherited from the natural $\textnormal{SO}(4)$ action, it is:
\be
\textnormal{Sym}(\mathscr{X}_\Gamma)=\textnormal{SU}(2)_R\times \Lambda_\Gamma\,.
\ee
Here $\Lambda_\Gamma$ is the subgroup of $\textnormal{SU}(2)_L$ conjugating $\Gamma$ to itself. For A-type singularities $\Lambda\equiv \Lambda_{\Z_N}=\textnormal{U}(1)\ltimes \Z_2$ consisting of diagonal matrices in $\textnormal{SU}(2)_L$ and complex conjugation\footnote{Complex conjugation corresponds to the outer automorphism of the A-type Lie algebras associated to the ADE singularity via the McKay correspondence. For D- and E-type singularities $\Lambda$ consists solely of the outer automorphisms of the respective Lie algebras.}.

The geometries $X$ we consider contain an  ADE locus whose normal geometry is twisted by symmetries $\textnormal{U}(1)\subset \Lambda_{\Z_N}$.  We refer to this twist as a $\Lambda$-twist and characterize it as follows. A-type ADE singularities take the form $xy=z^N$ for some integer $N\geq 2$. When viewing the ADE locus $\mathscr{S}\setminus \mathscr{S}_0$ as a family of ADE singularities $x,y,z$ become local functions on $\mathscr{S}\setminus \mathscr{S}_0$. As the symmetry subgroup $\Lambda$ commutes with $\Gamma=\Z_N$ the coordinates $x,y,z$ individually twist to sections of a line bundle. If $x$ twists to a section of $\mathcal{L}$, then $y$ and $z$ are sections of $\mathcal{L}^{-1}$ and the trivial line bundle respectively. The $\Lambda$-twist is specified by $\mathcal{L}$ which is determined by its connection $A_\Lambda$ with curvature $F_\Lambda$.

How does this twist enter our physical considerations? The smooth compact manifold $\partial^2 X^\circ$ always takes the form of a fibration with fiber $S^3/\Z_N$ and base $\mathscr{K}=\partial \mathscr{S}$, i.e.,
\be
S^3/\Z_N ~\hookrightarrow~ \partial^2 X^\circ ~\rightarrow~ \mathscr{K}\,.
\ee
The $\Lambda$-twist specifies this fibration. Whenever the ADE singularity in M-theory admits a dual description as a D6-brane in IIA we further have
\be
A_\Lambda=C_1\,, \qquad dC_1=F_\Lambda\,,
\ee
for the RR 1-form field $C_1$ on the D6-brane worldvolume, i.e., the $\Lambda$-twist locally specifies a M-theory circle bundle. Further, the $\Lambda$-twist constrains how the fibration $\partial^2 X^\circ \rightarrow \mathscr{K}$ can be filled radially to the fibration $\partial X^\circ \rightarrow \textnormal{Cone}(\mathscr{K})=\mathscr{S}$. It specifies how the generic fiber degenerates to an exceptional fiber $\partial \mathscr{E}$ projecting to the tip of the cone $ \textnormal{Cone}(\mathscr{K})$. For the two examples picked out above we have:
\begin{itemize}
\item $X=\textnormal{Cone}(\mathbb{W}\P^3)$ with $\mathbb{W}\P^3$ a weighted projective space whose projective coordinates $z_1,\bar z_2,z_3,\bar z_4$ have weights $N,N,1,1$ respectively. The generic fiber $S^3/\Z_N$ collapses to $S^2=(S^3/\Z_N)/\textnormal{U}(1)$ where $\textnormal{U}(1)$ is the Hopf circle.
\item $X=\C^3/\Z_{2n}= \textnormal{Cone}(S^5/\Z_{2n})$ where the complex coordinates $z_1,z_2,z_3$ have weights $1,1,2n-2$ respectively. The generic fiber $S^3/\Z_2$ collapses to $(S^3/\Z_2)/\Z_n=S^3/\Z_{2n}$ where $\Z_2,\Z_{2n}$ are subgroups of the $\textnormal{U}(1)$ Hopf circle.
\end{itemize}
We now describe the geometry of a family of examples containing the above pair of examples in more detail and derive the long exact sequences which formed the starting point of our analysis in section \ref{sec:Illustrative}. Of course when $\mathscr{S}\setminus \mathscr{S}_0$ has multiple disconnected components (flavor branes) then considerations similar to the above apply.

\subsection{Fibrations and (Co)Homology}

We in turn analyze the topology of the fibrations $\pi_{\mathscr{S}}:\partial T_{\mathscr{S}}\rightarrow \mathscr{S}$ and  $\pi_{I}:\partial T_{\mathscr{S}}\rightarrow I$ of the two geometries:
\begin{itemize}
\item $X=\textnormal{Cone}(\mathbb{W}\P^3)$ with $\mathbb{W}\P^3$ a weighted projective space whose projective coordinates $z_1,\bar z_2,z_3,\bar z_4$ have weights $N_1,N_1,N_2,N_2$ with $\gcd(N_1,N_2)=1$. This space contains up to two flavor branes.
\item $X=\C^3/\Z_{N}= \textnormal{Cone}(S^5/\Z_{N})$ where the complex coordinates $z_1,z_2,z_3$ have weights $k_1,k_2,k_3$ respectively. This space contains up to three flavor branes.
\end{itemize}

\subsubsection{$X=\textnormal{Cone}(\mathbb{W}\P^3)$}

We consider two cases.

\paragraph{Case 1: $X=\textnormal{Cone}(\mathbb{W}\P^3)$ and $(N_1,N_2)=(N,1)$}\mbox{}\medskip

First, consider the fibration $\pi_{\mathscr{S}}:\partial T_{\mathscr{S}}\rightarrow \mathscr{S}$. The generic fibers projecting to the codimension 4 ADE locus $\R^3 \setminus \{0\}= \mathscr{S}\setminus \mathscr{S}_0$ are lens spaces $S^3/\Z_{N}$. We claim that at the origin the exceptional fiber is $\partial \mathscr{E}=S^2$ and that $S^2$ is related to the generic fiber by the Hopf projection $S^3/\Z_{N}\rightarrow S^2$. The starting point in establishing $\partial \mathscr{E}=S^2$ is the characterization
\be
\partial \mathscr{E}=\partial X^\circ|_{\textnormal{retract}}\,,
\ee
derived on general grounds near \eqref{eq:Retract}. As such, we first describe a parametrization of the link $\partial X=\mathbb{W}\mathbb{P}^3$ geared towards excising the singularities supported at $\mathbb{P}^1_{12}$ and, subsequently, describing the deformation retraction.

Choosing the weights of a $\textnormal{U}(1)$-action appropriately we clearly have $\mathbb{W}\P^3=S^7/\!\:\textnormal{U}(1)$. With this we consider the fibration\footnote{This is the natural generalization of the 2-torus fibration of the 3-sphere upon replacing the complex plane $\C$ with the quaternions $\mathbb{H}$.} $S^7\rightarrow I= [-1,+1]$ with generic fiber $S^3_+\times S^3_{-}$. At the edges $\pm1$ of the base interval the spheres $S^3_{\pm}$ collapse and the generic fiber degenerates to $S^3_{\mp}$ respectively. The torus action respects this fibration and we obtain the fibration $\mathbb{W}\P^3\rightarrow I= [-1,+1]$ with generic fiber $S^3_+\times S^3_{-}/\textnormal{U}(1)$ where the quotient identifies the Hopf circles of $S^3_{\pm}$. The exceptional fibers are now $S^3_{\mp}/\textnormal{U}(1)=S^2_{\mp}$ respectively. One of these 2-spheres, for example $S^2_-$, supports the ADE singularities and upon excision the remaining space can be deformation retracted onto the other 2-sphere $S^2_+$, hence topologically $\partial \mathscr{E}=S^2$.

Next, we discuss the fibration $\pi_{I}:\partial T_{\mathscr{S}}\rightarrow I$ which follows by flipping the radial shells $S^2\subset \R^3$ of the ADE locus of the base for $\pi_{\mathscr{S}}$ into the fiber. The generic fiber at radius $r\neq 0$ is therefore $S^3_+\times S^3_{-}/\textnormal{U}(1)$. At $r=0$ one of the 3-spheres collapses, again resulting in the exceptional fiber $\partial \mathscr{E}=S^2$.\medskip

\paragraph{Case 2: $X=\textnormal{Cone}(\mathbb{W}\P^3)$ and $N_1,N_2>1$}\mbox{}\medskip

First, consider the fibration $\pi_{\mathscr{S}}:\partial T_{\mathscr{S}}\rightarrow \mathscr{S}$. The generic fibers projecting to the two codimension 4 ADE loci $\R^3_{N_i}\setminus \{0\}\subset \mathscr{S}$ are lens spaces $S^3/\Z_{N_i}$. We claim that at the origin the exceptional fiber is a real 5-dimensional manifold $\partial \mathscr{E}$, which is circle fibered as
\be\label{eq:CircleBundle}
S^1~\hookrightarrow~ \partial \mathscr{E}~\rightarrow~ S^2_1\times S^2_2\,, \qquad e=N_2\textnormal{vol}_{S^2_1}-N_1\textnormal{vol}_{S^2_2}\,,
\ee
where $e$ denotes the Euler class of the fibration. Via the Gysin sequence one then straightforwardly computes the (co)homology groups\footnote{Let us also determine explicit representatives for generator of the homology groups $H_n(\partial \mathscr{E})$. First, $H_1(\partial \mathscr{E})$ is generated by the circle fiber. Studying the Gysin sequence we find
\be
\Big\langle \frac{N_2}{g}S^2_1+\frac{N_1}{g}S^2_2\Big\rangle= H_2(\partial \mathscr{E})\,.
\ee
For this consider the image in the Gysin sequence of $H^3(\partial \mathscr{E})$ and apply Poincar\'e duality. Here we mean there exists a 2-cycles in $H_2(\partial \mathscr{E})$ which projects to the base cycle $ (N_2/g)S^2_1+(N_1/g)S^2_2$ and, to keep notation light, we have not given this 2-cycle a new name. Similarly, one determines $H_3(\partial \mathscr{E})\cong \Z_g\oplus \Z$ as, respectively,
\be \label{eq:Coolformula}
\Big\langle \frac{N_1}{g}S^3_1/\Z_{N_1}-\frac{N_2}{g}S^3_2/\Z_{N_2}\Big\rangle\oplus \langle l S^3_1/\Z_{N_1}+k S^3_1/\Z_{N_2} \rangle = H_3(\partial \mathscr{E})\,.
\ee
Here the lens spaces are the fibers of $S^3/\Z_{N_i}\hookrightarrow \partial \mathscr{E}\rightarrow S^2_i$. From this presentation we also have that there is a $g$-fold covering of the base. We denote this 2-cycle of $\partial \mathscr{E}$ as $gS^2_i$ for which we have $gS^2_i \cdot S^3/\Z_{N_j}=g\delta_{ij}$. This gives the check that indeed the free, torsional generators in degree 2 and 3, respectively, do not intersect, as it must be by general theory. Next, $k,l$ are integers such that $kN_1+lN_2=g$. This follows from the Gysin sequence, as the matrix, which appears in a basis change,
\be
\lb \begin{array}{c c}
 N_1/g& -N_2/g \\ l & k
\end{array}
\rb
\ee
must have have determinant $\pm1$, otherwise the new basis is `too coarse', only spanning an integral sublattice. Note we can always redefine generators to their negative, hence we have chosen without restricting ourselves $+1$. The generators of $H_0,H_5$ are the point and $\partial \mathscr{E}$ respectively.}
\be\label{eq:NormalGeo} \ba
H_n(\partial \mathscr{E})&\cong \{\Z,\Z_g,\Z,\Z\oplus\Z_g,0,\Z \} \\
H^n(\partial \mathscr{E})&\cong \{\Z,0,\Z\oplus\Z_g,\Z,\Z_g,\Z \}\,.
\ea \ee
All of this follows immediately from the discussion of case 1 by noting that if both $S^2_\pm$ support singularities then $\partial X^\circ$ deformation retracts to $S^3_+\times S^3_{-}/\textnormal{U}(1)$ which is described as above.

We are interested in the case $g=1$, for which all torsion groups trivialize
\be\label{eq:NormalGeoG1} \ba
H_n(\partial \mathscr{E})&\cong \{\Z,0,\Z,\Z,0,\Z \} \\
H^n(\partial \mathscr{E})&\cong \{\Z,0,\Z,\Z,0,\Z \}\,.
\ea \ee
from our discussion of generators in the general case, we see that generating (co)cycles in degree $2,3$ pair to (the top class) 1 via (the cup product) intersection. This determines the ring structure completely.

Next, we discuss the fibration $\pi_{I}:\partial T_{\mathscr{S}}\rightarrow I$ which follows by essentially doubling the discussion of case 1. The generic fiber at $r\neq 0$ is the disjoint union of a pair of 5-manifolds
\be
\partial \mathscr{E}\sqcup\partial \mathscr{E}\,.
\ee
At $r=0$ these are identified to give a single copy of $\partial \mathscr{E}$. Topologically $\partial T_{\mathscr{S}}$ is therefore simply the cylinder $\partial T_{\mathscr{S}}=\R \times \partial  \mathscr{E}$.

\paragraph{The $\Lambda$-twist}\mbox{}\medskip

Next, let us characterize the $\Lambda$-twist. \medskip

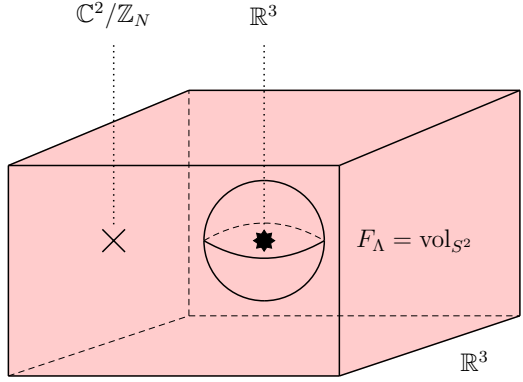
\begin{figure}
\centering
\scalebox{0.8}{
\begin{tikzpicture}
	\begin{pgfonlayer}{nodelayer}
		\node [style=none] (0) at (-4.25, 0.25) {};
		\node [style=none] (1) at (-1.25, 1.5) {};
		\node [style=none] (2) at (1.25, 0.25) {};
		\node [style=none] (3) at (4.25, 1.5) {};
		\node [style=none] (4) at (-4.25, -3.25) {};
		\node [style=none] (5) at (-1.25, -2.25) {};
		\node [style=none] (6) at (1.25, -3.25) {};
		\node [style=none] (7) at (4.25, -2.25) {};
		\node [style=Star] (8) at (0, -1) {};
		\node [style=none] (9) at (-1, -1) {};
		\node [style=none] (10) at (1, -1) {};
		\node [style=none] (11) at (0, 0) {};
		\node [style=none] (12) at (0, -2) {};
		\node [style=NodeCross] (13) at (-2.5, -1) {};
		\node [style=none] (14) at (-2.5, -0.75) {};
		\node [style=none] (15) at (0, -0.75) {};
		\node [style=none] (16) at (0, 2.25) {};
		\node [style=none] (17) at (-2.5, 2.25) {};
		\node [style=none] (18) at (-2.5, 2.75) {$\C^2/\Z_N$};
		\node [style=none] (19) at (0, 2.75) {$\mathbb{R}^3$};
		\node [style=none] (20) at (3.5, -3) {$\R^3$};
		\node [style=none] (21) at (2.5, -1) {$F_\Lambda=\textnormal{vol}_{S^2}$};
	\end{pgfonlayer}
	\begin{pgfonlayer}{edgelayer}
		\filldraw[fill=red!20, draw=red!20]  (-4.25, 0.25) -- (-1.25, 1.5) -- (4.25, 1.5) -- (4.25, -2.25) -- (1.25, -3.25) -- (-4.25, -3.25) -- cycle;
		\draw [style=ThickLine] (0.center) to (1.center);
		\draw [style=ThickLine] (3.center) to (1.center);
		\draw [style=ThickLine] (3.center) to (2.center);
		\draw [style=ThickLine] (2.center) to (0.center);
		\draw [style=ThickLine] (7.center) to (6.center);
		\draw [style=ThickLine] (6.center) to (4.center);
		\draw [style=ThickLine, bend left=45] (11.center) to (10.center);
		\draw [style=ThickLine, bend left=45] (10.center) to (12.center);
		\draw [style=ThickLine, bend left=45] (12.center) to (9.center);
		\draw [style=ThickLine, bend left=45] (9.center) to (11.center);
		\draw [style=ThickLine, bend right] (9.center) to (10.center);
		\draw [style=ThickLine] (0.center) to (4.center);
		\draw [style=ThickLine] (2.center) to (6.center);
		\draw [style=ThickLine] (3.center) to (7.center);
		\draw [style=DottedLine] (17.center) to (14.center);
		\draw [style=DottedLine] (16.center) to (15.center);
		\draw [style=DashedLineThin, bend left] (9.center) to (10.center);
		\draw [style=DashedLineThin] (4.center) to (5.center);
		\draw [style=DashedLineThin] (5.center) to (1.center);
		\draw [style=DashedLineThin] (5.center) to (7.center);
	\end{pgfonlayer}
\end{tikzpicture}}
\caption{Sketch of the local model $T_{\mathscr{S}}$ for case 1 with $\mathscr{S}=\R^3$ and $\mathscr{X}_f=\C^2/\Z_N$ and $\mathscr{E}=\R^3$ and the Euler class $F_\Lambda$ specifying the $\Lambda$-twist for the tube $T_{\mathscr{S}}$ within the $G_2$-holonomy cone $X=\textnormal{Cone}(\mathbb{W}\P^3)$.
}
\label{fig:FibrationsG2}
\end{figure}

\noindent {\bf Case 1:} Here, we can construct an interesting $\textnormal{U}(1)$-bundle from the above by collecting all Hopf circles in the generic fibers $\partial \mathscr{X}_f=S^3/\Z_N$ over $\R^3\setminus \{0\}=\mathscr{S}\setminus \mathscr{S}_0$. From the geometry we see that this $\textnormal{U}(1)$-bundle has unit Euler class when restricted to a 2-sphere linking the origin\footnote{This is easily understood in the dual IIA frame which consists of a stack of $N$ D6-branes supersymmetrically intersecting a single D6-brane. In our discussion we have centered coordinates on the $N$ D6-branes and the Euler class is counting the number of transverse D6-branes intersecting this stack at the origin.}. The connection $A_\Lambda$, defined on the ADE locus $\R^3\setminus \{0\}$, has curvature 2-form satisfying
\be
\frac{d F_\Lambda}{2\pi}= \delta(x) \, \textnormal{vol}_{\R^3}\,,
\ee
where $x\in \R^3$. The topology of the boundary of tubular neighbourhood $T_{\mathscr{S}}$ is completely specified by the 4-plet $(\mathscr{S},\partial \mathscr{X}_f, \partial \mathscr{E}, A_\Lambda)$ which we depict in figure \ref{fig:FibrationsG2}.\medskip

\noindent {\bf Case 2:} In this case the above structure simply doubles. There are two components to $\mathscr{S}\setminus \mathscr{S}_0$, each a copy of $\R^3\setminus \{0\}$ similar to above. To each we associated a $\textnormal{U}(1)$-bundle which we can characterize by two connections $ A_\Lambda^{(N_i)}$ with curvatures
\be
\frac{d F_\Lambda^{(N_1)}}{2\pi}= \delta(x) N_2 \, \textnormal{vol}_{\R^3_{N_1}}\,, \qquad d \frac{F_\Lambda^{(N_2)}}{2\pi}= -\delta(x) N_1 \, \textnormal{vol}_{\R^3_{N_2}}\,.
\ee
 The topology of $\partial T_{\mathscr{S}}$ is still characterized by the 4-plet $(\mathscr{S},\partial \mathscr{X}_f, \partial \mathscr{E}, A_\Lambda)$ where $\mathscr{X}_f$ now makes reference to the pair of singularity models $\C^2/\Z_{N_i}$, respectively over the two components of $\mathscr{S}\setminus \mathscr{S}_0$, and $A_\Lambda$ describes their twisting as specified by the pair of connections $ A_\Lambda^{(N_i)}$ on each of these components. These two fibrations then glue along one exceptional fiber.

\paragraph{The Long Exact Sequence in Relative (Co)Homology}\mbox{}\medskip

It will also be useful to determine long exact sequences in relative (co)homology of the pair $(\partial X^\circ,\partial^2 X^\circ)$. We do so for case 1. Here $\partial X^\circ$ has a single boundary component given by $\partial T_{\mathscr{K}}=\partial^2 X^\circ=S^3/\Z_N \rtimes S^2$ with (co)homology groups \eqref{eq:NormalGeoG1}. Together with the fact that $\partial X^\circ$ deformation retracts to $S^2$ we have the homology sequence:
\begin{equation}\label{eq:RHcircG2}
    \begin{array}{c||cccccc}
&   H_n(\partial T_{\mathscr{K}}) &  & H_n(\partial X^\circ) &  &   H_n(\partial X^\circ, \partial T_{\mathscr{K}}) &      \\[0.4em] \hline \hline \\[-0.9em]
	       n=6 ~&   0  & \rightarrow &0 & \rightarrow &  \mathbb{Z} & \rightarrow  \\
       n=5 ~&   \Z  & \rightarrow &0 & \rightarrow & 0 & \rightarrow  \\
      n=4 ~&  0 & \rightarrow & 0 & \rightarrow &  \Z & \rightarrow  \\
      n=3 ~&  \Z & \rightarrow &0& \rightarrow &0 & \rightarrow  \\
         n=2~ &  \Z  &\rightarrow & \Z & \rightarrow& 0&\xrightarrow[]{}  \\
       n=1~ &  0  &\rightarrow & 0 &\rightarrow &  0 &\rightarrow  \\
        n=0 ~&   \mathbb{Z}  &\rightarrow &  \Z &\rightarrow & 0  &
    \end{array}
\end{equation}
The cohomology sequence is:
\begin{equation}
    \begin{array}{c||cccccc}
& H^n(\partial X^\circ, \partial T_{\mathscr{K}}) &  & H^n(\partial X^\circ) &  &   H^n(\partial T_{\mathscr{K}})  &      \\[0.4em] \hline \hline \\[-0.9em]
       n=0 ~&   0  & \rightarrow & \Z & \rightarrow &  \mathbb{Z} & \rightarrow  \\
      n=1 ~& 0  & \rightarrow & 0 & \rightarrow &  0 & \xrightarrow[]{}  \\
      n=2 ~&   0  & \rightarrow &\Z & \rightarrow &\Z& \rightarrow  \\
         n=3~ &  0  &\rightarrow &0& \rightarrow& \Z &\rightarrow  \\
       n=4~ &    \Z  &\rightarrow &0&\rightarrow &0&\rightarrow  \\
        n=5 ~&  0 &\rightarrow & 0 &\rightarrow & \Z  & \rightarrow  \\
               n=6 ~&   \Z  & \rightarrow &0 & \rightarrow &  0 & \rightarrow  \\
    \end{array}
\end{equation}

\subsubsection{$X=\textnormal{Cone}(S^5/\Z_N)=\C^3/\Z_N$}

We begin by discussing some orbifold data of $X=\C^3/\Z_N$. Let $\Z_N\cong \Gamma \subset \textnormal{SU}(3)$ act faithfully on $\C^3$ as generated by
\be
(z_1,z_2,z_3) \mapsto (\omega^{m_1} z_1,\omega^{m_2} z_2,\omega^{m_3} z_3)
\ee
where $\omega$ is a primitive $N$-th root of unity and the integers $m_i$ satisfy $m_1+m_2+m_3=0$. Introduce $g_i=\gcd(m_i,N)$, then the faithfulness of the action implies that the $g_i$ are pairwise coprime. The singular locus of $\C^3/\Gamma$ consists of up to 3 cones of codimension 4 A-type ADE singularities. The apexes of these 3 cones coincide and support a codimension 6 singularity. The cones are cut out by setting two of three coordinates of $\C^3$ to vanish and are parametrized by the third coordinate $z_i$ modulo $\Gamma$. The $i$-th singular locus is characterized by the short exact sequence
\be
0~\rightarrow~ \Gamma_{\text{fix},i}\cong \Z_{g_i} ~\rightarrow~ \Gamma \cong \Z_N ~\rightarrow~ \Gamma/\Gamma_{\text{fix},i} \cong \Z_{N/g_i}~\rightarrow~ 0
\ee
where $\Gamma_{\text{fix},i}$ is the subgroup of $\Gamma$ folding the singularity. The ADE singularity is therefore of type $A_{g_i-1}$ and supported on the real 2-dimensional cone, parametrized by $z_i$,
\be
\mathscr{S}_{\Gamma_{\text{fix},i}}=\C/(\Gamma/\Gamma_{\text{fix},i})\subset \C^3/\Gamma\,.
\ee
Traversing once around the origin of $\mathscr{S}_{\Gamma_{\text{fix},i}}$ the normal geometry is glued back with twist $\Gamma/\Gamma_{\text{fix},i}$ which has a well-defined action on $\mathbb{C}^2/\Gamma_{\text{fix},i}$.

We now consider the asymptotic $\partial X= S^5/\Gamma$. The cones supporting ADE singularities intersect $\partial X$ in a circle $\mathscr{K}_i=S_i^1/(\Gamma/\Gamma_{\text{fix},i})$ where $S^1_i$ is thought of as parametrized by the argument of $z_i$. The total singular locus of $\partial X$ is
\be
\mathscr{K}\equiv \cup_i \mathscr{K}_i
\ee
consisting of up to three circles. We denote by $| \mathscr{K}|$ the number of circles. We further define $\partial X^\circ\equiv \partial X\setminus T_\mathscr{K}$ where $T_\mathscr{K}$ is an open tubular neighborhood of $\mathscr{K}$ in $\partial X$. With this $\partial X^\circ$ is 5-dimensional manifold with boundary. The boundary of $\partial T_{\mathscr{K}}$ has up to three connected components denoted $\partial T_{K_i}$. Each component takes the form of a lens space fibered over a circle, we write $\partial T_{\mathscr{K}_i} =S^3/ \Gamma_{\text{fix},i} \rtimes \mathscr{K}_i$
where traversing the circle $\mathscr{K}_i$ the fiber $S^3/\Gamma_{\text{fix},i}$ is twisted by $\Gamma/\Gamma_{\text{fix},i}$. Finally, we introduce the subgroup of $\Gamma_{\text{fix}}\subset \Gamma$ generated by all elements with fixed points in $S^5$ and straightforwardly we have
\be
\Gamma_{\text{fix}}=\prod_i \Gamma_{\text{fix},i} \cong \Z_{g_1g_2g_3}\,.
\ee

Let us consider the case $|\mathscr{K}|=1$. We denote the nontrivial $\Gamma_{\text{fix},i}$ by $\Gamma_{\text{fix}}\cong \Z_g$. In this case the fibration $X\rightarrow \mathscr{S}$ is straightforward. Take for example $\mathscr{S}$ to be cut out by $z_1,z_2=0$. Then $X\rightarrow \mathscr{S}$ is realized by projecting onto the third coordinate. The normal geometry jumps from $\mathbb{C}^2/\Gamma_{\text{fix}}$ to $\mathbb{C}^2/\Gamma$ with $\Gamma\cong \Z_N$. In this case
the fibrations $\pi_{\mathscr{S}}:\partial T_{\mathscr{S}}\rightarrow \mathscr{S}$ and  $\pi_I:\partial T_{\mathscr{S}}\rightarrow \R_{r\geq 0}$ are inferred straightforwardly via restriction.

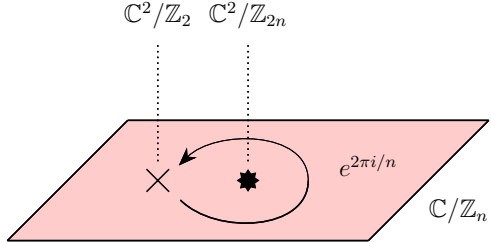
\begin{figure}
\centering
\scalebox{0.8}{
\begin{tikzpicture}
	\begin{pgfonlayer}{nodelayer}
		\node [style=none] (0) at (-4, -1) {};
		\node [style=none] (1) at (-2, 1) {};
		\node [style=none] (2) at (2, -1) {};
		\node [style=none] (3) at (4, 1) {};
		\node [style=Star] (4) at (0, 0) {};
		\node [style=NodeCross] (5) at (-1.5, 0) {};
		\node [style=none] (6) at (-1.5, 0.25) {};
		\node [style=none] (7) at (-1.5, 2.25) {};
		\node [style=none] (8) at (0, 0.25) {};
		\node [style=none] (9) at (0, 2.25) {};
		\node [style=none] (10) at (-1.125, -0.325) {};
		\node [style=none] (11) at (-1.15, 0.25) {};
		\node [style=none] (13) at (1, 0) {};
		\node [style=none] (14) at (3.5, -0.5) {$\C/\Z_n$};
		\node [style=none] (15) at (-1.5, 2.75) {$\C^2/\Z_2$};
		\node [style=none] (16) at (0, 2.75) {$\C^2/\Z_{2n}$};
		\node [style=none] (18) at (2, 0.25) {$e^{2\pi i/n}$};
	\end{pgfonlayer}
	\begin{pgfonlayer}{edgelayer}
		\filldraw[fill=red!20, draw=red!20]  (-4, -1) -- (-2, 1) -- (4, 1) -- (2, -1) -- cycle;
		\draw [style=ThickLine] (1.center) to (3.center);
		\draw [style=ThickLine] (3.center) to (2.center);
		\draw [style=ThickLine] (2.center) to (0.center);
		\draw [style=ThickLine] (0.center) to (1.center);
		\draw [style=DottedLine] (7.center) to (6.center);
		\draw [style=DottedLine] (9.center) to (8.center);
		\draw [style=ArrowLineRight, in=45, out=90] (13.center) to (11.center);
		\draw [style=ThickLine, in=-90, out=-45] (10.center) to (13.center);
		\draw [style=ThickLine] (1.center) to (3.center);
		\draw [style=ThickLine] (3.center) to (2.center);
		\draw [style=ThickLine] (2.center) to (0.center);
		\draw [style=ThickLine] (0.center) to (1.center);
		\draw [style=DottedLine] (7.center) to (6.center);
		\draw [style=DottedLine] (9.center) to (8.center);
		\draw [style=ArrowLineRight, in=45, out=90] (13.center) to (11.center);
		\draw [style=ThickLine, in=-90, out=-45] (10.center) to (13.center);
	\end{pgfonlayer}
\end{tikzpicture}}
\caption{We depict $\mathscr{S}=\C/\Z_n$ and $\mathscr{X}_f=\C^2/\Z_2$ and $\mathscr{E}=\C^2/\Z_{2n}$ and $\textnormal{Hol}(A_\Lambda)=e^{2\pi i/n}$ for the Calabi-Yau cone $X=\C^3/\Z_{2n}$.
}
\label{fig:FibrationsCY}
\end{figure}

The fibration $\pi_{\mathscr{S}}:\partial T_{\mathscr{S}}\rightarrow \mathscr{S}$ has a generic lens space fiber $S^3/\Gamma_{\text{fix}}$ and an exceptional fiber $\partial \mathscr{E}=S^3/\Gamma$. See figure \ref{fig:FibrationsCY} for the example with weights $(1,1,2n-2)$. The fibration $\pi_I:\partial T_{\mathscr{S}}\rightarrow \R_{r\geq 0}$ has as generic fiber a circle worth of lens space $S^3/H$ and as exceptional fiber also $\partial \mathscr{E}=S^3/\Gamma$.

In the general case with $|\mathscr{K}|>1$ we do not have a better characterization than $\partial \mathscr{E}=\partial X^\circ|_{\textnormal{retract}}$. Away from the codimension 6 singularity the fibrations $\pi_{\mathscr{S}},\pi_I$ have a disjoint union of $|\mathscr{K}|$ fibers with similar structure as in the $|\mathscr{K}|=1$ case.

\paragraph{The $\Lambda$-twist}\mbox{}\medskip

Next, let us characterize the $\Lambda$-twist. Consider the $k$-th flavor brane fixed by the subgroup $\Gamma_{\text{fix},k}\subset \Gamma$. Transporting the local model of the singularity $\mathscr{X}_f^{(k)}\cong \C^2/\Gamma_{\text{fix},k}$ once around the origin of the base $\C/(\Gamma/\Gamma_{\text{fix},k})$ it is glued back to itself twisted by $\Gamma/\Gamma_{\text{fix},k}\cong \Z_{N/g_k}$. This is the $\Lambda$-twist which here takes the form of an $\exp(2\pi i m_k /N)$ monodromy. The connection $A_\Lambda^{(k)}$, defined on $\C/(\Gamma/\Gamma_{\text{fix},k})\setminus \{0\}$  is therefore flat with curvature
\be
F_\Lambda^{(k)}= \frac{i}{2} m_k\delta(z_3) dz_3 \wedge d\bar z_3\,,
\ee
where $dz_3=d a+ id b$ with $a= \textnormal{re} \,z_3$ and $b= \textnormal{im} \,z_3$ and thus $F_\Lambda^{(k)}=m_k \delta(a)\delta(b) d  a \wedge d  b$. Indeed, the holonomy of $A_\Lambda^{(k)}$ over the circle $\ell^1_k=S^1/(\Gamma/\Gamma_{\text{fix},k})$ of the cone $\C/(\Gamma/\Gamma_{\text{fix},k})$ is
\be
\textnormal{Hol}(A_\Lambda^{(k)})= \exp \lb 2\pi i  \int_{\ell^1_k} A_{\Lambda}^{(k)} \rb =\exp \lb 2\pi i \int_{\C/(\Gamma/\Gamma_{\text{fix},k})}F_\Lambda \rb = \exp \lb \frac{2\pi i m_k}{N} \rb\,.
\ee

\paragraph{Boundary (Co)Homology}\mbox{}\medskip

We begin by computing the (co)homology of $\partial^2 X^\circ$. The smooth compact space $\partial^2 X^\circ$ is a disjoint union of $|\mathscr{K}|$ spaces. We therefore focus on one component and consider the $k$-th component $\partial^2 X^\circ_k$ associated with the $k$-th flavor brane. We compute
\be \label{eq:CohoGroups5DSCFT} \ba
H_n(\partial^2 X^\circ_k)\cong\{\Z,\Z_{g_k}\oplus \Z,\Z_{g_k},\Z,\Z\}\,,\\
H^n(\partial^2 X^\circ_k)\cong\{\Z, \Z,\Z_{g_k},\Z\oplus \Z_{g_k},\Z\}\,.
\ea \ee
It then follows
\be
H_n(\partial^2 X^\circ)=\bigoplus_{k=1}^{|\mathscr{K}|}H_n(\partial^2 X_k^\circ)\,, \qquad H^n(\partial^2 X^\circ)=\bigoplus_{k=1}^{|\mathscr{K}|}H^n(\partial^2 X_k^\circ)\,.
\ee
The groups \eqref{eq:CohoGroups5DSCFT} immediately follow from the fact that the homology groups of a space fibered over a circle $M\rightarrow S^1$ with fiber $F$ and monodromy homology mappings $f_n:H_n(F)\rightarrow H_n(F)$ are determined from the short exact sequence (obtained as subsequences of a Mayer-Vietoris long exact sequence)
\be
0~\rightarrow~\textnormal{coker}\!\lb f_n-1\rb ~\rightarrow~H_n(M)~\rightarrow~\textnormal{ker}\!\lb f_{n-1}-1\rb~\rightarrow~0\,.
\ee
In our case, only the Hopf circle experiences a monodromic shift, however these shifts do not alter its homology class, hence $f_n=1$ in all degrees $n$. The maps $f_n$ do not determine the cohomology ring in general, note however, that we can continuously decrease the discrete shift to zero, `untwisting' the monodromy in the process. The cohomology ring is preserved under such deformations, we learn that the cohomology ring of $\partial^2 X^\circ_k$ is identical to that of the direct product space $S^1\times S^3/\Gamma_{\text{fix},k}$.

We next compute the (co)homology of $\partial X=S^5/\Gamma$. To begin, following \cite{Kawasaki1973CohomologyOT}, we take a more general perspective (which will help when considering linkings, setting $n=2$ later) and note that the orbifold
\be
\mathbb{C}^{n+1}\:\!\!/\Gamma\,, \quad \Gamma\cong \Z_N\,, \quad (z_1,\dots,z_n)\sim (\omega^{m_1}z_1,\dots,\omega^{m_{n+1}}z_{n+1} )\,,
\ee
with $\omega$ a primitive $N^{th}$ root of unity and $m=(m_1,\dots,m_{n+1})\in \Z^{n+1}_{\geq 0}$ an integer weight vector, naturally occurs as a patch of the weighted projective space
\be
\mathbb{WCP}^{\;\!n+1}_{N,m_1,\dots,m_{n+1}}\,,
\ee
with homogenous coordinates $[z_0:z_1:\dots :z_{n+1}]$. For this, observe that $z_0=0$ cuts out the weighted projective space $\mathbb{WCP}^{\;\!n}_{m_1,\dots,m_{n+1}}$. Then, taking the set theoretic difference we find
\be
\mathbb{WCP}^{\;\!n+1}_{N,m_1,\dots,m_{n+1}} -\mathbb{WCP}^{\;\!n}_{m_1,\dots,m_{n+1}} = \mathbb{C}^{n+1}\:\!\!/\Gamma\,.
\ee
This is equivalent to noting that, in the construction of $\mathbb{WCP}^{\;\!n+1}_{N,m_1,\dots,m_{n+1}}$ as a CW complex with exactly one cell in even degrees, we can arrange for the final cell to be modelled on the orbifold $ \mathbb{C}^{n+1}/\Gamma$, glued to $\mathbb{WCP}^{\;\!n}_{m_1,\dots,m_{n+1}}$. Consequently, the weighted projective space $\mathbb{WCP}^{\;\!n+1}_{N,m_1,\dots,m_{n+1}}$ can be presented as the gluing
\be \label{eq:gluing}
\mathbb{WCP}^{\;\!n+1}_{N,m_1,\dots,m_{n+1}}=\mathbb{C}^n\:\!\!/\Gamma \cup_{S^{2n+1}/\Gamma}\mathscr{L}^{\:\!n+1}_{m_1,\dots,m_{n+1}}
\ee
where $\mathscr{L}^{\:\!n+1}_{m_1,\dots,m_{n+1}}$ is topologically a copy of the partial resolution of $\mathbb{C}^{n+1}/\Gamma$ obtained by resolving the tip to the (less) singular space $\mathbb{WCP}^{\;\!n}_{m_1,\dots,m_{n+1}}$.


Given weight vectors $M=(N,m)$ and $m=(m_1,\dots,m_{n+1})$ there exist integers $L_M^{k},L_m^{k}$ such that the cohomology rings of the two weighted projective spaces are generated as
\be \label{eq:WCPCohoRing}\ba
H^*(\mathbb{WCP}_M^{\;\!n+1})&=\Z\langle 1, L_M^1 u,  L_M^2 u^2,\dots,  L_M^{n+1} u^{n+1}\rangle\,, \\
H^*(\mathbb{WCP}_m^{\;\!n})&=\Z\langle 1, L_m^1 u,  L_m^2 u^2,\dots,  L_m^{n} u^{n}\rangle\,,
\ea \ee
and, in even degree with $u$ of degree 2, taking the quotient we find the lens space cohomology ring
\be\label{eq:screening}
H^*(S^{2n+1}/\Gamma)=\frac{\Z\langle 1, L_m^1 u,  L_m^2 u^2,\dots,  L_m^{n} u^{n}\rangle}{\Z\langle  L_M^1 u,  L_M^2 u^2,\dots,  L_M^{n+1} u^{n+1}\rangle}\,,
\ee
with the quotient is in fixed degree. In odd degree this is supplemented by $H^{2n+1}(S^{2n+1}/\Gamma)\cong \Z$. Further, it implicitly also contains various linking forms by formally extending the cup product to degree $n+1$ and taking the quotient by $ L_M^{n+1} u^{n+1}$ at face value. The integers $L^M_k$ and $L^m_k$ are given by
\be
L^b_k=\textnormal{lcm}\lb \lbbb \frac{b_{i_0}\dots b_{i_k}}{\textnormal{gcd}(b_{i_0},\dots, b_{i_k})}\rbbb_{0\leq b_{i_0}< \dots < b_{i_k}\leq |b|}\rb\,,
\ee
where $b$ is either of the weight vectors $M,m$ of length $|b|=n+2,n+1$ respectively.

For example, consider the case $n=2$ with $S^5/\Gamma$. Then we have the torsional groups
\be
H^2(S^5/\Gamma)\cong \Z_{L_M^1/L_m^1}\cong (\Gamma/\Gamma_{\text{fix}})^\vee \,,\qquad H^4(S^5/\Gamma)\cong \Z_{L_M^2/L_m^2}\cong \Gamma\,.
\ee
Here, it follows from $m_1+m_2+m_3=0$ mod $N$ and the assumption of a faithfully acting group action that $\gcd(m_1,m_2,m_3,N)=1$ and $\gcd(m_i,m_j,N)=1$. This implies $L^2_m=m_1m_2m_3$ and $L^2_M=Nm_1m_2m_3$ and $L_M^2/L_m^2=N=|\Gamma|$. We also immediately compute the cup product and linking pairing, and their combination into a triple product
\be
\ba \label{eq:tripleprodandmore}
\cup\,:&\quad H^2(S^5/\Gamma)\times H^2(S^5/\Gamma) ~\rightarrow~H^4(S^5/\Gamma) \\
&\quad (rL_m^1 u,sL_m^1 u)\cong (r,s) ~\mapsto~ rs(L_m^1)^2u^2\cong rs(L_m^1)^2/L_{m}^2\,, \\[0.75em]
\textnormal{Link}\,:&\quad H^2(S^5/\Gamma)\times H^4(S^5/\Gamma)~\rightarrow~\Q/\Z \\
&\quad(rL_m^1 u,sL_m^2 u)\cong (r,s)~\mapsto~ rs L_m^1L_m^2u^3\cong rs(L_m^1L_m^2/L_M^3)\,,\\[0.75em]
\textnormal{Triple}\,:&\quad  H^2(S^5/\Gamma)\times H^2(S^5/\Gamma)\times H^2(S^5/\Gamma)~\rightarrow~\Q/\Z\\
&\quad(rL_m^1 u,sL_m^1 u,tL_m^1 u)\cong (r,s,t)~\mapsto~ rst(L_m^1)^3u^3\cong rst(L_m^1)^3/L_M^3\,.
\ea\ee
which rewritten on generators given in \eqref{eq:screening} and using the above isomorphisms become
\begin{alignat}{3}
\cup\,:&\quad (\Gamma/\Gamma_{\text{fix}})^\vee \times (\Gamma/\Gamma_{\text{fix}})^\vee ~\rightarrow~ \Gamma \\
&\quad (1,1) ~\mapsto~ (L_m^1)^2/L_{m}^2\,, \\[0.75em]
\textnormal{Link}\,:&\quad (\Gamma/\Gamma_{\text{fix}})^\vee \times \Gamma ~\rightarrow~\Q/\Z \\
&\quad (1,1)~\mapsto~ L_m^1L_m^2/L_M^3\,,\\[0.75em]
\textnormal{Triple}\,:&\quad (\Gamma/\Gamma_{\text{fix}})^\vee \times (\Gamma/\Gamma_{\text{fix}})^\vee \times (\Gamma/\Gamma_{\text{fix}})^\vee~\rightarrow~\Q/\Z\\
&\quad  (1,1,1)~\mapsto~ (L_m^1)^3/L_M^3\,.
\end{alignat}

For the concrete example $\mathbb{C}^3/\Z_{2n}$ with weights $(1,1,2n-2)$, i.e.,  $M=(2n,1,1,2n-2)$ and $m=(1,1,2n-2)$, which gives
\be\ba
l^1_m&=2n-2\,, && l^2_m=2n-2\,,&& \\
L^1_M&=n(2n-2) \,,&&  L^2_M=2n(2n-2)\,,&&L_{M}^3=2n(2n-2)\,,
\ea \ee
we compute, now making the abelian groups $(\Gamma/\Gamma_{\text{fix}})^\vee \cong \Z_n$ and $\Gamma\cong \Z_{2n}$ explicit
\begin{alignat}{3}\label{eq:Products}
\cup\,:&\quad \Z_n \times \Z_n ~\rightarrow~ \Z_{2n} \\
&\quad (1,1) ~\mapsto~ 2n-2=-2 \quad \textnormal{mod 2}n\,, \\[0.75em]
\textnormal{Link}\,:&\quad \Z_n \times \Z_{2n} ~\rightarrow~\Q/\Z \\
&\quad (1,1)~\mapsto~ (2n-2)^2/(2n(2n-2))=-1/n \quad \textnormal{mod 1}\,,\\[0.75em]
\textnormal{Triple}\,:&\quad \Z_n \times\Z_n \times\Z_n ~\rightarrow~\Q/\Z\\
&\quad  (1,1,1)~\mapsto~(2n-2)^3/(2n(2n-2))=2/n \quad \textnormal{mod 1}\,.
\end{alignat}

\paragraph{Long Exact Sequences in Relative (Co)Homology}\mbox{}\medskip

We now compute some long exact sequence. We begin by discussing the smooth manifold with boundary $\partial X^\circ$. By Poincar\'e-Lefschetz duality we have an isomorphism
\be
H_k(\partial X^\circ)\cong H^{5-k}(\partial X^\circ, \partial^2 X^\circ)\,.
\ee
From here we have via excision
\be
H^{5-k}(\partial X^\circ, \partial^2 X^\circ) \cong H^{5-k}(\partial X, \mathscr{K})\,.
\ee
Overall studying the above groups we will come to an understanding of the (co)homology of $\partial X^\circ$ with the ultimate goal of determining the relative cohomology sequence of the pair $(\partial X^\circ, \partial^2 X^\circ)$.

We begin computing the long exact sequence in relative cohomology of $(\partial X,\mathscr{K})$:
\begin{equation}\label{eq:RCH}
    \begin{array}{c||cccccc}
& H^k(\partial X, \mathscr{K}) &  & H^k(\partial X) &  &   H^k(\mathscr{K})  &      \\[0.4em] \hline \hline \\[-0.9em]
       k=0 ~&   0  & \rightarrow & \Z & \rightarrow &  \mathbb{Z}^{| \mathscr{K}|} & \rightarrow  \\
      k=1 ~&   \Z^{{| \mathscr{K}|}-1}  & \rightarrow & 0 & \rightarrow &  \mathbb{Z}^{| \mathscr{K}|} & \xrightarrow[]{~\rho~}  \\
      k=2 ~&   \Z^{| \mathscr{K}|}  & \rightarrow & (\Gamma/\Gamma_{\text{fix}})^\vee & \rightarrow &0& \rightarrow  \\
         k=3~ &  0  &\rightarrow & 0 & \rightarrow& 0 &\rightarrow  \\
       k=4~ &    \Gamma  &\rightarrow & \Gamma &\rightarrow &0 &\rightarrow  \\
        k=5 ~&   \mathbb{Z}  &\rightarrow & \mathbb{Z} &\rightarrow & 0  & \rightarrow
    \end{array}
\end{equation}
It immediately follows from the above Poincar\'e-Lefschetz and excision arguments that
\be
H_k(\partial X^\circ)\cong  \{\Z, \Gamma, 0, \Z^{| \mathscr{K}|}, \Z^{{| \mathscr{K}|}-1} \}\,.
\ee
Here we introduced the map $\rho : \Z^{| \mathscr{K}|}\rightarrow \Z^{| \mathscr{K}|}$ with $\textnormal{coker}(\rho)=(\Gamma/\Gamma_{\text{fix}})^\vee$. We will encounter various geometric incarnations of the map $\rho$ and conflate these in notation.

Finally let us address how we determined $H^2(\partial X,\mathscr{K})$. We can dualize the relative sequence \eqref{eq:RCH} to homology. The result is
\begin{equation}\label{eq:RH}
    \begin{array}{c||cccccc}
&   H_k(\mathscr{K}) &  & H_k(\partial X) &  &   H_k(\partial X, \mathscr{K}) &      \\[0.4em] \hline \hline \\[-0.9em]
       k=5 ~&   0  & \rightarrow & \Z & \rightarrow &  \mathbb{Z} & \rightarrow  \\
      k=4 ~&   0 & \rightarrow & 0 & \rightarrow &  0 & \rightarrow  \\
      k=3 ~&  0 & \rightarrow & \Gamma^\vee & \rightarrow & \Gamma^\vee & \rightarrow  \\
         k=2~ &  0  &\rightarrow & 0 & \rightarrow& \Z^{| \mathscr{K}|} &\xrightarrow[]{~\rho~}  \\
       k=1~ &    \Z^{| \mathscr{K}|}  &\rightarrow & \Gamma/\Gamma_{\text{fix}} &\rightarrow &\Z^{{| \mathscr{K}|}-1} &\rightarrow  \\
        k=0 ~&   \mathbb{Z}^{| \mathscr{K}|}  &\rightarrow & \mathbb{Z} &\rightarrow & 0  &
    \end{array}
\end{equation}
and the problem is mapped onto determining the relative homology group $H_2(\partial X,\mathscr{K})$ which sits in
\be
0~\rightarrow~H_2(\partial X,K)~ \xrightarrow[]{~\rho~}~\Z^{| \mathscr{K}|} \rightarrow \Gamma/\Gamma_{\text{fix}} ~\rightarrow~0
\ee
and therefore $\rho$ maps between free groups. Further, via excision and Poincar\'e-Lefschetz duality, as above, we determine
\be
H^k(\partial X^\circ)\cong  \{\Z, 0, \Gamma^\vee, \Z^{| \mathscr{K}|}, \Z^{{| \mathscr{K}|}-1} \}\,.
\ee

Next we compute the long exact sequence in relative homology of the pair $(\partial X^\circ,\partial^2X^\circ)$. We have $\partial^2X^\circ =\partial T_{\mathscr{K}}$ and the sequence reads:
\begin{equation}\label{eq:RHcirc}
    \begin{array}{c||cccccc}
&   H_k(\partial T_{\mathscr{K}}) &  & H_k(\partial X^\circ) &  &   H_k(\partial X^\circ, \partial T_{\mathscr{K}}) &      \\[0.4em] \hline \hline \\[-0.9em]
       k=5 ~&   0  & \rightarrow &0 & \rightarrow &  \mathbb{Z} & \rightarrow  \\
      k=4 ~&   \Z^{| \mathscr{K}|} & \rightarrow & \Z^{{| \mathscr{K}|}-1} & \rightarrow &  0 & \rightarrow  \\
      k=3 ~&  \Z^{| \mathscr{K}|} & \rightarrow &\Z^{| \mathscr{K}|} & \rightarrow & \Gamma^\vee & \rightarrow  \\
         k=2~ &  H^\vee  &\rightarrow & 0 & \rightarrow& \Z^{| \mathscr{K}|} &\xrightarrow[]{~\rho'~}  \\
       k=1~ &    \Z^{| \mathscr{K}|} \oplus H  &\rightarrow & \Gamma &\rightarrow &\Z^{{| \mathscr{K}|}-1} &\rightarrow  \\
        k=0 ~&   \mathbb{Z}^{| \mathscr{K}|}  &\rightarrow & \mathbb{Z} &\rightarrow & 0  &
    \end{array}
\end{equation}
Here $\rho'$ followed by a projection to $\Z^{| \mathscr{K}|}$ has cokernel isomorphic to $\Gamma/\Gamma_{\text{fix}}$. However, the initial $\rho'$ has cokernel $\Gamma$. Similarly, let us record the respective sequence in relative cohomology:
\begin{equation}
    \begin{array}{c||cccccc}
& H^k(\partial X^\circ, \partial T_{\mathscr{K}}) &  & H^k(\partial X^\circ) &  &   H^k(\partial T_{\mathscr{K}})  &      \\[0.4em] \hline \hline \\[-0.9em]
       k=0 ~&   0  & \rightarrow & \Z & \rightarrow &  \mathbb{Z}^{| \mathscr{K}|} & \rightarrow  \\
      k=1 ~&   \Z^{{| \mathscr{K}|}-1}  & \rightarrow & 0 & \rightarrow &  \mathbb{Z}^{| \mathscr{K}|} & \xrightarrow[]{~\rho^\vee~}  \\
      k=2 ~&   \Z^{| \mathscr{K}|} & \rightarrow & \Gamma^\vee & \rightarrow &\Gamma_{\text{fix}}^\vee& \rightarrow  \\
         k=3~ &  0  &\rightarrow & \Z^{{| \mathscr{K}|}} & \rightarrow& \Z^{| \mathscr{K}|}\oplus \Gamma_{\text{fix}} &\rightarrow  \\
       k=4~ &    \Gamma  &\rightarrow & \Z^{{| \mathscr{K}|}-1} &\rightarrow &\Z^{| \mathscr{K}|} &\rightarrow  \\
        k=5 ~&   \mathbb{Z}  &\rightarrow & 0 &\rightarrow & 0  & \rightarrow
    \end{array}
\end{equation}

Finally we identify the generator of $H^2(\partial X^\circ)$. The long exact Mayer-Vietoris sequence for the covering $\partial X=\partial X^\circ \cup T_{\mathscr{K}}$ contains the restriction map $H^2(\partial X)\rightarrow H^2(\partial X^\circ)$. From here, via the parametrization introduced in \eqref{eq:screening}, we deduce
\be \label{eq:IdentifyCircGenerator}\ba
H^2(\partial X^\circ)&=\Big\langle \frac{1}{|\Gamma_{\text{fix}}|} L^1_m u \Big\rangle \Big/ \langle L^1_M u\rangle\,.
\ea \ee

\subsection{Triple Products and Anomalies}

We now discuss various geometric triple products which go on to determine coefficients in the symmetry theories we compute in section \ref{sec:Illustrative}.

\subsubsection{$X=\textnormal{Cone}(\mathbb{WP}^3)$}

We begin by reformulating results of \cite{Witten:2001uq} in symmetry TFT formalism. Initially, the 4D physical boundary conditions will be set by a single chiral superfield $\Phi$ with flavor symmetry $\mathfrak{u}_1$ as engineered in M-theory on the $G_2$-holonomy cone $X=\textnormal{Cone}(\P^3)$. This cone exhibits a single isolated codimension-7 singularity at its apex and therefore lies outside of the class of theories studied throughout this paper, nonetheless, it will be instructive to analyze. From here, we turn to more general cases with singular link.

\paragraph{ Geometries with Smooth Links: Review}\mbox{}\medskip

The symmetry theory for M-theory on $X=\textnormal{Cone}(\P^3)$ derives via the reduction of the 11D supergravity Chern-Simons term on the $\P^3$ link of the geometry following \cite{Apruzzi:2021nmk}. In addition, one-derivative terms contribute, as derived in \cite{Baume:2023kkf, GarciaEtxebarria:2024fuk}. The starting point for the former is
\be\label{eq:11DStart}
-\frac{2\pi}{6}\int \frac{C_3}{2\pi}\cup \frac{G_4}{2\pi}\cup \frac{G_4}{2\pi}+\frac{2\pi}{48}\int \frac{C_3}{2\pi} \cup (p_2-p_1^2/4) \,,
\ee
more precisely, its uplift to differential cohomology. Here $p_n$ denotes the $n$-th Pontryagin class.

The cohomology ring of $\P^n$ is simply $H^*(\P^n)= \Z[u]/u^{n+1}$ with $u\equiv u_2$ in degree $2$. To proceed with the reduction, one next makes the expansion
\be \label{eq:Expansion}
{G}_4={F}_2 \cup {v}_2 +\dots
\ee
where $F_2=dA_1$ is the field strength of the $\mathfrak{u}_1$ flavor symmetry and evaluates the integral over $\P^3$. These computations were completed in \cite{Witten:2001uq} and the resulting contribution to the 5D symmetry theory action is
\be
\mathcal{S}_{d+1}\supset- \frac{2\pi}{6} \int \frac{A_1}{2\pi}\cup \frac{F_2}{2\pi}\cup \frac{F_2}{2\pi}+ \frac{2\pi}{24}\int \frac{A_1}{2\pi}\cup p_1\,.
\ee
The exterior derivative of the integrand is precisely the anomaly polynomial of an unit charge chiral multiplet in 4D. Anomaly inflow is now formulated by noting that bulk gauge transformations $A_1\rightarrow A_1+df_0$ lead to boundary terms which are required to cancel via the physical boundary conditions. These boundary terms match the $\mathfrak{u}_1^3$ and $\mathfrak{u}_1\!\cdot p_1$ anomalies of a chiral superfield and are therefore consistent with the codimension-7 singularity at the cone apex supporting a chiral superfield.

\paragraph{ Geometries with Singular Links}\mbox{}\medskip

Next, we turn to discuss the case of a chiral superfield $\Phi$ transforming in the fundamental representation the flavor algebra $\mathfrak{u}_N=\mathfrak{su}_N\oplus \mathfrak{u}_1$, as discussed in section \ref{sec:BiFund}. First we naively repeat the above computation of the $\mathfrak{u}_1$ self-anomaly and then rephrase it with respect to the pair $\partial^2X^\circ, \partial X^\circ$. For this we require the cohomology ring of the relevant weighted projective space $\mathbb{W}\C\P^3$ with weights $N,N,1,1$, the link of the geometry. The underlying groups are isomorphic to those of ordinary weighted projective space, however, the ring structure differs, it is given in \eqref{eq:WCPCohoRing}. Here the two non-trivial cup products are given by
\be \label{eq:RingStructure}
v_2\cup v_2 =v_4\,, \qquad v_2\cup v_4 = N v_6\,.
\ee
This results in the symmetry theory contribution
\be \label{eq:AnomalyU1Sing}
-\frac{2\pi N}{6}\int \frac{A_1}{2\pi}\cup \frac{F_2}{2\pi}\cup \frac{F_2}{2\pi}\,,
\ee
which is, as expected, the anomaly of $N$ chiral superfields.

The class $v_2$ generating $H^2(\partial X)$ restricts to the generator of $H^2(\partial X^\circ)\cong H^2(\partial^2 X^\circ)$. Similarly, from long exact sequences we learn we have that  $v_4$ generating $H^4(\partial X)$ is $\partial v_3'$ with $v_3'$ generating $H^3(\partial^2 X^\circ)$. It follows that the triple product above is reproduced by the pair of pairings
\be\ba
P_{22}^3\,:\quad &~\,~H^2(\partial X^\circ) \times  H^2(\partial X^\circ) \!\!\!\! &&\rightarrow~ H^3(\partial^2 X^\circ)\,,\\
\cup\,:\quad &H^3(\partial^2 X^\circ)\times H^2(\partial^2 X^\circ) \!\!\!\! &&\rightarrow~ H^5(\partial^2 X^\circ)\,,
\ea \ee
where the latter is simply the cup product. The triple product setting the $ \mathfrak{u}_1^3$ self-anomaly is therefore
\be \ba
H^2(\partial X^\circ) \times  H^2(\partial X^\circ)\times H^2(\partial X^\circ) ~&\rightarrow~ H^6(\partial X^\circ,\partial^2 X^\circ)\,,\\
\Z \times  \Z \times \Z ~&\rightarrow~\Z \,,\\
(a ,b , c) ~&\rightarrow~Nabc \,,\\
(1 ,1 , 1) ~&\rightarrow~N \,,\\
\ea \ee
where we used the isomorphism $ H^5(\partial^2 X^\circ)\cong H^6(\partial X^\circ,\partial^2 X^\circ)$ and $H^2(\partial^2X^\circ)\cong H^2(\partial X^\circ)$. We immediately see, by virtue of the last group being a relative cohomology group, i.e., the cocycle returns zero applied to chains in $\partial^2 X^\circ$, we are correctly describing a bulk effect.

Let us briefly describe the mapping $P_{22}^3$. We have $v_2\cup v_2=0$ as an element of $H^4(\partial X^\circ)=0$. By Poincar\'e-Lefschetz duality we learn that $H^4(\partial X^\circ)\cong H_2(\partial X^\circ,\partial^2 X^\circ)$, i.e., $v_2\cup v_2$ could be dual to a 2-cycle in $\partial^2 X^\circ$. Applying Poincar\'e duality to this 2-cycle maps us to $H^3(\partial^2 X^\circ)$.

The terms involving Pontryagin classes are more subtle to analyze and we refer it to future work. They receives localized contributions from the singular locus beyond integral cohomology contributions.

\subsubsection{$X=\textnormal{Cone}(S^5/\Z_N)=\C^3/\Z_N$}
\label{sec:AnomalyTriple5DSCFTs}

Given the orbifold $S^5/\Gamma$ we computed in \eqref{eq:tripleprodandmore} the triple product
\be \ba
\textnormal{Triple}\,:\quad
 (\Gamma/\Gamma_{\text{fix}})^\vee \times (\Gamma/\Gamma_{\text{fix}})^\vee  \times (\Gamma/\Gamma_{\text{fix}})^\vee ~&\rightarrow~\Q/\Z \\
 (r,s,t)~&\mapsto~ rst(L_m^1)^3/L_M^3\,.
\ea \ee
where $H^2(S^5/\Gamma)\cong  (\Gamma/\Gamma_{\text{fix}})^\vee$. However, the natural linking pairing
\be\label{eq:linkingform}\ba
 \textnormal{Link} \,: \quad H^4(\partial X)\times H^2(\partial X^\circ)~ &\rightarrow~ \Q/\Z \\
\Gamma \times \Gamma^\vee ~&\rightarrow~  \Q/\Z\,.
\ea\ee
maps from larger groups. We can therefore extend the triple pairing derived from the cohomology ring of $\partial X$, by replacing one of the $H^2(\partial X)$ with $H^2(\partial X^\circ)$, resulting in
\be\label{eq:better} \ba
T\,:\quad
 (\Gamma/\Gamma_{\text{fix}})^\vee \times (\Gamma/\Gamma_{\text{fix}})^\vee  \times \Gamma^\vee ~&\rightarrow~\Q/\Z \\
 (r,s,t)~&\mapsto~ rst(L_m^1)^3/L_M^3|\Gamma_{\text{fix}}|\,,
\ea \ee
which we evaluated using \eqref{eq:IdentifyCircGenerator}. Overall, we learn of a finer product. See also \eqref{eq:Anomaly} where we encountered a lift of this product.

For example, when $M=(2n,1,1,2n-2)$ and $m=(1,1,2n-2)$, then we compute
\be \ba
T_{2n,1,1,2n-2}\,:\quad
 (\Gamma/H)^\vee \times (\Gamma/H)^\vee  \times \Gamma^\vee ~&\rightarrow~\Q/\Z \\[0.5em]
 (r,s,t)~&\mapsto~ \frac{(2n-2)^3}{4n(2n-2)}rst=\frac{rst}{n}\,.
\ea \ee

\newpage

\bibliographystyle{utphys}
\bibliography{TopOpM}

\end{document}